\documentclass[letterpaper, 12pt]{article}
\pdfoutput=1
\usepackage{filecontents}
\usepackage{silence}
\usepackage{amsmath}
\usepackage{amssymb}
\usepackage{amsfonts}

\usepackage{comment}
\usepackage{slashed}

\usepackage{hyperref}
\hypersetup{bookmarksnumbered}

\newcommand{\tr}{\operatorname{tr}}
\newcommand{\Tr}{\operatorname{Tr}}

\newcommand{\be}{\begin{equation}}
\newcommand{\ee}{\end{equation}}
\newcommand{\ba}{\begin{align}}
\newcommand{\ea}{\end{align}}
\newcommand{\bes}{\begin{equation*}}
\newcommand{\ees}{\end{equation*}}
\newcommand{\bas}{\begin{align*}}
\newcommand{\eas}{\end{align*}}
\newcommand{\ie}{\begin{equation}\begin{aligned}}
\newcommand{\fe}{\end{aligned}\end{equation}}

\renewcommand{\title}[1]{\vbox{\center\LARGE{#1}}\vspace{5mm}}
\renewcommand{\author}[1]{\vbox{\center#1}\vspace{5mm}}
\newcommand{\address}[1]{\vbox{\center\em#1}}

\numberwithin{equation}{section}
\allowdisplaybreaks

\begin{document}

\pagenumbering{roman} 
\begin{titlepage}

\begin{center}

\hfill \\
\hfill \\
\vskip 2.25cm

\title{Localization and Non-Renormalization in \\ Chern-Simons Theory}

\author{Yale Fan}

\address{Department of Physics, Princeton University, Princeton, NJ 08544, USA}

\end{center}

\vskip 1cm

\abstract
{
We revisit and clarify some aspects of perturbative renormalization in pure Chern-Simons theory by means of a localization principle associated with an underlying supersymmetry.  This perspective allows the otherwise perturbative one-loop shifts to be interpreted as nonperturbative consequences of a non-renormalization theorem, while providing a unified understanding of their origin (particularly in the case of Wilson lines).  We illustrate this approach explicitly for $SU(2)$ Chern-Simons theory in flat space, on Seifert manifolds, and on a solid torus.
}

\end{titlepage}
\pagenumbering{arabic} 

\setcounter{tocdepth}{3}
\tableofcontents

\section{Introduction}

The goal of this paper is to argue that certain properties of three-dimensional Chern-Simons theory can be understood in a unified way by regarding the theory as an effective description of an $\mathcal{N} = 2$ supersymmetric completion.  To an optimist, such a viewpoint might represent a particular instance of a more general program of using supersymmetry to elucidate aspects of quantum field theories without manifest supersymmetry.

The application of supersymmetry to topological field theories is far from new.  For instance, both the topological invariance and semiclassical exactness of observables in Witten-type (cohomological) TQFTs have long been recognized as consequences of a fermionic BRST symmetry \cite{Birmingham:1991ty}.  After a suitable topological twist, gauge-fixed Chern-Simons theory itself furnishes an example of a Witten-type TQFT \cite{Kallen:2011ny}.  The BRST supersymmetry is a restatement of the underlying general covariance of the theory: the subtraction of ghost degrees of freedom guarantees the absence of excited states.  By contrast, our approach relies on a further auxiliary supersymmetry.  The relevant fermions obey the spin-statistics theorem.  At finite Yang-Mills coupling, they result in an infinite tower of states with equal numbers of bosonic and fermionic degrees of freedom, which make no net contribution to supersymmetric observables.  However, they have the additional effect of shifting the number of vacuum states.  We will argue that this shift, combined with the localization principle afforded by the auxiliary fermionic symmetry, provides a natural framework in which to understand some features of correlation functions in bosonic Chern-Simons theory that are obscure from the point of view of perturbation theory.

It has long been understood that induced Chern-Simons terms are one-loop exact because higher-order corrections (via an expansion in $\hbar\sim 1/k$) cannot, in general, respect the quantization condition on the level \cite{Chen:1992ee, Witten:1999ds} (see \cite{Coleman:1985zi} for a diagrammatic proof in the abelian case, and \cite{Closset:2012vg, Closset:2012vp} for a modern perspective).  One manifestation of this fact is that quantum observables in pure Chern-Simons theory with simple gauge group $G$ and level $k > 0$, possibly involving Wilson loops in irreducible representations of $G$ labeled by highest weights $\lambda$, are naturally viewed as functions not of the ``bare'' parameters (suitably defined), but of
\begin{equation}
k\to k + h, \quad \lambda\to \lambda + \rho
\label{shifts}
\end{equation}
where $h$ is the dual Coxeter number and $\rho$ is the Weyl vector of $G$.  For example, when $G = SU(2)$, the shifts read $k\to k + 2$ and $j\to j + 1/2$, and the latter appears at the level of representation theory in the $SU(2)$ Weyl character
\begin{equation}
\chi_j(\theta) = \sum_{m=-j}^j e^{im\theta} = \frac{\sin[(j + 1/2)\theta]}{\sin(\theta/2)},
\label{su2Weylcharacter}
\end{equation}
which (up to a $j$-independent prefactor) takes the form of a sum over $m = \pm(j + 1/2)$, as familiar from equivariant localization formulas.  These shifts can be thought of as quantum corrections.  While $\lambda$, unlike $k$, does not appear in the bulk Lagrangian, the associated shift similarly lends itself to a Lagrangian point of view via an auxiliary system attached to the Wilson line, obtained by quantizing the coadjoint orbit of $\lambda$.

It is likewise well-known that correlation functions in pure $\mathcal{N} = 2$ and $\mathcal{N} = 0$ Chern-Simons coincide up to a shift of the above form:\footnote{The 3D Lorentzian spin group $SL(2, \mathbb{R})$ has Majorana representations, while the Euclidean version $SU(2)$ does not; hence the minimal amount of SUSY in three Euclidean dimensions is that associated with a single two-component complex spinor, and Euclidean supersymmetric partition functions can only be calculated for $\mathcal{N}\geq 2$.  3D $\mathcal{N}\geq 2$ theories are precisely those whose holomorphy properties allow them to be constrained by non-renormalization theorems \cite{Aharony:1997bx}.} in the Chern-Simons action for an $\mathcal{N} = 2$ vector multiplet at level $k + h$, all superpartners of the gauge field (real scalars $\sigma, D$ and a gaugino $\lambda$ -- not to be confused with the weight $\lambda$ of the previous paragraph) are auxiliary, and performing the Gaussian path integral over these fields leads to an effective $\mathcal{N} = 0$ Chern-Simons action at level $k$.  In practice, this can be understood for sufficiently large $k$ by regulating the $\mathcal{N} = 2$ theory with an irrelevant Yang-Mills term (so that the path integral converges absolutely), which introduces a scale that masses up all fields, and then integrating out the gaugino in the Wilsonian sense.

However, to make a merely perturbative analogy between $\mathcal{N} = 0$ and $\mathcal{N} = 2$ Chern-Simons theory is slightly misleading.  While the renormalized parameters in \eqref{shifts} are one-loop exact, general observables in the $\mathcal{N} = 0$ theory are not, reflecting the fact that Chern-Simons theory is conventionally formulated as a Schwarz-type rather than a Witten-type TQFT.  The real power of supersymmetry lies in its ability to explain how the shifts \eqref{shifts} persist nonperturbatively in a wide class of observables.  Enhancing both the 3D Chern-Simons action and the 1D coadjoint orbit action for Wilson loops with $\mathcal{N} = 2$ supersymmetry gives one access to a localization argument that ensures that correlation functions depend only on the bare couplings appearing in the respective actions.  This is a sort of non-renormalization principle.  These two supersymmetrizations are not independent, as there exists a precise map between fields in the bulk and fields on the line.  The supersymmetric, coupled 3D-1D path integral can be evaluated exactly, and after adjusting for parity anomalies\footnote{We are abusing terminology here: by this, we simply mean the trading of a parity-violating fermion mass for a parity-violating Chern-Simons term.  The induced Chern-Simons terms that we obtain from integrating out massive fermions will always be properly quantized, so we will not encounter any actual parity anomalies (the situation is different when $\mathcal{N} = 1$ \cite{Witten:1999ds}).} from integrating out the auxiliary fermions (in 3D and in 1D), we immediately deduce the exact result in the corresponding bosonic theory, including the famous shifts.  In this way, a one-loop supersymmetric localization computation reproduces an all-loop result in the bosonic theory.  This line of reasoning leads to a conceptually simpler explanation for \eqref{shifts} than that originally obtained from anomalies in the coherent state functional integral \cite{Elitzur:1989nr}.

The non-renormalization of the level in $\mathcal{N}\geq 2$ Chern-Simons theory is often acknowledged in the localization literature (such as when performing supersymmetric tests of non-supersymmetric dualities \cite{Kapustin:2010mh, Aharony:2015mjs}), but the non-renormalization of the weight(s) is seldom mentioned.  This omission may make the latter point seem pedantic, but it is in fact essential for a consistent mapping of line operators between the bosonic theory and its $\mathcal{N} = 2$ cousin.

Making the above statements precise requires fixing unambiguous physical definitions of the ``bare'' parameters $k$ and $\lambda$ -- for example, via the coefficient of the two-point function in the associated 2D current algebra and canonical quantization of the coadjoint orbit theory, respectively.\footnote{An intrinsically bulk definition of $k$ is as follows.  For positive integer $k$, the Hilbert space of Chern-Simons theory with simply connected $G$ on a Riemann surface $\Sigma$ is isomorphic to $H^0(\mathcal{M}, \mathcal{L}^k)$ where $\mathcal{M}$ is the moduli space of flat $G$-connections on $\Sigma$ and $\mathcal{L}$ is the basic line bundle over $\mathcal{M}$ in the sense of having positive curvature and that all other line bundles over $\mathcal{M}$ take the form $\mathcal{L}^n$ for some integer $n$ \cite{Witten:1999ds}.  For example, for simple, connected, simply connected $G$ and $\Sigma = T^2$, $\mathcal{M}$ is a weighted projective space of complex dimension $\operatorname{rank} G$ and $\mathcal{L} = \mathcal{O}(1)$ (whose sections are functions of degree one in homogeneous coordinates on $\mathcal{M}$).  In the $\mathcal{N} = 1$ and $\mathcal{N} = 2$ settings, fermions have the effect of tensoring $\mathcal{L}^k$ with $K^{1/2}$ or $K$ to give $\mathcal{L}^{k - h/2}$ or $\mathcal{L}^{k - h}$, respectively, where $K = \mathcal{L}^{-h}$ is the canonical bundle of $\mathcal{M}$.  Note that these fermions effectively implement the metaplectic correction in geometric quantization \cite{Mikhaylov:2014aoa, Schottenloher:2008zz}.}  Having done so, the shifts in $k$ and $\lambda$ arise in a unified fashion from jointly supersymmetrizing the 3D bulk theory and the 1D coadjoint orbit theory, giving rise to three equivalent descriptions of the same theory:
\begin{enumerate}
\item The bosonic Chern-Simons theory has level $k$ and Wilson loops
\begin{equation}
\Tr_\lambda P\exp\left(i\oint A_\mu dx^\mu\right).
\end{equation}
\item The supersymmetric Chern-Simons theory has level $k + h$ and Wilson loops
\begin{equation}
\Tr_\lambda P\exp\left[i\oint (A_\mu dx^\mu - i\sigma ds)\right].
\label{superloop}
\end{equation}
\item The coadjoint orbit description of half-BPS Wilson loops coupled to the bulk supersymmetric theory has level $k + h$ and weight $\lambda + \rho$ from the start; these parameters are not renormalized.  The trace in \eqref{superloop} is replaced by an appropriate supertrace in a 1D theory containing an auxiliary complex fermion $\psi$.  In the standard presentation of a supersymmetric Wilson loop, the fermion $\psi$ has already been integrated out.
\end{enumerate}
One would in principle expect to be able to match \emph{all} observables between these descriptions, not only those that are protected (BPS) and hence calculable using supersymmetric localization, because the path integral over the auxiliary fermions and the scalar $D$ can be performed exactly (shifting $(k + h, \lambda + \rho)\to (k, \lambda)$ and setting $\sigma = 0$, respectively).  The main limitation of our analysis is that we are able to demonstrate this equivalence only for correlation functions of Wilson loops that are BPS with respect to the bulk supersymmetry (for which the integration contour implicit in \eqref{superloop} is subject to certain constraints).

The shifts \eqref{shifts} can be thought of as fundamentally representation-theoretic in nature, with the correspondence between 3D Chern-Simons theory with compact $G$ and 2D RCFT placing them in a physical setting: Wilson loops encode Weyl characters of $G$, and quantizing the theory on various manifolds makes contact with the representation theory of the corresponding affine Kac-Moody algebra.  Many of the relevant statements regarding character formulas and their associated Weyl shifts have been known since the early days of equivariant localization and index theorems, with the notion of hidden supersymmetry being a common thread.  For a sample of the relevant literature, see the reviews \cite{Szabo:1996md, Pestun:2016zxk} and references therein.  Part of our aim is to review some of these old localization results in light of new ones, while emphasizing that in the supersymmetric context, the essential mechanism for the shifts is identical in 3D and in 1D.

The essence of the 1D localization argument can be seen in the prototypical system of a massless charged particle on $S^2$ in the field of a magnetic monopole, which we refer to as the ``monopole problem.''  Indeed, part of our discussion involves giving a slightly more modern formulation of the treatment of the monopole problem in \cite{Stone:1988fu}, while embedding it into Chern-Simons theory.  In \cite{Stone:1988fu}, it is shown using a hidden supersymmetry that the semiclassical approximation to the path integral for the monopole problem is exact: rather than taking the zero-mass limit, one can introduce a fermionic superpartner so that the contributions of all excited states to the partition function cancel regardless of the mass.  The upshot is a derivation of the Weyl character formula for $SU(2)$ from supersymmetric quantum mechanics, which provides a physical interpretation of the Duistermaat-Heckman formula.  The same strategy of localizing an apparently purely bosonic theory has many modern incarnations: see, for example, \cite{Stanford:2017thb}.

Passing to 3D, exact results for Chern-Simons theory have been obtained by a variety of methods: aside from surgery and 2D CFT, these include abelianization \cite{Blau:1993tv, Blau:2006gh}, nonabelian localization \cite{Beasley:2005vf, Beasley:2009mb}, and supersymmetric localization \cite{Kallen:2011ny}.\footnote{Abelianization is a common theme in Chern-Simons theory: it reduces to an abelian theory in the semi\-classical limit \cite{Witten:1988hf}, to an abelian effective quantum mechanics problem in canonical quantization \cite{Elitzur:1989nr}, and to an abelian matrix model in localization.}  Our goal is to explain why the supersymmetric localization approach provides a structural understanding of these exact results.

In \cite{Blau:1993tv}, Chern-Simons theory on $\Sigma\times S^1$ was reduced to an abelian $BF$-type theory on $\Sigma$ where the role of $B$ is played by a compact scalar (in this way, $k$ cannot be scaled away, and one obtains a sum over integrable representations at level $k$).  In \cite{Blau:2006gh}, the technique of abelianization was extended to Chern-Simons theory on nontrivial circle bundles.  The final abelianized expression for the partition function, obtained by integrating over all connections in the 2D $BF$ theory on the base, takes the form of an integral over the Cartan of $G$ and incorporates the shift in $k$.  If one had first integrated over $B$, one would have recovered the result of \cite{Beasley:2005vf} obtained by nonabelian localization, which involves not only an integral over the Cartan, but also over the moduli space of flat connections on $M^3$ to which the Chern-Simons path integral localizes.  In \cite{Beasley:2009mb}, the techniques of \cite{Beasley:2005vf} were extended to compute the expectation values of Wilson loops along the $U(1)$ fibers.

Our approach involves introducing an auxiliary fermionic symmetry with the aid of generalized Killing spinors, allowing the localization procedure to be carried out on arbitrary Seifert manifolds.  The underlying geometric structure that makes this possible is a transversely holomorphic foliation, or THF \cite{Closset:2012ru, Closset:2013vra}.  It is worth contrasting this approach with that of \cite{Kallen:2011ny}, which avoids assuming the existence of Killing spinors by using a contact structure to define the requisite fermionic symmetry.  A contact structure exists on any compact, orientable three-manifold.  It is, locally, a one-form $\kappa$ for which $\kappa\wedge d\kappa\neq 0$; a metric can always be chosen for which $\kappa\wedge d\kappa$ is the corresponding volume form, i.e., such that $\ast 1 = \kappa\wedge d\kappa$ and $\ast\kappa = d\kappa$.  The dual vector field $v$ such that $\iota_v\kappa = 1$ and $\iota_v d\kappa = 0$ is known as the Reeb vector field.  It was found in \cite{Kallen:2011ny} that to carry out the localization, the corresponding Reeb vector field must be a Killing vector field, which restricts this approach to Seifert manifolds (as in \cite{Beasley:2009mb}); this approach was generalized in \cite{Ohta:2012ev} to Chern-Simons theories with matter.  Therefore, while the geometric basis for our approach differs from that for the cohomological localization of \cite{Kallen:2011ny, Ohta:2012ev}, the domain of applicability is the same.  Our focus, however, is different: the compensating level shift from auxiliary fermions was ignored in \cite{Kallen:2011ny}, noted in \cite{Ohta:2012ev}, and essential in neither.

We begin by reviewing some background material and setting our conventions in Sections \ref{N0CS} and \ref{wilsonandcoadjoint}.  We then carry out the analysis for Wilson lines very explicitly for $G = SU(2)$ in Section \ref{N2CS} (we comment briefly on the generalization to arbitrary $G$ at the end of the paper).  Using the description of these lines as 1D $\mathcal{N} = 2$ sigma models, we compute the effective action for fermions at both zero and finite temperature, canonically quantize the system, and present the localization argument in 1D.

In Section \ref{couplingtobulk}, we show how to embed this story in bulk 3D $\mathcal{N} = 2$ Chern-Simons theory.  Crucially, while we expect $\mathcal{N} = 0$ and $\mathcal{N} = 2$ Chern-Simons to be equivalent by integrating out the extra fields in the vector multiplet, the equivalence only holds if we take into account \emph{both} the shift of the level and the weight (as discussed further in Section \ref{matching}).

In Section \ref{3Dloc}, we describe how to generalize the aforementioned analysis of a Wilson line in flat space, either straight or wrapping a compact direction, to various classes of compact three-manifolds.  We also give some examples of the observables that we can compute.  Both the $\mathcal{N} = 0$ and $\mathcal{N} = 2$ theories are topological, so their observables are metric-independent.  In the $\mathcal{N} = 0$ case, the introduction of a metric is usually regarded as a ``necessary evil'' for the purposes of gauge-fixing and regularization.  In the $\mathcal{N} = 2$ case, the metric plays a more essential role in computing observables because it determines which observables are compatible with supersymmetry and therefore accessible to localization techniques.  Seifert loops (i.e., Wilson loops along the Seifert fiber direction) can give different knots depending on the choice of Seifert fibration.  For instance, depending on the choice of Seifert fibration on $S^3$, the half-BPS sector can contain Wilson loop configurations with the topology of Hopf links or torus links \cite{Beasley:2009mb}.

We review in Appendix \ref{quantization} the necessary elements of the quantization of Chern-Simons theory to which we refer throughout the paper.  In Appendix \ref{surgvsloc}, we comment on SUSY as an alternative to surgery computations in some situations.

\section{\texorpdfstring{$\mathcal{N} = 0$}{N = 0} Chern-Simons Theory} \label{N0CS}

Let $M^3$ be a compact, oriented three-manifold and let $G$ be a simple, compact, connected, simply connected Lie group.  The latter two assumptions on $G$ ensure that any principal $G$-bundle $P$ over $M^3$ is trivial, so that the Chern-Simons gauge field $A$ is a connection on all of $P$.  It then suffices to define the Lorentzian $\mathcal{N} = 0$ $G_{k > 0}$ Chern-Simons action by
\begin{equation}
S_\text{CS} = \frac{k}{4\pi}\int_{M^3} \Tr\left(AdA - \frac{2i}{3}A^3\right).
\label{N0SCS}
\end{equation}
We normalize the trace such that the norm squared of the longest root is two (for example, when $G = SU(N)$, the trace is taken in the fundamental representation and $k$ is integrally quantized).  In more general settings (e.g., $G$ non-simply connected), the quantization of $k$ would depend on additional data, such as whether we choose a spin structure on $M^3$ \cite{Dijkgraaf:1989pz}.

In flat space, we work in Lorentzian signature, except when computing the supersymmetric index in Section \ref{indexshift}.  In curved space (Section \ref{3Dloc}), we work in Euclidean signature.  In flat Minkowski space, we have the $\mathcal{N} = 2$ Lagrangians
\begin{align}
\mathcal{L}_\text{CS}|_{\mathbb{R}^{1, 2}} &= \frac{k}{4\pi}\Tr\left[\epsilon^{\mu\nu\rho}\left(A_\mu\partial_\nu A_\rho - \frac{2i}{3}A_\mu A_\nu A_\rho\right) - 2i\lambda\bar{\lambda} - 2D\sigma\right], \label{LCS} \\
\mathcal{L}_\text{YM}|_{\mathbb{R}^{1, 2}} &= \frac{1}{g^2}\Tr\left(-\frac{1}{4}F_{\mu\nu}F^{\mu\nu} - \frac{1}{2}D_\mu\sigma D^\mu\sigma + \frac{1}{2}D^2 + i\bar{\lambda}\gamma^\mu D_\mu\lambda - i\bar{\lambda}[\sigma, \lambda]\right). \label{LYM}
\end{align}
These are written in the convention where the generators $T^a$ are Hermitian, which we use throughout this paper.\footnote{Writing $\lambda = (\eta + i\tilde{\eta})/\sqrt{2}$ with $\eta, \tilde{\eta}$ real adjoint Majorana fermions reproduces the $\mathcal{N} = 2$ expressions of \cite{Kao:1995gf}.  WLOG, we may take $k > 0$ because time reversal (equivalently, spacetime orientation reversal in Euclidean signature) flips the overall sign of \eqref{LCS}, i.e., the sign of the bosonic Chern-Simons term, the sign of the gaugino mass term, \emph{and} the sign of the pseudoscalar $\sigma$.}

\subsection{Perturbation Theory}

The level of the pure $\mathcal{N} = 2$ CS theory whose correlation functions reproduce those of the corresponding $\mathcal{N} = 0$ theory is $k_{\mathcal{N} = 2} = k_{\mathcal{N} = 0} + h$ ($k_{\mathcal{N} = 0} > 0$ by assumption).  This we refer to as the ``fermionic shift.''  A quick way to justify this shift in flat space is as follows.  Consider, in generality, some fermions in a representation $R$ of $G$, minimally coupled to the gauge field and with a negative real mass term:
\begin{equation}
\frac{i}{g^2}\Tr(\bar{\lambda}\gamma^\mu D_\mu\lambda - m\lambda\bar{\lambda}) = \frac{i}{g^2}(\bar{\lambda}^i)^\alpha(\delta^{ij}(\gamma^\mu)_\alpha{}^\beta\partial_\mu - i(\gamma^\mu)_\alpha{}^\beta A_\mu^a(T_R^a)^{ij} + m\delta^{ij}\delta_\alpha{}^\beta)\lambda_\beta^j,
\end{equation}
where $i, j = 1, \ldots, \dim R$ and $D_\mu = \partial_\mu - iA_\mu^a T_R^a$.  These $\dim R$ complex fermions can be thought of as $2\dim R$ Majorana fermions for $R$ real.  Ignoring the vacuum energy, the only terms in the one-loop effective action $iS_\text{eff}[A, m]$ that survive the IR limit (in which $m\to\infty$ and the external momenta $p\to 0$) are those quadratic and cubic in $A$.  In this limit, the parity-odd parts of these terms are
\begin{equation}
-\frac{i}{8\pi}\frac{m}{|m|}\Tr(T_R^a T_R^b)\int d^3 x\, \epsilon^{\mu\nu\rho}\left(A_\mu^a\partial_\nu A_\rho^b + \frac{1}{3}f^{acd}A_\mu^b A_\nu^c A_\rho^d\right).
\end{equation}
Keeping the parity-even parts leads to a linearly divergent mass term for $A_\mu$, which can be regularized by subtracting the parity-even effective action at $m = 0$ \cite{Redlich:1983kn, Redlich:1983dv}.  Recall that $\Tr(T_R^a T_R^b)$ is scaled up relative to $\Tr(T_\text{fund}^a T_\text{fund}^b)$ by $\smash{\frac{C(R)}{C(\text{fund})}}$ where $C(R)$ is the Dynkin index of $R$.  In our conventions, $C(\text{fund}) = 1/2$ and $C(\text{adj}) = h$, where the latter follows from our normalization of long roots.  Hence $S_\text{eff}[A, m]$ for two Majorana fermions in $R = \text{adj}$ is $S_\text{CS}$ at level $-h\operatorname{sign}(m)$.\footnote{In the case of an \emph{odd} number of Majorana fermions, this is the basic mechanism of the parity anomaly, wherein gauge invariance requires that the UV Lagrangian contain parity-violating local counterterms to compensate for the gauge non-invariance of the fermion determinant.  This requirement holds regardless of whether the fermions themselves have bare masses in the UV.}

Now consider the sum of the Lagrangians \eqref{LCS} and \eqref{LYM}.  The resulting theory has a mass gap of $m = kg^2/2\pi$.  At large $k$ ($m\gg g^2$), we may integrate out all massive superpartners of the gauge field.  Assuming unbroken supersymmetry, the result is the low-energy effective theory of zero-energy supersymmetric ground states.  Of course, the fact that integrating out $\lambda$ induces $\mathcal{L}_\text{CS}^{\mathcal{N} = 0}$ at level $-h$ (among other interactions), along with the assumption that $\mathcal{N} = 2$ SUSY is preserved quantum-mechanically, is only a heuristic justification for the renormalization of the coefficient of $\mathcal{L}_\text{CS}^{\mathcal{N} = 2}$ to $k - h$.  This expectation is borne out by computing the one-loop perturbative renormalization of couplings \cite{Kao:1995gf}.

The fermionic shift discussed above is entirely separate from any ``bosonic shift'' that might arise from gauge dynamics (as found in, e.g., \cite{Pisarski:1985yj}, which effectively integrates out the topologically massive $W$-boson).  Such a shift does not affect the number of vacuum states.  Indeed, it is an artifact of regularization scheme: in the YM-CS regularization (which preserves supersymmetry, and which we use throughout this paper), the IR level is shifted by $+h$ relative to the bare level, while dimensional regularization yields no such shift \cite{Chen:1992ee}.  It is, nonetheless, a convenient conceptual slogan that $k$ is renormalized to $k + h$ at one loop in $\mathcal{N} = 0$ YM-CS, so that $k$ is not renormalized in $\mathcal{N}\geq 2$ YM-CS.  The important point is that for $\mathcal{N}\geq 2$ supersymmetry, integrating out the gauginos in the 3D YM-CS Lagrangian yields a shift of $-h$, which is twice the shift of $-h/2$ in the $\mathcal{N} = 1$ case \cite{Kao:1995gf}.

Given a precise physical definition of the level $k$, such as those presented in the introduction, a more substantive ``bosonic'' shift of the form mentioned above is that exhibited by correlation functions of $\mathcal{N} = 0$ Chern-Simons theory as functions of $k$.  This can already be seen in the semiclassical limit \cite{Witten:1988hf}.  At large $k$, we may expand \eqref{N0SCS} to quadratic order around a flat connection $A_0$.  The semiclassical path integral evaluates to its classical value weighted by the one-loop contribution $e^{i\pi\eta(A_0)/2}T(A_0)$ where $T(A_0)$ is the Ray-Singer torsion of $A_0$ (a topological invariant).  The APS index theorem implies that the relative $\eta$-invariant
\begin{equation}
\frac{1}{2}(\eta(A_0) - \eta(0)) = \frac{h}{\pi}I(A_0),
\end{equation}
where $I(A_0)\equiv \frac{1}{k}S_\text{CS}(A_0)$, is a topological invariant.  The large-$k$ partition function is then
\begin{equation}
Z = e^{i\pi\eta(0)/2}\sum_\alpha e^{i(k + h)I(A_0^{(\alpha)})}T(A_0^{(\alpha)}),
\end{equation}
where the sum (assumed finite) runs over gauge equivalence classes of flat connections.  This is how the shift $k\to k + h$, which persists in the full quantum answer, appears perturbatively.  The phase $\eta(0)$ depends on the choice of metric.  However, given a trivialization of the tangent bundle of $M^3$, the gravitational Chern-Simons action $I_\text{grav}(g)$ has an unambiguous definition, and upon adding a counterterm $\frac{\dim G}{24}I_\text{grav}(g)$ to the action, the resulting large-$k$ partition function is a topological invariant of the framed, oriented three-manifold $M^3$ \cite{Witten:1988hf}.

Thus a framing of $M^3$ fixes the phase of $Z$.  Aside from the framing anomaly of $M^3$ itself, there exists a framing ambiguity of links within it.  This framing ambiguity appears in the computation of Wilson loop expectation values because the conventional regularization of overlapping integrals of fields along the loop involves a choice of self-linking number, which is not a topologically invariant notion.  This point will be important in our application: the supersymmetric framing of a BPS Wilson loop differs from the canonical framing, when it exists, because the point splitting must be performed with respect to another BPS loop \cite{Kapustin:2009kz}.

To make concrete the utility of supersymmetry in light of these perturbative considerations, take as an example $\mathcal{N} = 0$ $SU(2)_k$ on $S^3$.  A typical observable in this theory receives contributions from all loops.  For example, the full nonperturbative result for the partition function is
\begin{equation}
Z(S^3) = \sqrt{\frac{2}{k + 2}}\sin\left(\frac{\pi}{k + 2}\right).
\label{ZN0}
\end{equation}
Suppose we were to compute the logarithm of this quantity (the free energy on $S^3$) in perturbation theory as the sum of connected vacuum bubbles, without recourse to 2D conformal field theory.  Expanding around the trivial flat connection, the one-loop factor is simply the large-$k$ limit of the exact result:
\begin{equation}
Z_\text{1-loop} = \exp(\bigcirc) = \frac{\sqrt{2}\pi}{(k + 2)^{3/2}}.
\end{equation}
The reconstruction of the exact result from summing trivalent graphs is far from obvious, regardless of whether the expansion parameter is $k^{-1}$ or $(k + 2)^{-1}$ (the necessity of doing perturbation theory in the renormalized level has historically been a point of contention in the literature; for a review of early references on large-$k$ asymptotics of Chern-Simons invariants, see \cite{Marino:2002fk}).  On the other hand, a one-loop supersymmetric localization computation in $\mathcal{N} = 2$ $SU(2)_{k+2}$ on $S^3$ (with the level adjusted to account for the fermionic shift, suitably generalized to curved space) handily yields the all-loop non-supersymmetric result \eqref{ZN0}, up to a framing phase given in Appendix \ref{surgvsloc}.  The bulk of our discussion will focus on more complicated observables that include Wilson loops.

\subsection{Beyond Perturbation Theory}

As known since \cite{Witten:1988hf}, there exist completely general nonperturbative techniques for computing observables in the $\mathcal{N} = 0$ theory, and thus checks of any results obtained via supersymmetry.  These techniques rely on essentially two ingredients.  The first is the fact that $Z(\Sigma\times_K S^1) = \Tr_{\mathcal{H}_\Sigma}(K)$, where the mapping torus $\Sigma\times_K S^1$ is obtained by identifying the ends of the cylinder $\Sigma\times [0, 1]$ by a diffeomorphism $K$ of $\Sigma$.  The second is the fundamental surgery formula
\begin{equation}
Z(\tilde{M}; R_i) = \sum_j K_i{}^j Z(M; R_j),
\end{equation}
where $M$ contains an arbitrary Wilson loop in the representation $R_i$ (possibly trivial) and $\tilde{M}$ is the result of gluing a tubular neighborhood of this loop back into $M$ with a diffeomorphism $K$ on its boundary.  Topologically equivalent surgeries on three-manifolds may have different effects on framing.

To give a few examples of nonperturbative results computed by these means (stated in the canonical framing), consider $G_{k > 0}$ on $S^3$.  Let $S_{ij}$ be the representation of the modular transformation $S$ on $T^2$ in the Verlinde basis for $\mathcal{H}_{T^2}$.  Then
\begin{equation}
Z(S^3) = S_{00} = \frac{1}{(k + h)^{\operatorname{rank} G/2}}\left(\frac{\operatorname{vol} \Lambda_W}{\operatorname{vol} \Lambda_R}\right)^{1/2}\prod_{\alpha > 0} 2\sin\left(\frac{\pi\alpha(\rho)}{k + h}\right),
\label{ZS3}
\end{equation}
while for an unknotted Wilson loop in an irreducible representation $R_i$,
\begin{equation}
\langle W\rangle = \frac{Z(S^3; R_i)}{Z(S^3)} = \frac{S_{0i}}{S_{00}} = \prod_{\alpha > 0} \frac{\sin(\pi\alpha(\lambda + \rho)/(k + h))}{\sin(\pi\alpha(\rho)/(k + h))}.
\label{WS3}
\end{equation}
Here, $\alpha$ runs over positive roots and $\lambda$ is the highest weight of $R_i$.  As $k\to\infty$, $Z(S^3)\sim k^{-\dim G/2}$ and $\langle W\rangle\to \dim R_i$, the latter of which justifies the nomenclature ``quantum dimension.''  The expressions in terms of $S$-matrix elements were deduced in \cite{Witten:1988hf}, while the explicit formulas in \eqref{ZS3} and \eqref{WS3} are consequences of the Weyl denominator and character formulas \cite{Marino:2005sj}.\footnote{The result for $Z(S^3)$ follows from consistency between two different ways of gluing together two copies of a solid torus $D^2\times S^1$: one trivially to get $S^2\times S^1$, and another with an $S$ transformation on the boundary to get $S^3$.  More generally, by inserting Wilson lines in these solid tori, one obtains the expectation value of the Hopf link as a normalized $S$-matrix element.

For any $G$, the modular transformation $T$ is represented in the Verlinde basis by a diagonal matrix with $T_i{}^i = e^{2\pi i(h_i - c/24)}$ where $h_i$ is the conformal weight of the primary field in the representation $R_i$ and $c$ is the central charge of $\smash{\widehat{G}_k}$.}  In particular, for $SU(2)_k$,
\begin{equation}
S_{ij} = \sqrt{\frac{2}{k + 2}}\sin\left[\frac{(2i + 1)(2j + 1)\pi}{k + 2}\right]
\label{Smatrix}
\end{equation}
where $i, j$ label the spins of the corresponding representations (thus giving \eqref{ZN0}), and for an unknotted Wilson loop in the spin-$j$ representation,
\begin{equation}
\langle W\rangle = \frac{S_{0j}}{S_{00}} = \frac{q^{j + 1/2} - q^{-(j + 1/2)}}{q^{1/2} - q^{-1/2}} = \frac{\sin((2j + 1)\pi/(k + 2))}{\sin(\pi/(k + 2))}
\end{equation}
where $q = e^{2\pi i/(k + 2)}$.

In some observables, highest weights of integrable representations of the $G_k$ theory appear not due to explicit Wilson loop insertions, but rather because they are summed over.  Indeed, the shift in $\lambda$ already appears in the partition function on $\Sigma\times S^1$, which computes the dimension of the Hilbert space of the Chern-Simons theory on $\Sigma$ and hence the number of conformal blocks in the corresponding 2D RCFT.  The answer is famously given by the Verlinde formula, which for arbitrary compact $G$, reads \cite{Blau:1993tv}
\begin{equation}
\dim V_{g, k} = (|Z(G)|(k + h)^{\operatorname{rank} G})^{g-1}\sum_{\lambda\in \Lambda_k} \prod_\alpha (1 - e^{2\pi i\alpha(\lambda + \rho)/(k + h)})^{1-g}
\label{Verlinde}
\end{equation}
where $g$ is the genus of $\Sigma$ and $\Lambda_k$ denotes the set of integrable highest weights of $\smash{\widehat{G}_k}$.  For the $SU(2)_k$ WZW model, it becomes
\begin{equation}
\dim V_{g, k} = \left(\frac{k + 2}{2}\right)^{g-1}\sum_{m=0}^k \sin\left[\frac{(m + 1)\pi}{k + 2}\right]^{2-2g},
\end{equation}
where the RHS reduces to $k + 1$ for $g = 1$.  While our focus is on Wilson loops, it turns out that the appearance of $\lambda + \rho$ in $Z(\Sigma\times S^1)$ comes ``for free'' in our approach, without the need to adjust for any 1D fermionic shifts, which is consistent with the fact that the weights in \eqref{Verlinde} are not associated with Wilson loops.  This fact has already been appreciated in prior literature, as we briefly review in Section \ref{3Dloc}.

\section{Wilson Loops and Coadjoint Orbits} \label{wilsonandcoadjoint}

\subsection{The Orbit Method}

A central ingredient in our analysis is the fact that a Wilson loop over a curve $\gamma$ in $M^3$ is a path integral for a 1D Chern-Simons theory whose classical phase space is a coadjoint orbit of $G$, with the corresponding representation $R$ arising by the orbit method \cite{Witten:1988hf}.  We will be interested in the case of compact $G$, where this construction is also known as Borel-Weil-Bott quantization.  The philosophy is that one can eliminate both the trace and the path ordering from the definition of a Wilson loop in a nonabelian gauge theory at the cost of an additional path integral over all gauge transformations along $\gamma$.

To make this description explicit, we draw from the exposition of \cite{Beasley:2009mb}.  We would like to interpret a Wilson loop as the partition function of a quantum-mechanical system on $\gamma$ with time-dependent Hamiltonian.  In the Hamiltonian formalism, this is a matter of writing
\begin{equation}
W_R(\gamma) = \Tr_R P\exp\left(i\oint_\gamma A\right) = \Tr_{\mathcal{H}} T\exp\left(-i\oint_\gamma H\right)
\end{equation}
where the Hilbert space $\mathcal{H}$ is the carrier space of the representation $R$, $H$ generates translations along $\gamma$, and the time evolution operator is the holonomy of the gauge field.  In the path integral formalism, this becomes
\begin{equation}
W_R(\gamma) = \int DU\, e^{iS_\lambda(U, A|_\gamma)}
\label{Upathintegral}
\end{equation}
where $U$ is an auxiliary bosonic field on $\gamma$, $\lambda$ is the highest weight of $R$, and the restriction of the bulk gauge field $A|_\gamma$ is a background field in the (operator-valued) path integral over $U$.  Since the definition of a Wilson loop is independent of any metric on $\gamma$,\footnote{This is not true of its supersymmetric counterparts.} it is not surprising that the action $S_\lambda$ will turn out to describe a topological sigma model.

The Borel-Weil-Bott theorem identifies the irreducible representation $R$ with the space of holomorphic sections of a certain line bundle over the coadjoint orbit $\mathcal{O}_\lambda\subset \mathfrak{g}^\ast$ of $\lambda$, which (in the generic case) is isomorphic to the flag manifold $G/T$ where $T$ is a maximal torus of $G$.  In physical terms, it states that $R$ is the Hilbert space obtained by quantizing $\mathcal{O}_\lambda$.  We are therefore led to consider the quantum mechanics of a particle on $\mathcal{O}_\lambda$ given by a 1D sigma model of maps $U : S^1\to \mathcal{O}_\lambda$, where the compact worldline is identified with $\gamma\subset M^3$.  To ensure that $\mathcal{O}_\lambda$ (rather than $T^\ast\mathcal{O}_\lambda$) appears as the classical phase space, the action for $U$ must be first-order in the time derivative along $S^1$.  Moreover, on general grounds, it should be independent of the metric on $S^1$.

There is an essentially unique choice of action that fulfills these wishes.  For convenience, we identify $\lambda$ via the Killing form as an element of $\mathfrak{g}$ rather than $\mathfrak{g}^\ast$, so that $\mathcal{O}_\lambda\subset \mathfrak{g}$ is the corresponding adjoint orbit (henceforth, we shall not be careful to distinguish $\mathfrak{g}$ and $\mathfrak{g}^\ast$).  We assume that $\lambda$ is a regular weight, so that $\mathcal{O}_\lambda\cong G/G_\lambda$ where $G_\lambda\cong T$.  The (left-invariant) Maurer-Cartan form $\theta$ is a distinguished $\mathfrak{g}$-valued one-form on $G$ that satisfies $d\theta + \theta\wedge\theta = 0$.  We obtain from it two natural forms on $G$, namely the real-valued presymplectic one-form $\Theta_\lambda$ and the coadjoint symplectic two-form $\nu_\lambda$:
\begin{equation}
\theta = g^{-1}dg\in \Omega^1(G)\otimes \mathfrak{g}, \qquad \Theta_\lambda = i\Tr(\lambda\theta)\in \Omega^1(G), \qquad \nu_\lambda = d\Theta_\lambda\in \Omega^2(G).
\label{forms}
\end{equation}
Both $\Theta_\lambda$ and $\nu_\lambda$ descend to forms on $\mathcal{O}_\lambda$.  The weight $\lambda$ naturally determines a splitting of the roots of $G$ into positive and negative, positive roots being those having positive inner product with $\lambda$.  Endowing $\mathcal{O}_\lambda$ with the complex structure induced by this splitting makes $\mathcal{O}_\lambda$ a K\"ahler manifold, with K\"ahler form $\nu_\lambda$ of type $(1, 1)$.\footnote{This is usually phrased as a choice of Borel subalgebra $\mathfrak{b}\supset \mathfrak{t}$, so that the coadjoint orbit is isomorphic to $G_\mathbb{C}/B$ where $B$ is the corresponding Borel subgroup and the roots of $B$ are defined to be the positive roots of $G$; then representations are labeled by their lowest weights.  We instead adhere to the ``highest weight'' conventions of \cite{Beasley:2009mb}.}  Now consider the action
\begin{equation}
S_\lambda(U) = \oint_{S^1} U^\ast(\Theta_\lambda) = \oint_{S^1} (\Theta_\lambda)_m\frac{dU^m}{d\tau}\, d\tau.
\end{equation}
The second expression (written in local coordinates $U^m$ on $\mathcal{O}_\lambda$) is indeed first-order in derivatives, so that the solutions to the classical EOMs are constant maps $U$, as desired.

To be concrete, we may think of $U$ as parametrizing gauge transformations.  Using the isomorphism $G/G_\lambda\xrightarrow{\sim} \mathcal{O}_\lambda$ given by $gG_\lambda\mapsto g\lambda g^{-1}$, we lift $U$ to a map $g : S^1\to G$, so that
\begin{equation}
S_\lambda(U) = i\oint_{S^1} \Tr(\lambda g^{-1}dg).
\label{Uactiong}
\end{equation}
From \eqref{Uactiong}, we see very explicitly that the canonical symplectic form $\nu_\lambda$ on $\mathcal{O}_\lambda$, given in \eqref{forms}, takes the form $d\pi_g\wedge dg$ where the components of $g$ are canonical coordinates.  The fact that $\lambda\in \mathfrak{g}$ is quantized as a weight of $G$ implies that \eqref{Uactiong} is independent of the choice of lift from $\mathcal{O}_\lambda$ to $G$.  Namely, $g$ is only determined by $U$ up to the right action of $G_\lambda$; under a large gauge transformation $g\mapsto gh$ where $h : S^1\to G_\lambda$, the integrand of \eqref{Uactiong} changes by $d\Tr(\lambda\log h)$ and the action changes by an integer multiple of $2\pi$.\footnote{From the geometric quantization point of view, the quantization of $\lambda$ is necessary for the existence of a prequantum line bundle $\mathcal{L}(\lambda)$ over $\mathcal{O}_\lambda$, with curvature $\nu_\lambda$.  Each $\lambda$ in the weight lattice gives a homomorphism $\rho_\lambda : T\to U(1)$, which can be used to construct an associated line bundle $\mathcal{L}(\lambda) = G\times_{\rho_\lambda} \mathbb{C}$ over $G/T$, so that the Hilbert space is the space of holomorphic sections of $\mathcal{L}(\lambda)$.  Then $\Theta_\lambda$ is a connection on $\mathcal{L}(\lambda)$.}  Thus $\Theta_\lambda$ descends (up to exact form) to $\mathcal{O}_\lambda$.  The path integral \eqref{Upathintegral} is over all maps $U$ in $L\mathcal{O}_\lambda$, or equivalently, over all maps $g$ in $LG/LG_\lambda$ (accounting for the gauge redundancy).

To couple \eqref{Uactiong} to the bulk gauge field, we simply promote $dg$ to $d_A g = dg - iA|_\gamma\cdot g$:
\begin{equation}
S_\lambda(U, A|_\gamma) = i\oint_{S^1} \Tr(\lambda g^{-1}d_A g).
\end{equation}
Prescribing the correct gauge transformations under $G\times T$ (with $T$ acting on the right and $G$ acting on the left), the 1D Lagrangian transforms by the same total derivative as before.

The first-order action \eqref{Uactiong}, in the absence of a background gauge field, can be thought of as describing the IR limit of a charged particle on $\mathcal{O}_\lambda$ in a magnetic field $\nu_\lambda$.  In complete analogy to 3D Chern-Simons theory, the irrelevant two-derivative kinetic terms have the effect of renormalizing $\lambda$ to $\lambda + \rho$ at one loop, and upon supersymmetrizing the theory, the fermion effective action provides a compensating shift by $-\rho$.\footnote{As in 3D, the effect of these fermions can be compared to that of the metaplectic correction in geometric quantization, which states that wavefunctions should not be viewed as sections of $\mathcal{L}(\lambda)$, but rather as half-densities valued in $\mathcal{L}(\lambda)$, meaning that they belong to $\mathcal{L}(\lambda)\otimes K^{1/2}\cong \mathcal{L}(\lambda - \rho)$ where $K^{1/2}$ is a square root of the canonical bundle of $\mathcal{O}_\lambda$ \cite{Mikhaylov:2014aoa}.}  We will substantiate this interpretation for $G = SU(2)$ in exhaustive detail.

\subsection{Wilson/'t Hooft Loops in Chern-Simons Theory} \label{loopsinCS}

While the coadjoint representation of a Wilson loop holds in any gauge theory, it is especially transparent in Chern-Simons theory, where it can be derived straightforwardly via a surgery argument \cite{Moore:1989yh}.  Consider Chern-Simons on $S^1\times \mathbb{R}^2$, where the Wilson line wraps the $S^1$ at a point on the $\mathbb{R}^2$.  Cutting out a small tube around $\gamma$ and performing a gauge transformation $\tilde{g}$, the action changes by\footnote{By Appendix \ref{gaugetransformations}, varying the bulk action gives a boundary term of $-\frac{ik}{4\pi}d(A\tilde{g}^{-1}d\tilde{g})$; the Pontryagin density term does not contribute because $G$ is assumed simply connected.  By Appendix \ref{boundaryconditions}, specifying nonzero $A_t$ on the boundary requires adding a boundary term of $\frac{k}{4\pi}\int_{\partial M^3} d^2 x \Tr(A_t A_\phi)$ to the action, whose variation under a $\phi$-dependent gauge transformation $\tilde{g}$ gives another contribution of $\smash{-\frac{ik}{4\pi}\int_{\partial M^3} d^2 x \Tr(A_t\tilde{g}^{-1}\partial_\phi\tilde{g})}$.}
\begin{equation}
\Delta S = -\frac{ik}{2\pi}\int_{\partial M^3} \Tr(A\tilde{g}^{-1}d\tilde{g}).
\end{equation}
Set $\tilde{g} = e^{i\alpha\phi}$ where $e^{2\pi i\alpha} = 1$ (this gauge transformation is singular along the loop; $t$ is the coordinate along $\gamma$ and $\phi$ the coordinate around it).  To define a gauge-invariant operator, average over $\tilde{g}\to g\tilde{g}$ and $A\to gAg^{-1} - idgg^{-1}$ where $g = g(t)$, whereupon this becomes
\begin{equation}
\Delta S = ik\int_\gamma \Tr(\alpha g(\partial_t - iA_t)g^{-1})\, dt,
\end{equation}
where we have performed the $\phi$ integral and shrunk the boundary to a point.  Finally, replace $g$ by $g^{-1}$.  Hence $k\alpha$ must be quantized as a weight $\lambda$.\footnote{We have corrected the transformation rule $\tilde{g}\to g\tilde{g}g^{-1}$ and a spurious factor of $\frac{1}{2}$ in \cite{Moore:1989yh}.}  This derivation illustrates that Wilson and 't Hooft loops are equivalent in pure Chern-Simons theory.

To summarize, consider a bulk theory with gauge group $G$ and the 1D Lagrangian
\begin{equation}
\mathcal{L}_\text{1D} = i\Tr[\lambda g^{-1}(\partial_t - iA)g]
\label{Lforg}
\end{equation}
where $g\in G$, $A\equiv A|_\gamma$, and $\lambda\in \mathfrak{t}$ (properly, $\lambda\in \mathfrak{t}^\ast$).  Since $\lambda$ is Hermitian in our conventions, the factor of $i$ ensures that the coadjoint orbit action is real.  The Lagrangian \eqref{Lforg} transforms by a total derivative under $t$-dependent $G\times T$ gauge transformations
\begin{equation}
g\to h_\ell gh_r, \quad A\to h_\ell Ah_\ell^{-1} - i\partial_t h_\ell h_\ell^{-1},
\label{GtimesT}
\end{equation}
namely $i\Tr(\lambda\partial_t\log h_r)$, where $h_\ell$ is the restriction of a $G$-gauge transformation in the bulk and $h_r\in T$.  Hence $\lambda$ is quantized to be a weight of $G$.  The $T$-gauge symmetry restricts the degrees of freedom in $g$ to $G/T$.  Quantizing $g$ in this Lagrangian leads to the Wilson line.

Strictly speaking, the global symmetry of the model \eqref{Uactiong} that we gauge to obtain \eqref{Lforg} is $G/Z(G)$, since the center is already gauged.  This should be contrasted with the global symmetry $G\times G/Z(G)$ of a particle on a group manifold with the usual kinetic term $\Tr((g^{-1}\dot{g})^2)$, which consists of isometries of the bi-invariant Killing metric on $G$.

\section{Wilson Loops in \texorpdfstring{$\mathcal{N} = 2$}{N = 2} Chern-Simons Theory} \label{N2CS}

We now show that properly defining half-BPS Wilson loops in $\mathcal{N} = 2$ Chern-Simons theory ensures that their weights are not renormalized, in direct parallel to the non-renormalization of the bulk Chern-Simons level.  This involves enhancing the sigma model of the previous section with 1D $\mathcal{N} = 2$ supersymmetry in a way compatible with bulk 3D $\mathcal{N} = 2$ supersymmetry.

\subsection{Shift from Line Dynamics}

\subsubsection{\texorpdfstring{$\mathcal{N} = 2$}{N = 2} Coadjoint Orbit}

We work in Lorentzian 1D $\mathcal{N} = 2$ superspace with coordinates $(t, \theta, \theta^\dag)$ (see Appendix \ref{1DN2}).  Implicitly, we imagine a quantum-mechanical system on a line embedded in $\mathbb{R}^{1, 2}$, but we will not need to pass to 3D until the next section.  Our primary case study is $G = SU(2)$.  We first construct, without reference to the 3D bulk, an $SU(2)$-invariant and supersymmetric coadjoint orbit Lagrangian from the 1D $\mathcal{N} = 2$ chiral superfield
\begin{equation}
\Phi = \phi + \theta\psi - i\theta\theta^\dag\dot{\phi}
\end{equation}
descending from bulk super gauge transformations and the 1D $\mathcal{N} = 2$ vector superfields
\begin{equation}
V_i = a_i + \theta\psi_i - \theta^\dag\psi_i^\dag + \theta\theta^\dag A_i
\end{equation}
obtained from restrictions of the bulk fields to the Wilson line, which extends along the 0 direction in flat space.  Here, $i = 1, 2, 3$ label the $\mathfrak{su}(2)$ components in the $\vec{\sigma}/2$ basis; $\phi$ is a complex scalar and $\psi$ is a complex fermion; $a_i, A_i$ are real scalars and $\psi_i$ are complex fermions; and the relevant SUSY transformations are given in \eqref{1Dsusyvector} and \eqref{1Dsusychiral}.

We begin by writing \eqref{Lforg} in a form more amenable to supersymmetrization, namely in terms of a complex scalar $\phi$ whose two real degrees of freedom come from those in $g\in G = SU(2)$ minus those in $h\in T = U(1)$.  Along with its conjugate $\phi^\dag$, it parametrizes the phase space $SU(2)/U(1)\cong \mathbb{CP}^1$.  Take $\lambda = -j\sigma_3$ with $j\in \frac{1}{2}\mathbb{Z}_{\geq 0}$, which fixes a Cartan; then
\begin{equation}
g = \left(\begin{array}{cc} a & b \\ -\bar{b} & \bar{a} \end{array}\right), \quad |a|^2 + |b|^2 = 1
\end{equation}
is subject to a $U(1)$ gauge redundancy $g\sim ge^{i\theta\sigma_3}$.  We identify variables via the Hopf map $SU(2)\to S^2$, followed by stereographic projection:
\begin{equation}
\phi = -\frac{a}{\bar{b}}.
\end{equation}
This map respects the chosen $U(1)$ gauge equivalence: $(a, b)\to (ae^{i\theta}, be^{-i\theta})$.  Let us gauge-fix the $U(1)$ action on the right by taking $b = r$ real.  Since $|a|^2 + r^2 = 1$, $r$ is only determined by $a$ up to a sign (reflecting the ambiguity in the action of $SU(2)$ on $S^2$).  Note that the gauge fixing breaks down when $|a| = 1$ ($r = 0$).  Accounting for the sign ambiguity, we have
\begin{equation}
\phi = -\frac{a}{\pm\sqrt{1 - |a|^2}} \implies a = \mp\frac{\phi}{\sqrt{1 + |\phi|^2}}, \mbox{ } r = \pm\frac{1}{\sqrt{1 + |\phi|^2}}.
\label{gtophi}
\end{equation}
The relative minus sign is important for ensuring equivariance of the map from $a$ to $\phi$ with respect to the action of $SU(2)$.  Let us fix the overall sign to ``$(a, r) = (+, -)$.''  This is a one-to-one map between the interior of the unit disk $|a| < 1$ and the $\phi$-plane that takes the boundary of the disk to the point at infinity.  To couple the $\phi$ degrees of freedom to the gauge field, we work in the basis $\vec{\sigma}/2$, so that
\begin{equation}
A = \frac{1}{2}\left(\begin{array}{cc} A_3 & A_1 - iA_2 \\ A_1 + iA_2 & -A_3 \end{array}\right)
\end{equation}
where the three $\mathfrak{su}(2)$ components $A_{1, 2, 3}$ are real (note that $A$ has only one spacetime component).  Then the non-supersymmetric 1D coadjoint orbit Lagrangian \eqref{Lforg} can be written as $\mathcal{L}_\text{1D} = j\mathcal{L}$ where $\mathcal{L} = \mathcal{L}_0 + \mathcal{L}_A$ and
\begin{align}
\mathcal{L}_0 &= \frac{i}{j}\Tr(\lambda g^{-1}\partial_t g) = \frac{i(\phi\dot{\phi}^\dag - \phi^\dag\dot{\phi})}{1 + |\phi|^2}, \label{L0} \\
\mathcal{L}_A &= \frac{1}{j}\Tr(\lambda g^{-1}Ag) = -\left[\frac{(A_1 + iA_2)\phi + (A_1 - iA_2)\phi^\dag - A_3(1 - |\phi|^2)}{1 + |\phi|^2}\right]. \label{LA}
\end{align}
Note that with Hermitian generators, the Killing form given by $\Tr$ is positive-definite.

By promoting $\phi$ to $\Phi$, we find that the supersymmetric completion of $\mathcal{L}_0$ (the coadjoint orbit Lagrangian with vanishing background gauge field, i.e., the pullback of the presymplectic one-form for $SU(2)$) is
\begin{equation}
\tilde{\mathcal{L}}_0 = \int d^2\theta\, K = \frac{i(\phi\dot{\phi}^\dag - \phi^\dag\dot{\phi})}{1 + |\phi|^2} - \frac{\psi^\dag\psi}{(1 + |\phi|^2)^2}, \quad K\equiv \log(1 + |\Phi|^2).
\label{K}
\end{equation}
We have covered $\mathbb{CP}^1$ with the standard patches having local coordinates $\Phi$ and $1/\Phi$, so that $K$ is the K\"ahler potential for the Fubini-Study metric in the patch containing the origin.

To gauge $\tilde{\mathcal{L}}_0$ in a supersymmetric way and thereby obtain the supersymmetric completion of $\mathcal{L}$ requires promoting the $A_i$ to $V_i$, which is more involved.  Having eliminated the integration variable $g$ in favor of $\phi$, let us denote by $g$ what we called $h_\ell$ in \eqref{GtimesT}.  Writing finite and infinitesimal local $SU(2)$ transformations as
\begin{equation}
g = \left(\begin{array}{cc} a & b \\ -\bar{b} & \bar{a} \end{array}\right)\sim \left(\begin{array}{cc} 1 + \frac{i\epsilon_3}{2} & \frac{i\epsilon_1 + \epsilon_2}{2} \\ \frac{i\epsilon_1 - \epsilon_2}{2} & 1 - \frac{i\epsilon_3}{2} \end{array}\right),
\label{infinitesimal}
\end{equation}
finite and infinitesimal gauge transformations take the form
\begin{align}
A\to gAg^{-1} - i\dot{g}g^{-1} &\Longleftrightarrow \delta_{SU(2)}A_i = \epsilon_{ijk}A_j\epsilon_k + \dot{\epsilon}_i, \label{gaugeA} \\
\Phi\to \frac{a\Phi + b}{-\bar{b}\Phi + \bar{a}} &\Longleftrightarrow \delta_{SU(2)}\Phi = \epsilon_i X_i, \mbox{ } (X_1, X_2, X_3)\equiv \frac{1}{2}(i(1 - \Phi^2), 1 + \Phi^2, 2i\Phi),
\end{align}
where the holomorphic $SU(2)$ Killing vectors $X_i$ satisfy $[X_i\partial_\Phi, X_j\partial_\Phi] = \epsilon_{ijk}X_k\partial_\Phi$.  Then
\begin{equation}
\delta_{SU(2)}K = \epsilon_i(\mathcal{F}_i + \bar{\mathcal{F}}_i), \quad (\mathcal{F}_1, \mathcal{F}_2, \mathcal{F}_3)\equiv \frac{1}{2}(-i\Phi, \Phi, i)
\end{equation}
(any purely imaginary $\mathcal{F}_3$ would do, but our choice leads to the ``canonical'' Noether currents transforming in the adjoint representation).  To implement the Noether procedure, we promote the real $\epsilon_i$ to complex chiral superfields $\Lambda_i$:
\begin{equation}
\delta_{SU(2)}\Phi = \Lambda_i X_i.
\end{equation}
The corresponding change in $\tilde{\mathcal{L}}_0$ can be read off from
\begin{equation}
\delta_{SU(2)}K = \Lambda_i\mathcal{F}_i + \bar{\Lambda}_i\bar{\mathcal{F}}_i - i(\Lambda_i - \bar{\Lambda}_i)J_i
\label{deltaK}
\end{equation}
where the $SU(2)$ Noether currents (Killing potentials) are the real superfields
\begin{equation}
J_i = \frac{iX_i\Phi^\dag}{1 + |\Phi|^2} - i\mathcal{F}_i \implies (J_1, J_2, J_3) = \frac{1}{2}\left(-\frac{\Phi + \Phi^\dag}{1 + |\Phi|^2}, -\frac{i(\Phi - \Phi^\dag)}{1 + |\Phi|^2}, \frac{1 - |\Phi|^2}{1 + |\Phi|^2}\right),
\label{Noether}
\end{equation}
which satisfy $J_i^2 = 1/4$ and
\begin{equation}
\delta_{SU(2)}J_i = -\frac{1}{2}\epsilon_{ijk}(\Lambda_j + \bar{\Lambda}_j)J_k + i(\Lambda_j - \bar{\Lambda}_j)J_j J_i - \frac{i}{4}(\Lambda_i - \bar{\Lambda}_i).\footnote{Under $\Phi\to 1/\Phi$, we have $J_1 + iJ_2\leftrightarrow J_1 - iJ_2$ and $J_3\to -J_3$.  The difference between $\mathcal{F}_3 = i/2$ and $\mathcal{F}_3 = 0$ ($J_3$ and $J_3 - 1/2$) is a $U(1)$ Chern-Simons term in the third component of the gauge field, which is singled out by our conventions for the maximal torus (Chern-Simons terms for simple gauge groups do not exist in 1D).  The $J_i$ are only defined up to additive constants, but for nonabelian gauge groups, these constants can be fixed by choosing the $J_i$ to transform in the adjoint representation; in our case,
\[
X_i\partial_\Phi J_j - X_j\partial_\Phi J_i = (X_i\partial_\Phi + \bar{X}_i\partial_{\Phi^\dag})J_j = \epsilon_{ijk}J_k.
\]
For each $U(1)$ factor of the gauge group, there is one undetermined constant (corresponding to an FI term).}
\label{deltaJ}
\end{equation}
This generalizes $\delta_{SU(2)}J_i = -\epsilon_{ijk}\epsilon_j J_k$ for real $\epsilon_i$.  Now, if we could find a counterterm $\Gamma$ such that $\delta_{SU(2)}\Gamma = i(\Lambda_i - \bar{\Lambda}_i)J_i$, then we would be done: the supersymmetric completion of $\mathcal{L}$ would be the minimally gauged supersymmetric $\mathbb{CP}^1$ model $\smash{\tilde{\mathcal{L}} = \tilde{\mathcal{L}}_0 + \tilde{\mathcal{L}}_A}$ where
\begin{equation}
\tilde{\mathcal{L}}_A = \int d^2\theta\, \Gamma, \quad \delta_{SU(2)}\Gamma = i(\Lambda_i - \bar{\Lambda}_i)J_i.
\end{equation}
Note that $\tilde{\mathcal{L}}$ is invariant under local $SU(2)$ because, in light of \eqref{deltaK}, the total variation of $K + \Gamma$ takes the form of a K\"ahler transformation.  There exists a standard procedure for constructing such a $\Gamma$ \cite{Wess:1992cp}, which we review in Appendix \ref{higherGamma}.  Its exact form is
\begin{equation}
\Gamma = 2\int_0^1 d\alpha\, e^{i\alpha V_i O_i}V_j J_j
\end{equation}
where $O_i = X_i\partial_\Phi - \bar{X}_i\partial_{\Phi^\dag}$.  For our purposes, it suffices to work in Wess-Zumino gauge, where the bulk vector superfield is nilpotent of degree three ($V_\text{3D}^3 = 0$) and its restriction to the line is nilpotent of degree two ($V_\text{1D}^2 = 0$): namely, $V_i = \theta\theta^\dag A_i$.  In this gauge, we have $\Gamma = 2V_i J_i$, so that $\tilde{\mathcal{L}}$ reduces to the non-manifestly supersymmetric Lagrangian $\tilde{\mathcal{L}}_0 + \mathcal{L}_A$.  In arbitrary gauge, $\tilde{\mathcal{L}}$ contains terms of arbitrarily high order in the dimensionless bottom component of $V$.\footnote{With $V = V_i T_i$ and $\Lambda = \Lambda_i T_i$ where $T_i = \sigma_i/2$, we have in Wess-Zumino gauge that a 1D super gauge transformation truncates to
\begin{align}
e^{2V}\to e^{i\Lambda}e^{2V}e^{-i\bar{\Lambda}} &\Longleftrightarrow V\to V + \frac{i}{2}(\Lambda - \bar{\Lambda}) - \frac{i}{2}[V, \Lambda + \bar{\Lambda}] \nonumber \\
&\Longleftrightarrow \delta_{SU(2)}V_i = \frac{i}{2}(\Lambda_i - \bar{\Lambda}_i) + \frac{1}{2}\epsilon_{ijk}V_j(\Lambda_k + \bar{\Lambda}_k) \label{gaugeV}
\end{align}
(note that the order of chiral and antichiral parameters is opposite to that in 3D due to our conventions for 1D $\mathcal{N} = 2$ superspace).  Wess-Zumino gauge is preserved under super gauge transformations with parameters $\Lambda_i = \epsilon_i - i\theta\theta^\dag\dot{\epsilon}_i$ where $\epsilon_i\in \mathbb{R}$ (i.e., where the lowest component of $\Lambda$ is real and the fermionic component vanishes).  For such $\Lambda_i$, \eqref{gaugeV} is precisely equivalent to \eqref{gaugeA}.}

An important point is the following.  There are two standard ways of geometrizing the action of $SU(2)$ on $S^2$, both of which can be found in the literature.  These two conventions differ by signs, leading to slightly different $SU(2)$ Noether currents.  First, the action of $SU(2)$ on $S^2$ descends from the adjoint action of $SU(2)$ on $\mathfrak{su}(2)\cong \mathbb{R}^3$, which preserves the Killing form (hence $S^2\subset \mathbb{R}^3$).  This convention is used in, e.g., \cite{Wess:1992cp}, corresponding to
\begin{equation}
(J_1, J_2, J_3)_\text{other} = \frac{1}{2}\left(\frac{\Phi + \Phi^\dag}{1 + |\Phi|^2}, -\frac{i(\Phi - \Phi^\dag)}{1 + |\Phi|^2}, -\frac{1 - |\Phi|^2}{1 + |\Phi|^2}\right)
\end{equation}
(with the relative sign of $J_2$ reversed relative to our \eqref{Noether}).  Second, $SU(2)$ acts on $\mathbb{CP}^1$ by linear fractional transformations.  We use the latter convention unless stated otherwise.  For further details, see Appendix \ref{sigmadetails}.

\subsubsection{Effective Action}

To compute the effective action generated by integrating out $\psi$, we add an $SU(2)$-invariant kinetic term for $\psi$ (with an implicit dimensionful coefficient) as a UV regulator:
\begin{equation}
\mathcal{L}' = \int d^2\theta\, K' = -\frac{i(\psi^\dag\dot{\psi} - \dot{\psi}^\dag\psi) + 4\dot{\phi}\dot{\phi}^\dag}{(1 + |\phi|^2)^2} - \frac{2i(\dot{\phi}^\dag\phi - \phi^\dag\dot{\phi})\psi^\dag\psi}{(1 + |\phi|^2)^3}, \quad K'\equiv \frac{D^\dag\Phi^\dag D\Phi}{(1 + |\Phi|^2)^2}.
\label{Kprime}
\end{equation}
Note that since $D\Phi = \psi - 2i\theta^\dag\dot{\phi} + i\theta\theta^\dag\dot{\psi}$ transforms in the same way under $SU(2)$ as its bottom component $\psi$, $K'$ is automatically invariant under global $SU(2)$.  We want to gauge $K'$.  With chiral superfield gauge transformation parameters, we have (note $DX_i = 2\mathcal{F}_i D\Phi$)
\begin{equation}
\delta_{SU(2)}K' = -i(\Lambda_i - \Lambda_i^\dag)J_i' - i(D\Lambda_i I_i - D^\dag\Lambda_i^\dag I_i^\dag)
\label{deltaKprime}
\end{equation}
where $J_i'$ are the bosonic Noether currents associated to $K'$ and the $I_i$ are fermionic:
\begin{equation}
J_i' = -2K' J_i, \quad I_i = \frac{iX_i(D\Phi)^\dag}{(1 + |\Phi|^2)^2}.
\label{Kprimecurrents}
\end{equation}
There exists a counterterm $\Gamma'$ satisfying
\begin{equation}
\delta_{SU(2)}\Gamma' = i(\Lambda_i - \bar{\Lambda}_i)J_i' + i(D\Lambda_i I_i - D^\dag\bar{\Lambda}_i\bar{I}_i),
\label{Gammaprimecondition}
\end{equation}
which takes the form
\begin{equation}
\int d^2\theta\, \Gamma' = \frac{2\psi^\dag\psi}{(1 + |\phi|^2)^2}\left[\frac{(A_1 + iA_2)\phi + (A_1 - iA_2)\phi^\dag - A_3(1 - |\phi|^2)}{1 + |\phi|^2}\right] + \cdots
\end{equation}
in Wess-Zumino gauge, such that the Lagrangian
\begin{equation}
\tilde{\mathcal{L}}' = \int d^2\theta\, (K' + \Gamma')\equiv \mathcal{L}_\psi - \frac{4\dot{\phi}\dot{\phi}^\dag}{(1 + |\phi|^2)^2} + \cdots
\end{equation}
(written in Wess-Zumino gauge) is invariant under local $SU(2)$, where
\begin{equation}
\mathcal{L}_\psi = -\frac{i(\psi^\dag\dot{\psi} - \dot{\psi}^\dag\psi)}{(1 + |\phi|^2)^2} - \frac{2\mathcal{L}\psi^\dag\psi}{(1 + |\phi|^2)^2}
\end{equation}
is itself invariant under local $SU(2)$ (we construct $\Gamma'$ in Appendix \ref{higherGammap} using a general pres\-cription for the full nonlinear gauging of supersymmetric sigma models with higher-derivative terms).  Thus the ``${\cdots}$'' in $\tilde{\mathcal{L}}'$ contains only dimension-two terms not involving $\psi$, namely the couplings to $A_i$ necessary to make the two-derivative term in $\phi$ invariant under local $SU(2)$.  Making the scale $\mu$ of the higher-dimension terms explicit, consider
\begin{equation}
\tilde{\mathcal{L}}_\text{tot} = j\tilde{\mathcal{L}} - \frac{1}{2\mu}\tilde{\mathcal{L}}'\equiv j\mathcal{L} + \psi^\dag\mathcal{D}\psi + \frac{2\dot{\phi}\dot{\phi}^\dag}{\mu(1 + |\phi|^2)^2} + \cdots,
\end{equation}
where we have integrated by parts.  Performing the path integral over $\psi$ generates the one-loop effective action
\begin{equation}
\tr\log\mathcal{D} = \pm\frac{i}{2}\int dt\, \mathcal{L},
\label{effectiveaction}
\end{equation}
as derived in Appendix \ref{effactiondetails}.  The regularization-dependent sign is fixed to ``$-$'' by canonical quantization, leading to a shift $j\to j - 1/2$.  The ``${\cdots}$'' terms in $\tilde{\mathcal{L}}'$ decouple at low energies ($\mu\to\infty$).

The full component-wise Lagrangian $\tilde{\mathcal{L}}'$ in Wess-Zumino gauge is $\tilde{\mathcal{L}}' = \mathcal{L}_\psi - 4\mathcal{L}_\phi$ where
\begin{equation}
\mathcal{L}_\phi\equiv F - \frac{1}{2}(A_1 - iA_2)F_- - \frac{1}{2}(A_1 + iA_2)F_+ - A_3 F_3 - \frac{1}{4}\mathcal{L}_A^2 + \frac{1}{4}A_i^2
\end{equation}
and we have defined
\begin{equation}
F = \frac{\dot{\phi}\dot{\phi}^\dag}{(1 + |\phi|^2)^2}, \quad F_+ = F_-^\dag = \frac{-i(\dot{\phi} + \phi^2\dot{\phi}^\dag)}{(1 + |\phi|^2)^2}, \quad F_3 = \frac{i(\phi\dot{\phi}^\dag - \phi^\dag\dot{\phi})}{(1 + |\phi|^2)^2}.
\end{equation}
Note that $\mathcal{L}_A = 2A_i J_i$ where $J_i$ denotes the lowest component.  One can check that $\mathcal{L}_\phi$, hence $\tilde{\mathcal{L}}'$, is invariant under local $SU(2)$.  We have $\delta_{SU(2)}\mathcal{L}_\psi = \delta_{SU(2)}\mathcal{L}_\phi = 0$ exactly (not up to total derivatives), which is a consequence of the fact that $\delta_{SU(2)}\Gamma'$ cancels $\delta_{SU(2)}K'$ exactly.

\subsection{Shift from Canonical Quantization}

Canonical quantization of the $\mathcal{N} = 2$ quantum mechanics provides another perspective on the shift in $j$.  Here, we set $A_i = 0$, whence
\begin{equation}
\tilde{\mathcal{L}}|_{A_i = 0} = \tilde{\mathcal{L}}_0, \quad \tilde{\mathcal{L}}'|_{A_i = 0} = \mathcal{L}',
\end{equation}
so that the full Lagrangian is $\smash{\tilde{\mathcal{L}}_\text{tot}|_{A_i = 0}} = j\tilde{\mathcal{L}}_0 - \smash{\frac{1}{2\mu}}\mathcal{L}' = L_B + L_F$ where $L_B$ and $L_F$ describe 1D sigma models with $S^2$ target space:
\begin{align}
L_B &= \frac{ij(\phi\dot{\phi}^\dag - \phi^\dag\dot{\phi})}{1 + |\phi|^2} + \frac{i\alpha}{2}\left(\frac{\dot{\phi}}{\phi} - \frac{\dot{\phi}^\dag}{\phi^\dag}\right) + \frac{2\dot{\phi}\dot{\phi}^\dag}{\mu(1 + |\phi|^2)^2}, \label{LB} \\
L_F &= -\left[j - \frac{i(\phi\dot{\phi}^\dag - \phi^\dag\dot{\phi})}{\mu(1 + |\phi|^2)}\right]\frac{\psi^\dag\psi}{(1 + |\phi|^2)^2} + \frac{i(\psi^\dag\dot{\psi} - \dot{\psi}^\dag\psi)}{2\mu(1 + |\phi|^2)^2}.
\end{align}
For later convenience, we have added a total derivative, parametrized by $\alpha\in \mathbb{R}$, to $L_B$.  Its meaning is as follows: $L_B$ describes an electrically charged particle on $S^2$ in the field of a magnetic monopole of charge $\propto j$ at the center, with the scale $\mu\in \mathbb{R}_{\geq 0}$ (the spectral gap) proportional to its inverse mass and $\alpha$ parametrizing the longitudinal gauge of the monopole vector potential.  We define the gauges $S$, $E$, and $N$ by setting $\alpha = (0, j, 2j)$, respectively.  We refer to $L_B$ as the ``bosonic system'' and to $L_B + L_F$ as the corresponding ``supersymmetric system.''  We now summarize the results of quantizing the theories $L_B$ and $L_B + L_F$: details are given in Appendix \ref{canonicalquantization}.  As when computing the effective action, we use the $\mu$-suppressed kinetic terms as a technical aid; they have the effect of enlarging the phase space.

\subsubsection{Bosonic System}

As a warmup, consider $L_B$ alone.  At finite $\mu$, the phase space is $(2 + 2)$-dimensional and the quantum Hamiltonian can be written as
\begin{equation}
H_j = \frac{\mu}{2}(\vec{L}^2 - j^2) = \frac{\mu}{2}(\ell(\ell + 1) - j^2).
\label{Hj}
\end{equation}
Here, $\smash{\vec{L}}^2 = \frac{1}{2}(L_+ L_- + L_- L_+) + L_3^2$ and we have defined the operators
\begin{align}
L_+ &= -\phi^2\frac{\partial}{\partial\phi} - \frac{\partial}{\partial\phi^\dag} + \frac{2j|\phi|^2 + \alpha(1 - |\phi|^2)}{2|\phi|^2}\phi, \nonumber \\
L_- &= \frac{\partial}{\partial\phi} + (\phi^\dag)^2\frac{\partial}{\partial\phi^\dag} + \frac{2j|\phi|^2 + \alpha(1 - |\phi|^2)}{2|\phi|^2}\phi^\dag, \label{Lquantumwithmu} \\
L_3 &= \phi\frac{\partial}{\partial\phi} - \phi^\dag\frac{\partial}{\partial\phi^\dag} - (j - \alpha), \vphantom{\frac{\phi^\dag}{2}} \nonumber
\end{align}
which satisfy $[L_3, L_\pm] = \pm L_\pm$, $[L_+, L_-] = 2L_3$.  The spectrum is constrained to $\ell\geq j$ by an $L_3$ selection rule,\footnote{Considering the matrix element $\langle\theta = 0|L_3|\ell, m'\rangle$ shows that $\langle\theta = 0|\ell, m'\rangle = 0$ for $m'\neq -j$ (note that while the position eigenstate $|\theta = 0\rangle$ is $\varphi$-independent in the gauge $\alpha = 0$, it acquires a phase factor of $e^{-i\alpha\varphi}$ for general $\alpha$).  In particular, since $|\theta, \varphi\rangle$ is related to $|\theta = 0\rangle$ by a rotation, $\langle\theta, \varphi|\ell, m\rangle = 0$ unless $\ell\geq j$.} with each level $\ell$ appearing once; the eigenfunctions of the associated generalized angular momentum are monopole spherical harmonics.  As $\mu\to\infty$, all states except those with $\ell = j$ decouple (add $-j\mu/2$ to $H_j$).  Rather than taking the decoupling limit $\mu\to\infty$ in $L_B$, which projects out all but the spin-$j$ states, setting $j = 0$ yields the rigid rotor.  Its Hamiltonian is given in terms of the Laplace-Beltrami operator $\Delta_{S^2}$, whose spectrum is $-\ell(\ell + 1)$ with degeneracy $2\ell + 1$ for $\ell\geq 0$.

The bosonic theory with $\mu = \infty$ ($L_B = j\mathcal{L}_0$, in $S$ gauge) is the well-known Wess-Zumino term for quantization of spin.  The action computes the solid angle enclosed by a trajectory on the sphere, and the Dirac quantization condition requires that the coefficient $j$ be a half-integer.  Quantizing the compact phase space $S^2$ yields $2j + 1$ states $|j, m\rangle$, all eigenstates of $L_3$.  Indeed, at $\mu = \infty$, the phase space is $(1 + 1)$-dimensional and we can write
\begin{equation}
L_+ = -\phi^2\partial_\phi + (2j - \alpha)\phi, \quad L_- = \partial_\phi + \frac{\alpha}{\phi}, \quad L_3 = \phi\partial_\phi - (j - \alpha).
\label{Lquantumnomu}
\end{equation}
The wavefunctions are $\phi^{-\alpha}, \ldots, \phi^{2j - \alpha}$, the eigenvalues range from $-j$ to $j$ in integer steps, and $\smash{\vec{L}}^2 = j(j + 1)$.

\subsubsection{Supersymmetric System}

For $L_B + L_F$, let us keep $\mu$ finite (work in the full phase space) and set $\alpha = 0$.  Write
\begin{equation}
\chi = \frac{\psi}{\sqrt{\mu}(1 + |\phi|^2)},
\label{chi}
\end{equation}
which satisfies $\{\chi, \chi^\dag\} = 1$ upon quantization.  The supercharges are represented by differential operators as
\begin{equation}
Q = \psi\left(\frac{\partial}{\partial\phi} - \frac{(j + 1/2)\phi^\dag}{1 + |\phi|^2}\right), \quad Q^\dag = \psi^\dag\left(-\frac{\partial}{\partial\phi^\dag} - \frac{(j - 1/2)\phi}{1 + |\phi|^2}\right),
\label{Qquantum}
\end{equation}
which are adjoints with respect to the Fubini-Study measure.  The Hamiltonian is
\begin{equation}
H' = \frac{1}{2}\{Q, Q^\dag\} = H_{j + \chi^\dag\chi - 1/2} + j\mu\chi^\dag\chi - \frac{\mu}{2}(j - 1/2) = \frac{\mu}{2}(\vec{L}_f^2 - (j + 1/2)(j - 1/2))
\label{Hp}
\end{equation}
where $\vec{L}_f = \vec{L}|_{j + \chi^\dag\chi - 1/2}$.  On the Hilbert space $(L^2(S^2, \mathbb{C})\otimes |0\rangle)\oplus (L^2(S^2, \mathbb{C})\otimes \chi^\dag|0\rangle)$,
\begin{align}
H' &= \left(\begin{array}{cc} H_{j - 1/2} - \mu(j - 1/2)/2 & 0 \\ 0 & H_{j + 1/2} + \mu(j + 1/2)/2 \end{array}\right) \\
&= \frac{\mu}{2}\left(\begin{array}{cc} \ell_b(\ell_b + 1) - (j - 1/2)(j + 1/2) & 0 \\ 0 & \ell_f(\ell_f + 1) - (j - 1/2)(j + 1/2) \end{array}\right) \label{Hpmatrix}
\end{align}
where $\ell_b\geq j - 1/2$ and $\ell_f\geq j + 1/2$.  There are $2j$ bosonic ground states at $\ell_b = j - 1/2$.  This fixes the sign of the previous path integral calculation.  As a further check, the quantum representations of the fermionic monopole angular momenta $(L_f)_i$ are presented in \eqref{Lfquantum}.  Their classical counterparts \eqref{Lfclassical} reduce to the classical $L_i$ with $j - 1/2$ as $\mu\to\infty$.

\subsection{Shift from 1D Supersymmetric Index} \label{indexshift}

To make contact with bulk Wilson loops, we compute both the non-supersymmetric twisted partition function and the flavored Witten index
\begin{equation}
I_{\mathcal{N} = 0} = \Tr(e^{-\beta H}e^{izL_3}), \quad I_{\mathcal{N} = 2} = \Tr[(-1)^F e^{-\beta H'}e^{iz(L_f)_3}]
\label{indices}
\end{equation}
by working semiclassically in the Euclidean path integral.  Let
\begin{align}
L_{B, E} &= \frac{j(\phi\dot{\phi}^\dag - \phi^\dag\dot{\phi})}{1 + |\phi|^2} + \frac{\alpha}{2}\left(\frac{\dot{\phi}}{\phi} - \frac{\dot{\phi}^\dag}{\phi^\dag}\right) + \frac{2\dot{\phi}\dot{\phi}^\dag}{\mu(1 + |\phi|^2)^2}, \label{LBE} \\
L_{F, E} &= \left[j + \frac{\phi\dot{\phi}^\dag - \phi^\dag\dot{\phi}}{\mu(1 + |\phi|^2)}\right]\frac{\psi^\dag\psi}{(1 + |\phi|^2)^2} + \frac{\psi^\dag\dot{\psi} - \dot{\psi}^\dag\psi}{2\mu(1 + |\phi|^2)^2}
\end{align}
denote the Euclideanized versions of $L_B$ and $L_F$, with dots denoting $\tau$-derivatives.  Then
\begin{equation}
I_{\mathcal{N} = 0} = \int D\phi^\dag D\phi\, e^{-\int_0^\beta d\tau\, L_{B, E}}, \quad I_{\mathcal{N} = 2} = \int D\phi^\dag D\phi D\psi^\dag D\psi\, e^{-\int_0^\beta d\tau\, (L_{B, E} + L_{F, E})},
\end{equation}
with boundary conditions twisted by $e^{izL_3}$ or $e^{iz(L_f)_3}$ as appropriate.  While both $I_{\mathcal{N} = 0}$ and $I_{\mathcal{N} = 2}$ are known from canonical quantization, our goal here is to introduce the localization argument via what amounts to a derivation of the Weyl character formula \eqref{su2Weylcharacter} as a sum of two terms coming from the classical saddle points with a spin-independent prefactor coming from the one-loop determinants.  For our precise normalization conventions in what follows, see Appendix \ref{quantummechanics}.

We first compute $I_{\mathcal{N} = 0}$ in the bosonic problem.  Set $\mu = \infty$ and work in the $E$ gauge (not to be confused with ``$E$ for Euclidean''), where
\begin{equation}
L_{B, E} = \frac{j(\phi\dot{\phi}^\dag - \phi^\dag\dot{\phi})}{1 + |\phi|^2} + \frac{j}{2}\partial_\tau\log\left(\frac{\phi}{\phi^\dag}\right).
\label{LBE-Egauge}
\end{equation}
We restrict the path integral to field configurations satisfying $\phi(\tau + \beta) = e^{iz}\phi(\tau)$, for which
\begin{equation}
\int_0^\beta d\tau\, \partial_\tau\log\left(\frac{\phi}{\phi^\dag}\right) = 2iz.
\end{equation}
With this restriction, the action is extremized when $\phi = \phi_\text{cl}\in \{0, \infty\}$ (the two fixed points of the $L_3$ action).  We see that $L_{B, E}|_0 = ijz/\beta$ and $L_{B, E}|_\infty = -ijz/\beta$.  First expand around $\phi_\text{cl} = 0$ with perturbation $\Delta$: $\phi = \phi_\text{cl} + \Delta = \Delta$, where $\Delta$ satisfies the twisted boundary condition.  Its mode expansion takes the form
\begin{equation}
\Delta = \frac{1}{\sqrt{\beta}}\sum_{n=-\infty}^\infty \Delta_n e^{i(2\pi n + z)\tau/\beta},
\end{equation}
from which we obtain simply
\begin{equation}
\int_0^\beta d\tau\, L_{B, E}|_{O(\Delta^2)} = j\int_0^\beta d\tau\, (\Delta\dot{\Delta}^\dag - \Delta^\dag\dot{\Delta}) = -\frac{2ij}{\beta}\sum_{n=-\infty}^\infty (2\pi n + z)|\Delta_n|^2.
\end{equation}
Thus the one-loop factor from expanding around $\phi_\text{cl} = 0$ is
\begin{equation}
Z_\text{1-loop}|_0 = \exp\left[-\sum_{n=-\infty}^\infty \log(2\pi n + z)\right] = \frac{e^{az + b}}{\sin(z/2)} = -\frac{e^{-iz/2}}{2i\sin(z/2)}
\label{abconstants}
\end{equation}
where the integration constants $a, b$ parametrize the counterterms by which different regularization schemes differ.  We present several ways to fix the values of $a$ and $b$ to those written above.  First, \eqref{abconstants} is the only choice consistent with canonical quantization.  Second, Hurwitz zeta function regularization yields
\begin{equation}
\sum_{n=-\infty}^\infty \log(An + B) = \log(1 - e^{2\pi iB/A}) \implies Z_\text{1-loop}|_0 = \frac{1}{1 - e^{iz}} = -\frac{e^{-iz/2}}{2i\sin(z/2)}.
\end{equation}
Third, performing free-field subtraction (normalizing the functional determinant, sans zero mode) at finite $\mu$ and then taking $\mu\to\infty$ yields the same answer.  Indeed, accounting for the $1/\mu$ term in \eqref{LBE}, the kinetic operator for bosonic fluctuations $\Delta$ is $-2(j\partial_\tau + \partial_\tau^2/\mu)$ where the eigenvalues of $\partial_\tau$ are $i(2\pi n + z)/\beta$, giving the regularized product
\begin{align}
Z_\text{1-loop}|_0 = \frac{1}{\det(j\partial_\tau + \partial_\tau^2/\mu)} = -\frac{\sinh(\beta\mu j/2)}{2i\sin(z/2)\sinh((\beta\mu j + iz)/2)}\xrightarrow{\beta\mu\to\infty} -\frac{e^{-(j/|j|)iz/2}}{2i\sin(z/2)}. \label{bosdeterminant}
\end{align}
Now note that taking $\phi\to 1/\phi$ leaves $L_{B, E}$ in $E$ gauge \eqref{LBE-Egauge} invariant (with the $1/\mu$ term in \eqref{LBE} being invariant by itself) while taking $z\to -z$ in the boundary condition for the path integral.  Hence
\begin{equation}
Z_\text{1-loop}|_\infty = (Z_\text{1-loop}|_0)|_{z\to -z} = \frac{e^{iz/2}}{2i\sin(z/2)},
\end{equation}
and it follows that
\begin{equation}
I_{\mathcal{N} = 0} = \sum_{0, \infty} e^{-\beta L_{B, E}}Z_\text{1-loop} = \frac{e^{i(j + 1/2)z} - e^{-i(j + 1/2)z}}{2i\sin(z/2)} = \frac{\sin((j + 1/2)z)}{\sin(z/2)}.
\label{IN0}
\end{equation}
This is, of course, a special case of the Duistermaat-Heckman formula for longitudinal rotations of $S^2$, with the contribution from each fixed point weighted by the appropriate sign.  As a consequence, the index is an even function of $z$ (invariant under the Weyl group $\mathbb{Z}_2$), as it must be, because the Hilbert space splits into representations of $SU(2)$.\footnote{That the index is even in $z$, as implied by canonical quantization, fixes potential multiplicative ambiguities in the path integral computation.  For example, regardless of $\alpha$ in \eqref{LBE}, $L_3$ in \eqref{Lquantumwithmu} satisfies $[L_3, \phi] = \phi$ and hence implements the same twisted boundary condition $\phi(\tau + \beta) = e^{iz}\phi(\tau)$.  However, to obtain an answer that is even in $z$ requires implementing the boundary condition using the operator $L_3|_{\alpha = j}$.  In this way, the constant shift in $L_3$ relative to $L_3|_{\alpha = j}$ gives an overall phase of $e^{-i(j - \alpha)z}$, which combines with the classical contributions $e^{-\beta L_{B, E}}|_0 = e^{-i\alpha z}$ and $e^{-\beta L_{B, E}}|_\infty = e^{i(2j - \alpha)z}$ to produce the gauge-independent result \eqref{IN0}.  To avoid this complication, we have chosen to work in the $E$ gauge from the beginning.}

We now compute $I_{\mathcal{N} = 2}$, keeping $\mu$ finite.  In the supersymmetric problem, the $E$ gauge corresponds to choosing the K\"ahler potential $\log(1 + |\Phi|^2) - \frac{1}{2}\log|\Phi|^2$, which is invariant under $\Phi\to 1/\Phi$.  In component fields, the Lagrangian is $L_{B, E} + L_{F, E}$ with $\alpha = j$.  Expanding in both bosonic fluctuations $\Delta$ and fermionic fluctuations $\Xi$ ($\psi = \psi_\text{cl} + \Xi = \Xi$) gives
\begin{equation}
(L_{B, E} + L_{F, E})|_{O(\Delta^2 + \Xi^2)} = j(\Delta\dot{\Delta}^\dag - \Delta^\dag\dot{\Delta} + \Xi^\dag\Xi) + \frac{2}{\mu}\dot{\Delta}\dot{\Delta}^\dag + \frac{1}{2\mu}(\Xi^\dag\dot{\Xi} - \dot{\Xi}^\dag\Xi).
\label{quadterms}
\end{equation}
The part of the Lagrangian quadratic in fluctuations, as written above, is supersymmetric by itself.\footnote{We have that
\[
\delta(\Delta\dot{\Delta}^\dag - \Delta^\dag\dot{\Delta} + \Xi^\dag\Xi) = -\partial_\tau(\epsilon\Xi\Delta^\dag + \epsilon^\dag\Xi^\dag\Delta), \quad \textstyle \delta[2\dot{\Delta}\dot{\Delta}^\dag + \frac{1}{2}(\Xi^\dag\dot{\Xi} - \dot{\Xi}^\dag\Xi)] = \partial_\tau(\epsilon\Xi\dot{\Delta}^\dag - \epsilon^\dag\Xi^\dag\dot{\Delta})
\]
under the (global) Euclidean SUSY variations $(\delta\Delta, \delta\Xi) = (\epsilon\Xi, 2\epsilon^\dag\dot{\Delta})$ and $(\delta\Delta^\dag, \delta\Xi^\dag) = (-\epsilon^\dag\Xi^\dag, -2\epsilon\dot{\Delta}^\dag)$.}  Twisted boundary conditions in the path integral are implemented by $(L_f)_3$, which satisfies $[(L_f)_3, \phi] = \phi$ and $[(L_f)_3, \psi] = \psi$.  The moding for the fermionic fluctuations
\begin{equation}
\Xi = \frac{1}{\sqrt{\beta}}\sum_{n=-\infty}^\infty \Xi_n e^{i(2\pi n + z)\tau/\beta}
\end{equation}
is integral because at $z = 0$, the insertion of $(-1)^F$ would require periodic boundary conditions for fermions on the thermal circle.  Hence the fermions contribute a factor of
\begin{equation}
\exp\left[\sum_{n=-\infty}^\infty \log\left(\frac{2\pi n + z}{\beta\mu} - ij\right)\right]
\end{equation}
to $Z_\text{1-loop}|_0$ (to obtain a nontrivial functional determinant, we cannot neglect the fermion kinetic term, which is why we have kept $\mu$ finite).  Hurwitz zeta function regularization alone does not suffice for taking the $\beta\mu\to\infty$ limit, so we instead perform free-field subtraction (divide by a fiducial functional determinant):
\begin{equation}
\det(j + \partial_\tau/\mu) = \prod_{n=-\infty}^\infty \frac{(2\pi n + z)/\beta\mu - ij}{2\pi n/\beta\mu - ij} = \frac{\sin((i\beta\mu j - z)/2)}{\sin(i\beta\mu j/2)} \xrightarrow{\beta\mu\to\infty} e^{(j/|j|)iz/2}.
\label{ferdeterminant}
\end{equation}
Taking $j$ positive, this reduces to a phase of $e^{iz/2}$.  By similar reasoning to that in the bosonic case, we conclude that
\begin{equation}
I_{\mathcal{N} = 2} = \frac{\sin(jz)}{\sin(z/2)}.
\label{IN2}
\end{equation}
Again, this is the only answer consistent with canonical quantization.  Thus in the supersymmetric theory, the one-loop shift of $j$ due to the bosons ($+1/2$) exactly cancels that due to the fermions ($-1/2$).

\subsubsection{Localization in 1D}

In both the bosonic and supersymmetric theories, direct comparison to canonical quantization shows that the semiclassical (one-loop) approximation for the index is exact.  It is natural to ask why this should be so, and supersymmetry provides an answer.  While the exactness in the bosonic case can only be heuristically justified by the Dirac quantization condition on $j$, it can be rigorously justified by appealing to the supersymmetric case.

In its most basic form, the localization principle starts from the fact that a Euclidean partition function deformed by a total variation of some nilpotent symmetry $\delta$ ($\delta^2 = 0$) of both the action and the measure is independent of the coefficient of this deformation:
\begin{equation}
Z(t) = \int \mathcal{D}\Phi\, e^{-S[\Phi] + t\delta V} \implies \frac{dZ(t)}{dt} = \int \mathcal{D}\Phi\, \delta\left(e^{-S[\Phi] + t\delta V}V\right) = 0.
\end{equation}
If the bosonic part of $\delta V$ is positive-semidefinite, then as $t\to\infty$, the path integral localizes to $\delta V = 0$.  For a given field configuration with $\delta V = 0$, one can compute a semiclassical path integral for fluctuations on top of this background, and then integrate over all such backgrounds to obtain the exact partition function.

In our case, the quadratic terms arising from perturbation theory are already $(Q + Q^\dag)$-exact, without the need to add any localizing terms.  Indeed, we compute that\footnote{Here, we again use that in Euclidean signature,
\[
\delta\phi = \delta_\epsilon\phi = \epsilon\psi, \quad \delta\phi^\dag = \delta_{\epsilon^\dag}\phi^\dag = -\epsilon^\dag\psi^\dag, \quad \delta\psi = \delta_{\epsilon^\dag}\psi = 2\epsilon^\dag\dot{\phi}, \quad \delta\psi^\dag = \delta_\epsilon\psi^\dag = -2\epsilon\dot{\phi}^\dag
\]
where $\delta\mathcal{O}\equiv [\epsilon Q + \epsilon^\dag Q^\dag, \mathcal{O}]$ and $\delta_{\epsilon, \epsilon^\dag}$ are Grassmann-even.}
\begin{align}
\delta(\delta(\phi^\dag\phi)) &= 2\epsilon^\dag\epsilon(\phi\dot{\phi}^\dag - \phi^\dag\dot{\phi} + \psi^\dag\psi), \\
\delta(\delta(\psi^\dag\psi)) &= 2\epsilon^\dag\epsilon(4\dot{\phi}\dot{\phi}^\dag + \psi^\dag\dot{\psi} - \dot{\psi}^\dag\psi).
\end{align}
Up to overall factors, these are precisely the quadratic expressions \eqref{quadterms} that we integrate over the fluctuations $\Delta, \Xi$ to compute the one-loop factors in the index $I_{\mathcal{N} = 2}$.  As we take the coefficient of either the $\delta(\delta(\phi^\dag\phi))$ term or the $\delta(\delta(\psi^\dag\psi))$ term to infinity, the original Lagrangian $L_{B, E} + L_{F, E}$ becomes irrelevant for the one-loop analysis, but since these terms have the same critical points as the original Lagrangian, the result of the localization analysis coincides with that of the original Lagrangian, proving that the path integral for the latter is one-loop exact.\footnote{Note that the bosonic part of the $\delta(\delta(\phi^\dag\phi))$ term is not positive-semidefinite; indeed, it is imaginary.  We are implicitly using a stationary phase argument.}  Furthermore, the final result is independent of the coefficient of either term.  This has a simple explanation: the regularized bosonic and fermionic functional determinants \eqref{bosdeterminant} and \eqref{ferdeterminant} have a product which is independent of $\beta\mu$, namely
\begin{equation}
\frac{\det(j + \partial_\tau/\mu)}{\det(j\partial_\tau + \partial_\tau^2/\mu)} = -\frac{1}{2i\sin(z/2)}.
\end{equation}
Hence the one-loop factor has the same limit whether $\beta\mu\to\infty$ or $\beta\mu\to 0$.\footnote{To complete the argument, one should check that the path integral measure is invariant under $Q$ (and/or $Q^\dag$).  While we have assumed that this measure reduces to $D\Delta^\dag D\Delta\, D\Xi^\dag D\Xi$ for fluctuations ($D$ here should not be confused with a superderivative), the full nonperturbative path integral measure (i.e., the supersymmetrized Fubini-Study measure) must be invariant under both SUSY and global $SU(2)$.}

\subsubsection{Finite Temperature}

We have shown in Lorentzian signature and at zero temperature that integrating out the fermions in the supersymmetric theory with isospin $J$ ($2J$ bosonic ground states) yields an effective bosonic theory with isospin $j = J - 1/2$ ($2j + 1$ bosonic ground states), which is consistent with the equality of $I_{\mathcal{N} = 0}(j)$ in \eqref{IN0} and $I_{\mathcal{N} = 2}(J)$ in \eqref{IN2}.

The index, however, is computed at finite temperature.  The temperature can only enter the effective action through the dimensionless combination $\beta\mu$, and this dependence must disappear in the limit $\mu\to\infty$.  Therefore, the statement of the preceding paragraph must be independent of temperature.  Let us show this directly at finite temperature by mimicking the index computation, thereby giving an alternative and cleaner derivation of \eqref{effectiveaction}.

We first perform a field redefinition $\psi' = \psi/(1 + |\phi|^2)$ (the associated Jacobian determinant cancels in regularization).  Integrating by parts then gives
\begin{equation}
L_{F, E} = \psi'^\dag\mathcal{D}\psi', \quad \mathcal{D}\equiv \frac{\partial_\tau + \mathcal{L}_{0, E}}{\mu} + j, \quad \mathcal{L}_{0, E}\equiv \frac{\phi\dot{\phi}^\dag - \phi^\dag\dot{\phi}}{1 + |\phi|^2}.
\end{equation}
In Euclidean signature, the eigenfunctions of $\mathcal{D}$ are simple:
\begin{equation}
f(\tau) = \exp\left[(\lambda - j)\mu\tau - \int^\tau d\tau'\, \mathcal{L}_{0, E}\right].
\end{equation}
With periodic (supersymmetric) boundary conditions for the fermions, the eigenvalues are
\begin{equation}
\lambda_n = j + \frac{2\pi in + \mathcal{A}}{\beta\mu}, \quad \mathcal{A} = \int_0^\beta d\tau\, \mathcal{L}_{0, E}, \quad n\in \mathbb{Z}.
\end{equation}
Free-field subtraction then gives
\begin{equation}
\frac{\det(\partial_\tau/\mu + j + \mathcal{L}_{0, E}/\mu)}{\det(\partial_\tau/\mu + j)} = \prod_{n=-\infty}^\infty \frac{j + (2\pi in + \mathcal{A})/\beta\mu}{j + 2\pi in/\beta\mu} = \frac{e^{-\mathcal{A}/2}(1 - e^{\mathcal{A} + \beta\mu j})}{1 - e^{\beta\mu j}}.
\end{equation}
Upon taking $\mu\to\infty$, this becomes $e^{(j/|j|)\mathcal{A}/2}$, whose exponent has the correct sign because the Euclidean action appears with a minus sign in the path integral.

Note that while this computation seemingly fixes the sign outright, our regularization crucially assumes a positive sign for $\mu$.  Moreover, different regularization schemes lead to different global anomalies in the effective action \cite{Dunne:1998qy, Elitzur:1985xj}.  For instance, using Hurwitz zeta function regularization \emph{before} free-field subtraction would give
\begin{equation}
\frac{1 - e^{\mathcal{A} + \beta\mu j}}{1 - e^{\beta\mu j}} \xrightarrow{\mu\to\infty} e^{(1 + j/|j|)\mathcal{A}/2}.
\end{equation}
These ambiguities can be phrased as a mixed anomaly between the ``charge conjugation'' symmetry taking $z\to -z$ and invariance under global gauge transformations $z\to z + 2\pi n$ for $n\in \mathbb{Z}$ \cite{Elitzur:1985xj} (as we will see shortly, $z$ can be interpreted as a background gauge field).  Indeed, in terms of the effective bosonic system, $I_{\mathcal{N} = 0}$ in \eqref{IN0} is invariant under $z\to z + 2\pi n$ for integer $j$ but picks up a sign of $(-1)^n$ for half-integer $j$.  On the other hand, in an alternate regularization where $I_{\mathcal{N} = 0}\to e^{i(j + 1/2)z}I_{\mathcal{N} = 0}$, $I_{\mathcal{N} = 0}$ is no longer even in $z$ but picks up a sign of $(-1)^n$ for all $j$.  To fix the sign of the shift unambiguously (i.e., such that the effective action computation is consistent with the index), we appeal to canonical quantization.  In other words, in the Hamiltonian formalism, we demand that the $SU(2)$ symmetry be preserved quantum-mechanically.

\subsubsection{Background Gauge Field}

The quantities \eqref{indices} are useful because the twisted index with vanishing background gauge field is in fact equivalent to the \emph{un}twisted index with \emph{arbitrary} constant background gauge field.  To see this, set $\mu = \infty$ for simplicity.  To restore the background gauge field, we simply take $L_B\to L_B + j\mathcal{L}_A$, or equivalently
\begin{equation}
L_{B, E}\to L_{B, E} - j\mathcal{L}_A,
\end{equation}
with $\mathcal{L}_A$ in \eqref{LA} (note that $\mathcal{L}_{A, E} = -\mathcal{L}_A$, where the gauge field is always written in Lorentzian conventions).  With $A_i = 0$, the bosonic index $I_{\mathcal{N} = 0}$ corresponds to the partition function for $L_{B, E}$ on $S^1$ with twisted boundary conditions implemented by the quantum operator $L_3$, whose classical expression is given in \eqref{Lclassical}.  Clearly, $I_{\mathcal{N} = 0}$ can also be viewed as a thermal partition function for a deformed Hamiltonian with periodic boundary conditions:
\begin{equation}
I_{\mathcal{N} = 0} = \Tr(e^{-\beta H_z}), \quad H_z\equiv H - \frac{izL_3}{\beta} = H + \frac{ijz}{\beta}\frac{1 - |\phi|^2}{1 + |\phi|^2}.
\end{equation}
This corresponds to a path integral with the modified Lagrangian
\begin{equation}
L_{B, E} + \frac{ijz}{\beta}\frac{1 - |\phi|^2}{1 + |\phi|^2}.
\end{equation}
Setting $z = i\beta A_3$, we recover precisely $(L_{B, E} - j\mathcal{L}_A)|_{A_1 = A_2 = 0}$, so we deduce from \eqref{IN0} that
\begin{equation}
\int D\phi^\dag D\phi\, e^{-\int_0^\beta d\tau\, (L_{B, E} - j\mathcal{L}_A)|_{A_1 = A_2 = 0}} = \frac{\sinh((j + 1/2)\beta A_3)}{\sinh(\beta A_3/2)},
\end{equation}
with periodic boundary conditions implicit.  But for a constant gauge field, we can always change the basis in group space to set $A_1 = A_2 = 0$: under a finite global $SU(2)$ transformation $\phi\to (a\phi + b)/(-\bar{b}\phi + \bar{a})$, the measure is invariant, the single-derivative Wess-Zumino term changes by a total derivative, and the Noether currents \eqref{Noether} rotate into each other.  Letting $|A| = \sqrt{\sum_i A_i^2}$ denote the norm in group space, we conclude that
\begin{equation}
\int D\phi^\dag D\phi\, e^{-\int_0^\beta d\tau\, (L_{B, E} - j\mathcal{L}_A)} = \frac{\sinh((j + 1/2)\beta|A|)}{\sinh(\beta|A|/2)}.\footnote{While we inferred this result from the $SU(2)$ symmetry of the twisted partition function, it can also be seen directly from a semiclassical analysis of the Euclidean Lagrangian.  Setting $\mu = \infty$ and imposing periodic boundary conditions, the critical points of $L_{B, E} - j\mathcal{L}_A$ occur at the constant values
\[
\phi_\text{cl} = \frac{A_3\pm |A|}{A_1 + iA_2} \implies (L_{B, E} - j\mathcal{L}_A)|_{\phi_\text{cl}} = \pm j|A|.
\]
Including the one-loop determinants around these classical contributions gives the expected answer.}
\end{equation}
Setting $L_{B, E}^\text{WZ}\equiv j\mathcal{L}_{0, E}$ and noting that $\Tr_j e^{\pm\beta A_i J_i} = \Tr_j e^{\beta|A|J_3}$, this result can be written more suggestively as
\begin{equation}
\int D\phi^\dag D\phi \exp\left[-\int_0^\beta d\tau\, (L_{B, E}^\text{WZ} - 2jA_i J_i)\right] = \Tr_j e^{-\beta A_i J_i}
\end{equation}
where on the left, the $J_i$ are interpreted as classical Noether currents and on the right, they are interpreted as quantum non-commuting matrices (the Hermitian generators of $SU(2)$ in the spin-$j$ representation).  Hence the path integral for the 1D quantum mechanics with constant background gauge field computes a Wilson loop of spin $j$ with constant gauge field along the $S^1$, i.e., the character of the spin-$j$ representation.  This identification holds even for arbitrary background gauge field because one can always choose a time-dependent gauge such that the gauge field is constant along the loop; the only invariant information is the conjugacy class of the holonomy around the loop.  Indeed, a Wilson loop can be thought of as a dynamical generalization of a Weyl character.

We can now be even more explicit about the relation between the standard path-ordered definition of a Wilson loop and the coadjoint orbit description, with path ordering identified with time ordering in the quantum mechanics on the line and noncommutativity arising as a quantum effect.  In this way, we derive Kirillov's character formula from the partition function of the quantum mechanics \cite{Noma:2009dq}.  Take the gauge field along the $S^1$ to be time-dependent and consider the path-ordered exponential
\[
P\equiv P\exp\left[\int_0^T dt\, (A + B(t))\right] = \sum_{n=0}^\infty \int_0^T dt_1\int_0^{t_1} dt_2\cdots \int_0^{t_{n-1}} dt_n\, (A + B(t_1))\cdots (A + B(t_n))
\]
where $A, B$ are matrices and $A$ is constant.  Observe that $P = P'$ where
\[
P' = \sum_{n=0}^\infty \int_0^T dt_1\int_0^{t_1} dt_2\cdots \int_0^{t_{n-1}} dt_n\, e^{(T - t_1)A}B(t_1)e^{(t_1 - t_2)A}B(t_2)\cdots e^{(t_{n-1} - t_n)A}B(t_n)e^{t_n A}
\]
because $P$ and $P'$ both satisfy the differential equation $f'(T) = (A + B(T))f(T)$ subject to the initial condition $f(0) = 1$.  Now consider a Euclideanized Wilson loop wrapping the $S^1$ and split the gauge field into a fiducial time-independent part and the remainder:
\begin{equation}
\Tr_j P\exp\left[-\int_0^\beta d\tau\, (A_i^c + A_i^\tau)J_i\right]\equiv \Tr_j P\exp\left[\int_0^\beta d\tau\, (A + \mathcal{A})\right]\equiv \sum_{n=0}^\infty P_n.
\end{equation}
One can view the terms $P_n$ as operator insertions inside
\begin{equation}
\int D\phi^\dag D\phi \exp\left[-\int_0^\beta d\tau\, (L_{B, E}^\text{WZ} + 2jA)\right] = \Tr_j e^{\beta A}
\end{equation}
(where the implicit $J_i$ are classical on the left and quantum on the right) as follows:
\begin{align*}
P_n &= \int_0^\beta d\tau_1\int_0^{\tau_1} d\tau_2\cdots \int_0^{\tau_{n-1}} d\tau_n \Tr_j(e^{(\beta - \tau_1)A}\mathcal{A}(\tau_1)e^{(\tau_1 - \tau_2)A}\mathcal{A}(\tau_2)\cdots e^{(\tau_{n-1} - \tau_n)A}\mathcal{A}(\tau_n)e^{\tau_n A}) \\
&= \int_0^\beta d\tau_1\int_0^{\tau_1} d\tau_2\cdots \int_0^{\tau_{n-1}} d\tau_n\int D\phi^\dag D\phi\left[\prod_{i=1}^n -2j\mathcal{A}(\tau_i)\right]\exp\left[-\int_0^\beta d\tau\, (L_{B, E}^\text{WZ} + 2jA)\right] \\
&= \frac{1}{n!}\int D\phi^\dag D\phi\left[\prod_{i=1}^n -2j\int_0^\beta d\tau_i\, \mathcal{A}(\tau_i)\right]\exp\left[-\int_0^\beta d\tau\, (L_{B, E}^\text{WZ} + 2jA)\right].
\end{align*}
In the last step, we have used that the $\mathcal{A}(\tau_i)$ are classical quantities inside the path integral.  Hence the sum exponentiates to
\begin{equation}
\Tr_j P\exp\left[-\int_0^\beta d\tau\, (A_i^c + A_i^\tau)J_i\right] = \int D\phi^\dag D\phi \exp\left[-\int_0^\beta d\tau\, (L_{B, E}^\text{WZ} - 2j(A_i^c + A_i^\tau)J_i)\right]
\end{equation}
where again, the $J_i$ on the left and right have different meanings.

The above arguments can be carried over wholesale to the supersymmetric index $I_{\mathcal{N} = 2}$, since the $(L_f)_i$ rotate into each other under global $SU(2)$.  The fermions modify the representation in which the trace is taken, and (as we will see) the fact that a particular linear combination of the bulk gauge field and the auxiliary scalar $\sigma$ appears in the quantum mechanics is reflected in the appearance of these fields in the supersymmetric path-ordered expression.

\section{Coupling to the Bulk} \label{couplingtobulk}

We now take a top-down approach to the quantum mechanics on the line by restricting the 3D $\mathcal{N} = 2$ multiplets to 1D $\mathcal{N} = 2$ multiplets closed under SUSY transformations that generate translations along the line, which we take to extend along the 0 direction in $\mathbb{R}^{1, 2}$ (as in the previous section, aside from Section \ref{indexshift}, we work in Lorentzian signature).  We thus identify the components of the 1D vector multiplet  with restrictions of the bulk fields; in principle, the 1D chiral multiplet $\Phi$ of the previous section descends from bulk super gauge transformations.

Our conventions for SUSY in $\mathbb{R}^{1, 2}$ are given in Appendix \ref{3DN2}.  The linear combination of supercharges that generates translations along the line is $\Omega\equiv (Q_1 + iQ_2)/\sqrt{2}$ (any choice $\Omega = c_1 Q_1 + c_2 Q_2$ with $|c_1|^2 = |c_2|^2 = 1/2$ and $c_1 c_2^\ast$ purely imaginary would suffice), which satisfies $\{\Omega, \Omega^\dag\} = -2P_0 = 2H$ for vanishing central charge.  Therefore, to restrict to the line, we choose the infinitesimal spinor parameter $\xi$ such that
\begin{equation}
\xi Q = \xi_1 Q_2 - \xi_2 Q_1 = \omega\Omega \implies (\xi_1, \xi_2) = \frac{1}{\sqrt{2}}(i\omega, -\omega)
\end{equation}
where $\omega$ is some fiducial Grassmann parameter (note that $\xi Q$ has suppressed spinor indices, while $\omega\Omega$ does not).  In terms of the linear representations of the supercharges on 3D and 1D $\mathcal{N} = 2$ superspace (\eqref{diffops3D} and \eqref{diffops1D}, respectively), we compute that for superfields whose only spacetime dependence is on the 0 direction, 1D $\mathcal{N} = 2$ SUSY transformations are implemented by $\xi\mathcal{Q} - \bar{\xi}\bar{\mathcal{Q}} = \omega\hat{Q} + \bar{\omega}\hat{Q}^\dag$ with $\theta = \frac{1}{\sqrt{2}}(\theta^1 - i\theta^2)$ and $\partial_\theta = \frac{1}{\sqrt{2}}(\partial_{\theta^1} + i\partial_{\theta^2})$.

\subsection{Linearly Realized SUSY on the Line}

With all auxiliary fields necessary to realize SUSY transformations linearly, a 3D $\mathcal{N} = 2$ vector multiplet ($V = V^\dag$) takes the form
\begin{equation}
\begin{aligned}
V &= C + \theta\chi - \bar{\theta}\bar{\chi} + \frac{1}{2}\theta^2(M + iN) - \frac{1}{2}\bar{\theta}^2(M - iN) - i\theta\bar{\theta}\sigma - \theta\gamma^\mu\bar{\theta}A_\mu \\
&\phantom{==} + i\theta^2\bar{\theta}\left(\bar{\lambda} - \frac{1}{2}\gamma^\mu\partial_\mu\chi\right) - i\bar{\theta}^2\theta\left(\lambda - \frac{1}{2}\gamma^\mu\partial_\mu\bar{\chi}\right) + \frac{1}{2}\theta^2\bar{\theta}^2\left(D - \frac{1}{2}\partial^2 C\right)
\end{aligned}
\label{3DN2vector}
\end{equation}
where $V = V^a T^a$, etc., and all bosonic components are real.  A 3D $\mathcal{N} = 2$ chiral multiplet ($\bar{D}_\alpha\Phi = 0$) takes the form
\begin{equation}
\Phi = A - i\theta\gamma^\mu\bar{\theta}\partial_\mu A - \frac{1}{4}\theta^2\bar{\theta}^2\partial^2 A + \sqrt{2}\theta\psi - \frac{i}{\sqrt{2}}\theta^2\bar{\theta}\gamma^\mu\partial_\mu\psi + \theta^2 F
\label{3DN2chiral}
\end{equation}
where the scalar components are complex.  Bulk (3D) SUSY acts on the vector and chiral multiplets as in \eqref{3Dsusyvector} and \eqref{3Dsusychiral}.  For $f$ any complex 3D fermion, it is convenient to set
\begin{equation}
f'\equiv \frac{f_1 + if_2}{\sqrt{2}}, \quad f''\equiv \frac{f_1 - if_2}{\sqrt{2}}.
\label{linenotation}
\end{equation}
We find that the 3D $\mathcal{N} = 2$ vector multiplet restricts to the following 1D $\mathcal{N} = 2$ multiplets:
\begin{itemize}
\item a 1D vector $\{-C, \chi', \sigma + A_0\}$,
\item a 1D chiral $\{(N + iM)/2, \lambda' - i\partial_0\bar{\chi}''\}$ (and its conjugate antichiral),
\item and a 1D chiral $\{(iD - \partial_0\sigma)/2, \partial_0\bar{\lambda}''\}$ (and its conjugate antichiral).
\end{itemize}
We find that the 3D $\mathcal{N} = 2$ chiral multiplet restricts to the following 1D $\mathcal{N} = 2$ multiplets:
\begin{itemize}
\item a 1D chiral $\{A, -\sqrt{2}\psi'\}$
\item and a 1D antichiral $\{F, -\sqrt{2}\partial_0\psi''\}$.
\end{itemize}
The above 1D $\mathcal{N} = 2$ multiplets transform according to \eqref{1Dsusyvector} and \eqref{1Dsusychiral} with $\epsilon = \omega$.  Note that $\chi, \lambda, \psi$ in 3D each restrict to two independent complex fermions in 1D.

\subsection{Nonlinearly Realized SUSY on the Line}

The most direct way to see how $\Phi$ in the coadjoint orbit Lagrangian arises from bulk super gauge transformations would be to perform the supersymmetric analogue of the derivation of Section \ref{loopsinCS} by cutting out a tubular neighborhood of the line and examining the effect of a bulk super gauge transformation on the resulting boundary (an action is induced on the line after integrating over all such transformations and taking the radius to zero).  For this derivation, it would not suffice to work in Wess-Zumino gauge.\footnote{The conditions on the chiral superfield transformation parameter $\Lambda$ to preserve Wess-Zumino gauge are that $A = -A^\ast$, $\psi = 0$, and $F = 0$, in which case the super gauge transformation $e^{2V}\to e^{\bar{\Lambda}}e^{2V}e^\Lambda$ reduces to an ordinary gauge transformation with parameter $-A$: $V\to V - i\theta\gamma^\mu\bar{\theta}\partial_\mu A + [V, A]$.  These conditions preclude the possibility of inducing fermions on the line.}  Therefore, let us not presuppose a gauge.  In superspace, the 3D $\mathcal{N} = 2$ Chern-Simons Lagrangian
\begin{equation}
\mathcal{L}_\text{CS} = \frac{k}{4\pi i}\int d^4\theta\int_0^1 dt\, \Tr[V\bar{D}_\alpha(e^{-2tV}D^\alpha e^{2tV})]
\end{equation}
is invariant under (linearly realized) SUSY and reduces to \eqref{LCS} in Wess-Zumino gauge.  To see the effect of a super gauge transformation (following \cite{Ivanov:1991fn}), consider more generally
\begin{equation}
\mathcal{L}_\text{CS} = \frac{k}{8\pi i}\int d^4\theta\int_0^1 dt\, \ell_\text{CS}, \quad \ell_\text{CS} = \Tr[(e^{-2V(t)}\partial_t e^{2V(t)})\bar{D}_\alpha(e^{-2V(t)}D^\alpha e^{2V(t)})]
\end{equation}
with boundary conditions $V(0) = 0$ and $V(1) = V$.\footnote{The integration can only be explicitly performed in the abelian case: upon integrating by parts,
\[
\ell_\text{CS} = 2\partial_t(V(t)\bar{D}_\alpha D^\alpha V(t)) \implies \mathcal{L}_\text{CS} = \frac{k}{4\pi}\int d^4\theta\, V\Sigma \xrightarrow{\text{WZ gauge}} \frac{k}{4\pi}(\epsilon^{\mu\nu\rho}A_\mu\partial_\nu A_\rho - 2i\lambda\bar{\lambda} - 2D\sigma)
\]
where $\Sigma = -i\epsilon^{\alpha\beta}\bar{D}_\alpha D_\beta V$ is the linear superfield associated to $V$.  In Wess-Zumino gauge \eqref{WZgauge},
\begin{gather*}
\Sigma = -2\sigma + 2\bar{\theta}\lambda - 2\theta\bar{\lambda} + 2i\theta\bar{\theta}D - \epsilon^{\mu\nu\rho}\theta\gamma_\rho\bar{\theta}F_{\mu\nu} + i\bar{\theta}^2\theta\gamma^\mu\partial_\mu\lambda - i\theta^2\bar{\theta}\gamma^\mu\partial_\mu\bar{\lambda} + \frac{1}{2}\theta^2\bar{\theta}^2\partial^2\sigma.
\end{gather*}
The superspace Lagrangian transforms by a total spinor derivative under $V\to V + \Phi + \bar{\Phi}$, by virtue of the relations $\{D_\alpha, \bar{D}_\beta\} = -2i\gamma_{\alpha\beta}^\mu\partial_\mu$ and hence $\{D^\alpha, \bar{D}_\alpha\} = 0$.}  Under a super gauge transformation $e^{2V(t)}\to e^{\bar{\Phi}(t)}e^{2V(t)}e^{\Phi(t)}$, we have $\ell_\text{CS}\to \ell_\text{CS} + \delta'\ell_\text{CS}$ where
\begin{equation}
\delta'\ell_\text{CS} = \Tr[D^\alpha(e^{-\bar{\Phi}(t)}\partial_t e^{\bar{\Phi}(t)}e^{2V(t)}\bar{D}_\alpha e^{-2V(t)}) + \bar{D}_\alpha(\partial_t e^{\Phi(t)}e^{-\Phi(t)}e^{-2V(t)}D^\alpha e^{2V(t)})].
\end{equation}
Obtaining an explicit expression for this total derivative (in particular, for $\Phi(t)$ when $V(t) = tV$) is prohibitively complicated.  Thus, rather than imitating the derivation of \cite{Moore:1989yh}, we will arrive at a bulk interpretation of the quantum-mechanical variables $\phi, \psi$ in Wess-Zumino gauge, which partially fixes ``super gauge'' while retaining the freedom to perform ordinary gauge transformations.  To this end, it is useful to work in terms of the correponding nonlinearly realized supersymmetry (SUSY') transformations.

In Wess-Zumino gauge, a 3D $\mathcal{N} = 2$ vector multiplet takes the form
\begin{equation}
V|_\text{WZ} = -i\theta\bar{\theta}\sigma - \theta\gamma^\mu\bar{\theta}A_\mu + i\theta^2\bar{\theta}\bar{\lambda} - i\bar{\theta}^2\theta\lambda + \frac{1}{2}\theta^2\bar{\theta}^2 D.
\label{WZgauge}
\end{equation}
Bulk (3D) SUSY' acts on the vector multiplet as
\begin{align}
\delta'\sigma &= -(\xi\bar{\lambda} - \bar{\xi}\lambda), \nonumber \\
\delta' A_\mu &= i(\xi\gamma_\mu\bar{\lambda} + \bar{\xi}\gamma_\mu\lambda), \nonumber \\
\delta'\lambda &= \textstyle -i\xi D - i\gamma^\mu\xi D_\mu\sigma - \frac{1}{2}\epsilon^{\mu\nu\rho}\gamma_\rho\xi F_{\mu\nu}, \label{3Dsusypvector} \\
\delta'\bar{\lambda} &= \textstyle i\bar{\xi}D + i\gamma^\mu\bar{\xi}D_\mu\sigma - \frac{1}{2}\epsilon^{\mu\nu\rho}\gamma_\rho\bar{\xi}F_{\mu\nu}, \nonumber \\
\delta' D &= -(\xi\gamma^\mu D_\mu\bar{\lambda} - \bar{\xi}\gamma^\mu D_\mu\lambda) + [\xi\bar{\lambda} + \bar{\xi}\lambda, \sigma] \nonumber
\end{align}
where $D_\mu(\cdot) = \partial_\mu(\cdot) - i[A_\mu, (\cdot)]$ and $F_{\mu\nu} = \partial_\mu A_\nu - \partial_\nu A_\mu - i[A_\mu, A_\nu]$.  Bulk (3D) SUSY' acts on a fundamental chiral multiplet as
\begin{align}
\delta' A &= -\sqrt{2}\xi\psi, \nonumber \\
\delta'\psi &= -\sqrt{2}\xi F + i\sqrt{2}\gamma^\mu\bar{\xi}D_\mu A + i\sqrt{2}\bar{\xi}\sigma A, \label{3Dsusypchiral} \\
\delta' F &= i\sqrt{2}\bar{\xi}\gamma^\mu D_\mu\psi - i\sqrt{2}\sigma\bar{\xi}\psi - 2i\bar{\xi}\bar{\lambda}A \nonumber
\end{align}
where $D_\mu(\cdot) = \partial_\mu(\cdot) - iA_\mu(\cdot)$.  SUSY' transformations close off shell into the algebra
\begin{equation}
[\delta_\zeta', \delta_\xi'](\cdot) = -2i(\xi\gamma^\mu\bar{\zeta} + \bar{\xi}\gamma^\mu\zeta)D_\mu(\cdot) - 2i(\xi\bar{\zeta} - \bar{\xi}\zeta)\sigma\cdot (\cdot)
\end{equation}
on gauge-covariant fields where, e.g., $\sigma\cdot (\cdot)\equiv [\sigma, (\cdot)]$ for $\sigma, F_{\mu\nu}, \lambda, \bar{\lambda}, D$ and $\sigma\cdot (\cdot)\equiv \sigma(\cdot)$ for $A, \psi, F$.  The above transformation laws and commutators can be obtained by dimensional reduction from 4D (set $\partial_3 = 0$).

The 3D SUSY' transformations restrict to the line as follows.  We again use the notation \eqref{linenotation}.  For the vector multiplet, defining the SUSY'-covariant derivative $D_0'(\cdot)\equiv D_0(\cdot) - i[\sigma, (\cdot)] = \partial_0(\cdot) - i[\sigma + A_0, (\cdot)]$, which satisfies $\delta' D_0'(\cdot) = D_0'\delta'(\cdot)$ and $D_0'\sigma = D_0\sigma$, we obtain the following (rather degenerate) restricted multiplets in 1D:
\begin{itemize}
\item a 1D vector $\{0, 0, \sigma + A_0\}$,
\item a 1D adjoint chiral $\{0, \lambda'\}$ (and its complex conjugate),
\item and a 1D adjoint chiral $\{(iD - D_0'\sigma)/2, D_0'\bar{\lambda}''\}$ (and its complex conjugate).
\end{itemize}
For a fundamental chiral multiplet, defining the SUSY'-covariant derivative $D_0'(\cdot)\equiv D_0(\cdot) - i\sigma(\cdot) = \partial_0(\cdot) - i(\sigma + A_0)(\cdot)$, which satisfies $\delta' D_0'(\cdot) = D_0'\delta'(\cdot)$, we obtain a single restricted multiplet in 1D, namely
\begin{itemize}
\item a 1D fundamental chiral $\{A, -\sqrt{2}\psi'\}$,
\end{itemize}
whose scalar component is associated with bulk gauge transformations.  All of the above 1D $\mathcal{N} = 2$ chiral multiplets transform according to \eqref{1Dsusypchiral} with $\epsilon = \omega$ and $D_0\to D_0'$.  Note that the putative 1D fundamental antichiral $\{F, -\sqrt{2}D_0'\psi''\}$ transforms according to
\begin{align*}
\delta' F &= -\bar{\omega}(-\sqrt{2}D_0'\psi'') - 2iA\bar{\omega}\bar{\lambda}', \\
\delta'(-\sqrt{2}D_0'\psi'') &= 2i\omega D_0' F,
\end{align*}
which is incompatible with 1D SUSY'.  On a 1D chiral multiplet, the 1D SUSY' algebra is realized as
\begin{equation}
[\delta_\eta', \delta_\epsilon'](\cdot) = -2i(\epsilon\eta^\dag + \epsilon^\dag\eta)D_0'(\cdot)
\end{equation}
for $(\cdot) = \phi, \psi$, while $\delta'$ acts trivially on a 1D vector multiplet in Wess-Zumino gauge.

One would expect to write a coupled 3D-1D action
\begin{equation}
S_\text{3D-1D} = \int d^3 x\, \mathcal{L}_\text{CS} + j\int dt\, \tilde{\mathcal{L}}
\end{equation}
that is both supersymmetric and gauge-invariant (under SUSY' and ordinary gauge transformations), with the transformation of the 1D action compensating for any boundary terms induced along the line in the transformation of the 3D action.  However, in Wess-Zumino gauge, $\mathcal{L}_\text{CS}$ in \eqref{LCS} has the following SUSY' variation:
\begin{equation}
\delta'\mathcal{L}_\text{CS} = \frac{k}{4\pi}\partial_\mu\Tr[i\epsilon^{\mu\nu\rho}(\xi\gamma_\nu\bar{\lambda} + \bar{\xi}\gamma_\nu\lambda)A_\rho + 2(\xi\gamma^\mu\bar{\lambda} - \bar{\xi}\gamma^\mu\lambda)\sigma].
\label{LCS-boundary}
\end{equation}
This induces a boundary term along the line only if the fields are singular as the inverse of the radial distance to the line.  Since they are not, it suffices to show that the 1D action is itself invariant under appropriately defined 1D SUSY' transformations.

\subsection{Nonlinearly Realized SUSY in the Sigma Model}

To carry out this last step, we specialize to $SU(2)$.  For the vector multiplet, the bulk and line variables are identified as $a_i = -C_i$, $\psi_i = \chi_i'$, $A_i = \sigma_i + (A_0)_i$.  The quantum mechanics $\tilde{\mathcal{L}}|_\text{WZ} = \tilde{\mathcal{L}}_0 + \mathcal{L}_A$ is invariant under the 1D SUSY' transformations
\begin{align}
\delta'\phi &= \epsilon\psi, \nonumber \\
\delta'\psi &= \textstyle -2i\epsilon^\dag(\dot{\phi} - \frac{i}{2}A_1(1 - \phi^2) - \frac{1}{2}A_2(1 + \phi^2) - iA_3\phi),
\end{align}
which satisfy the algebra
\begin{align}
[\delta_\eta', \delta_\epsilon']\phi &= \textstyle -2i(\epsilon\eta^\dag + \epsilon^\dag\eta)(\dot{\phi} - \frac{i}{2}A_- + \frac{i}{2}A_+\phi^2 - iA_3\phi), \nonumber \\
[\delta_\eta', \delta_\epsilon']\psi &= -2i(\epsilon\eta^\dag + \epsilon^\dag\eta)(\dot{\psi} + i(A_+\phi - A_3)\psi), \\
[\delta_\eta', \delta_\epsilon']\psi^\dag &= -2i(\epsilon\eta^\dag + \epsilon^\dag\eta)(\dot{\psi}^\dag - i(A_-\phi^\dag - A_3)\psi^\dag). \nonumber
\end{align}
The adjoint action of $SU(2)$ on its Lie algebra induces an action on $S^2$, which explains the appearance of the $SU(2)$ Killing vectors in $\delta'\psi$.  Explicitly, at the level of scalar components, the map between the adjoint (gauge parameter) chiral superfield $S = s + \theta f - i\theta\theta^\dag\dot{s} = S^a\sigma^a/2$ and the (scalar) $SU(2)/U(1)$ coset chiral superfield $\Phi = \phi + \theta\psi - i\theta\theta^\dag\dot{\phi}$ is
\begin{equation}
\frac{1}{|s|}\left(\begin{array}{c} s^1 \\ -s^2 \\ s^3 \end{array}\right)\leftrightarrow \left(\begin{array}{c} \sin\theta\cos\varphi \\ \sin\theta\sin\varphi \\ \cos\theta \end{array}\right)\leftrightarrow \phi = \frac{s^1 - is^2}{|s| - s^3}
\label{superfieldmap}
\end{equation}
by stereographic projection (note that this only makes sense for $s$ real).  In terms of angles,
\begin{equation}
e^{i\varphi} = \frac{s^1 - is^2}{\sqrt{|s|^2 - (s^3)^2}}, \quad \tan(\theta/2) = \sqrt{\frac{|s| - s^3}{|s| + s^3}}.
\end{equation}
Keep in mind that to translate between the adjoint action and linear fractional transformations, one must flip the sign of the second Killing vector: that is, one must identify $\vec{\sigma}/2$ with $(\vec{e}_1, -\vec{e}_2, \vec{e}_3)$.  The action of $SU(2)$ is as expected: writing $\epsilon = \epsilon^a\sigma^a/2$ and $s = s^a\sigma^a/2$, we have with $\epsilon^i$ infinitesimal that
\begin{equation}
g = 1 + i\epsilon \implies gsg^{-1} = s + i[\epsilon, s] \implies \delta_{SU(2)}s^i = \epsilon^{ijk}s^j\epsilon^k.
\end{equation}
Under the given map \eqref{superfieldmap}, this is equivalent to $\delta_{SU(2)}\phi = \epsilon_i x_i$.  Now we check that SUSY' acts correctly.  Na\"ively, we have for the components of $S$ that (with $A = A^a\sigma^a/2$)
\begin{align}
\delta' s &= \epsilon f, \nonumber \\
\delta' f &= -2i\epsilon^\dag(\dot{s} - i[A, s]),
\end{align}
but to make sense of SUSY' transformations for real $s$, we must take $f$ real and $\epsilon$ purely imaginary (though $S$ itself is not real):
\begin{align}
\delta' s &= i\epsilon f, \nonumber \\
\delta' f &= -2\epsilon(\dot{s} - i[A, s]),
\end{align}
where $\epsilon, f$ are real Grassmann variables.  In terms of chiral superfields, the desired map is
\begin{equation}
\Phi = \frac{S^1 - iS^2}{|S| - S^3} = \phi + \theta\psi - i\theta\theta^\dag\dot{\phi}.
\end{equation}
Upon substituting for $\delta' s^a$ and $\delta' f^a$, the $\delta'$ variations of $\phi = \phi(s^a, f^a)$ and $\psi = \psi(s^a, f^a)$ are
\begin{align}
\delta'\phi &= i\epsilon\psi, \nonumber \\
\delta'\psi &= \textstyle -2\epsilon(\dot{\phi} - \frac{i}{2}A^1(1 - \phi^2) - \frac{1}{2}A^2(1 + \phi^2) - iA^3\phi),
\end{align}
as expected (for our choice of $\epsilon$).  It would be interesting to clarify the interpretation of the coadjoint orbit theory as resulting from promoting the gauge transformation parameter in the group element $g$ to a chiral superfield ($g = e^{iS^a\sigma^a/2}$).

\section{Localization in 3D} \label{3Dloc}

\subsection{Overview}

We now examine how the understanding achieved for a straight line in $\mathbb{R}^{1, 2}$ can be extended to compact Euclidean spaces.  We will describe shortly the backgrounds to which our analysis generalizes, but some general considerations are as follows.

A supersymmetric field theory minimally coupled to a curved metric is invariant under variations with covariantly constant spinors.  Going beyond the minimal coupling paradigm, one can preserve supersymmetry by generalizing the spinor condition $\nabla_\mu\xi = 0$ in various ways.  It is convenient to start by assuming superconformal symmetry, which requires only the existence of conformal Killing spinors.  Under this assumption, we construct in Appendix \ref{3DN2} the curved-space SUSY' transformations \eqref{3Dsusypvector-M3} and \eqref{3Dsusypchiral-M3}, in which the eight independent superconformal symmetries associated to $\xi, \tilde{\xi}$ generate the 3D $\mathcal{N} = 2$ superconformal algebra $\mathfrak{osp}(2|2, 2)$.  These transformations turn out to be a special case of a more general set of transformations derived from the ``new minimal'' supergravity background (see \cite{Willett:2016adv}, which we follow closely in this section).  Of course, not all of the backgrounds that we are interested in are conformally flat: having derived the SUSY' transformations with conformal Killing spinor parameters, we restrict to those spinors that generate a suitable non-conformal subalgebra; the resulting transformations, for suitably generalized Killing spinors, pertain to all of the backgrounds that we consider.  The results coincide with the supergravity background perspective that we describe in the next subsection.

Having placed the theory supersymmetrically on a curved space, the next question is that of computability.  The fact that the Chern-Simons partition function on Seifert manifolds can be written as a matrix model is well-known from \cite{Marino:2002fk, Aganagic:2002wv} (see \cite{Marino:2005sj} for a review), and has been discussed in the framework of nonabelian localization in \cite{Beasley:2009mb, Beasley:2005vf}.  By now, the computation of observables in $\mathcal{N} = 2$ Chern-Simons theory via supersymmetric localization \cite{Kapustin:2009kz} is also a well-established technique.  The original approach of \cite{Kapustin:2009kz} applies to SCFTs (of which $\mathcal{N} = 2$ Chern-Simons-matter theories with no superpotential, and particularly pure $\mathcal{N} = 2$ Chern-Simons theory, are examples \cite{Gaiotto:2007qi}), but it can be generalized to non-conformal theories with a $U(1)_R$ symmetry \cite{Hama:2010av, Jafferis:2010un}.

The basic approach is as follows.  In Euclidean signature, we regard all fields as complexified and the path integration cycle as middle-dimensional.  To use $\delta V$ as a localizing term while preserving $\delta$-invariance of the theory (and correlation functions of $\delta$-closed operators), we choose $\delta$ to square to (if not zero) a bosonic symmetry under which $V$ is invariant up to total derivatives, and we choose the bosonic part of $\delta V$ to be positive-semidefinite to ensure convergence of the path integral.  On the backgrounds of interest to us, the Euclidean Yang-Mills action may be conveniently chosen to play the role of $\delta V$.  The localization locus $\mathcal{F}$, comprised of field configurations that contribute to the path integral in the $t\to\infty$ limit ($t$ being the coefficient of $\delta V$), is the intersection of the BPS field configurations with the saddle points of $\delta V$ (for gauge theories, we mean that the gauge-fixed action localizes to $\mathcal{F}$; properly, one would form an extended cochain complex with respect to the nilpotent $\delta + \delta_\text{BRST}$ \cite{Pestun:2007rz}).  To avoid a potential confusion about order of limits when deriving the absence of the shift in $\mathcal{N}\geq 2$ Chern-Simons theories \cite{Tanaka:2012nr}, one should integrate out the gauginos at any finite value for the coefficient of the localizing term, before taking it to infinity (which sets the gauginos to zero).

Passing to compact Euclidean three-manifolds allows us to compute correlation functions of nontrivially linked, mutually half-BPS Wilson loops.  On $S^3$, for example, we can access links whose components are fibers of the same Hopf fibration and are therefore unknots with mutual linking number one (the simplest example is the Hopf link).  One can also squash the $S^3$ to obtain the Berger sphere with $SU(2)\times U(1)$ isometry (where the Killing vector has closed orbits and points along the Seifert fiber) \cite{Hama:2011ea, Imamura:2011wg} or the ellipsoid with $U(1)\times U(1)$ isometry (where the Killing vector does not, generically, point along the fiber) \cite{Closset:2012ru}; on the latter background, one can compute expectation values of nontrivial torus knots \cite{Tanaka:2012nr, Kapustin:2013hpk}.  One can also consider lens spaces \cite{Gang:2009wy} and more general Seifert manifolds \cite{Kallen:2011ny, Ohta:2012ev, Closset:2016arn, Closset:2017zgf}.  With appropriate boundary conditions, localizing on a solid torus $D^2\times S^1$ \cite{Sugishita:2013jca, Yoshida:2014ssa} makes contact with supersymmetric analogues of the gluing and Heegaard decompositions usually encountered in the context of Chern-Simons theory \cite{Pasquetti:2011fj, Beem:2012mb, Fujitsuka:2013fga, Benini:2013yva}.  We examine the quantum mechanics on Wilson loops in these general backgrounds.

\subsection{Supergravity Background} \label{sugra}

For the sake of a unified presentation, we first review the relevant aspects of the background supergravity formalism of \cite{Festuccia:2011ws}, following \cite{Willett:2016adv, Closset:2012ru}.  The idea is that one can systematically formulate quantum field theories preserving some rigid supersymmetry as BPS configurations of off-shell supergravity theories to which they couple via a chosen multiplet containing both the supercurrent and the energy-momentum tensor.  For a given theory, different supercurrent multiplets lead to different off-shell supergravities, which have different rigid limits.

In the 3D $\mathcal{N} = 2$ context, this approach allows for the construction of a scalar supercharge by partially topologically twisting the $U(1)_R$ symmetry of the $\mathcal{N} = 2$ algebra.  Namely, suppose that $M^3$ admits a transversely holomorphic foliation (THF), which consists of a nowhere vanishing unit vector field $v^\mu$ and a complex structure $J$ on the two-dimensional leaves transverse to $v^\mu$ such that $\mathcal{L}_v J = 0$.  Then, very roughly speaking, one may twist the spatial rotations in the ``planes'' transverse to the ``time'' direction \cite{Closset:2012ru}.  This construction subsumes both the round sphere and squashed sphere backgrounds.  The relevant supergravity theory is ``new minimal'' supergravity, defined as the off-shell formulation of 3D supergravity that couples to the $\mathcal{R}$-multiplet of a 3D $\mathcal{N} = 2$ quantum field theory with a $U(1)_R$ symmetry.  For the supersymmetry algebra and multiplets resulting from the rigid limit of new minimal supergravity, see Section 6 of \cite{Closset:2012ru}.  The bosonic fields in new minimal supergravity are the metric $g_{\mu\nu}$, the $R$-symmetry gauge field $\smash{A_\mu^{(R)}}$, a two-form gauge field $B_{\mu\nu}$, and the central charge symmetry gauge field $C_\mu$.  It is convenient to let $H$ and $V_\mu$ denote the Hodge duals of the field strengths of $B_{\mu\nu}$ and $C_\mu$, respectively.

For 3D $\mathcal{N} = 2$ theories with a $U(1)_R$ symmetry, \cite{Closset:2012ru} classifies the backgrounds that preserve some supersymmetry.  In particular, to preserve two supercharges of opposite $R$-charge, the three-manifold $M^3$ must admit a nowhere vanishing Killing vector $K^\mu$.  If $K^\mu$ is real, then $M^3$ is necessarily an orientable Seifert manifold.  An example with $K^\mu$ complex is the $S^2\times S^1$ background of \cite{Imamura:2011su} relevant to computing the superconformal index (as opposed to the topologically twisted index of \cite{Benini:2015noa}).  We focus on the case of a real, nowhere vanishing Killing vector $K^\mu$, but we do not restrict the orbit to be a Seifert fiber.  Under these assumptions, it suffices to consider backgrounds with $V_\mu = 0$, so that the conditions for the existence of a rigid supersymmetry are
\begin{equation}
\begin{aligned}
(\nabla_\mu - iA_\mu^{(R)})\xi &= -\frac{1}{2}H\gamma_\mu\xi, \\
(\nabla_\mu + iA_\mu^{(R)})\tilde{\xi} &= -\frac{1}{2}H\gamma_\mu\tilde{\xi}.
\end{aligned}
\label{generalizedKS}
\end{equation}
These are the generalized Killing spinor equations, under which $\xi$ and $\tilde{\xi}$ have $R$-charges $\pm 1$, respectively.  The corresponding SUSY' transformations with $V_\mu = 0$ \cite{Willett:2016adv} are
\begin{align}
\delta'\sigma &= -(\xi\tilde{\lambda} - \tilde{\xi}\lambda), \nonumber \\
\delta' A_\mu &= i(\xi\gamma_\mu\tilde{\lambda} + \tilde{\xi}\gamma_\mu\lambda), \nonumber \\
\delta'\lambda &= \textstyle -i\xi(D - \sigma H) - i\gamma^\mu\xi D_\mu\sigma - \frac{i}{2}\sqrt{g}^{-1}\epsilon^{\mu\nu\rho}\gamma_\rho\xi F_{\mu\nu}, \vphantom{\tilde{\xi}} \label{V0susypvector} \\
\delta'\tilde{\lambda} &= \textstyle i\tilde{\xi}(D - \sigma H) + i\gamma^\mu\tilde{\xi}D_\mu\sigma - \frac{i}{2}\sqrt{g}^{-1}\epsilon^{\mu\nu\rho}\gamma_\rho\tilde{\xi}F_{\mu\nu}, \nonumber \\
\delta' D &= \textstyle -D_\mu(\xi\gamma^\mu\tilde{\lambda} - \tilde{\xi}\gamma^\mu\lambda) + [\xi\tilde{\lambda} + \tilde{\xi}\lambda, \sigma] + H(\xi\tilde{\lambda} - \tilde{\xi}\lambda) \nonumber
\end{align}
for the vector multiplet and
\begin{align}
\delta' A &= -\sqrt{2}\xi\psi, \nonumber \\
\delta'\psi &= \textstyle -\sqrt{2}\xi F + i\sqrt{2}\gamma^\mu\tilde{\xi}D_\mu A + i\sqrt{2}\tilde{\xi}\sigma A - i\sqrt{2}\Delta H\tilde{\xi}A, \label{V0susypchiral} \\
\delta' F &= \textstyle i\sqrt{2}D_\mu(\tilde{\xi}\gamma^\mu\psi) - i\sqrt{2}\sigma\tilde{\xi}\psi - 2i\tilde{\xi}\tilde{\lambda}A + i\sqrt{2}(\Delta - 2)H\tilde{\xi}\psi \nonumber
\end{align}
for a fundamental chiral multiplet of dimension $\Delta$ (here, the dimensions coincide with the $R$-charges, differing by a sign for antichiral multiplets).  The covariant derivative is now
\begin{equation}
D_\mu = \nabla_\mu - iA_\mu - irA_\mu^{(R)}
\label{DmuwithR}
\end{equation}
where $r$ is the $R$-charge of the field on which it acts.  The transformations \eqref{V0susypvector} and \eqref{V0susypchiral} furnish a representation of the algebra $\mathfrak{su}(1|1)$.  Taking into account the generalized Killing spinor equations \eqref{generalizedKS} and replacing
\begin{equation}
\nabla_\mu\xi\to D_\mu\xi = (\nabla_\mu - i\smash{A_\mu^{(R)}})\xi, \quad \nabla_\mu\tilde{\xi}\to D_\mu\tilde{\xi} = (\nabla_\mu + i\smash{A_\mu^{(R)}})\tilde{\xi}
\end{equation}
in the curved-space SUSY' transformations \eqref{3Dsusypvector-M3} and \eqref{3Dsusypchiral-M3} results in precisely the $V_\mu = 0$ SUSY' transformations above.  The latter transformations satisfy the same algebra as \eqref{3Dsusypvector-M3} and \eqref{3Dsusypchiral-M3}, given in Appendix \ref{3DN2}, but with parameters
\begin{equation}
U^\mu = 2i\xi\gamma^\mu\tilde{\xi}, \quad \epsilon_{\mu\nu} = 2H\sqrt{g}\epsilon_{\mu\nu\rho}\xi\gamma^\rho\tilde{\xi}, \quad \rho = 0, \quad \alpha = 2iH\xi\tilde{\xi}
\label{V0susypparameters}
\end{equation}
and $D_\mu$ as in \eqref{DmuwithR}.  We also have at our disposal
\begin{gather}
\delta_\xi'\delta_{\tilde{\xi}}'\Tr\left(\frac{1}{2}\lambda\tilde{\lambda} - iD\sigma\right) = \xi\tilde{\xi}\mathcal{L}_\text{YM}, \label{V0LYM} \\
\mathcal{L}_\text{YM} = \Tr\left[\frac{1}{4}F_{\mu\nu}F^{\mu\nu} + \frac{1}{2}D_\mu\sigma D^\mu\sigma - \frac{1}{2}(D - \sigma H)^2 - i\tilde{\lambda}\gamma^\mu D_\mu\lambda + i\tilde{\lambda}[\sigma, \lambda] + \frac{i}{2}H\lambda\tilde{\lambda}\right] \nonumber
\end{gather}
as a convenient localizing term, where we have omitted the Yang-Mills coupling.

If the generalized Killing spinor equations have at least one solution, then $M^3$ admits a THF.  The existence of solutions to both equations implies that $K^\mu\equiv \xi\gamma^\mu\tilde{\xi}$ is a nowhere vanishing Killing vector.  Assuming that $K^\mu$ is real, we can find local (``adapted'') coordinates $(\tilde{\psi}, z, \bar{z})$ such that $K = \partial_{\tilde{\psi}}$ and
\begin{equation}
ds^2 = (d\tilde{\psi} + a(z, \bar{z})\, dz + \bar{a}(z, \bar{z})\, d\bar{z})^2 + c(z, \bar{z})^2\, dz\, d\bar{z}
\label{standardmetric}
\end{equation}
where $a$ is complex and $c$ is real (following \cite{Willett:2016adv}, we have normalized the metric such that $|K|^2 = 1$, which does not affect results for supersymmetric observables \cite{Closset:2013vra}; see also \cite{Alday:2013lba}).  Coordinate patches are related by transformations of the form $\tilde{\psi}' = \tilde{\psi} + \alpha(z, \bar{z})$, $z' = \beta(z)$, $\bar{z}' = \bar{\beta}(\bar{z})$ with $\alpha$ real and $\beta$ holomorphic.  We choose the vielbein
\begin{equation}
e^1 = \frac{1}{2}c(z, \bar{z})(dz + d\bar{z}), \quad e^2 = \frac{i}{2}c(z, \bar{z})(dz - d\bar{z}), \quad e^3 = d\tilde{\psi} + a(z, \bar{z})\, dz + \bar{a}(z, \bar{z})\, d\bar{z},
\label{standardvielbein}
\end{equation}
for which the corresponding spin connection (determined from $de^a + \omega^a{}_b\wedge\smash{e^b} = 0$) is
\begin{equation}
\omega^{12} = -\omega^{21} = F_a e^3 + (\omega_\text{2D})^{12}, \quad \omega^{23} = -\omega^{32} = -F_a e^1, \quad \omega^{31} = -\omega^{13} = -F_a e^2
\end{equation}
where we have defined
\begin{equation}
F_a(z, \bar{z})\equiv \frac{i(\partial_{\bar{z}}a - \partial_z\bar{a})}{c^2}, \quad (\omega_\text{2D})^{12} = -(\omega_\text{2D})^{21} = -\frac{i}{c}(\partial_z c\, dz - \partial_{\bar{z}}c\, d\bar{z})
\end{equation}
with $\omega_\text{2D}$ being the spin connection associated to $e^1, e^2$ for the 2D metric $c^2\, dz\, d\bar{z}$.  Note that $F_a$ is independent of the choice of chart, while $\omega_\text{2D}$ is not.  We have on spinors that
\begin{equation}
\nabla_\mu = \partial_\mu - \frac{i}{2}F_a\gamma_\mu\cdot {} + i\left(F_a e_\mu^3 + \frac{1}{2}(\omega_\text{2D})_\mu^{12}\right)\gamma^3\cdot {}
\end{equation}
(cf.\ \eqref{nablamuonspinors}), where the dots indicate matrix multiplication rather than spinor contraction (see Appendix \ref{3DN2}).  Hence if we take
\begin{equation}
H = -iF_a, \quad A^{(R)} = -\left(F_a e^3 + \frac{1}{2}(\omega_\text{2D})^{12}\right),
\label{standardbkgd}
\end{equation}
then the generalized Killing spinor equations \eqref{generalizedKS} are solved by
\begin{equation}
\xi = x\left(\begin{array}{c} 1 \\ 0 \end{array}\right), \quad \tilde{\xi} = x\left(\begin{array}{c} 0 \\ 1 \end{array}\right)
\label{generalizedKSsolns}
\end{equation}
in a basis where $\gamma^a = x\sigma^a x^{-1}$ (here, as in the definition of $K^\mu$, we really mean the commuting spinors $\xi|_0$ and $\tilde{\xi}|_0$).  In particular, $\xi, \tilde{\xi}$ are constant in the chosen frame, and since $x\in SU(2)$, we have both $\tilde{\xi} = \xi^\dag$ and $\xi|_0\xi^\dag|_0 = 1$.  Regardless of basis, we have
\begin{equation}
K^a = (\xi|_0)\gamma^a(\tilde{\xi}|_0) = \left(\begin{array}{cc} 0 & -1 \end{array}\right)\gamma^a\left(\begin{array}{c} 0 \\ 1 \end{array}\right) = \delta^{a3},
\end{equation}
so that $K = \partial_{\tilde{\psi}}$.

\subsection{Localizing ``Seifert'' Loops}

We now describe how bulk $V_\mu = 0$ SUSY' restricts to BPS Wilson loops.  To summarize, our assumption that $M^3$ admits a real, nowhere vanishing Killing vector restricts it to be a Seifert manifold.  On any such manifold, it is possible to define a 3D $\mathcal{N} = 2$ supergravity background with $V_\mu = 0$, in which the Killing spinors take a simple form.  Namely, we work in local coordinates $(\tilde{\psi}, z, \bar{z})$ such that $K = \smash{\partial_{\tilde{\psi}}}$ and the metric takes the standard form \eqref{standardmetric}: upon choosing the frame \eqref{standardvielbein} and the background fields $H$ and $\smash{A^{(R)}}$ as in \eqref{standardbkgd}, the generalized Killing spinor equation $D_\mu\eta = -\frac{1}{2}H\gamma_\mu\eta$ (with $D_\mu$ as in \eqref{DmuwithR}) has solutions $\eta = \smash{\xi, \tilde{\xi}}$ of $R$-charge $\pm 1$ as in \eqref{generalizedKSsolns}.

However, the integral curves of the Killing vector field may not be compact.  Therefore, local coordinates adapted to the Killing vector do not necessarily define a Seifert fibration of $M^3$.  Thus the Wilson loops that we consider, while supported on the Seifert manifold $M^3$, are not necessarily Seifert loops.  The quotation marks in the title of this subsection serve to emphasize that the term ``Seifert loop'' (in the sense of \cite{Beasley:2009mb}) is a misnomer.

To begin, consider a Euclidean 3D $\mathcal{N} = 2$ Wilson loop along a curve $\gamma$ \cite{Gaiotto:2007qi, Kapustin:2013hpk}:
\begin{equation}
W = \Tr_R P\exp\left[i\oint_\gamma (A_\mu dx^\mu - i\sigma ds)\right] = \Tr_R P\exp\left[i\oint_\gamma d\tau\, (A_\mu\dot{x}^\mu - i\sigma|\dot{x}|)\right].
\label{3DN2loop}
\end{equation}
The BPS conditions following from \eqref{V0susypvector} take the same form on any background geometry:
\begin{equation}
n^\mu\gamma_\mu\xi - \xi = 0, \quad n^\mu\gamma_\mu\tilde{\xi} + \tilde{\xi} = 0,
\label{bpsconditions}
\end{equation}
with $n^\mu = \dot{x}^\mu/|\dot{x}|$ being the unit tangent vector to $\gamma$.  They are satisfied when $n^\mu = -K^\mu$.  Hence a BPS Wilson loop preserving both supercharges under consideration lies along an integral curve of $K^\mu$.\footnote{These conditions would be equivalent in Lorentzian signature: a Lorentzian 3D $\mathcal{N} = 2$ Wilson loop
\[
W = \Tr_R P\exp\left[i\oint_\gamma (A_\mu dx^\mu + \sigma ds)\right] = \Tr_R P\exp\left[i\oint_\gamma dt\, (A_\mu\dot{x}^\mu + \sigma|\dot{x}|)\right]
\]
is locally half-BPS in $\mathbb{R}^{1, 2}$ (using \eqref{3Dsusypvector}) if we choose
\[
-\frac{\dot{x}^\mu}{|\dot{x}|}\gamma_\mu\xi + i\xi = 0 \Longleftrightarrow -\frac{\dot{x}^\mu}{|\dot{x}|}\gamma_\mu\bar{\xi} - i\bar{\xi} = 0.
\]
If the line extends along the 0 direction, then these conditions reduce to $i\xi_1 = \xi_2$, as in Section \ref{couplingtobulk}.}

To determine how bulk SUSY' restricts to these BPS Wilson loops, note that even after demanding that the Killing spinors $\xi, \tilde{\xi}$ be properly normalized, we still have the freedom to introduce a relative phase between them (the overall phase is immaterial).  Therefore, let us keep $\xi$ as in \eqref{generalizedKSsolns}, with $K^\mu = \xi\gamma^\mu\xi^\dag$, and write
\begin{equation}
\tilde{\xi} = \rho x\left(\begin{array}{c} 0 \\ 1 \end{array}\right) = \rho\xi^\dag, \quad |\rho| = 1.
\end{equation}
The linear combinations of 3D fermions that appear in the 1D multiplets depend on the gamma matrix conventions.  For simplicity, we work in the basis $\gamma^a = \sigma^a$ ($a = 1, 2, 3$).  According to the above discussion, we fix $(n^1, n^2, n^3) = (0, 0, -1)$.  Restoring Grassmann parameters, we have
\begin{equation}
\left(\begin{array}{c} \xi_1 \\ \xi_2 \end{array}\right) = \left(\begin{array}{c} \omega \\ 0 \end{array}\right), \quad \left(\begin{array}{c} \tilde{\xi}_1 \\ \tilde{\xi}_2 \end{array}\right) = \left(\begin{array}{c} 0 \\ \rho\bar{\omega} \end{array}\right).
\end{equation}
To restrict the SUSY' transformations \eqref{V0susypvector} and \eqref{V0susypchiral}, we drop dependence on the 1 and 2 directions and consider only the component of the gauge field along the loop.  Along the loop, frame and spacetime indices are equivalent since $e_3^3 = 1$.  For the vector multiplet, it is convenient to define the 1D SUSY'-covariant derivative
\begin{equation}
D_3'(\cdot)\equiv \partial_3(\cdot) - i[A_3 + i\sigma, (\cdot)]
\label{susypcovariantvector}
\end{equation}
on \emph{both} scalars and spinors, which satisfies $\delta' D_3'(\cdot) = D_3'\delta'(\cdot)$ and $D_3'\sigma = D_3\sigma$.  Note that $D_3'(\cdot)$ and $D_3(\cdot) + [\sigma, (\cdot)]$ coincide on scalars, but not on spinors; note also that in 1D, we need not diffeomorphism-covariantize the derivative acting on spinors because the spin connection is trivial.  In our supergravity background and frame, we have on spinors that
\begin{equation}
\nabla_i = \partial_i - \frac{1}{2}H\gamma_i + iA_i^{(R)}\gamma_3 \implies \nabla_3\psi_{i_\perp} = \partial_3\psi_{i_\perp} - (-1)^{i_\perp}\left(\frac{1}{2}H - iA_3^{(R)}\right)\psi_{i_\perp}
\label{intermediate1}
\end{equation}
where $i_\perp = 1, 2$.  Moreover, it follows from the $V_\mu = 0$ SUSY' algebra that the gauginos $\lambda, \tilde{\lambda}$ have $R$-charges $\mp 1$, so \eqref{DmuwithR} and \eqref{intermediate1} give
\begin{equation}
D_3\lambda_1 = \partial_3\lambda_1 + \frac{1}{2}H\lambda_1 - i[A_3, \lambda_1], \quad D_3\tilde{\lambda}_2 = \partial_3\tilde{\lambda}_2 - \frac{1}{2}H\tilde{\lambda}_2 - i[A_3, \tilde{\lambda}_2].
\label{intermediate2}
\end{equation}
Specializing to our specific $\xi, \tilde{\xi}$, we obtain from \eqref{V0susypvector} (using \eqref{susypcovariantvector} and \eqref{intermediate2}) that
\begin{align}
\delta'\sigma &= -(\omega\tilde{\lambda}_2 + \rho\bar{\omega}\lambda_1), \nonumber \\
\delta' A_3 &= i(\omega\tilde{\lambda}_2 + \rho\bar{\omega}\lambda_1), \nonumber \\
\delta'\lambda_1 &= -i\omega D + i\omega D_3'\sigma + i\omega\sigma H, \vphantom{\tilde{\xi}} \\
\delta'\tilde{\lambda}_2 &= i\rho\bar{\omega}D + i\rho\bar{\omega}D_3'\sigma - i\rho\bar{\omega}\sigma H, \nonumber \\
\delta' D &= -(\omega D_3'\tilde{\lambda}_2 - \rho\bar{\omega}D_3'\lambda_1), \nonumber
\end{align}
with $\delta'\lambda_2 = \delta'\tilde{\lambda}_1 = 0$.  We thus obtain the following restricted multiplets in 1D:
\begin{itemize}
\item A 1D vector $\{0, 0, A_3 + i\sigma\}$ where $\delta'(A_3 + i\sigma) = 0$.
\item Two independent 1D adjoint chirals (not related by complex conjugation) $\{0, \lambda_2\}$ and $\{0, \tilde{\lambda}_1\}$ where $\delta'\lambda_2 = \delta'\tilde{\lambda}_1 = 0$.
\end{itemize}
The remaining fields do not comprise good multiplets.  Namely, we have:
\begin{itemize}
\item A putative 1D adjoint chiral $\{(D + D_3'\sigma)/2, -D_3'\tilde{\lambda}_2\}$ where
\begin{align*}
\delta'((D + D_3'\sigma)/2) &= \omega(-D_3'\tilde{\lambda}_2), \\
\delta'(-D_3'\tilde{\lambda}_2) &= -2i\rho\bar{\omega}D_3'((D + D_3'\sigma)/2) + i\rho\bar{\omega}H D_3'\sigma.
\end{align*}
\item A putative 1D adjoint antichiral $\{(D - D_3'\sigma)/2, iD_3'\lambda_1\}$ where
\begin{align*}
\delta'((D - D_3'\sigma)/2) &= -i\rho\bar{\omega}(iD_3'\lambda_1), \\
\delta'(iD_3'\lambda_1) &= 2\omega D_3'((D - D_3'\sigma)/2) - \omega HD_3'\sigma.
\end{align*}
\end{itemize}
These do not comprise good multiplets for two reasons.  First, the Euclidean SUSY' transformation rules of a 1D chiral are $\delta\phi = \epsilon\psi$, $\delta\psi = 2\epsilon^\dag\dot{\phi}$, and those of a 1D antichiral are $\delta\phi' = -\epsilon^\dag\psi'$, $\delta\psi' = 2\epsilon\dot{\phi}'$; hence the above transformation rules do not close for nonzero $H$.  Second, even for $H = 0$, it is impossible to choose the phase $\rho$ such that both sets of transformation rules close (if $\rho = i$, then the chiral closes while the antichiral does not, while if $\rho = -i$, then the opposite is true).  We will see that for a 3D chiral to restrict to a 1D chiral, we must choose $\rho = i$.  Indeed, consider a fundamental chiral multiplet of dimension $\Delta$.  The corresponding 1D SUSY'-covariant derivative is
\begin{equation}
D_3'(\cdot)\equiv \partial_3(\cdot) - i(A_3 + i\sigma)(\cdot),
\label{susypcovariantchiral}
\end{equation}
which satisfies $\delta' D_3'(\cdot) = D_3'\delta'(\cdot)$.  From the $V_\mu = 0$ SUSY' algebra, we see that $A, \psi, F$ have $R$-charges $-\Delta, 1 - \Delta, 2 - \Delta$, respectively, so that
\begin{equation}
\begin{aligned}
D_3 A &= (\partial_3 - iA_3 + i\Delta A_3^{(R)})A, \\
D_3\psi_1 &= \nabla_3\psi_1 - iA_3\psi_1 - i(1 - \Delta)A_3^{(R)}\psi_1
\end{aligned}
\end{equation}
with $\nabla_3\psi_1$ as in \eqref{intermediate1}.  Substituting our specific $\xi, \tilde{\xi}$ into \eqref{V0susypchiral} and using \eqref{susypcovariantchiral} then gives the restricted transformation rules
\begin{align}
\delta' A &= -\sqrt{2}\omega\psi_2, \nonumber \\
\delta'\psi_1 &= -\sqrt{2}\omega F, \nonumber \\
\delta'\psi_2 &= i\sqrt{2}\rho\bar{\omega}(D_3' + i\Delta A_3^{(R)} - \Delta H)A, \\
\delta' F &= i\sqrt{2}\rho\bar{\omega}(D_3' - i(2 - \Delta)A_3^{(R)} + H)\psi_1 + 2i\rho\bar{\omega}\tilde{\lambda}_1 A - i\sqrt{2}\Delta\rho\bar{\omega}\psi_1 H. \nonumber
\end{align}
Choosing both $\rho = i$ and $\Delta = 0$, we obtain a single restricted multiplet in 1D, namely a 1D fundamental chiral $\{A, -\sqrt{2}\psi_2\}$ where
\begin{equation}
\begin{aligned}
\delta' A &= \omega(-\sqrt{2}\psi_2), \\
\delta'(-\sqrt{2}\psi_2) &= 2\bar{\omega}D_3' A.
\end{aligned}
\end{equation}
The remaining transformation rules can be written as
\begin{align*}
\delta' F &= \bar{\omega}(-\sqrt{2}(D_3' - 2iA_3^{(R)} + H)\psi_1) - 2\bar{\omega}\tilde{\lambda}_1 A, \\
\delta'(-\sqrt{2}D_3'\psi_1) &= 2\omega D_3' F,
\end{align*}
which superficially resemble those of a 1D fundamental antichiral, but which do not close and do not have the correct relative sign.

The key point is that the transformation rules for the restricted 1D multiplets are independent of the supergravity background fields and take exactly the same form as in flat space.  The fields are \emph{a priori} complex, and $D_3'$ is not Hermitian because it involves a complexified gauge field.  After imposing reality conditions in the path integral, we want $\sigma$ purely imaginary, $A_3$ purely real, and $D$ purely imaginary; the fermions remain independent.

\subsubsection{Example: \texorpdfstring{$S^3$}{S3}}

Let us see how this setup works in the familiar setting of $S^3$, whose radius we take to be $\ell$.  This is a special case due to the high amount of symmetry, so we first make some comments on the geometry of $S^3$.  We coordinatize $S^3$ by an element $g\in SU(2)$, which admits both left and right actions of $SU(2)$.  Frame indices are identified with $\mathfrak{su}(2)$ indices in the basis $T^a = \sigma^a/2$.  The $\mathfrak{su}(2)$-valued left- and right-invariant one-forms are
\begin{equation}
\Omega_L\equiv g^{-1}dg = i(\Omega_L)^a T^a, \quad \Omega_R\equiv dgg^{-1} = i(\Omega_R)^a T^a
\end{equation}
where $(\Omega_L)^a\equiv (\Omega_L)_\mu^a\, dx^\mu$ and $(\Omega_R)^a\equiv (\Omega_R)_\mu^a\, dx^\mu$, which satisfy the Maurer-Cartan equations
\begin{equation}
d(\Omega_L)^a - \frac{1}{2}\epsilon^{abc}(\Omega_L)^b\wedge (\Omega_L)^c = 0, \quad d(\Omega_R)^a + \frac{1}{2}\epsilon^{abc}(\Omega_R)^b\wedge (\Omega_R)^c = 0.
\end{equation}
The bi-invariant Riemannian metric (i.e., the metric on $S^3$ induced by its embedding in $\mathbb{C}^2$ with coordinates $(a, b)$ via $g = (\begin{smallmatrix} a & b \\ -\bar{b} & \bar{a} \end{smallmatrix})$) is
\begin{equation}
ds^2 = \frac{\ell^2}{2}\Tr(dg\otimes dg^{-1}) = -\frac{\ell^2}{2}\Tr(\Omega_L^{\otimes 2}) = -\frac{\ell^2}{2}\Tr(\Omega_R^{\otimes 2}).
\end{equation}
In terms of Euler angles $\theta\in [0, \pi)$, $\phi\in [0, 2\pi)$, $\psi\in [0, 4\pi)$, we have
\begin{equation}
g = \begin{pmatrix} \cos\frac{\theta}{2}e^{i(\phi + \psi)/2} & i\sin\frac{\theta}{2}e^{i(\phi - \psi)/2} \\[5 pt] i\sin\frac{\theta}{2}e^{-i(\phi - \psi)/2} & \cos\frac{\theta}{2}e^{-i(\phi + \psi)/2} \end{pmatrix}, \mbox{ } ds^2 = \frac{\ell^2}{4}(d\theta^2 + d\phi^2 + d\psi^2 + 2\cos\theta\, d\phi\, d\psi).
\label{eulerangles}
\end{equation}
The frame one-forms are
\begin{equation}
e_L = \frac{\ell}{2}\Omega_L, \quad e_R = \frac{\ell}{2}\Omega_R,
\label{frameoneforms}
\end{equation}
which satisfy $e_\mu^a e_\nu^b\delta_{ab} = g_{\mu\nu}$.  With the canonical orientation $\sqrt{\det g}\epsilon_{\theta\phi\psi} = \ell^3\sin\theta/8$ (writing $\det g$ in full to avoid confusion with the $SU(2)$ coordinate $g$), the volume form is
\begin{equation}
\sqrt{\det g}\, d\theta\wedge d\phi\wedge d\psi = -(e_L)^1\wedge (e_L)^2\wedge (e_L)^3 = -(e_R)^1\wedge (e_R)^2\wedge (e_R)^3,
\end{equation}
giving $\operatorname{vol}(S^3) = 2\pi^2\ell^3$.  From the Maurer-Cartan equations, we read off the spin connection $\omega^{ab}\equiv \omega_\mu^{ab}\, dx^\mu$ in the two frames: $(\omega_L)^{ab} = \frac{1}{\ell}\epsilon^{abc}(e_L)^c$ and $(\omega_R)^{ab} = -\frac{1}{\ell}\epsilon^{abc}(e_R)^c$.  Hence on spinors (see \eqref{nablamuonspinors}), we have
\begin{equation}
\nabla_\mu|_\text{LI} = \partial_\mu + \frac{i}{2\ell}\gamma_\mu\cdot {}, \quad \nabla_\mu|_\text{RI} = \partial_\mu - \frac{i}{2\ell}\gamma_\mu\cdot {}
\end{equation}
in the left- and right-invariant frames, respectively.  The left- and right-invariant vector fields that generate the left and right actions of $SU(2)$ via $L^a g = -T^a g$ and $R^a g = gT^a$ are dual to $\Omega_{R, L}$ in that their components are given by the inverse vielbeins:
\begin{equation}
L_a = \frac{i\ell}{2}(e_R)_a^\mu\partial_\mu, \quad R_a = -\frac{i\ell}{2}(e_L)_a^\mu\partial_\mu,
\label{LandR}
\end{equation}
with normalizations chosen such that they satisfy the algebra $\mathfrak{su}(2)_\ell\oplus \mathfrak{su}(2)_r$:
\begin{equation}
[L_a, L_b] = i\epsilon_{abc}L_c, \quad [R_a, R_b] = i\epsilon_{abc}R_c.
\end{equation}
Their actions on $g$ imply that
\begin{equation}
L^a(\Omega_R)^b = i\epsilon^{abc}(\Omega_R)^c, \quad R^a(\Omega_L)^b = i\epsilon^{abc}(\Omega_L)^c, \quad L^a(\Omega_L)^b = R^a(\Omega_R)^b = 0.
\end{equation}
The four $\mathbb{C}$-linearly independent conformal Killing spinor fields on $S^3$ can be constructed by taking $\xi$ constant in the left-invariant frame or constant in the right-invariant frame:
\begin{equation}
\nabla_\mu\xi = \pm\frac{i}{2\ell}\gamma_\mu\cdot\xi = \mp\frac{i}{2\ell}\gamma_\mu\xi \Longleftrightarrow \xi' = \mp\frac{i}{2\ell}\xi,
\end{equation}
with two solutions for each sign.  Keeping in mind that \eqref{CKScondition} means $\nabla_\mu\xi = -\gamma_\mu\cdot\xi'$, we refer to spinors with $\xi' = -\smash{\frac{i}{2\ell}\xi}$ as ``positive'' and those with $\xi' = \smash{\frac{i}{2\ell}\xi}$ as ``negative.''  These conformal Killing spinors are genuine Killing spinors, for which $\xi'\propto \xi$ (on maximally symmetric spaces with nonzero curvature, there always exists a basis for the space of conformal Killing spinors consisting of Killing spinors).  For these spinors, the condition \eqref{extracondition} for closure of the superconformal algebra (involving $R = 6/\ell^2$ and $h = -9/4\ell^2$) is automatically satisfied.  Using \eqref{frameoneforms}, the Killing spinors can be written more explicitly as follows.  Let $\xi_0$ denote a constant spinor, and write $\gamma^a = x\sigma^a x^{-1}$ for constant $x\in SU(2)$ parametrizing the basis.  In the left-invariant frame, ``positive'' and ``negative'' Killing spinors satisfy
\begin{equation}
\partial_\mu\xi = 0, \quad \partial_\mu\xi = -\frac{i}{\ell}\gamma_\mu\cdot\xi = -\frac{i}{\ell}(e_L)_\mu^a\gamma^a\cdot\xi,
\end{equation}
respectively.  The first equation is solved by $\xi = \xi_0$, and the second by $\xi = xg^{-1}x^{-1}\cdot\xi_0$.  In the right-invariant frame, ``positive'' and ``negative'' Killing spinors satisfy
\begin{equation}
\partial_\mu\xi = \frac{i}{\ell}\gamma_\mu\cdot\xi = \frac{i}{\ell}(e_R)_\mu^a\gamma^a\cdot\xi, \quad \partial_\mu\xi = 0,
\end{equation}
respectively.  The first equation is solved by $\xi = xgx^{-1}\cdot\xi_0$, and the second by $\xi = \xi_0$.

By taking the spinor parameters $\xi, \tilde{\xi}$ of the superconformal algebra both to be positive or negative (hence constant in the left- or right-invariant frame), the dilatation parameter $\rho = \frac{2i}{3}\nabla_\mu(\xi\gamma^\mu\tilde{\xi})$ vanishes and we restrict to either of the non-conformal subalgebras
\[
\mathfrak{osp}(2|2)_\text{left}\times \mathfrak{su}(2)_\text{right}, \mbox{ } \mathfrak{su}(2)_\text{left}\times \mathfrak{osp}(2|2)_\text{right}\subset \mathfrak{osp}(2|2, 2),
\]
which are $S^3$ analogues of $\mathcal{N} = 2$ Poincar\'e supersymmetry in $\mathbb{R}^3$.  The Killing vector $\smash{\xi\gamma^\mu\tilde{\xi}}$ is likewise constant in the appropriate frame.  The supercharges generate $\mathfrak{osp}(2|2)$ (left or right), whose bosonic subalgebra contains the $\mathfrak{su}(2)$ (left or right) isometry and the $\mathfrak{u}(1)_R$.  The SUSY' transformations become
\begin{equation}
\begin{aligned}
\delta'\sigma &= -(\xi\tilde{\lambda} - \tilde{\xi}\lambda), \\
\delta' A_\mu &= i(\xi\gamma_\mu\tilde{\lambda} + \tilde{\xi}\gamma_\mu\lambda), \\
\delta'\lambda &= \textstyle -i\xi D - i\gamma^\mu\xi D_\mu\sigma - \frac{i}{2}\sqrt{g}^{-1}\epsilon^{\mu\nu\rho}\gamma_\rho\xi F_{\mu\nu}\mp \frac{1}{\ell}\sigma\xi, \vphantom{\tilde{\xi}} \\
\delta'\tilde{\lambda} &= \textstyle i\tilde{\xi}D + i\gamma^\mu\tilde{\xi}D_\mu\sigma - \frac{i}{2}\sqrt{g}^{-1}\epsilon^{\mu\nu\rho}\gamma_\rho\tilde{\xi}F_{\mu\nu}\pm \frac{1}{\ell}\sigma\tilde{\xi}, \\
\delta' D &= \textstyle -(\xi\gamma^\mu D_\mu\tilde{\lambda} - \tilde{\xi}\gamma^\mu D_\mu\lambda) + [\xi\tilde{\lambda} + \tilde{\xi}\lambda, \sigma]\mp \frac{i}{2\ell}(\xi\tilde{\lambda} - \tilde{\xi}\lambda)
\end{aligned}
\label{S3susypvector}
\end{equation}
for the vector multiplet and
\begin{equation}
\begin{aligned}
\delta' A &= -\sqrt{2}\xi\psi, \\
\delta'\psi &= \textstyle -\sqrt{2}\xi F + i\sqrt{2}\gamma^\mu\tilde{\xi}D_\mu A + i\sqrt{2}\tilde{\xi}\sigma A\pm \frac{\sqrt{2}\Delta}{\ell}\tilde{\xi}A, \\
\delta' F &= \textstyle i\sqrt{2}\tilde{\xi}\gamma^\mu D_\mu\psi - i\sqrt{2}\sigma\tilde{\xi}\psi - 2i\tilde{\xi}\tilde{\lambda}A\mp \frac{\sqrt{2}(\Delta - 1/2)}{\ell}\tilde{\xi}\psi
\end{aligned}
\label{S3susypchiral}
\end{equation}
for a fundamental chiral multiplet of dimension $\Delta$, with the top and bottom signs corresponding to $\xi, \tilde{\xi}$ positive and $\xi, \tilde{\xi}$ negative, respectively.  The non-conformal $\mathcal{N} = 2$ Yang-Mills Lagrangian is invariant under this restricted supersymmetry.  Indeed, we find that \eqref{V0LYM} holds, where $\mathcal{L}_\text{YM}|_{S^3}$ corresponds to taking $H = \pm i/\ell$ with the top and bottom signs as above:
\begin{equation}
\mathcal{L}_\text{YM}|_{S^3} = \Tr\left[\frac{1}{4}F_{\mu\nu}F^{\mu\nu} + \frac{1}{2}D_\mu\sigma D^\mu\sigma - \frac{1}{2}\left(D\mp \frac{i\sigma}{\ell}\right)^2 - i\tilde{\lambda}\gamma^\mu D_\mu\lambda + i\tilde{\lambda}[\sigma, \lambda]\mp \frac{1}{2\ell}\lambda\tilde{\lambda}\right].
\label{LYM-S3}
\end{equation}
\emph{A fortiori}, $\mathcal{L}_\text{YM}|_{S^3}$ is supersymmetric with respect to the restricted SUSY'.

Let us restrict to the subalgebra of positive (``left-invariant'') spinors and use as a localizing term $\mathcal{L}_\text{YM}|_{S^3}$ with $H = i/\ell$.  A Wilson loop with unit tangent vector $n^\mu$ is locally half-BPS if we choose $n^\mu$ such that \eqref{bpsconditions} holds (these are half-BPS rather than BPS conditions due to the extra symmetry of $S^3$).  Since $\xi, \tilde{\xi}$ are constant in the left-invariant frame, the loop can only be globally half-BPS if $n^\mu$ is a constant linear combination of the $(e_L)_a^\mu$: $n^\mu = n^a(e_L)_a^\mu$.  The positive spinors $\xi, \tilde{\xi}$ each belong to a two-complex-dimensional space of two-component complex spinors; the Wilson loop selects a one-complex-dimensional line inside each of these spaces, so it preserves a single complex supercharge.  In practice, one localizes with respect to a single real supercharge, which further restricts us to one of the two left-invariant Killing spinors selected by the Wilson loop.  Regardless of the basis for the gamma matrices, the property $\sigma_2\sigma_a^\ast\sigma_2 = -\sigma_a$ implies that the BPS conditions require that $\tilde{\xi}\propto \sigma_2\cdot\xi^\ast\propto \xi^\dag$.  Now define the Killing vector
\begin{equation}
K^\mu = (\xi\gamma^\mu\xi^\dag)|_0 = (\xi|_0)\gamma^\mu(\xi^\dag|_0).
\end{equation}
Normalizing $\xi$ by setting $\xi|_0\xi^\dag|_0 = 1$, we compute using Fierz identities for mutually commuting spinors (see Appendix \ref{3DN2}) that $K^\mu = (K^\mu)^\ast$, $K_\mu K^\mu = 1$, and
\begin{equation}
K^\mu\gamma_\mu\xi = -\xi, \quad K^\mu\gamma_\mu\xi^\dag = \xi^\dag.
\end{equation}
Thus for a properly normalized, positive Killing spinor $\xi$ with corresponding $K^\mu$, the BPS equation in $\xi$ for a Wilson loop with $n^\mu = -K^\mu$ is automatically satisfied, while the BPS equation in $\tilde{\xi}$ requires that $\tilde{\xi}\propto \xi^\dag$.  The supercharge with respect to which we localize is defined by the choice of $\xi$ (for which $n^\mu = -K^\mu$) and $\tilde{\xi} = 0$.

To determine how bulk SUSY' restricts to half-BPS Wilson loops, note that the left-invariant SUSY' algebra on $S^3$, corresponding to \eqref{S3susypvector} and \eqref{S3susypchiral} with the top sign, takes the same form as the $V_\mu = 0$ SUSY' algebra with parameters \eqref{V0susypparameters}, but with $H = i/\ell$ and $\smash{A_\mu^{(R)}} = 0$.  We work in the basis $\gamma^a = \sigma^a$.  Any constant $n^a$ defines a family of Wilson loops; WLOG, we fix $(n^1, n^2, n^3) = (0, 0, -1)$, giving the normalized Killing spinors $(\xi_1, \xi_2) = (\omega, 0)$ and $(\tilde{\xi}_1, \tilde{\xi}_2) = (0, \rho\bar{\omega})$, where we have fixed a convenient phase for $\xi$ and let $\rho$ denote the relative phase between $\tilde{\xi}, \xi^\dag$.  The corresponding Killing vector is $(K^1, K^2, K^3) = (0, 0, 1)$.  The rest of the analysis proceeds in the same way as for a general Seifert manifold, but with $H = i/\ell$ and $\smash{A_\mu^{(R)}} = 0$: we obtain the same restricted multiplets.\footnote{Choosing $(n^1, n^2, n^3) = (0, 0, 1)$ would change the 1D gauge field to $A_3 - i\sigma$, with 1D SUSY' transformations modified such that $\delta'(A_3 - i\sigma) = 0$.}

Let us see how the general construction of Section \ref{sugra} reduces to the known one in the case of $S^3$.  In a form that makes the Hopf fibration manifest \cite{Drukker:2012sr}, the metric \eqref{eulerangles} on $S^3$ is
\begin{equation}
ds^2 = \frac{\ell^2}{4}(d\theta^2 + \sin^2\theta\, d\phi^2 + (d\psi + \cos\theta\, d\phi)^2),
\end{equation}
where the first two terms are the round metric on $S^2$ of radius $\ell/2$.\footnote{The integral curves of Killing vectors on $S^3$ are great circles corresponding to fibers of the Hopf fibration $S^3\to S^2$.  The fiber over a given point on $S^2$ is the locus with fixed $(\theta, \phi)$ and arbitrary $\psi$.  These are great circles because such circles have $ds = \ell\, d\psi/2$ and hence circumference $2\pi\ell$, and when parametrized by arc length $s$, they are clearly integral curves of $(e_L)_3^\mu\partial_\mu = \frac{2}{\ell}\partial_\psi$ with unit tangent vector.}  Defining the dimensionful variable $\tilde{\psi} = \ell\psi/2$, so that $\tilde{\psi}/\ell\in [0, 2\pi)$, we have $K = \smash{\partial_{\tilde{\psi}}}$.  Stereographic projection gives the relation $z = \ell e^{i\phi}/\tan(\theta/2)$ between complex coordinates $z, \bar{z}$ and spherical coordinates $(\theta, \phi)$ on $S^2$.  To go from the patch containing the origin to the patch containing $\infty$ on $S^2$, we simply take $z' = \ell^2/z$ ($\tilde{\psi}$ does not transform).  These correspond to adapted coordinates \eqref{standardmetric} on $S^3$ with
\[
a = \frac{i\ell}{4z}\left(\frac{1 - |z|^2/\ell^2}{1 + |z|^2/\ell^2}\right), \mbox{ } c = \frac{1}{1 + |z|^2/\ell^2}\implies H = -\frac{i}{\ell}, \mbox{ } A^{(R)} = -\frac{d\tilde{\psi}}{\ell} - \frac{i}{4}\left(\frac{dz}{z} - \frac{d\bar{z}}{\bar{z}}\right).
\]
Despite that the resulting SUSY' algebra takes the same form as in the right-invariant frame, our standard frame \eqref{standardvielbein} for the adapted coordinates \eqref{standardmetric} is \emph{neither} the left- nor the right-invariant frame on $S^3$.  Indeed, in the $L$ and $R$ frames, we may choose $A^{(R)} = 0$.  To reproduce the analysis in the left- or right-invariant frame directly from adapted coordinates, one can work in toroidal rather than Hopf coordinates, as we do on $S_b^3$.\footnote{By contrast, \cite{Closset:2013vra} uses the following adapted Hopf coordinates for the round sphere:
\[
ds^2 = \frac{dz\, d\bar{z}}{(1 + |z|^2/\ell^2)^2} + \left(d\tilde{\tau} + \frac{i}{2\ell}\frac{\bar{z}\, dz - z\, d\bar{z}}{1 + |z|^2/\ell^2}\right)^2,
\]
where $\tilde{\tau}/\ell\in [0, 2\pi)$ and $K = \smash{\partial_{\tilde{\tau}}}$.  To go from the patch containing the origin to the patch containing $\infty$, we take $z' = \ell^2/z$ and $\tilde{\tau}' = \tilde{\tau} - \frac{i\ell}{2}\log\frac{\bar{z}}{z}$.}

Finally, we recall the computation of Wilson loop expectation values in $\mathcal{N} = 2$ Chern-Simons theory on $S^3$ in the left-invariant frame, where $\mathcal{L}_\text{CS}|_{S^3}$ is as in \eqref{LCS-curved} and $\mathcal{L}_\text{YM}|_{S^3}$ is given in \eqref{LYM-S3}.  For left-invariant $\xi, \tilde{\xi}$, the BPS equations are
\begin{equation}
D - \frac{i\sigma}{\ell} = 0, \quad D_\mu\sigma\pm \frac{1}{2}\sqrt{g}\epsilon_{\mu\nu\rho}F^{\nu\rho} = 0
\label{S3BPSintermediate}
\end{equation}
where the top and bottom signs correspond to setting $\delta'\lambda = 0$ or $\delta'\tilde{\lambda} = 0$ in \eqref{S3susypvector}, respectively.  Since we localize with respect to a supercharge with $\tilde{\xi} = 0$, we would in principle impose only $\delta'\lambda = 0$; however, seeing as $\mathcal{L}_\text{YM}|_{S^3}$ is both $\delta_\xi$- and $\delta_{\tilde{\xi}}$-exact, and the operators that we localize preserve both of the corresponding supercharges, the BPS conditions really require both vanishing conditions to hold.  Hence we see that the solutions to \eqref{S3BPSintermediate}, namely
\begin{equation}
F_{\mu\nu} = 0, \quad D_\mu\sigma = 0, \quad D - \frac{i\sigma}{\ell} = 0,
\label{S3BPSequations}
\end{equation}
are precisely the minima (zeros) of \eqref{LYM-S3}, and \emph{a fortiori} saddle points thereof.  In other words, the localization locus coincides with the BPS locus.  In fact, one sees directly that the localization locus is simply the zero locus of the Yang-Mills action because all other saddle points are infinitely suppressed in the limit of zero Yang-Mills coupling.

The BPS equations require that $A_\mu$ be pure gauge, and since $S^3$ is simply connected, the zero modes $V_0$ of the vector multiplet fields are given by
\begin{equation}
A_\mu = 0, \quad \sigma = -i\ell D = \sigma_0, \quad \lambda = \tilde{\lambda} = 0
\end{equation}
(we write $\sigma_0\equiv \hat{\sigma}_0/\ell$ for constant $\hat{\sigma}_0\in \mathfrak{g}$).  The partition function can be evaluated in the $t\to\infty$ limit by expanding in transverse fluctuations to the $V_0$ in field space.  It reduces to an integral over the finite-dimensional space of zero modes:
\begin{equation}
Z = \int_{\mathfrak{g}}\, d\hat{\sigma}_0\, e^{-S_\text{CS}[V_0]}Z_\text{1-loop}^\text{vector}[\hat{\sigma}_0]\equiv \int_{\mathfrak{g}}\, d\hat{\sigma}_0\, Z_\text{cl}[\hat{\sigma}_0]Z_\text{1-loop}^\text{vector}[\hat{\sigma}_0],
\end{equation}
where the classical contribution is
\[
S_\text{CS}[V_0] = \frac{k}{2\pi}\int d^3 x\, \sqrt{g}\Tr(i\sigma_0^2/\ell) = k\pi i\Tr\hat{\sigma}_0^2.
\]
WLOG, we may fix $\hat{\sigma}_0$ in a Cartan subalgebra $\mathfrak{t}\subset \mathfrak{g}$ (which introduces a Jacobian factor), and fixing the residual Weyl symmetry gives
\begin{equation}
Z = \frac{1}{|\mathcal{W}|}\int_{\mathfrak{t}}\, da\, \left|\prod_\alpha \alpha(\hat{\sigma}_0)\right|Z_\text{cl}[\hat{\sigma}_0]Z_\text{1-loop}^\text{vector}[\hat{\sigma}_0]
\end{equation}
where the real scalars $a_i$ ($i = 1, \ldots, \operatorname{rank} G$) parametrize $\mathfrak{t}$, $\alpha$ ranges over all roots of $\mathfrak{g}$, and $\hat{\sigma}_0$ is a function of the $a_i$.  The integration measure, including the Vandermonde determinant $|{\prod_\alpha \alpha(\hat{\sigma}_0)}|$, is manifestly Weyl-invariant.  Making a Cartan decomposition of the transverse (divergenceless) component of the gauge field and of the fermions, we find that only the components in the direction of the root spaces contribute nontrivially to $Z_\text{1-loop}^\text{vector}[\hat{\sigma}_0]$.  By the standard spectral analysis,\footnote{This is aided by the $L^a$ and $R^a$ in \eqref{LandR}, which generate the $SU(2)_\text{left}\times SU(2)_\text{right}$ isometry group of $S^3$.  The scalar Laplacian is $\nabla^2 = -\frac{4}{\ell^2}L_a L^a = -\frac{4}{\ell^2}R_a R^a$.  The Dirac operator can be written as
\[
-i\gamma^\mu\nabla_\mu|_\text{LI} = \frac{2}{\ell}S^a(2R_a + S_a), \quad -i\gamma^\mu\nabla_\mu|_\text{RI} = -\frac{2}{\ell}S^a(2L_a + S_a),
\]
where $S^a\equiv \frac{1}{2}\gamma^a$ are the generators of $SU(2)_\text{spin}$ and matrix multiplication (as opposed to spinor contraction) is understood.} one finds, up to an overall power of $\ell$,
\begin{equation}
Z_\text{1-loop}^\text{vector}[\hat{\sigma}_0] = \prod_\alpha \frac{2\sinh(\pi\alpha(\hat{\sigma}_0))}{\alpha(\hat{\sigma}_0)}.
\end{equation}
The full partition function of $\mathcal{N} = 2$ $G_k$ on $S^3$ is therefore (up to a dimensionful constant)
\begin{equation}
Z = \frac{1}{|\mathcal{W}|}\int da\, e^{-k\pi i\Tr\hat{\sigma}_0^2}\prod_\alpha 2\sinh(\pi\alpha(\hat{\sigma}_0))\equiv \frac{1}{|\mathcal{W}|}\int da\, e^{-k\pi i\Tr(a^2)}\operatorname{det}_\text{Ad} 2\sinh(\pi a).
\end{equation}
Half-BPS Wilson loops simply give insertions of $\Tr_R(e^{2\pi a})$:
\begin{equation}
\langle W_{R_1}\cdots W_{R_n}\rangle = \frac{1}{|\mathcal{W}|Z}\int da\, e^{-k\pi i\Tr(a^2)}\Tr_{R_1}(e^{2\pi a})\cdots \Tr_{R_n}(e^{2\pi a})\operatorname{det}_\text{Ad} 2\sinh(\pi a).
\label{matrixmodelS3}
\end{equation}
The Weyl character formula leaves us with a sum of Gaussian integrals, which can be evaluated to reproduce the $\mathcal{N} = 0$ result, up to a framing phase and a level shift.

\subsubsection{Example: \texorpdfstring{$S_b^3$}{Sb3}}

The backgrounds that we consider include $U(1)$ fibrations over arbitrary Riemann surfaces $\Sigma$, where the integral curves of $K$ are the Seifert fibers.  When $\Sigma$ has genus zero, the base has additional isometries (this comment applies also to genus one), allowing for squashed sphere backgrounds where $K$ generates an isometry that does not point along the Seifert fiber.  Consider parametrizing the round sphere by toroidal coordinates, which manifest $S^3$ as a torus fibered over a closed interval \cite{Willett:2016adv, Drukker:2012sr}.  As in the previous section, we regard $S^3$ as the locus $(u, v)\in \mathbb{C}^2$ satisfying $|u|^2 + |v|^2 = 1$, but rather than parametrizing $u, v$ using Euler angles, we use coordinates $\chi\in [0, \pi/2]$ and $\phi_1, \phi_2\in [0, 2\pi)$:
\begin{equation}
u = \cos\chi e^{i\phi_1}, \quad v = \sin\chi e^{i\phi_2}, \quad ds^2 = \ell^2(d\chi^2 + \cos^2\chi\, d\phi_1^2 + \sin^2\chi\, d\phi_2^2).
\label{toroidal}
\end{equation}
This metric clearly admits two independent $U(1)$ isometries.  More generally, consider the following squashed-sphere metric, which preserves a $U(1)\times U(1)$ subgroup of the $SU(2)\times SU(2)$ isometry group of $S^3$:
\begin{equation}
ds^2 = f(\chi)^2\, d\chi^2 + \ell_1^2\cos^2\chi\, d\phi_1^2 + \ell_2^2\sin^2\chi\, d\phi_2^2.
\label{Sb3metric}
\end{equation}
Here, $\ell_1, \ell_2 > 0$ and $f$ is a smooth, positive function on $[0, \pi/2]$ satisfying $f(0) = \ell_2$ and $f(\pi/2) = \ell_1$ to avoid conical singularities along the $\phi_1$ and $\phi_2$ circles, respectively.  The squashing parameter of this geometry that enters supersymmetric observables is $b\equiv \sqrt{\ell_1/\ell_2}$.  For example, the metric on $S_b^3$ induced by its embedding as the locus $(u, v)\in \mathbb{C}^2$ satisfying $b^{-2}|u|^2 + b^2|v|^2 = 1$ where $u = b\cos\chi e^{i\phi_1}$ and $v = b^{-1}\sin\chi e^{i\phi_2}$ takes the above form, with
\[
f(\chi) = \sqrt{\ell_1^2\sin^2\chi + \ell_2^2\cos^2\chi}
\]
(cf.\ \cite{Hama:2011ea, Kapustin:2013hpk}).  A generic Killing vector is a linear combination of the $U(1)$ generators: $K = \alpha\partial_{\phi_1} + \beta\partial_{\phi_2}$.  To use the 3D $\mathcal{N} = 2$ supergravity background described above, we take
\begin{equation}
|K|^2 = \alpha^2\ell_1^2\cos^2\chi + \beta^2\ell_2^2\sin^2\chi = 1 \Longleftrightarrow K = \ell_1^{-1}\partial_{\phi_1} + \ell_2^{-1}\partial_{\phi_2}
\label{Sb3Killing}
\end{equation}
and define local coordinates ($z = x + iy$)
\begin{equation}
x = \int^\chi \frac{f(\chi')}{\sin\chi'\cos\chi'}\, d\chi', \quad y = \ell_1\phi_1 - \ell_2\phi_2, \quad \tilde{\psi} = \ell_1\phi_1\cos^2\chi + \ell_2\phi_2\sin^2\chi,
\end{equation}
in terms of which the metric \eqref{Sb3metric} on $S_b^3$ can be written in the standard form \eqref{standardmetric} with
\begin{equation}
a = 2\sin\chi\cos\chi(\ell_1\phi_1 - \ell_2\phi_2)\, d\chi, \quad c = \sin\chi\cos\chi.
\end{equation}
Note that by (locally) inverting the relation between $x$ and $\chi$, we can write $a = yF(x)\, dx$ for some $F$ and $c = G(x)$ for some $G$; hence $a, c$ are independent of $\tilde{\psi}$ and $d\tilde{\psi}$, as required.  In addition, we see that $K = \partial_{\tilde{\psi}}$.  In these coordinates, the background supergravity formalism \eqref{standardvielbein} yields the vielbein
\begin{equation}
e^1 = f(\chi)\, d\chi, \,\,\, e^2 = -\sin\chi\cos\chi(\ell_1\, d\phi_1 - \ell_2\, d\phi_2), \,\,\, e^3 = \ell_1\cos^2\chi\, d\phi_1 + \ell_2\sin^2\chi\, d\phi_2.
\label{Sb3vielbein}
\end{equation}
We compute that
\begin{equation}
F_a = -\frac{1}{f(\chi)}, \quad (\omega_\text{2D})^{12} = -(\omega_\text{2D})^{21} = \frac{\cos^2\chi - \sin^2\chi}{f(\chi)}(\ell_1\, d\phi_1 - \ell_2\, d\phi_2).
\end{equation}
The corresponding background field configuration is
\begin{equation}
H = \frac{i}{f}, \quad V_\mu = 0, \quad A^{(R)} = \frac{\ell_1\, d\phi_1 + \ell_2\, d\phi_2}{2f}.
\label{Sb3bkgd}
\end{equation}
We may fix a gauge in which $A^{(R)}$ is regular everywhere (i.e., such that when the $\phi_i$ circle shrinks, the coefficient of $d\phi_i$ vanishes):
\begin{equation}
A^{(R)} = \frac{1}{2}\left(\frac{\ell_1}{f} - 1\right)d\phi_1 + \frac{1}{2}\left(\frac{\ell_2}{f} - 1\right)d\phi_2.
\end{equation}
Performing the corresponding $R$-symmetry gauge transformation on the Killing spinors in \eqref{generalizedKSsolns}, we have in the chosen frame and gauge that
\begin{equation}
\xi = e^{i(\phi_1 + \phi_2)/2}\left(\begin{array}{c} 1 \\ 0 \end{array}\right), \quad \tilde{\xi} = e^{-i(\phi_1 + \phi_2)/2}\left(\begin{array}{c} 0 \\ 1 \end{array}\right).
\label{Sb3Killingspinors}
\end{equation}
For generic $b$, the metric in $\tilde{\psi}, z, \bar{z}$ coordinates does not define an $S^1$ fibration because the coordinate $\tilde{\psi}$ is not periodic: namely, we see from \eqref{Sb3Killing} that the integral curves of $K$ do not close on the tori at $\chi\neq 0, \pi/2$ unless $b^2 = \ell_1/\ell_2$ is rational.  If $b^2 = m/n$ with $m, n$ relatively prime integers, then the integral curves for $\chi\neq 0, \pi/2$ are $(n, m)$ torus knots wrapping the $\phi_1$ cycle $n$ times and the $\phi_2$ cycle $m$ times.\footnote{In general, the $(n, m)$ torus link has $\operatorname{gcd}(n, m)$ components.}  These curves have length
\[
2\pi(\ell_1^2 n^2\cos^2\chi + \ell_2^2 m^2\sin^2\chi)^{1/2} = 2\pi\sqrt{\ell_1\ell_2 mn},
\]
regardless of $\chi$.  On the other hand, at $\chi = 0, \pi/2$, the vector field $K$ is singular and its integral curves are circles (regardless of $b$) of lengths $2\pi\ell_1$ and $2\pi\ell_2$, respectively.  One can insert supersymmetric Wilson loops along these knots or circles.

The round-sphere limit is given by $f(\chi) = \ell_1 = \ell_2 = \ell$.  In this limit, we have $H = i/\ell$ and $V_\mu = \smash{A_\mu^{(R)}} = 0$, and both the generalized Killing spinor equations and the $V_\mu = 0$ SUSY' transformations reduce to those for the left-invariant frame on $S^3$.  Note that the $\mathfrak{su}(1|1)$ algebra contains half of the Killing spinors that generate $\mathfrak{osp}(2|2)_\text{left}$ on $S^3$; the existence of two additional left-invariant Killing spinors is due to the extra symmetry of $S^3$.  Recall that on $S^3$, the available spinors in $\mathfrak{osp}(2|2)_\text{left}$ are halved by the BPS condition for a half-BPS Wilson loop; it then suffices to use one of the two remaining supercharges for localization.

As on $S^3$, the vector multiplet localization locus on $S_b^3$ can be read off from $\mathcal{L}_\text{YM}$ \eqref{V0LYM}:
\begin{equation}
\sigma = D/H = \sigma_0\equiv \hat{\sigma}_0/\ell,
\end{equation}
with $\hat{\sigma}_0\in \mathfrak{g}$ constant and all other fields vanishing (here, $\ell\equiv \sqrt{\ell_1\ell_2}$ and $H = i/f$).  Unlike on $S^3$, the spectra of the relevant differential operators are in general infeasible to compute exactly.  Nonetheless, cohomological \cite{Hama:2011ea, Hosomichi:2014hja} or index theory \cite{Drukker:2012sr, Alday:2013lba} arguments that identify and keep only unpaired bosonic and fermionic eigenmodes can be used to extract the one-loop determinants in this situation (similar cancellations arise due to BRST symmetry in topological field theories \cite{Blau:1993tv}).  It suffices to show that for a chiral multiplet of $R$-charge $r$ transforming in the representation $R$ of $G$,
\begin{equation}
Z_\text{1-loop}^\text{chiral}[\hat{\sigma}_0] = \prod_{\rho\in R} Z_\text{chiral}^r[\rho(\hat{\sigma}_0)], \quad Z_\text{chiral}^r[\rho(\hat{\sigma})]\equiv s_b(iQ(1 - r)/2 - \rho(\hat{\sigma}))
\end{equation}
where the product is taken over the weights $\rho$ of $R$, $Q = b + b^{-1}$, and $s_b$ is the double sine function.  The corresponding result for the vector multiplet then follows from a standard Higgsing argument \cite{Willett:2016adv, Benini:2015noa}.  First observe that (up to constant factors)
\begin{equation}
Z_\text{chiral}^r[\rho(\hat{\sigma})]Z_\text{chiral}^{2-r}[-\rho(\hat{\sigma})] = 1
\label{Higgsing1}
\end{equation}
because two chirals that may be coupled through a superpotential mass term do not contribute to the partition function.\footnote{In this case, the formula reduces to the identity $s_b(x)s_b(-x) = e^{i\pi(1 - Q^2/2)/6}$.}  Now suppose that we have fixed $\hat{\sigma}_0$ to lie in a Cartan subalgebra of $\mathfrak{g}$, which incurs a Vandermonde determinant $\mathcal{V}[\hat{\sigma}_0]$.  Write
\begin{equation}
\mathcal{V}[\hat{\sigma}_0]Z_\text{1-loop}^\text{vector}[\hat{\sigma}_0]\equiv \tilde{Z}_\text{1-loop}^\text{vector}[\hat{\sigma}_0] = \prod_\alpha \tilde{Z}_\text{vector}[\alpha(\hat{\sigma}_0)]
\end{equation}
where the product is taken over roots $\alpha$ of $G$.  As usual, the Cartan components of the transverse vector multiplet modes contribute only constant factors.  The contribution of a mode in the direction of a root $\alpha$ is determined by the Higgs mechanism to be
\begin{equation}
\tilde{Z}_\text{vector}[\alpha(\hat{\sigma})]Z_\text{chiral}^0[-\alpha(\hat{\sigma})] = 1
\label{Higgsing2}
\end{equation}
(again, up to constant factors), since a massive vector multiplet contributes trivially to the partition function: the massless chiral multiplet that is eaten has no flavor or $R$-charges, so that its VEV breaks no global symmetries.  From \eqref{Higgsing1} and \eqref{Higgsing2}, we infer that
\begin{equation}
\tilde{Z}_\text{vector}[\alpha(\hat{\sigma})] = Z_\text{chiral}^2[\alpha(\hat{\sigma})] \implies \tilde{Z}_\text{1-loop}^\text{vector}[\hat{\sigma}_0] = \prod_\alpha Z_\text{chiral}^2[\alpha(\hat{\sigma}_0)].
\end{equation}
Specializing to $S_b^3$, this gives\footnote{Here, we use $s_b(x) = \prod_{p, q = 0}^\infty \frac{pb + qb^{-1} + Q/2 - ix}{pb + qb^{-1} + Q/2 + ix}$ and zeta function regularization (see, e.g., \cite{Hama:2011ea}).}
\begin{equation}
\tilde{Z}_\text{1-loop}^\text{vector}[\hat{\sigma}_0] = \prod_\alpha s_b(-iQ/2 - \alpha(\hat{\sigma}_0)) = \prod_{\alpha > 0} 4\sinh(\pi b\alpha(\hat{\sigma}_0))\sinh(\pi b^{-1}\alpha(\hat{\sigma}_0)).
\end{equation}
When $b = 1$, this reduces to the expected result on $S^3$ (up to phases):
\begin{equation}
\tilde{Z}_\text{1-loop}^\text{vector}[\hat{\sigma}_0] = \prod_\alpha 2\sinh(\pi\alpha(\hat{\sigma}_0))\Longleftrightarrow Z_\text{1-loop}^\text{vector}[\hat{\sigma}_0] = \prod_\alpha \frac{2\sinh(\pi\alpha(\hat{\sigma}_0))}{\alpha(\hat{\sigma}_0)}.
\end{equation}
On $S_b^3$, as on $S^3$, the classical contribution is $Z_\text{cl}[\hat{\sigma}_0] = e^{-S_\text{CS}[V_0]}$ where
\[
S_\text{CS}[V_0] = \frac{k}{2\pi\ell^2}\Tr(\hat{\sigma}_0^2)\int d^3 x\, \sqrt{g}H = \frac{2\pi i\ell_1\ell_2 k}{\ell^2}\Tr(\hat{\sigma}_0^2)\int_0^{\pi/2} d\chi\, \cos\chi\sin\chi = k\pi i\Tr(\hat{\sigma}_0^2).
\]
Thus the partition function $Z = \frac{1}{|\mathcal{W}|}\int_{\mathfrak{t}}\, d\hat{\sigma}_0\, Z_\text{cl}[\hat{\sigma}_0]\tilde{Z}_\text{1-loop}^\text{vector}[\hat{\sigma}_0]$ of pure Chern-Simons theory is
\begin{equation}
Z_{S_b^3} = \frac{1}{|\mathcal{W}|}\int_{\mathfrak{t}}\, d\hat{\sigma}_0\, e^{-k\pi i\Tr\hat{\sigma}_0^2}\prod_{\alpha > 0} 4\sinh(\pi b\alpha(\hat{\sigma}_0))\sinh(\pi b^{-1}\alpha(\hat{\sigma}_0)).
\end{equation}
Again by \eqref{bpsconditions}, BPS Wilson loops lie along integral curves of $K^\mu$ and thus correspond to insertions of
\[
W[\hat{\sigma}_0] = \Tr_R\exp\left(\frac{\hat{\sigma}_0}{\ell}\oint_\gamma ds\right) = \sum_{\rho\in R} \begin{cases} e^{2\pi b\rho(\hat{\sigma}_0)} & \text{if $\chi = 0$}, \\ e^{2\pi b^{-1}\rho(\hat{\sigma}_0)} & \text{if $\chi = \pi/2$}, \\ e^{2\pi\sqrt{mn}\rho(\hat{\sigma}_0)} & \text{if $b^2 = m/n$, $\gcd(m, n) = 1$, $\chi\neq 0, \pi/2$} \end{cases}
\]
in the partition function.  Clearly, this matrix model reduces to that for $S^3$ when $b = 1$.  One can use this matrix model to calculate the Jones polynomial for torus knots and torus links \cite{Tanaka:2012nr} (see also \cite{Beasley:2009mb}).  To do so, one must account for supersymmetric framing: the self-linking number at generic $\chi$ is given by the linking number of nested torus knots, while circular Wilson loops at $\chi = 0, \pi/2$ have fractional self-linking number (as seen from the phase relative to the expectation value of an unknot in bosonic Chern-Simons).  Explicit formulas for torus knots in $S^3$ can be found in \cite{Marino:2005sj}.

\subsubsection{Non-Example: (Squashed) Lens Spaces}

One might wish to apply the formalism of Section \ref{sugra} to more general Seifert manifolds (for useful references, see \cite{Beasley:2005vf, Beasley:2009mb, Kallen:2011ny}).  Lens spaces provide a nice class of examples: these are Seifert manifolds that admit infinitely many distinct Seifert fibrations.

Two convenient definitions of the lens space $L(p, q)$, for relatively prime integers $p, q$, are as follows.  First, it is the quotient of $S^3$, namely the locus $(u, v)\in \mathbb{C}^2$ where $|u|^2 + |v|^2 = 1$, by the free $\mathbb{Z}_p$ action
\[
(u, v)\mapsto (e^{2\pi i/p}u, e^{-2\pi iq/p}v).
\]
Second, it is the result of gluing two solid tori by a homeomorphism $g : T^2\to T^2$ of their boundaries, specified up to isotopy by its $SL(2, \mathbb{Z})$ action on the homology of $T^2$: $g = (\begin{smallmatrix} m & n \\ p & q \end{smallmatrix})$.  More generally, one obtains a space $L_b(p, q)$ that is topologically $L(p, q)$ by starting with a metric \eqref{Sb3metric} for $S_b^3$ and imposing the identifications
\begin{equation}
(\chi, \phi_1, \phi_2)\sim (\chi, \phi_1 + 2\pi/p, \phi_2 - 2\pi q/p).
\label{Zpsquashed}
\end{equation}
As examples, $S^3$ and $S^2\times S^1$ are equivalent to $L(1, 1)$ and $L(0, 1)$, respectively (note that $L(1, 0)$ is also $S^3$, as can be seen from the gluing definition).  Aside from the case of $L(p, -1)$ considered in \cite{Closset:2017zgf}, however, lens spaces cannot be written as Seifert manifolds in such a way that the base is a smooth Riemann surface, without orbifold points (the analysis of \cite{Ohta:2012ev} is also restricted to Seifert manifolds with a smooth base).

The only case in which it is possible to localize on a squashed lens space $L_b(p, q)$ using the $V_\mu = 0$ supergravity background \eqref{Sb3bkgd} on $S_b^3$ is that of $L_b(p, 1)$, because only in this case does the $\mathbb{Z}_p$ action \eqref{Zpsquashed} preserve the Killing spinors \eqref{Sb3Killingspinors}.  More general $L_b(p, q)$ would require a different supergravity background.  For example, localization on $L(p, -1)$ is performed in \cite{Closset:2017zgf} using a background in which the $R$-symmetry gauge field has a holonomy around the nontrivial cycle, leading to integrally quantized $R$-charges.

On $L_b(p, 1)$, the values of the supergravity background fields and the BPS equations are the same as on $S_b^3$.  However, the localization locus is different because $\pi_1(L_b(p, 1)) = \mathbb{Z}_p$, so flat connections on $L_b(p, 1)$ are labeled by $g\in G$ satisfying $g^p = 1$, modulo conjugation.  Fixing $g$ to lie in a maximal torus, we may write
\begin{equation}
g = e^{2\pi i\mathfrak{m}/p}, \quad \mathfrak{m}\in \frac{\Lambda_W^\vee}{p\Lambda_W^\vee}
\end{equation}
where $\Lambda_W^\vee\subset \mathfrak{t}$ is the coweight lattice of $G$ and $\mathfrak{t}$ is the chosen Cartan subalgebra.  The remaining BPS equations in \eqref{S3BPSequations} impose that
\begin{equation}
\sigma = -i\ell D = \sigma_0\equiv \hat{\sigma}_0/\ell, \quad [\hat{\sigma}_0, \mathfrak{m}] = 0,
\end{equation}
with $\hat{\sigma}_0\in \mathfrak{g}$ constant; the latter condition requires that $\hat{\sigma}_0\in \mathfrak{t}$.  Thus the space of BPS configurations is $(\mathfrak{g}\times \Lambda_W^\vee/p\Lambda_W^\vee)/\mathcal{W}$; i.e., fixing $\hat{\sigma}_0\in \mathfrak{g}$ selects a Cartan subalgebra $\mathfrak{t}$, after which one has a choice of $\mathfrak{m}$ and a residual Weyl symmetry.  The classical contribution from the $\mathcal{N} = 2$ Chern-Simons action in the holonomy-$\mathfrak{m}$ sector is $Z_\text{cl}[\hat{\sigma}_0] = e^{-S_\text{CS}[V_0]}$ where
\[
S_\text{CS}[V_0] = S_\text{CS}[\hat{\sigma}_0, \mathfrak{m}] = \frac{k\pi i}{p}\Tr(\hat{\sigma}_0^2 - \mathfrak{m}^2).
\]
The first term is the contribution from the scalars, which is identical to that on $S_b^3$ (or on $S^3$) up to a division by $p$; the second term is the contribution from the flat connection \cite{Gang:2009wy, Alday:2012au}.  The one-loop determinants are obtained by keeping only $\mathbb{Z}_p$-covariant modes on $S_b^3$.  In terms of a suitably modified double sine function, we have \cite{Willett:2016adv}
\begin{align}
Z_\text{chiral}^r[\rho(\hat{\sigma}_0), \mathfrak{m}] &= s_b^{(p)}(iQ(1 - r)/2 - \rho(\hat{\sigma}_0); \rho(\mathfrak{m})), \\
s_b^{(p)}(x; k) &\equiv \prod_{(\ast)} \frac{mb + nb^{-1} + Q/2 - ix}{mb + nb^{-1} + Q/2 + ix} \nonumber
\end{align}
where $(\ast)$ means $m, n\geq 0$ and $m - n\equiv k\text{ (mod $p$)}$.\footnote{The results for $Z_\text{cl}[\hat{\sigma}_0]$ and $Z_\text{chiral}^r[\rho(\hat{\sigma}_0), \mathfrak{m}]$ need to be dressed by nontrivial signs to ensure factorization of the $L_b(p, 1)$ partition function of general $\mathcal{N} = 2$ theories into holomorphic blocks (see \cite{Willett:2016adv} and references therein).}  With holonomy $\mathfrak{m}$, the Vandermonde determinant appearing in
\begin{equation}
\mathcal{V}[\hat{\sigma}_0, \mathfrak{m}]Z_\text{1-loop}^\text{vector}[\hat{\sigma}_0, \mathfrak{m}] = \tilde{Z}_\text{1-loop}^\text{vector}[\hat{\sigma}_0, \mathfrak{m}] = \prod_\alpha Z_\text{chiral}^2[\alpha(\hat{\sigma}_0), \mathfrak{m}]
\end{equation}
is $\mathcal{V}[\hat{\sigma}_0, \mathfrak{m}] = \smash{|{\prod_{\alpha(\mathfrak{m}) = 0} \alpha(\hat{\sigma}_0)}|}$ because the generators of $\mathfrak{g}$ with $\alpha(\mathfrak{m})\neq 0$ are broken.  Upon summing over holonomies (gauge bundles), the partition function of $\mathcal{N} = 2$ Chern-Simons theory on $L_b(p, 1)$ is (by similar simplifications as for $S_b^3$)
\begin{equation}
Z_{L_b(p, 1)} = \frac{1}{|\mathcal{W}|}\sum_{\mathfrak{m}}\int_{\mathfrak{t}}\, d\hat{\sigma}_0\, e^{-\frac{k\pi i}{p}\Tr(\hat{\sigma}_0^2 - \mathfrak{m}^2)}\prod_{\alpha > 0}\prod_\pm 2\sinh(\pi b^{\pm 1}\alpha(\hat{\sigma}_0\pm i\mathfrak{m})/p).
\end{equation}
One can obtain from this matrix model the expectation value of a Wilson loop wrapping the non-contractible cycle, corresponding to the generator of $\pi_1$ (see \cite{Gang:2009wy} and references therein).

Now suppose we want to localize Wilson loops on $L_b(p, 1)$ with $b^2$ rational, so that the orbits of the Killing vector close into torus knots.  For any Wilson loop along a torus knot in $S_b^3$, one can easily compute the expectation value of its image in $L_b(p, 1)$ by writing down the appropriate matrix model.  One could in principle take a different approach to localizing Wilson loops on $L_b(p, q)$, which is more akin to that in \cite{Closset:2017zgf}: instead of viewing $L_b(p, q)$ as a quotient of $S_b^3$, exhibit it directly as a Seifert fibration and consider loops wrapping the circle fibers (for rational $b^2$, the $L_b(p, q)$ are Seifert fibrations over $S^2$ with two singular fibers).  This was accomplished in \cite{Closset:2017zgf} for the case of round $L(p, -1)$.  Moreover, whereas the lens spaces $L(p, 1), L(p, -1), L(p, p - 1)$ are all homeomorphic, the latter two are defined by the same $\mathbb{Z}_p$ action and therefore have the same induced THF.

As for the special case of $S^2\times S^1 = L(0, 1)$, our $V_\mu = 0$ background (with $K$ generating translations along the $S^1$) computes the topologically twisted index \cite{Benini:2015noa} with a negative unit of $R$-symmetry flux through the $S^2$.  Taking the $S^1$ to have circumference $\beta$ and the $S^2$ to have radius $\ell$, we have
\[
ds^2 = \beta^2\, dt^2 + \ell^2(d\theta^2 + \sin^2\theta\, d\phi^2)
\]
where $t\sim t + 1$.  Putting $z = 2\ell e^{i\phi}/\tan(\theta/2)$ and following the standard recipe \eqref{standardvielbein} (with $\tilde{\psi} = \beta t$), we choose the vielbein $e^3 = \beta\, dt$ and
\begin{align*}
e^1 = -\ell(\cos\phi\, d\theta + \sin\theta\sin\phi\, d\phi), \quad e^2 = \ell(\sin\phi\, d\theta - \sin\theta\cos\phi\, d\phi).
\end{align*}
We then compute that $H = 0$ and $A^{(R)} = \cos^2(\theta/2)\, d\phi$.  Note that our vielbein differs from that of \cite{Benini:2015noa} (namely, $e^1 = \ell\, d\theta$ and $e^2 = \ell\sin\theta\, d\phi$) in the $1, 2$ directions, and our $A^{(R)}$ differs from theirs (namely, $A^{(R)} = \frac{1}{2}\cos\theta\, d\phi$) by an $R$-symmetry gauge transformation.  One can place Wilson loops along the $S^1$ over any point on the $S^2$.

\subsubsection{Non-Example: 3D Supersymmetric Index}

The above framework is well-suited to the computation of the 3D supersymmetric index, which has already been discussed from various points of view in the literature and which we will therefore not describe in any great detail.  Due to its significance in Chern-Simons theory (for which it yields the ground-state degeneracy), we will simply review the known results for the index while paying special attention to signs, as a prelude to our subsequent localization computation on a solid torus (which gives a partial result for the wavefunctions of Chern-Simons theory).

In the case of finite ground-state degeneracy, the Witten index can be computed as the Euclidean partition function on $T^3$ with periodic boundary conditions for fermions along each $S^1$ to preserve supersymmetry:
\begin{equation}
I(k)\equiv \Tr_{\mathcal{H}_{T^2}} (-1)^F.
\end{equation}
The supersymmetric index of pure Chern-Simons theory suggests that supersymmetry is spontaneously broken when $|k| < h/2$ for $\mathcal{N} = 1$ \cite{Witten:1999ds} and when $|k| < h$ for $\mathcal{N} = 2$ \cite{Intriligator:2013lca}.  The standard mnemonic of simply accounting for the level shift after integrating out gauginos at large $|k|$ and plugging the result into the genus-one Verlinde formula applies, but the derivation for arbitrary $k$ requires care.  Namely, for $k\geq 0$, the pure $G_k$ theory has index
\[
I_{\mathcal{N} = 1}(k) = \begin{cases} J_G(k - h/2) & \text{for } k\geq h/2, \\ 0 & \text{for } k < h/2, \end{cases} \qquad I_{\mathcal{N} = 2}(k) = \begin{cases} J_G(k - h) & \text{for } k\geq h, \\ 0 & \text{for } k < h, \end{cases}
\]
where $J_G(k')$ is the number of primary operators in the $G_{k'}$ WZW model.  The indices for $k < 0$ are related to those for $k > 0$ as follows:
\begin{equation}
I_{\mathcal{N} = 1}(-k) = (-1)^r I_{\mathcal{N} = 1}(k), \qquad I_{\mathcal{N} = 2}(-k) = I_{\mathcal{N} = 2}(k),
\end{equation}
where $r = \operatorname{rank} G$.  The reason is that the sign of the operator $(-1)^F$ is potentially ambiguous in finite volume.  From a microscopic point of view (i.e., using a Born-Oppenheimer approximation in the regime $g^2 k\ll 1/r$, $r$ being the size of the $T^2$), the $\mathcal{N} = 1$ index can be computed by quantizing $r$ pairs of fermion zero modes $\eta_\pm^a$, where $a = 1, \ldots, r$ (as in \cite{Witten:1999ds}).  There are two choices $|\Omega_\pm\rangle$ for the zero-mode Fock vacuum, depending on which chirality ($\pm$) we choose for the creation or annihilation operators.  They are related by
\begin{equation}
|\Omega_+\rangle = \prod_{a=1}^r \eta_+^a|\Omega_-\rangle.
\end{equation}
Taking $k\to -k$ is akin to a reversal of spacetime orientation and hence exchanges $|\Omega_+\rangle$ with $|\Omega_-\rangle$.  But fixing the statistics ($(-1)^F$-eigenvalue) of one of these vacua also fixes that of the other to be $(-1)^r$ times this value.  This explains the sign in the $\mathcal{N} = 1$ case.  On the other hand, when $\mathcal{N} = 2$, there are $2r$ pairs of fermion zero modes $\eta_\pm^a, \tilde{\eta}_\pm^a$ and four possibilities for the zero-mode Fock vacuum: $|\Omega_\pm\rangle\otimes |\smash{\tilde{\Omega}_{\pm'}}\rangle$, where $\pm$ and $\pm'$ are independent pairs of signs.  Regardless of which vacuum state one chooses, orientation reversal does not change its statistics under $(-1)^F$.  This explains the absence of the sign in the $\mathcal{N} = 2$ case.

For example, consider $SU(N)$, for which $h = N$ and
\begin{equation}
J_{SU(N)}(k') = \binom{N + k' - 1}{N - 1}.
\end{equation}
Note that while this function gives the dimension of the space of $SU(N)_{k'}$ conformal blocks only when $k'\geq 0$, it has an analytic continuation to all $k'$.  Hence we may write succinctly
\begin{equation}
I_{\mathcal{N} = 1}(k) = J_{SU(N)}(k - N/2) = \frac{1}{(N - 1)!}\prod_{j=-N/2+1}^{N/2-1} (k - j),
\end{equation}
which indeed vanishes when $|k| < N/2$ and satisfies $I_{\mathcal{N} = 1}(-k) = (-1)^{N-1}I_{\mathcal{N} = 1}(k)$.  On the other hand, we have that
\begin{equation}
I_{\mathcal{N} = 2}(k) = J_{SU(N)}(|k| - N) = \frac{1}{(N - 1)!}\prod_{j=1}^{N-1} (|k| - j) \text{ for } k\neq 0, \quad I_{\mathcal{N} = 2}(0) = 0,
\end{equation}
which vanishes when $|k| < N$ and satisfies $I_{\mathcal{N} = 2}(-k) = I_{\mathcal{N} = 2}(k)$.

Supersymmetric localization has the benefit of quantitatively justifying why the irrelevant Yang-Mills term does not affect the number of vacua and hence the index, even beyond the regime where semiclassical reasoning applies: it is $Q$-exact.  The genus-zero Verlinde formula ($Z(S^2\times S^1) = 1$ for all $G, k$) was computed by localization in \cite{Benini:2015noa}, where the Verlinde algebra was also obtained from correlation functions of Wilson loops wrapping the $S^1$ above arbitrary points on the $S^2$.  The Verlinde formula for $\Sigma\times S^1$ in arbitrary genus $g$ was later computed by localization in \cite{Benini:2016hjo, Closset:2016arn} using a background $R$-symmetry flux of $g - 1$ through $\Sigma$, which imposes a quantization condition of $q_R(g - 1)\in \mathbb{Z}$ on the $R$-charges; in particular, in the case $g = 1$ where no twisting is necessary, this computation reproduces the result of \cite{Blau:1993tv} for $Z_{T^3} = \dim\mathcal{H}_{T^2}$.  One can also compute by these means correlation functions of Wilson loops over arbitrary points on $\Sigma$, leading to the algebra of Wilson loops in arbitrary genus \cite{Closset:2016arn}, which dimensionally reduces to the twisted chiral ring of a 2D $(2, 2)$ theory on $\Sigma$ and generalizes the results of \cite{Kapustin:2013hpk}.  This approach was further generalized to nontrivial circle bundles \cite{Closset:2017zgf}.  An alternative approach to the supersymmetric index of $\mathcal{N} = 2$ Chern-Simons theory in arbitrary genus, in the spirit of \cite{Kallen:2011ny}, is presented in \cite{Ohta:2012ev}.

\subsection{Localization on Solid Torus}

So far, we have considered various compact manifolds, but in all cases, the half-BPS sector includes a very limited subset of the possible loop observables.  In pursuit of greater generality, it is natural to ask whether the localization argument can be applied to a basic ``building block'' that can then be used to assemble Wilson loop configurations of more complicated topologies.  Since every closed, orientable, connected three-manifold can be obtained from $S^3$ by surgery on knots \cite{Marino:2005sj, Guadagnini:1994cc}, the natural candidate for such a building block is a solid torus.  Here, we take only the most tentative possible step toward making this program precise by localizing on a solid torus dressed by a Wilson loop: we do not address the gluing procedure at all, much less how to make it supersymmetric.  This constructive approach has appeared in the supersymmetric context in the guise of holomorphic blocks \cite{Pasquetti:2011fj, Beem:2012mb}, and also as a gluing formula of a suggestively similar form to surgery in pure Chern-Simons theory \cite{Closset:2017zgf}; however, both of these incarnations depend on more than topological data.

Localization on a solid torus is a simple application of the results of \cite{Sugishita:2013jca}.  Their calculations fit easily into our framework for Seifert manifolds, as we now describe.  We restrict our attention to Dirichlet boundary conditions, as considered there.  These boundary conditions eliminate half of the fermions and, with the appropriate choice of localizing term, do not require introducing boundary degrees of freedom.  For pure $\mathcal{N} = 2$ gauge theory without matter, we can preserve two supersymmetries on the solid torus, which gives us precisely the 1D $\mathcal{N} = 2$ supersymmetry required for cancellation of the line shift.

The solid torus is constructed by starting with toroidal coordinates \eqref{toroidal} on round $S^3$ and restricting to $\chi\in [0, \chi_0]$ with $\chi_0 < \pi/2$.  In the standard supergravity background \eqref{Sb3bkgd} and frame \eqref{Sb3vielbein} (with $f = \ell_1 = \ell_2 = \ell$), we may choose a gauge in which $A^{(R)} = 0$ and the Killing spinors are as in \eqref{Sb3Killingspinors}.  Seeing as $\gamma^\chi = \frac{1}{\ell}\gamma^1$ in our frame, we have
\begin{equation}
-\ell e^{-i(\phi_1 + \phi_2)}\gamma^\chi\xi = \tilde{\xi}
\label{property}
\end{equation}
(recall that $-\gamma^\chi\xi = \gamma^\chi\cdot\xi$).  Up to a slight difference in conventions, this is precisely the condition satisfied by Killing spinors preserved by the boundary conditions of \cite{Sugishita:2013jca} on the solid torus.  Specifically, the boundary conditions that we wish to impose are
\begin{equation}
A_{\tilde{\mu}}| = a_{\tilde{\mu}}, \quad \sigma| = \sigma_0, \quad \ell e^{-i(\phi_1 + \phi_2)}\gamma^\chi\lambda| = \tilde{\lambda}|
\label{torusbcs}
\end{equation}
where $\tilde{\mu}\in \{\phi_1, \phi_2\}$ denotes the directions tangent to the boundary $T^2$, ``$|$'' denotes restriction to the boundary $\chi = \chi_0$, and $a_{\tilde{\mu}}$ and $\sigma_0$ are constants.  The fields $A_\chi$ and $D$ are left free.  These boundary conditions and the property \eqref{property} of the Killing spinors imply that
\begin{equation}
\xi\tilde{\lambda}| = \tilde{\xi}\lambda|, \quad \xi\gamma^\chi\tilde{\lambda}| = \tilde{\xi}\gamma^\chi\lambda|, \quad \xi\gamma^{\tilde{\mu}}\tilde{\lambda}| = -\tilde{\xi}\gamma^{\tilde{\mu}}\lambda|.
\end{equation}
Given these properties and \eqref{property}, one easily sees that the boundary conditions are compatible with the relevant SUSY' transformations, which are given by \eqref{S3susypvector} with the top sign (note, however, that we are \emph{not} working in the left-invariant frame).

The boundary terms in the SUSY' variation of the curved-space Chern-Simons action \eqref{LCS-curved} are given by \eqref{LCS-curved-boundary}.  For the Yang-Mills action on the solid torus, we write the fermion kinetic terms symmetrically between $\lambda$ and $\tilde{\lambda}$ (following \cite{Sugishita:2013jca}) to ensure that the boundary conditions kill surface terms without the need to introduce a compensating boundary action.  Namely, we use as a localizing term
\begin{equation}
\mathcal{L}_\text{(YM)}|_{S^3}\equiv \mathcal{L}_\text{YM}|_{S^3, \, \tilde{\lambda}\gamma^\mu D_\mu\lambda\to \frac{1}{2}(\tilde{\lambda}\gamma^\mu D_\mu\lambda + \lambda\gamma^\mu D_\mu\tilde{\lambda})},
\end{equation}
where $\mathcal{L}_\text{YM}|_{S^3}$ is defined by choosing the top sign in \eqref{LYM-S3} (the parentheses in (YM) are a mnemonic for symmetrization).  One computes that the corresponding boundary terms are
\begin{align*}
\delta'\mathcal{L}_\text{(YM)}|_{S^3} &= \frac{1}{2}\nabla_\mu\Tr\bigg[\frac{1}{2}\sqrt{g}^{-1}\epsilon^{\mu\nu\rho}F_{\nu\rho}(\xi\tilde{\lambda} + \tilde{\xi}\lambda) + i\sqrt{g}^{-1}\epsilon^{\mu\nu\rho}D_\nu\sigma(\xi\gamma_\rho\tilde{\lambda} - \tilde{\xi}\gamma_\rho\lambda) \\
&\hspace{1 cm} + iF^{\mu\nu}(\xi\gamma_\nu\tilde{\lambda} + \tilde{\xi}\gamma_\nu\lambda) - D^\mu\sigma(\xi\tilde{\lambda} - \tilde{\xi}\lambda) + \left(D - \frac{i\sigma}{\ell}\right)(\xi\gamma^\mu\tilde{\lambda} - \tilde{\xi}\gamma^\mu\lambda)\bigg].
\end{align*}
By the boundary conditions \eqref{torusbcs}, all of the Chern-Simons and Yang-Mills boundary terms vanish.  Finally, one can also show along the lines of \cite{Sugishita:2013jca} that the Yang-Mills action written in this way remains $Q$-exact with these boundary conditions, again without the need to add a compensating boundary action (see also the useful summary of boundary terms in \cite{Drukker:2012sr}).

The BPS equations are the same as on $S^3$, \eqref{S3BPSequations}.  Given the boundary conditions \eqref{torusbcs}, one can choose a gauge in which the saddle points of the Yang-Mills action are given by $A_\chi = 0$, $A_{\tilde{\mu}} = a_{\tilde{\mu}}$, $\sigma = \sigma_0$ where the constants $a_{\tilde{\mu}}$ and $\sigma_0$ satisfy $[a_{\tilde{\mu}}, \sigma_0] = 0$.  Moreover, regularity of the gauge field at $\chi = 0$ requires $a_{\phi_2} = 0$; as long as $\chi_0 < \pi/2$, $a_{\phi_1}$ can be nonzero, and it is only this component on which a localized Wilson loop depends.  Namely, a BPS Wilson loop along a curve $\gamma$ at fixed $\chi\in [0, \chi_0]$ localizes as follows:
\begin{equation}
W = \Tr_R P\exp\left[i\oint_\gamma d\tau\, (A_\mu\dot{x}^\mu - i\sigma|\dot{x}|)\right]\to \langle W\rangle = \Tr_R e^{2\pi i\ell(a_{\phi_1} - i\sigma_0)}.
\label{localizedloop}
\end{equation}
For $\chi\in (0, \chi_0]$, $\gamma$ is an unknot represented as a $(1, 1)$ torus knot.  At the core of the torus ($\chi = 0$), $\gamma$ is a $(1, 0)$ torus knot.  For any $\chi$, $\gamma$ has length $2\pi\ell$.  Because the value of $\sigma$ at the saddle point is fixed rather than integrated over, the expectation value of a Wilson loop is trivial to compute: both the classical contribution and the one-loop determinants cancel in the normalized expectation value.

Note that $\chi$ parametrizes the time direction in Hamiltonian quantization on $T^2$.  Hence we would like to interpret \eqref{localizedloop} as the wavefunction of a state in the Hilbert space on the boundary $T^2$.  The latter is more properly identified with the \emph{unnormalized} expectation value of a Wilson loop threading the torus, which is obtained by dressing \eqref{localizedloop} with factors of $Z_\text{cl}$ and $Z_\text{1-loop}^\text{vector}$ from \cite{Sugishita:2013jca}: we have
\[
Z_\text{cl} = e^{-\frac{ik}{2\pi\ell}\Tr(\sigma_0)^2 V(\chi_0)}
\]
where $V(\chi_0) = 2\pi^2\ell^3\sin^2\chi_0$ is the volume of the torus and
\[
Z_\text{1-loop}^\text{vector} = \prod_{\alpha > 0}\prod_{n\in \mathbb{Z}} (n - \ell\alpha(a_{\phi_1} - i\sigma_0)) = \prod_{\alpha > 0} 2\sinh(\pi\ell\alpha(ia_{\phi_1} + \sigma_0))
\]
up to an overall dimensionful factor, where we have used zeta function regularization.  Apart from the classical contribution, the boundary gauge field and $\sigma$ appear in the expected supersymmetric combination.  The localized Wilson loop trivially gives a character, but its interpretation as a wavefunction is unclear.  One obvious shortcoming is that with the half-BPS Dirichlet boundary conditions used for localization, one cannot obtain an arbitrary wavefunctional of the boundary fields from the path integral on the torus; rather, it is evaluated at particular boundary values of the gauge field.  It is tempting to compare this result to that obtained from holomorphic (coherent state) quantization \cite{Elitzur:1989nr, Bos:1989kn}.  There, one obtains affine (Weyl-Kac) characters at level $k$, which are both Weyl-invariant and modular-invariant.  Recall that an affine character is a combination of a Virasoro character and an ordinary Weyl character: for $\smash{\widehat{G}_k}$ with vacuum representation of highest weight $\lambda$,
\begin{equation}
\chi_\lambda^{(k)}(\tau, \theta^i) = q^{-c/24}\tr_\lambda^{(k)}(q^{L_0}e^{i\theta^i H^i})
\end{equation}
where the $H^i$ are Cartan generators.  However, to arrive at an answer that is a function (rather than a functional) of the constant $\phi_1$-component, we have effectively chosen a real polarization \cite{Elitzur:1989nr}; moreover, the localization calculation implicitly fixes a complex structure on the $T^2$ rather than allowing it to be arbitrary.\footnote{The flat metric on the boundary $T^2$ is $ds^2 = \ell^2\cos^2\chi_0\, d\phi_1^2 + \ell^2\sin^2\chi_0\, d\phi_2^2$, from which we identify the modular parameter as $i\operatorname{max}(\tan^2\chi_0, \cot^2\chi_0)$ up to a modular transformation (as usual, choose $\operatorname{Im}\tau > 0$).}  Hence we cannot hope to reproduce the Kac-Moody part of the affine character, which depends on the modular parameter $\tau$; only the Weyl character for compact $G$ is visible in this calculation.  What we \emph{can} do is isolate the shift in the highest weight $\lambda$, because the Weyl character already fixes $\lambda$.  This is enough for our purposes.  The constant Cartan element in the supersymmetric boundary condition for the gauge field is identified with the Cartan parameter of the Hodge decomposition on the torus \eqref{flatonT2}, which determines the Cartan angles that appear in the Weyl character.  Note that after fixing a gauge, we must still add a boundary term to make the variational principle well-defined in our real polarization (see Appendix \ref{boundaryconditions}); this term vanishes on the localization locus because it involves the product of both components of the gauge field along the boundary torus.

To be slightly more precise, the wavefunctions \eqref{ansatzT2} obtained via holomorphic quantization are functionals of $A_z$; restricting them to functions of $A_z = a_z$ where $a_z$ is a constant in the Cartan subalgebra shows that a basis for the physical wavefunctions is given by Weyl-Kac characters at level $k$, labeled by distinct $\lambda\in \Lambda_W/(\mathcal{W}\ltimes k\Lambda_R)$:
\begin{equation}
\psi_\lambda(a_z)\equiv e^{-\frac{k\operatorname{Im}\tau}{\pi}\Tr a_z^2}\chi_\lambda^{(k)}(\bar{\tau}, u), \quad u\equiv -\frac{\operatorname{Im}\tau}{\pi}a_z.
\label{wavefunctionsT2}
\end{equation}
For constant $a_z$, one might na\"ively make a change of variables to interpret the wavefunctions as functions of the coordinates $a_1$.  However, it is precisely the passage from holomorphic to real polarization that involves nontrivial Jacobians and leads to the famous shifts \cite{Elitzur:1989nr}.  Instead of simply setting $A_z = a_z$, one should integrate out the modes that are not constant and in the Cartan to obtain an effective wavefunction for $a_z$, leading to an effective wavefunction in $a_1$ that coincides with the na\"ive one up to the famous shifts (see Appendix \ref{holomorphicpol}).

Finally, let us comment on the broader context.  For a large class of 3D $\mathcal{N} = 2$ theories that preserve a $U(1)_R$ symmetry, the squashed lens space partition functions (including those on squashed spheres and various supersymmetric indices) factorize as
\begin{equation}
Z_{M^3}(m_\alpha) = \sum_\alpha B_\alpha(x_\alpha; q)\tilde{B}_\alpha(\tilde{x}_\alpha; \tilde{q})
\label{factorization}
\end{equation}
where $\alpha$ labels vacua of the mass-deformed theory on $T^2$, $x_\alpha, \tilde{x}_\alpha$ are $U(1)$ flavor symmetry fugacities, $(q, \tilde{q}) = (e^{2\pi ib^2}, e^{2\pi ib^{-2}})$, and the holomorphic blocks $B_\alpha, \tilde{B}_\alpha$ are intrinsic to the theory but independent of $M^3$ ($Z_{M^3}$ depends on $M^3$ only through how the blocks are glued together) \cite{Pasquetti:2011fj, Beem:2012mb}.  More precisely,
\begin{equation}
B_\alpha(x_\alpha; q) = Z_{D^2\times_q S^1}(x_\alpha; \alpha)
\end{equation}
where the theory on the ``Melvin cigar'' $D^2\times_q S^1$ (the $D^2$ being fibered over the $S^1$ with holonomy $q$) is topologically twisted such that the partition function is independent of the metric of the $D^2$, with $\alpha$ determined by boundary conditions.  If $M^3$ admits a Heegaard splitting into two solid tori glued together by an element of $SL(2, \mathbb{Z})$, and if both pieces can be deformed to a Melvin cigar in a $Q$-exact manner, then we have the desired factorization \eqref{factorization}.  The $M^3$ partition function can be exhibited directly in factorized form through Higgs branch localization \cite{Fujitsuka:2013fga, Benini:2013yva}, and the individual blocks can be computed by localization on the solid torus $D^2\times S^1$ \cite{Yoshida:2014ssa}.

The computation of \cite{Yoshida:2014ssa} does not fit neatly into our framework for Seifert manifolds because it uses both a different metric and different boundary conditions than \cite{Sugishita:2013jca}.  Nonetheless, it is also found in \cite{Yoshida:2014ssa} that the $\mathcal{N} = 2$ Yang-Mills action can be written in a $Q$-exact form without surface terms.  On the other hand, the $\mathcal{N} = 2$ Chern-Simons action is invariant under neither SUSY nor gauge transformations in the presence of a boundary.  Two proposals are given in \cite{Yoshida:2014ssa} for maintaining gauge invariance, namely that the compensating boundary action should contain either a chiral 2D $(0, 2)$ theory (namely, $(0, 2)$ matter multiplets to cancel gauge anomalies) or a trivially supersymmetric gauged chiral WZW model obtained by viewing gauge parameters as physical fields on the boundary.\footnote{In fact, one can consider possibilities intermediate between these two extremes, involving a chiral WZW model in which only a subgroup $H\subset G$ is gauged, along with $(0, 2)$ matter multiplets coupled to the $H$-gauge field.}  The second option makes direct contact with the Chern-Simons wavefunction computed in holomorphic polarization: the K\"ahler potential in the inner product \eqref{innerproduct} of coherent states appears from the localization point of view as a boundary term necessary to preserve half-BPS boundary conditions (again, the localization computation selects a constant gauge field, for which the wavefunction becomes a Weyl-Kac character).  However, as in holomorphic quantization, one obtains a character in the representation $\lambda$ by taking Weyl-invariant linear combinations of generalized theta functions rather than by directly evaluating the path integral with a Wilson loop in the representation $\lambda$; it is not clear how such a Wilson loop would fit into the approach of \cite{Yoshida:2014ssa}.\footnote{The computation of \cite{Yoshida:2014ssa} can be viewed as that of a half-index with Neumann boundary conditions for the vector multiplet.  For a discussion of how to recover Weyl-Kac characters from the half-index with Dirichlet $(0, 2)$ boundary conditions in $\mathcal{N} = 2$ Chern-Simons theories, see \cite{Dimofte:2017tpi}.}

\section{Matching \texorpdfstring{$\mathcal{N} = 0$}{N = 0} and \texorpdfstring{$\mathcal{N} = 2$}{N = 2} Line Operators} \label{matching}

So far, we have explained the quantum-mechanical non-renormalization of the weight only for certain classes of BPS observables in pure $\mathcal{N} = 2$ Chern-Simons, which can be computed via localization on three-manifolds that admit a real, nowhere vanishing Killing vector.  This amounts to an explanation of the renormalization of the weight for a similarly restricted set of observables in the corresponding $\mathcal{N} = 0$ theory.  The correspondence is as follows: embed the $\mathcal{N} = 0$ theory in an $\mathcal{N} = 2$ theory with the appropriate level (``integrate in'' auxiliary fields), and then for those links that are deformable to a BPS configuration, deform them to said configuration and enrich them with $\sigma$ as in \eqref{3DN2loop}.  This operation is trivial at the level of the functional integral.  Clearly, this procedure is not possible for arbitrary links, even though all observables are explicitly computable by topological means.  While we cannot explain the non-renormalization of the weight for completely general observables, we have sketched how one might approach a more general understanding by localizing on a solid torus.  In this section, we make some further comments on the correspondence between $\mathcal{N} = 0$ and $\mathcal{N} = 2$ observables.

\subsection{(Non-)Renormalization}

To substantiate the claim that the natural UV completion of Chern-Simons theory should have $\mathcal{N} = 2$ supersymmetry, it is (as mentioned in the introduction) important to fix unambiguous definitions of the level $k$ and the representation $\lambda$.  Throughout, we have used the canonical definition that the $k$ in $\mathcal{N} = 0$ $G_k$ Chern-Simons theory is the level of the corresponding 2D WZW model, where it appears in correlation functions and has a precise physical meaning.  Relative to this definition, the level that appears in the Chern-Simons braiding matrix with parameter $q$ is $k + h$.\footnote{Monodromy matrices in Chern-Simons theory follow from $R$-matrices in braid representations, and by ``braiding matrix,'' we mean the half-monodromy matrix \cite{Guadagnini:1994cc}.}  This shift is independent of regularization scheme, i.e., the question of how the renormalized coupling depends on the bare coupling.  Said differently, our $k\equiv k_\text{phys}$ is what determines the dimension of the Hilbert space and changes sign under time reversal, while $k_\text{phys} + h$ is what appears in correlation functions.  The relation of $k_\text{phys}$ to some UV parameter $k_\text{bare}$ (e.g., via $k_\text{phys} = k_\text{bare} + h$ or $k_\text{phys} = k_\text{bare}$) is a question of regularization scheme and not physically meaningful.

On the other hand, $\lambda$ determines the conjugacy class of the holonomy around a Wilson loop to be $e^{-2\pi i\lambda/k}$, as measured by another loop that links it.  This relation, derived from the classical EOMs (see Appendix \ref{realpol}), receives quantum corrections.  For example, in the case of $SU(2)$ (and using our convention for $\lambda$ from Section \ref{N2CS}), the classical and quantum holonomies are $e^{2\pi ij\sigma_3/k}$ and $e^{2\pi i(j + 1/2)\sigma_3/(k + 2)}$, respectively.  To interpret the statement that ``$\lambda$ is not renormalized'' in the $\mathcal{N} = 2$ setting, it should be kept in mind that Wilson loops are typically written not in terms of the \emph{bare} $\lambda$, but rather in terms of an effective $\lambda$ that corresponds to having integrated out the fermions along the line.\footnote{One can compare the supersymmetric case to the complex (analytically continued) case, where there are also no shifts.  Assuming the standard integration cycle over real connections, the Chern-Simons path integral is oscillatory: the level shift can be attributed to a Wick rotation in the space of connections that renders the integral absolutely convergent (see \cite{Mikhaylov:2014aoa} and references therein).  Since analytically continued Chern-Simons theory requires no further regularization, it is free of the attendant shift ambiguities (which is fortunate because, for instance, the lack of a Killing form of definite signature means that deforming by an irrelevant Yang-Mills term no longer gives rise to a sensible quantum field theory).}

\subsection{3D Point of View}

For completeness, let us comment on how the differences between the bosonic and supersymmetric theories bear on the mapping of line operators between the two theories.  These subtleties do not affect our conclusions.

An obvious difference is that the $\mathcal{N} = 2$ theory contains extra bulk fields, as well as both supersymmetric and non-supersymmetric line operators.  Wilson loops in $\mathcal{N} = 0$ CS are functions of the gauge field $A$, while Wilson loops in $\mathcal{N} = 2$ CS are (schematically) functions of the combination $A + \sigma$.  A collection of the former loops in an arbitrary smooth configuration and a collection of the latter loops in the same configuration have identical correlation functions in the respective theories, up to an appropriate identification of parameters.  This is true even if the configuration is not BPS from the point of view of the latter theory, hence not calculable using localization, as one sees by integrating out $\sigma$.  Moreover, in the latter theory, correlation functions of non-intersecting loops not involving local operators constructed from the extra bulk fields are independent of whether the loop operators are written as functions of $A$ or of $A + \sigma$.  However, such correlation functions can still have contact terms with integrated local operators; these contact terms differ for $A$ and $A + \sigma$ loops.  The Schwinger-Dyson equation for $A$ says that both $\mathcal{N} = 0$ and $\mathcal{N} = 2$ loops have contact terms with the equation of motion for $A$.  The Schwinger-Dyson equation for $\sigma$ says that only $\mathcal{N} = 2$ loops have contact terms with the auxiliary scalar $D$ in the $\mathcal{N} = 2$ vector multiplet.  At finite Yang-Mills coupling, the $A$ and $\sigma$ EOMs involve fermionic sources, but these irrelevant terms are $Q$-exact, so do not affect correlation functions of BPS loops.  If one is only interested in correlators of non-intersecting loops, as we are, then these issues are not relevant.  For a related discussion of the loop equation for BPS Wilson loops, see Section 4 of \cite{Drukker:1999zq}.

To properly define the localizing term requires both a metric and a spin structure (the latter because the fermions become dynamical at finite Yang-Mills coupling), neither of which seem to be necessary to define the bosonic theory (which, for $G$ connected and simply connected, is independent of spin structure \cite{Dijkgraaf:1989pz}).\footnote{The Killing spinor equations \eqref{generalizedKS} require only a spin$^c$ structure, which exists on any orientable three-manifold \cite{Closset:2012ru}.}  But recall that a metric is already needed both to regularize and to gauge-fix the bosonic theory \cite{Witten:1988hf}.  Moreover, recall that computing observables in $\mathcal{N} = 0$ Chern-Simons requires choosing a framing of $M^3$, which automatically fixes a spin structure: every orientable three-manifold is spin, hence parallelizable, and a spin structure is specified by a homotopy class of trivializations of the tangent bundle over the one-skeleton that extends over the two-skeleton.\footnote{More precisely, only a two-framing, or a framing of $TM^3\oplus TM^3$, is required to define the phase of the partition function.  Every three-manifold admits a canonical two-framing.  In fact, every Seifert fibration $\pi : M^3\to \Sigma$ also determines a two-framing on $M^3$, which in general differs from the canonical one \cite{Beasley:2005vf}.}  Therefore, even at finite Yang-Mills coupling, the regularized pure $\mathcal{N} = 2$ Chern-Simons theory does not depend on any additional geometric data beyond that required to compute observables in the $\mathcal{N} = 0$ theory.\footnote{There are at least two qualifications to this statement.  First, while the non-topological localizing terms are $Q$-exact, the metric still enters into the computation of observables in a more essential way in the $\mathcal{N} = 2$ theory because BPS Wilson loops must lie along isometries.  However, just as in the $\mathcal{N} = 0$ theory, smoothly deforming links in the (non-manifestly topological) $\mathcal{N} = 2$ theory leaves correlation functions unchanged (to see this, set the coefficient of the localizing term to zero and integrate out the extra bulk fields).  Second, the localization procedure determines the framing of knots, which is additional data beyond the framing of the three-manifold (the latter cancels in normalized expectation values).}

Finally, we reviewed in Section \ref{loopsinCS} the well-known fact that in bosonic Chern-Simons theory, Wilson and vortex ('t Hooft) loops are equivalent \cite{Moore:1989yh}, the latter being defined by their holonomy.  The same is true in $\mathcal{N} = 2$ Chern-Simons theory \cite{Kapustin:2012iw, Drukker:2012sr, Hosomichi:2014hja}, where a vortex loop is defined by a vector multiplet in a singular BPS configuration.  In this case, vortex loops entail nontrivial background profiles for $\sigma, D$, but the path integral with appropriate boundary conditions and boundary actions completely decouples between these fields and $A$.  Hence, modulo the exceptional situations discussed above for Wilson loops, supersymmetric vortex loops in pure Chern-Simons are equivalent to their non-supersymmetric counterparts.  Note that in \emph{abelian} theories, vortex loops for gauge (rather than flavor) connections are trivial in the sense that up to an overall factor (the classical contribution from the Chern-Simons action), a vortex loop insertion simply results in an imaginary shift of the Coulomb branch parameters that is integrated over in the matrix model and can therefore be absorbed into a redefinition of the integration contour.

\subsection{A Quasi-2D Point of View}

We now show that there exists a one-to-one correspondence between line operators in the bosonic and supersymmetric theories that is clear only if we take into account both shifts.  Given our assumptions on $G$, this correspondence can be viewed as a restatement of well-known facts about 2D rational conformal field theory.

As can be seen in canonical quantization, the distinct Wilson lines in pure Chern-Simons theory are in one-to-one correspondence with the ground states of the theory on a (spatial) torus.  To explain what ``distinct'' means, we must identify the precise equivalence classes of Wilson lines that map to these ground states.  $SU(2)_k$ Chern-Simons on a torus has $k + 1$ ground states labeled by half-integers $j = 0, \ldots, k/2$.  These can equivalently be viewed as the $k + 1$ primary operators in the $SU(2)_k$ WZW model, where the truncation of $SU(2)$ representations to integrable representations is a selection rule that follows from combining two different $\mathfrak{su}(2)$ subalgebras of $\smash{\widehat{SU}(2)_k}$.  From the 3D point of view, however, a Wilson line can carry any $SU(2)$ representation (half-integer $j$).  To respect the 2D truncation, all such lines fall into equivalence classes labeled by the basic lines $j = 0, \ldots, k/2$.  The equivalence relations turn out to be a combination of Weyl conjugation and translation \cite{Elitzur:1989nr}:
\begin{equation}
j\sim -j, \quad j\sim j + k.
\label{N0relations}
\end{equation}
As reviewed in Appendix \ref{realpol}, the story is similar for general $G$.  Line operators are subject to equivalence relations given by the action of the affine Weyl group at level $k$ ($\mathcal{W}\ltimes k\Lambda_R^\vee$ where $\Lambda_R^\vee$ is the coroot lattice of $G$), whose fundamental domain we refer to as an affine Weyl chamber or a Weyl alcove and which contains all inequivalent weights (corresponding to integrable representations of $\smash{\widehat{G}_k}$).\footnote{The equivalence classes of Wilson lines in abelian Chern-Simons can be found in Appendix C of \cite{Seiberg:2016rsg}.}

Now consider the correlation functions of these lines.  Two basic observables of $SU(2)_k$ Chern-Simons on $S^3$ are the expectation value of an unknotted spin-$j$ Wilson loop and the expectation value of two Wilson loops of spins $j, j'$ in a Hopf link:
\begin{equation}
\langle W_j\rangle_{\mathcal{N} = 0} = \frac{S_{0j}}{S_{00}}, \quad \langle W_j W_{j'}\rangle_{\mathcal{N} = 0} = \frac{S_{jj'}}{S_{00}}.
\label{basiccorrelators}
\end{equation}
Recall that the modular $S$-matrix of $SU(2)_k$ is given by \eqref{Smatrix} in a basis of integrable representations.  The formulas \eqref{basiccorrelators} apply only to Wilson loops with $j$ within the restricted range $0, \ldots, k/2$.  Indeed, \eqref{Smatrix} is not invariant under the equivalence relations \eqref{N0relations}.  Nonetheless, let us na\"ively extend these formulas to arbitrary $j, j'$.  The first positive value of $j$ for which $\langle W_j\rangle = 0$ is that immediately above the truncation threshold: $j = (k + 1)/2$.  More generally, from \eqref{basiccorrelators}, it is clear that a line of spin $j$ and a line of spin $j + k + 2$ have identical correlation functions, while lines with $j = n(k/2 + 1) - 1/2$ for any integer $n$ vanish identically.  Here, one should distinguish the \emph{trivial} line $j = 0$, which has $\langle W_0\rangle = 1$ and trivial braiding with all other lines, from \emph{nonexistent} lines, which have $\langle W_j\rangle = 0$ and vanishing correlation functions with all other lines.  On the other hand, a line with $j$ and a line with $j + k/2 + 1$ have the same expectation value and braiding, \emph{up to a sign}.  In other words, at the level of correlation functions, $SU(2)_k$ Wilson lines are antiperiodic with period $k/2 + 1$.

An analogous antiperiodicity phenomenon holds for arbitrary simple, compact $G$.  In the WZW model, the fusion rule eigenvalues (computed from the $S$-matrix elements) are equal to the finite Weyl characters of $G$, evaluated on some special Cartan angles that respect the truncation of the relevant representations \cite{Verlinde:1988sn}.  For example, in $SU(2)_k$, $\smash{\lambda_\ell^{(n)}} = S_{\ell n}/S_{0n}$ is the Weyl character $\chi_\ell(\theta)$ in \eqref{su2Weylcharacter} evaluated at $\theta/2 = (2n + 1)\pi/(k + 2)$ for $n = 0, \ldots, k/2$, chosen such that the Weyl character of spin $\ell = (k + 1)/2$ vanishes.

The (anti)periodicity of $S$ under $j\to j + (k + 2)/2$ can be understood in terms of the renormalized parameters $K = k + 2$ and $J = j + 1/2$.\footnote{See also \cite{Mikhaylov:2014aoa}.  I thank V. Mikhaylov for correspondence on this point.}  In the $\mathcal{N} = 2$ theory, a $J$ Wilson line has holonomy $e^{2\pi iJ\sigma_3/K}$, so the equivalence relations are
\begin{equation}
J\sim -J, \quad J\sim J + K \quad \Longleftrightarrow \quad j\sim -1 - j, \quad j\sim j + k + 2.
\label{N2relations}
\end{equation}
The inequivalent values of $j$ are $-1/2, \ldots, (k + 1)/2$.  The extremal values $j = -1/2$ and $j = (k + 1)/2$ correspond to identically zero line operators, and the remaining values are the same as in the $\mathcal{N} = 0$ formulation.  In other words, in contrast to \eqref{basiccorrelators} for $\mathcal{N} = 0$ $SU(2)_k$ on $S^3$, we have for $\mathcal{N} = 2$ $SU(2)_K$ on $S^3$ that
\begin{equation}
\langle W_J\rangle_{\mathcal{N} = 2} = \frac{S_{\frac{1}{2}J}}{S_{\frac{1}{2}\frac{1}{2}}}, \quad \langle W_J W_{J'}\rangle_{\mathcal{N} = 2} = \frac{S_{JJ'}}{S_{\frac{1}{2}\frac{1}{2}}}, \quad S_{JJ'}\equiv \sqrt{\frac{2}{K}}\sin\left[\frac{(2J)(2J')\pi}{K}\right],
\end{equation}
where the \emph{bare} $J$ must satisfy $J\geq 1/2$ for supersymmetry not to be spontaneously broken.  In the supersymmetric theory, labeling lines by $J = 1/2, \ldots, (k + 1)/2$, the $J = 0$ line does not exist due to the vanishing Grassmann integral over the zero modes of the fermion in the $\mathcal{N} = 2$ coadjoint orbit sigma model.  The conclusion is that the $\mathcal{N} = 2$ theory has the same set of independent line operators as the $\mathcal{N} = 0$ theory.  In the $\mathcal{N} = 2$ formulation, the $S$-matrix $S_{jj'}$ is explicitly invariant under the equivalence relations \eqref{N2relations}.

For general $G$, let $\Lambda = \lambda + \rho$ and $K = k + h$.  Then $\Lambda$, modulo the action of the affine Weyl group at level $K$, takes values in an affine Weyl chamber at level $K$.  Those $\lambda = \Lambda - \rho$ for $\Lambda$ at the boundary of the chamber correspond to nonexistent lines, while those for $\Lambda$ in the interior are in one-to-one correspondence with weights in the affine Weyl chamber at level $k$ (for further details, see Appendix \ref{holomorphicpol}).

It would be interesting to understand both shifts from an intrinsically 2D point of view.  The shift in $k$ is, as in the 3D case, transparent; the shift in $\lambda$ is less so.  As the ring of line operators in Chern-Simons with compact gauge group is the fusion ring of the corresponding WZW model, one would like to translate the equivalence between $(J, K)$ and $(j, k)$ into an equivalence between ordinary and super WZW models.  One can impose half-BPS boundary conditions to obtain both $(1, 1)$ and $(0, 2)$ WZW models on the boundary of bulk $\mathcal{N} = 2$ CS \cite{Faizal:2016skd}.  It appears that $(1, 1)$ is relevant to the level shift while $(0, 2)$ is relevant to holomorphic blocks \cite{Yoshida:2014ssa}.  It is well-known that after a suitable redefinition of the super Kac-Moody currents, the $(1, 1)$ super WZW model at level $k$ (for compact, connected, simply connected $G$) is equivalent to a bosonic WZW model at level $k - h$ (with central charge $(1 - \frac{h}{k})\dim G$) plus decoupled free Majorana fermions in the adjoint representation (with central charge $\frac{1}{2}\dim G$), resulting in a super Virasoro algebra with central charge $\hat{c} = \frac{2}{3}c = (1 - \frac{2h}{3k})\dim G$ \cite{DiVecchia:1984nyg, Fuchs:1988gm}.  On the other hand, just as pure $\mathcal{N} = 2$ Chern-Simons is the bosonic theory plus some decoupled auxiliary fields, the corresponding $(0, 2)$ WZW model in 2D is the bosonic WZW model plus some decoupled fields; its left-moving sector is simply a non-supersymmetric chiral WZW model.

Finally, it is interesting to note that similar truncations of Wilson loop representations exist due to quantum relations in $\mathcal{N} = 2$ Chern-Simons-matter theories, despite that we do not expect an equivalence to a WZW model in this case \cite{Kapustin:2013hpk}.

\section{Discussion}

Using $SU(2)$ as a case study, we have supersymmetrized the coadjoint orbit quantum mechanics on a Wilson line in flat space from both intrinsically 1D and 3D points of view, providing several complementary ways to understand the shift in the representation $j$.  We have described how to extend this understanding to certain compact Euclidean manifolds.  For some classes of observables in Chern-Simons theory, the existence of an auxiliary supersymmetry lends itself not only to conceptual unity, but also to increased computability.

For arbitrary simple groups, one has both generic and degenerate coadjoint orbits, corresponding to quotienting $G$ by the maximal torus $T$ or by a Levi subgroup $L\supset T$ (see \cite{Bernatska:2008} and references therein).  For example, the gauge group $SU(N + 1)$ has for a generic orbit the phase space $SU(N + 1)/U(1)^N$, a flag manifold with real dimension $N^2 + N$ (corresponding to a regular weight); on the other hand, the most degenerate orbit is $SU(N + 1)/(SU(N)\times U(1))\cong S^{2N + 1}/S^1\cong \mathbb{CP}^N$, which has $2N$ real dimensions and a simple K\"ahler potential (corresponding to a weight that lies in the most symmetric part of the boundary of the positive Weyl chamber).\footnote{A regular weight $\lambda$ satisfies $\lambda(\alpha)\neq 0$ for all roots $\alpha$; otherwise, the coadjoint orbit is isomorphic to $G/L$ where the Lie algebra of the Levi subgroup $L$ is that of $T$ adjoined with all roots $\alpha$ such that $\lambda(\alpha) = 0$.}  The quantization of the phase space $\mathbb{CP}^N$ is well-known and can be made very explicit in terms of coherent states (see, e.g., Section 5 of \cite{Szabo:1996md}).  The Fubini-Study metric for $\mathbb{CP}^N$ follows from covering the manifold with $N + 1$ patches with the K\"ahler potential in each patch being the obvious generalization of that for $SU(2)$.  In principle, one can carry out a similar analysis with $SU(N + 1)$ Killing vectors.  We will not attempt the full analysis in this paper.  We simply remark that in general, the shift of a fundamental weight by the Weyl vector (half-sum of positive roots, or sum of fundamental weights) is no longer a fundamental weight, so one would need a qualitatively different sigma model than the original to describe the coadjoint orbit of the shifted weight.  An option is not to work in local coordinates at all, along the lines of \cite{Beasley:2009mb} (however, this approach seems harder to supersymmetrize).

Finally, perhaps this story is more natural in a setting with twice as much supersymmetry (3D $\mathcal{N} = 4$), where one has the option of twisting spatial rotations by either of the $SU(2)$ $R$-symmetry groups, allowing for the construction of $1/4$-BPS Wilson or vortex loops supported on arbitrary curves \cite{Assel:2015oxa}.

\section*{Acknowledgements}

I thank B. Le Floch, S. Giombi, I. Klebanov, V. Mikhaylov, S. Pufu, P. Putrov, N. Seiberg, H. Verlinde, B. Willett, E. Witten, and I. Yaakov for helpful discussions, correspondence, and (in some cases) comments on a preliminary draft.  I would like to give a special thanks to N. Seiberg for suggesting this line of investigation.  I am also grateful to M. Dedushenko, S. Pufu, and R. Yacoby for collaboration on related work.  I am supported in part by the NSF GRFP under Grant No.\ DGE-1656466 and by the Graduate School at Princeton University.

\appendix

\section{Conventions} \label{conventions}

\subsection{Gauge Transformations} \label{gaugetransformations}

With Hermitian generators ($[T^a, T^b] = if^{abc}T^c$ with $f^{abc}$ real and $\Tr(T^a T^b) = \frac{1}{2}\delta^{ab}$), we have
\begin{equation}
\mathcal{L}_\text{CS}^g = \epsilon^{\mu\nu\rho}\Tr\left(A_\mu^g\partial_\nu A_\rho^g - \frac{2i}{3}A_\mu^g A_\nu^g A_\rho^g\right), \quad \mathcal{L}_\text{CS}\equiv \mathcal{L}_\text{CS}^\text{id}.
\label{N0LCS}
\end{equation}
Writing $g = e^{i\theta^a T^a}$, we have $A_\mu^g = gA_\mu g^{-1} - i\partial_\mu gg^{-1} = (A_\mu^a + D_\mu\theta^a + \cdots)T^a$ where $D_\mu\theta^a = \partial_\mu\theta^a + f^{abc}A_\mu^b\theta^c$ and
\begin{equation}
\mathcal{L}_\text{CS}^g = \mathcal{L}_\text{CS} - i\epsilon^{\mu\nu\rho}\partial_\mu\Tr(A_\nu g^{-1}\partial_\rho g) - \frac{1}{3}\epsilon^{\mu\nu\rho}\Tr[(\partial_\mu gg^{-1})(\partial_\nu gg^{-1})(\partial_\rho gg^{-1})].
\label{gaugechange}
\end{equation}
In coordinate-free notation, $A^g = gAg^{-1} - idgg^{-1}$ and
\begin{equation}
\Tr\left[A^g dA^g - \frac{2i}{3}(A^g)^3\right] = \Tr\left[AdA - \frac{2i}{3}A^3 - id(Ag^{-1}dg) - \frac{1}{3}(dgg^{-1})^3\right].
\end{equation}

\subsection{Quantum Mechanics} \label{quantummechanics}

We fix the normalization of the partition function, for the purpose of computing the 1D supersymmetric index, as in the following two elementary examples.\footnote{In both cases, we use $\prod_{n=1}^\infty (4\pi^2 n^2)^{\pm 1} = e^{\pm 2(\zeta(0)\log(2\pi) - \zeta'(0))} = 1$.}

For a (real) bosonic oscillator, we have
\begin{equation}
L_E = \frac{1}{2}\dot{x}^2 + \frac{1}{2}\omega^2 x^2 \implies Z = \int Dx\, \exp\left[-\frac{1}{2}\int_0^\beta d\tau\, (\dot{x}^2 + \omega^2 x^2)\right]
\end{equation}
where periodic boundary conditions are implicit.  The corresponding mode expansion is
\begin{equation}
x = \frac{1}{\sqrt{\beta}}\sum_{n=-\infty}^\infty x_n e^{2\pi in\tau/\beta},
\end{equation}
where reality requires that $x_n = x_{-n}^\ast$.  Write $x_n = (a_n + ib_n)/\sqrt{2}$ for $n\neq 0$.  Since $x$ has mass dimension $-1/2$, the modes have mass dimension $-1$.  Hence the correct dimensionless path integration measure is
\begin{equation}
\int Dx = \int_{-\infty}^\infty \frac{dx_0}{\sqrt{2\pi}\beta}\left(\prod_{n=1}^\infty \int_{-\infty}^\infty \frac{da_n}{\sqrt{2\pi}\beta}\int_{-\infty}^\infty \frac{db_n}{\sqrt{2\pi}\beta}\right),
\end{equation}
with a denominator of $\sqrt{2\pi}$ for each real mode.  It follows that $Z = [2\sinh(\beta\omega/2)]^{-1}$.

For a (complex) fermionic oscillator, we have
\begin{equation}
L_E = \psi^\dag\dot{\psi} + \omega\psi^\dag\psi \implies Z' = \int D\psi^\dag D\psi\, \exp\left[-\int_0^\beta dt\, (\psi^\dag\dot{\psi} + \omega\psi^\dag\psi)\right]
\end{equation}
where $Z'$ is the twisted partition function and periodic boundary conditions are implicit.  The corresponding mode expansion is
\begin{equation}
\psi = \frac{1}{\sqrt{\beta}}\sum_{n=-\infty}^\infty \psi_n e^{2\pi int/\beta}.
\end{equation}
Since $\psi$ and $\psi^\dag$ are dimensionless, the modes have mass dimension $-1/2$ and $d(\text{modes})$ have mass dimension $1/2$.  Hence the correct dimensionless path integration measure is
\begin{equation}
\int D\psi^\dag D\psi = \int \prod_{n=-\infty}^\infty (\sqrt{\beta}\, d\psi_n^\dag)(\sqrt{\beta}\, d\psi_n),
\end{equation}
from which we obtain $Z' = 2\sinh(\beta\omega/2)$.

Combining the two examples above results in a simple supersymmetric system with one real bosonic and one real fermionic DOF on shell, whose twisted partition function (Witten index) is $ZZ' = 1$.

\subsection{1D \texorpdfstring{$\mathcal{N} = 2$}{N = 2}} \label{1DN2}

We work in Lorentzian\footnote{To Euclideanize the following, take $\tau = it$ and $iS = -S_E$.} 1D $\mathcal{N} = 2$ superspace with coordinates $(t, \theta, \theta^\dag)$ and write SUSY transformations in terms of a complex spinor parameter $\epsilon$.  The representations of the supercharges as differential operators on superspace and the supercovariant derivatives are
\begin{gather}
\hat{Q} = \partial_\theta + i\theta^\dag\partial_t, \qquad \hat{Q}^\dag = \partial_{\theta^\dag} + i\theta\partial_t, \label{diffops1D} \\
D = \partial_\theta - i\theta^\dag\partial_t, \qquad D^\dag = \partial_{\theta^\dag} - i\theta\partial_t.
\end{gather}
The nonvanishing anticommutators are
\begin{equation}
\{\hat{Q}, \hat{Q}^\dag\} = -\{D, D^\dag\} = 2i\partial_t.
\end{equation}
A general superfield takes the form
\begin{equation}
\Xi(t, \theta) = \phi(t) + \theta\psi(t) + \theta^\dag\chi(t) + \theta\theta^\dag F(t)
\end{equation}
with SUSY acting as $\delta\Xi = (\epsilon\hat{Q} + \epsilon^\dag\hat{Q}^\dag)\Xi$, or in components as
\begin{align}
\delta\phi &= \epsilon\psi + \epsilon^\dag\chi, \nonumber \\
\delta\psi &= -i\epsilon^\dag\dot{\phi} + \epsilon^\dag F, \nonumber \\
\delta\chi &= -i\epsilon\dot{\phi} - \epsilon F, \nonumber \\
\delta F &= -i\epsilon\dot{\psi} + i\epsilon^\dag\dot{\chi}.
\end{align}
All components of $\Xi$ are complex.  The vector multiplet satisfies
\begin{equation}
V = V^\dag \implies V(t, \theta) = \phi(t) + \theta\psi(t) - \theta^\dag\psi^\dag(t) + \theta\theta^\dag F(t).
\end{equation}
The SUSY transformations of its components $\phi$ (real), $\psi$ (complex), and $F$ (real) are
\begin{align}
\delta\phi &= \epsilon\psi - \epsilon^\dag\psi^\dag, \nonumber \\
\delta\psi &= -i\epsilon^\dag\dot{\phi} + \epsilon^\dag F, \nonumber \\
\delta\psi^\dag &= i\epsilon\dot{\phi} + \epsilon F, \nonumber \\
\delta F &= -i\epsilon\dot{\psi} - i\epsilon^\dag\dot{\psi}^\dag. \label{1Dsusyvector}
\end{align}
Here, $F$ plays the role of the non-dynamical gauge field, so that the 1D Chern-Simons action $\int F$ is automatically both gauge- and SUSY-invariant.  The chiral multiplet satisfies
\begin{equation}
D^\dag\Phi = 0 \implies \Phi(t, \theta) = \phi(t) + \theta\psi(t) - i\theta\theta^\dag\dot{\phi}(t).
\end{equation}
The SUSY transformations of its components $\phi$ (complex) and $\psi$ (complex) are
\begin{align}
\delta\phi &= \epsilon\psi, \nonumber \\
\delta\psi &= -2i\epsilon^\dag\dot{\phi}. \label{1Dsusychiral}
\end{align}
To integrate over superspace, we use $d^2\theta\equiv d\theta^\dag d\theta$.\footnote{Given a vector multiplet $V = \phi + \theta\psi - \theta^\dag\psi^\dag + \theta\theta^\dag F$, one can write the top component as the middle component of the fermionic chiral superfield
\[
D^\dag V = -\psi^\dag - \theta(F + i\dot{\phi}) + i\theta\theta^\dag\dot{\psi}^\dag
\]
(thus writing a $D$-term as an $F$-term), or as the bottom component of the real superfield
\[
\frac{1}{2}(D^\dag D - DD^\dag)V = F - i\theta\dot{\psi} - i\theta^\dag\dot{\psi}^\dag - \theta\theta^\dag\ddot{\phi}
\]
(a 1D analogue of a linear multiplet).  Thus, for instance, the superfield equation of motion in an $\mathcal{N} = 2$ quantum mechanics described by K\"ahler potential $K$ is $D^\dag\partial_\Phi K = 0$.}

The 1D SUSY' transformations are derived as follows.  In Wess-Zumino gauge, we have $V|_\text{WZ} = \theta\theta^\dag F$, which transforms under SUSY to $V' = V|_\text{WZ} + \delta V|_\text{WZ}$.  To preserve Wess-Zumino gauge, we choose the compensatory super gauge transformation parameter $\Phi = i\Lambda$ such that $\delta V|_\text{WZ} + \frac{1}{2}(\Phi + \Phi^\dag)$ is $O(\theta\theta^\dag)$:
\begin{equation}
e^{2V'}\to e^\Phi e^{2V'}e^{\Phi^\dag} \Longleftrightarrow V'\to V|_\text{WZ},
\end{equation}
which means that $\delta'$ acts trivially on the vector multiplet.  For the chiral multiplet, only the transformation rule for $\psi$ is modified by taking $\partial_0\to D_0$:
\begin{align}
\delta'\phi &= \epsilon\psi, \nonumber \\
\delta'\psi &= -2i\epsilon^\dag D_0\phi, \label{1Dsusypchiral}
\end{align}
where the gauge field appearing in $D_0$ is the single nonzero component of $V|_\text{WZ}$.

\subsection{3D \texorpdfstring{$\mathcal{N} = 2$}{N = 2}} \label{3DN2}

We largely follow the conventions of \cite{Intriligator:2013lca}.  Regardless of spacetime signature, spinor indices are raised and lowered on the left by $\epsilon^{\alpha\beta} = -\epsilon_{\alpha\beta}$ ($\epsilon^{12}\equiv 1$), where $\psi\chi\equiv \epsilon^{\alpha\beta}\psi_\alpha\chi_\beta$.  This convention requires that we distinguish matrix multiplication (``$\cdot$'') from spinor contraction (no symbol), which differ by a sign:
\begin{equation}
(\gamma^a\xi)_\alpha = -(\gamma^a)_\alpha{}^\beta\xi_\beta \Longleftrightarrow \gamma^a\xi = -\gamma^a\cdot \xi.\footnote{We also sometimes use ``$\cdot$'' to denote multiplication in the appropriate representation of the gauge group: for example, $[D_\mu, D_\nu]({\cdots}) = -iF_{\mu\nu}\cdot ({\cdots})$.}
\end{equation}
The basic Fierz identities, written with a free spinor index (taken to be lower by default), follow from the basis-independent properties $\tr(\gamma^\mu\gamma^\nu) = 2g^{\mu\nu}$ and $\tr\gamma^\mu = 0$ and take the same form for both commuting and anticommuting spinors:
\begin{equation}
\xi(\zeta\lambda) + \zeta(\lambda\xi) + \lambda(\xi\zeta) = 0, \quad 2\lambda(\xi\zeta) + \xi(\zeta\lambda) + \gamma^\mu\xi(\zeta\gamma_\mu\lambda) = 0.\footnote{For matrix-valued $\lambda$ (whose components $\lambda^{1, 2}$ are not nilpotent), we have $[\xi\gamma^\mu\lambda, \zeta\gamma_\mu\lambda] = [\xi\lambda, \zeta\lambda]$.}
\end{equation}
By default, our spinors are anticommuting; the notation ``$|_0$'' applied to a Grassmann-odd spinor denotes its Grassmann-even version.  Spinors satisfy $\psi\chi = \pm\chi\psi$, $\psi\gamma^\mu\chi = \mp\chi\gamma^\mu\psi$, $\overline{\psi\chi} = \mp\bar{\psi}\bar{\chi}$ depending on whether they anticommute (top sign) or commute (bottom sign).

Spinors that would be conjugate (e.g., $\lambda, \bar{\lambda}$) in Lorentzian signature are independent in Euclidean signature (e.g., $\lambda, \tilde{\lambda}$) because the \textbf{2} of $SL(2, \mathbb{R})$ is complex while the \textbf{2} of $SU(2)$ is pseudoreal.  Lower and upper indices denote the fundamental and antifundamental, respectively.  In Lorentzian signature, as in 1D, we use $x^\ast, \bar{x}, x^\dag$ interchangeably to denote the complex conjugate of $x$.  In Euclidean signature, we use bars and stars interchangeably to denote complex conjugation, while daggers denote Hermitian conjugation: $(\psi^\dag)^\alpha = (\psi_\alpha)^\ast$ (the combination $\xi^\dag\psi$ is $SU(2)$-invariant).  Explicitly,
\begin{gather*}
\begin{pmatrix} \xi_1 \\ \xi_2 \end{pmatrix} = \begin{pmatrix} a \\ b \end{pmatrix} \implies \begin{pmatrix} \xi^1 & \xi^2 \end{pmatrix} = \begin{pmatrix} b & -a \end{pmatrix}, \mbox{ } \begin{pmatrix} \xi^{\dag 1} & \xi^{\dag 2} \end{pmatrix} = \begin{pmatrix} a^\ast & b^\ast \end{pmatrix}, \mbox{ } \begin{pmatrix} \xi^\dag_1 \\ \xi^\dag_2 \end{pmatrix} = \begin{pmatrix} -b^\ast \\ a^\ast \end{pmatrix}.
\end{gather*}

\subsubsection{Lorentzian}

We use Lorentzian signature only in flat space.  In $\mathbb{R}^{1, 2}$ (with signature ${-}{+}{+}$), the 3D gamma matrices are
\begin{equation}
(\gamma^{\mu = 0, 1, 2})_{\alpha\beta} = (-1, \sigma^1, \sigma^3)_{\alpha\beta}, \quad \epsilon^{\alpha\beta}\gamma_{\gamma\alpha}^\mu\gamma_{\beta\delta}^\nu = \eta^{\mu\nu}\epsilon_{\gamma\delta} - \epsilon^{\mu\nu\rho}(\gamma_\rho)_{\gamma\delta}.
\end{equation}
With lowered indices, these matrices are real and symmetric, so that $\theta\gamma^\mu\bar{\theta}$ is real.  We take $\epsilon^{012} = 1$; note that the identity $\epsilon^{\mu\nu\lambda}\epsilon^{\rho\sigma}{}_\lambda = \eta^{\mu\sigma}\eta^{\nu\rho} - \eta^{\mu\rho}\eta^{\nu\sigma}$ differs by a sign relative to the Euclidean case.  The 3D $\mathcal{N} = 2$ algebra is
\begin{equation}
\{Q_\alpha, \bar{Q}_\beta\} = 2\gamma_{\alpha\beta}^\mu P_\mu + 2i\epsilon_{\alpha\beta}Z, \quad \{Q_\alpha, Q_\beta\} = 0.
\end{equation}
Equivalently,
\begin{equation}
\{Q_\pm, \bar{Q}_\pm\} = P_2\pm iP_1, \quad \{Q_\pm, \bar{Q}_\mp\} = -P_0\mp Z
\end{equation}
where $Q_\pm = \frac{1}{2}(Q_1\pm iQ_2)$ and $\bar{Q}_\pm = \overline{Q_\mp} = \frac{1}{2}(\bar{Q}_1\pm i\bar{Q}_2)$.  The representations of the supercharges as differential operators on superspace and the supercovariant derivatives are
\begin{gather}
\mathcal{Q}_\alpha = \frac{\partial}{\partial\theta^\alpha} - i\gamma_{\alpha\beta}^\mu\bar{\theta}^\beta\partial_\mu, \qquad \bar{\mathcal{Q}}_\alpha = -\frac{\partial}{\partial\bar{\theta}^\alpha} + i\theta^\beta\gamma_{\beta\alpha}^\mu\partial_\mu, \label{diffops3D} \\
D_\alpha = \frac{\partial}{\partial\theta^\alpha} + i\gamma_{\alpha\beta}^\mu\bar{\theta}^\beta\partial_\mu, \qquad \bar{D}_\alpha = -\frac{\partial}{\partial\bar{\theta}^\alpha} - i\theta^\beta\gamma_{\beta\alpha}^\mu\partial_\mu.
\end{gather}
We abbreviate $\bar{\partial}_\alpha\equiv \partial/\partial\bar{\theta}^\alpha$, $\partial_\beta\equiv \partial/\partial\theta^\beta$ and define $\int d^4\theta\, \theta^2\bar{\theta}^2 = 1$.

In component form, the SUSY transformations of the 3D $\mathcal{N} = 2$ vector and chiral multiplets \eqref{3DN2vector} and \eqref{3DN2chiral} are
\begin{align}
\delta C &= -(\xi\chi - \bar{\xi}\bar{\chi}), \nonumber \\
\delta\chi_\alpha = \smash{\overline{\delta\bar{\chi}_\alpha}} &= i\bar{\xi}_\alpha\sigma + \gamma_{\alpha\beta}^\mu\bar{\xi}^\beta(A_\mu + i\partial_\mu C) - \xi_\alpha(M + iN), \nonumber \\
\delta(M + iN) = \smash{\overline{\delta(M - iN)}} &= -2i(\bar{\xi}\bar{\lambda} - \bar{\xi}\gamma^\mu\partial_\mu\chi), \nonumber \\
\delta\sigma &= -(\xi\bar{\lambda} - \bar{\xi}\lambda), \label{3Dsusyvector} \\
\delta A_\mu &= i(\xi\gamma_\mu\bar{\lambda} + \bar{\xi}\gamma_\mu\lambda - \xi\partial_\mu\chi - \bar{\xi}\partial_\mu\bar{\chi}), \nonumber \\
\delta\lambda_\alpha = \smash{\overline{\delta\bar{\lambda}_\alpha}} &= -i\xi_\alpha D - i\gamma_{\alpha\beta}^\mu\xi^\beta\partial_\mu\sigma - \epsilon^{\mu\nu\rho}(\gamma_\rho)_{\alpha\beta}\xi^\beta\partial_\mu A_\nu, \nonumber \\
\delta D &= -(\xi\gamma^\mu\partial_\mu\bar{\lambda} - \bar{\xi}\gamma^\mu\partial_\mu\lambda) \nonumber
\end{align}
(following from $\delta_\xi V = (\xi\mathcal{Q} - \bar{\xi}\bar{\mathcal{Q}})V$) and
\begin{align}
\delta A &= -\sqrt{2}\xi\psi, \nonumber \\
\delta\psi_\alpha &= -\sqrt{2}\xi_\alpha F + i\sqrt{2}\gamma_{\alpha\beta}^\mu\bar{\xi}^\beta\partial_\mu A, \label{3Dsusychiral} \\
\delta F &= i\sqrt{2}\bar{\xi}\gamma^\mu\partial_\mu\psi \nonumber
\end{align}
(following from $\delta_\xi\Phi = (\xi\mathcal{Q} - \bar{\xi}\bar{\mathcal{Q}})\Phi$).  The 3D SUSY' transformations are derived as follows.  Under SUSY, $V|_\text{WZ}$ transforms into $V'\equiv V|_\text{WZ} + \delta V|_\text{WZ}$.  To preserve Wess-Zumino gauge, we choose $\Lambda$ such that $\delta V|_\text{WZ} + \frac{1}{2}(\Lambda + \bar{\Lambda})$ is $O(\theta\bar{\theta})$ and set the lowest component of $\Lambda$ (the parameter of ordinary gauge transformations) to zero.  With these choices, and to first order in $\xi, \bar{\xi}$, the super gauge transformation $e^{2V'}\to e^{\bar{\Lambda}}e^{2V'}e^\Lambda$ truncates to
\begin{equation}
V'\to V' + \frac{1}{2}(\Lambda + \bar{\Lambda}) + \frac{1}{2}[V', \Lambda - \bar{\Lambda}],
\end{equation}
from which we read off \eqref{3Dsusypvector}.  SUSY' also modifies the chiral multiplet transformation laws by terms involving vector multiplet fields, so that $\Phi + \delta\Phi\to e^{-\Lambda}(\Phi + \delta\Phi)$ and $\bar{\Phi} + \delta\bar{\Phi}\to (\bar{\Phi} + \delta\bar{\Phi})e^{-\bar{\Lambda}}$, from which \eqref{3Dsusypchiral} follows.

\subsubsection{Euclidean}

We start with flat Euclidean space $\mathbb{R}^3$, passing to Euclidean signature via $\tau = it$ and $iS = -S_E$ ($i\int dt\, L = -\int d\tau\, L_E$).  We set $\partial_0\equiv \partial_\tau = -i\partial_t$, and similarly for other tensors.  Thus
\begin{equation}
(\gamma^{\mu = 0, 1, 2})_{\alpha\beta} = (-i, \sigma^1, \sigma^3)_{\alpha\beta}, \quad \epsilon^{\alpha\beta}\gamma_{\gamma\alpha}^\mu\gamma_{\beta\delta}^\nu = \delta^{\mu\nu}\epsilon_{\gamma\delta} - i\epsilon^{\mu\nu\rho}(\gamma_\rho)_{\gamma\delta}.
\end{equation}
In particular, with lowered indices, the gamma matrices are no longer real.  We often write $\mu = 1, 2, 3$ in place of $\mu = 0, 1, 2$ in Euclidean signature.

The Euclidean SUSY' transformations of a vector multiplet are
\begin{align}
\delta'\sigma &= -(\xi\tilde{\lambda} - \tilde{\xi}\lambda), \nonumber \\
\delta' A_\mu &= i(\xi\gamma_\mu\tilde{\lambda} + \tilde{\xi}\gamma_\mu\lambda), \nonumber \\
\delta'\lambda &= \textstyle -i\xi D - i\gamma^\mu\xi D_\mu\sigma - \frac{i}{2}\epsilon^{\mu\nu\rho}\gamma_\rho\xi F_{\mu\nu}, \vphantom{\tilde{\xi}} \label{3Dsusypvector-E} \\
\delta'\tilde{\lambda} &= \textstyle i\tilde{\xi}D + i\gamma^\mu\tilde{\xi}D_\mu\sigma - \frac{i}{2}\epsilon^{\mu\nu\rho}\gamma_\rho\tilde{\xi}F_{\mu\nu}, \nonumber \\
\delta' D &= -(\xi\gamma^\mu D_\mu\tilde{\lambda} - \tilde{\xi}\gamma^\mu D_\mu\lambda) + [\xi\tilde{\lambda} + \tilde{\xi}\lambda, \sigma] \nonumber
\end{align}
(compare to \eqref{3Dsusypvector}), and those of a fundamental chiral multiplet are
\begin{align}
\delta' A &= -\sqrt{2}\xi\psi, \nonumber \\
\delta'\psi &= -\sqrt{2}\xi F + i\sqrt{2}\gamma^\mu\tilde{\xi}D_\mu A + i\sqrt{2}\tilde{\xi}\sigma A, \label{3Dsusypchiral-E} \\
\delta' F &= i\sqrt{2}\tilde{\xi}\gamma^\mu D_\mu\psi - i\sqrt{2}\sigma\tilde{\xi}\psi - 2i\tilde{\xi}\tilde{\lambda}A \nonumber
\end{align}
(compare to \eqref{3Dsusypchiral}).  Writing $\delta' = \delta_\xi' + \delta_{\tilde{\xi}}'$ with $\smash{(\delta_\xi')^2 = (\delta_{\tilde{\xi}}')^2 = 0}$, the SUSY' algebra is
\begin{equation}
[\delta_{\xi_1}', \delta_{\xi_2}'](\cdot) = [\delta_{\tilde{\xi}_1}', \delta_{\tilde{\xi}_2}'](\cdot) = 0, \quad [\delta_\xi', \delta_{\tilde{\xi}}'](\cdot) = 2i\xi\gamma^\mu\tilde{\xi}D_\mu(\cdot) + 2i\xi\tilde{\xi}\sigma\cdot (\cdot).
\label{3Dsusypalgebra-E}
\end{equation}
The Lagrangians of interest are
\begin{align}
\mathcal{L}_\text{CS}|_{\mathbb{R}^3} &= \frac{k}{4\pi i}\Tr\left[\epsilon^{\mu\nu\rho}\left(A_\mu\partial_\nu A_\rho - \frac{2i}{3}A_\mu A_\nu A_\rho\right) - 2\lambda\tilde{\lambda} + 2iD\sigma\right], \\
\mathcal{L}_\text{YM}|_{\mathbb{R}^3} &= \frac{1}{g^2}\Tr\left(\frac{1}{4}F_{\mu\nu}F^{\mu\nu} + \frac{1}{2}D_\mu\sigma D^\mu\sigma - \frac{1}{2}D^2 - i\tilde{\lambda}\gamma^\mu D_\mu\lambda + i\tilde{\lambda}[\sigma, \lambda]\right),
\end{align}
where the path integration contour is over purely imaginary $D$.  We find that
\begin{equation}
\frac{1}{g^2}\delta_\xi'\delta_{\tilde{\xi}}'\Tr\left(\frac{1}{2}\lambda\tilde{\lambda} - iD\sigma\right) = \xi\tilde{\xi}\mathcal{L}_\text{YM}|_{\mathbb{R}^3}.
\end{equation}
The Euclidean Yang-Mills action is in fact both $Q$-exact and $\tilde{Q}$-exact because in this case, the commutator of $\smash{\delta_\xi', \delta_{\tilde{\xi}}'}$ in \eqref{3Dsusypalgebra-E} is the sum of a total derivative and a commutator, the latter of which vanishes inside the trace.

In curved Euclidean space, making appropriate modifications by the metric gives, e.g.,
\begin{equation}
\epsilon^{\alpha\beta}\gamma_{\gamma\alpha}^\mu\gamma_{\beta\delta}^\nu = g^{\mu\nu}\epsilon_{\gamma\delta} - i\sqrt{g}^{-1}\epsilon^{\mu\nu\rho}(\gamma_\rho)_{\gamma\delta}
\end{equation}
($\epsilon^{\mu\nu\rho}$ always denotes the Levi-Civita symbol; we make the corresponding tensor explicit by writing $\sqrt{g}^{-1}\epsilon^{\mu\nu\rho}$ or $\sqrt{g}\epsilon_{\mu\nu\rho}$).  We derive the curved-space SUSY' transformations from the bottom up, following \cite{Doroud:2012xw}.  One can construct the superconformal algebra on a curved space that admits a conformal Killing spinor by diffeomorphism- and Weyl-covariantizing the corresponding flat-space Poincar\'e supersymmetry (or SUSY') algebra, in addition to replacing the constant supersymmetry parameters by conformal Killing spinors.  The supplementary terms required for Weyl covariance depend on both the Weyl weights of the fields and the spacetime dimension.  The conformal Killing spinor (twistor spinor) condition is
\begin{equation}
\nabla_\mu\xi = \gamma_\mu\xi' \Longleftrightarrow \xi' = \frac{1}{3}\gamma^\mu\nabla_\mu\xi,
\label{CKScondition}
\end{equation}
with the intuition being that the spin-$3/2$ component of $\nabla_\mu\xi$ vanishes, and only the spin-$1/2$ component given by $\xi'$ survives.  The subalgebra of the superconformal algebra that generates only isometries (not conformal transformations) of the curved space is the curved-space analogue of the flat-space super-Poincar\'e algebra.

First recall some basic facts about Weyl transformations.  On an oriented Riemannian manifold, we define an orthonormal frame by $g_{\mu\nu} = e_\mu^a e_\nu^b\delta_{ab}$ and denote by $\smash{\omega_\mu^{ab} = -\omega_\mu^{ba}}$ the minimal spin connection (such that the vielbein $e_\mu^a$ is covariantly constant), which defines the diffeomorphism-covariant derivative $\nabla_\mu$ on tensors carrying internal indices (we denote the gauge-covariant derivative by $D_\mu$).  Under local $SO(d)$ Lorentz transformations $\Lambda_{ab} = \delta_{ab} + \epsilon_{ab} + O(\epsilon^2)$ with $\epsilon_{ab} = -\epsilon_{ba}$, we have
\begin{equation}
\omega_\mu\to \Lambda\omega_\mu\Lambda^{-1} - \partial_\mu\Lambda\Lambda^{-1}, \quad \psi\to S(\Lambda)\psi\equiv \smash{e^{\epsilon^{ab}\sigma_{ab}}}\psi,
\end{equation}
where $\sigma^{ab} = \frac{1}{8}[\gamma^a, \gamma^b]$ for consistency with $S(\Lambda)^{-1}\gamma^a S(\Lambda) = \Lambda^a{}_b\gamma^b$.  Hence on spinors,
\begin{equation}
\nabla_\mu\psi = (\partial_\mu + \omega_\mu^{ab}\sigma_{ab})\psi.
\label{nablamuonspinors}
\end{equation}
Under a Weyl transformation $\smash{e_\mu^a\to e^\Omega e_\mu^a}$, we have
\begin{equation}
\omega_\mu^{ab}\to \omega_\mu^{ab} + e_\mu^a\partial^b\Omega - e_\mu^b\partial^a\Omega.
\end{equation}
The Weyl weight $w(\varphi)$ of a field is the charge appearing in $\varphi\to e^{-w\Omega}\varphi$.  Our convention is that tensors with lower coordinate indices have zero Weyl weight (with raised indices, they gain a Weyl weight due to the metric) while tensors with tangent space indices generally have nonzero Weyl weight, with the difference being due to the transformation of the vielbein.  The opposite is true for the metric, which is not a dynamical field: Weyl transformations act nontrivially on $g_{\mu\nu}$ and leave the tangent space metric unchanged.  In particular, whether the gamma matrices are Weyl-invariant depends on whether they carry tangent space indices (yes) or coordinate indices (no).  Thus for a scalar $\varphi$ and fermion $\psi$ of Weyl weight $w$,
\begin{align}
\partial_\mu\varphi &\to e^{-w\Omega}(\partial_\mu - w\partial_\mu\Omega)\varphi, \label{scalarweyl} \\
\gamma^\mu\nabla_\mu\psi &\to e^{-(w + 1)\Omega}\gamma^\mu\nabla_\mu\psi + e^{-(w + 1)\Omega}\left(\frac{d - 1}{2} - w\right)\gamma^\mu\partial_\mu\Omega\psi. \label{fermionweyl}
\end{align}
Note that $w(\gamma_\mu) = -1$ and $w(g_{\mu\nu}) = -2$ (hence $w(\sqrt{g}) = -d$).

For the vector multiplet, our conventions fix the following Weyl weights:
\[
w(\sigma) = 1, \mbox{ } w(A_\mu) = 0, \mbox{ } w(\xi) = w(\tilde{\xi}) = -1/2, \mbox{ } w(\lambda) = w(\tilde{\lambda}) = 3/2, \mbox{ } w(D) = 2.\footnote{The conformal Killing spinor condition $\nabla_\mu\xi = \gamma_\mu\xi'$ is Weyl-covariant because Weyl transformations map conformal Killing spinors to conformal Killing spinors: $\xi'\to e^{-\Omega/2}(\xi' + \frac{1}{2}\gamma^\mu\partial_\mu\Omega\xi)$.}
\]
Aside from $A_\mu$ (which has $[A_\mu] = 1$), these coincide with the scaling dimensions.  In $d = 3$, we have from \eqref{scalarweyl} and \eqref{fermionweyl} that
\begin{equation}
w\left(\gamma^\mu\xi D_\mu\sigma + \frac{2}{3}\sigma\gamma^\mu\nabla_\mu\xi\right) = \frac{3}{2}, \quad w\left(\xi\gamma^\mu D_\mu\tilde{\lambda} - \frac{1}{3}\tilde{\lambda}\gamma^\mu\nabla_\mu\xi\right) = 2,
\label{weylcovariantvector}
\end{equation}
and similarly for $(\xi, \tilde{\lambda})\leftrightarrow (\tilde{\xi}, \lambda)$.  To Weyl-covariantize \eqref{3Dsusypvector-E}, we replace the terms $\gamma^\mu\xi D_\mu\sigma$ and $\xi\gamma^\mu D_\mu\tilde{\lambda}$ with the expressions of well-defined Weyl weight in \eqref{weylcovariantvector} (and similarly for $(\xi, \tilde{\lambda})\leftrightarrow (\tilde{\xi}, \lambda)$), leading to the curved-space SUSY' transformations for a vector multiplet:
\begin{align}
\delta'\sigma &= -(\xi\tilde{\lambda} - \tilde{\xi}\lambda), \nonumber \\
\delta' A_\mu &= i(\xi\gamma_\mu\tilde{\lambda} + \tilde{\xi}\gamma_\mu\lambda), \nonumber \\
\delta'\lambda &= \textstyle -i\xi D - i\gamma^\mu\xi D_\mu\sigma - \frac{i}{2}\sqrt{g}^{-1}\epsilon^{\mu\nu\rho}\gamma_\rho\xi F_{\mu\nu} - \frac{2i}{3}\sigma\gamma^\mu\nabla_\mu\xi, \vphantom{\tilde{\xi}} \label{3Dsusypvector-M3} \\
\delta'\tilde{\lambda} &= \textstyle i\tilde{\xi}D + i\gamma^\mu\tilde{\xi}D_\mu\sigma - \frac{i}{2}\sqrt{g}^{-1}\epsilon^{\mu\nu\rho}\gamma_\rho\tilde{\xi}F_{\mu\nu} + \frac{2i}{3}\sigma\gamma^\mu\nabla_\mu\tilde{\xi}, \nonumber \\
\delta' D &= \textstyle -(\xi\gamma^\mu D_\mu\tilde{\lambda} - \tilde{\xi}\gamma^\mu D_\mu\lambda) + [\xi\tilde{\lambda} + \tilde{\xi}\lambda, \sigma] + \frac{1}{3}(\tilde{\lambda}\gamma^\mu\nabla_\mu\xi - \lambda\gamma^\mu\nabla_\mu\tilde{\xi}) \nonumber
\end{align}
where $D_\mu(\cdot) = \nabla_\mu(\cdot) - i[A_\mu, (\cdot)]$ and $F_{\mu\nu} = \partial_\mu A_\nu - \partial_\nu A_\mu - i[A_\mu, A_\nu]$.  Now consider a chiral multiplet with non-canonical scaling dimensions
\[
([A], [\psi], [F]) = (\Delta, \Delta + 1/2, \Delta + 1),
\]
which coincide with the Weyl weights.  In $d = 3$, we have from \eqref{scalarweyl} and \eqref{fermionweyl} that
\begin{equation}
w\left(\gamma^\mu\tilde{\xi}D_\mu A + \frac{2\Delta}{3}A\gamma^\mu\nabla_\mu\tilde{\xi}\right) = \Delta + \frac{1}{2}, \quad w\left(\tilde{\xi}\gamma^\mu D_\mu\psi - \frac{2\Delta - 1}{3}\psi\gamma^\mu\nabla_\mu\tilde{\xi}\right) = \Delta + 1.
\label{weylcovariantchiral}
\end{equation}
To Weyl-covariantize \eqref{3Dsusypchiral-E}, we replace $\gamma^\mu\tilde{\xi}D_\mu A$ and $\tilde{\xi}\gamma^\mu D_\mu\psi$ with the expressions in \eqref{weylcovariantchiral}, leading to the curved-space SUSY' transformations for a fundamental chiral multiplet:
\begin{align}
\delta' A &= -\sqrt{2}\xi\psi, \nonumber \\
\delta'\psi &= \textstyle -\sqrt{2}\xi F + i\sqrt{2}(\gamma^\mu\tilde{\xi}D_\mu A + \frac{2\Delta}{3}A\gamma^\mu\nabla_\mu\tilde{\xi}) + i\sqrt{2}\tilde{\xi}\sigma A, \label{3Dsusypchiral-M3} \\
\delta' F &= \textstyle i\sqrt{2}(\tilde{\xi}\gamma^\mu D_\mu\psi - \frac{2\Delta - 1}{3}\psi\gamma^\mu\nabla_\mu\tilde{\xi}) - i\sqrt{2}\sigma\tilde{\xi}\psi - 2i\tilde{\xi}\tilde{\lambda}A \nonumber
\end{align}
where $D_\mu(\cdot) = \nabla_\mu(\cdot) - iA_\mu(\cdot)$.

Closure of the 3D $\mathcal{N} = 2$ superconformal algebra (i.e., the curved-space SUSY' algebra with conformal Killing spinor parameters) on the auxiliary fields requires a refinement of the conformal Killing spinor condition \cite{Hama:2010av, Jafferis:2010un, Marino:2011nm}, namely that $\xi, \tilde{\xi}$ satisfy both \eqref{CKScondition} and
\begin{equation}
\gamma^\mu\gamma^\nu\nabla_\mu\nabla_\nu\xi = h\xi \Longleftrightarrow 3\gamma^\mu\nabla_\mu\xi' = h\xi
\label{extracondition}
\end{equation}
for some scalar function $h$.  Using $[\nabla_\mu, \nabla_\nu]\psi = R_{\mu\nu}{}^{ab}\sigma_{ab}\psi$ on fermions and $R_{\mu\nu\rho\sigma}\gamma^\mu\gamma^\nu\gamma^\rho\gamma^\sigma = -2R$, we deduce that $h = -3R/8$.  Explicitly, the algebra is as follows.  For $\xi, \tilde{\xi}$ satisfying \eqref{CKScondition}, define the parameters
\begin{equation}
\begin{aligned}
U^\mu\equiv 2i\xi\gamma^\mu\tilde{\xi}, \quad \epsilon_{\mu\nu}\equiv 2\sqrt{g}\epsilon_{\mu\nu\rho}(\tilde{\xi}\nabla^\rho\xi - \xi\nabla^\rho\tilde{\xi}), \\
\rho\equiv \frac{2i}{3}\nabla_\mu(\xi\gamma^\mu\tilde{\xi}), \quad \alpha\equiv \frac{2i}{3}(\nabla_\mu\xi\gamma^\mu\tilde{\xi} - \xi\gamma^\mu\nabla_\mu\tilde{\xi})
\end{aligned}
\end{equation}
(note that $\nabla^{[\mu}U^{\nu]} = 2\epsilon^{\mu\nu}$).  For the vector multiplet, $[\delta_{\xi_1}', \delta_{\xi_2}'] = [\delta_{\tilde{\xi}_1}', \delta_{\tilde{\xi}_2}'] = 0$ on all fields and
\begin{align}
[\delta_\xi', \delta_{\tilde{\xi}}']\sigma &= U^\mu D_\mu\sigma + \rho\sigma, \nonumber \\
[\delta_\xi', \delta_{\tilde{\xi}}']A_\mu &= 2D_\mu(\xi\tilde{\xi}\sigma) + U^\nu F_{\nu\mu}, \nonumber \\
[\delta_\xi', \delta_{\tilde{\xi}}']F_{\mu\nu} &= U^\rho D_\rho F_{\mu\nu} + 2i\xi\tilde{\xi}[\sigma, F_{\mu\nu}] + \epsilon_{\mu\rho}F^\rho{}_\nu - \epsilon_{\nu\rho}F^\rho{}_\mu + 2\rho F_{\mu\nu}, \nonumber \\
[\delta_\xi', \delta_{\tilde{\xi}}']\lambda &= \textstyle U^\mu D_\mu\lambda + 2i\xi\tilde{\xi}[\sigma, \lambda] + \epsilon_{\mu\nu}\sigma^{\mu\nu}\lambda + \frac{3}{2}\rho\lambda - \alpha\lambda, \\
[\delta_\xi', \delta_{\tilde{\xi}}']\tilde{\lambda} &= \textstyle U^\mu D_\mu\tilde{\lambda} + 2i\xi\tilde{\xi}[\sigma, \tilde{\lambda}] + \epsilon_{\mu\nu}\sigma^{\mu\nu}\tilde{\lambda} + \frac{3}{2}\rho\tilde{\lambda} + \alpha\tilde{\lambda}, \nonumber \\
[\delta_\xi', \delta_{\tilde{\xi}}']D &= \textstyle U^\mu D_\mu D + 2i\xi\tilde{\xi}[\sigma, D] + 2\rho D + \frac{2i}{3}\sigma(\xi\gamma^\mu\gamma^\nu\nabla_\mu\nabla_\nu\tilde{\xi} - \tilde{\xi}\gamma^\mu\gamma^\nu\nabla_\mu\nabla_\nu\xi). \nonumber
\end{align}
The last term vanishes given \eqref{extracondition}.  Thus on gauge-covariant fields, the nonvanishing commutators are a sum of translation, gauge transformation, Lorentz rotation, dilatation, and $R$-rotation (the dimensions of all component fields and $R$-charges of $\lambda, \tilde{\lambda}$ can be read off from the above commutators).\footnote{Under a Lorentz transformation $F_{\mu\nu}\to \Lambda_\mu{}^\alpha\Lambda_\nu{}^\beta F_{\alpha\beta}$ where $\Lambda_{\alpha\beta} = g_{\alpha\beta} + \epsilon_{\alpha\beta} + O(\epsilon^2)$ with $\epsilon_{\alpha\beta} = -\epsilon_{\beta\alpha}$, the change in $F_{\mu\nu}$ is $\epsilon_{\mu\alpha}F^\alpha{}_\nu - \epsilon_{\nu\alpha}F^\alpha{}_\mu$.  The change in $\lambda$ under a rotation is $\epsilon^{\alpha\beta}\sigma_{\alpha\beta}\lambda$.}  For a fundamental chiral multiplet whose bottom component has scaling dimension $\Delta$, we have $\smash{[\delta_{\xi_1}', \delta_{\xi_2}'] = [\delta_{\tilde{\xi}_1}', \delta_{\tilde{\xi}_2}'] = 0}$ on all fields except for
\begin{equation}
[\delta_{\tilde{\xi}_1}', \delta_{\tilde{\xi}_2}']F = \frac{4\Delta}{3}(\tilde{\xi}_1\gamma^\mu\gamma^\nu\nabla_\mu\nabla_\nu\tilde{\xi}_2 - \tilde{\xi}_2\gamma^\mu\gamma^\nu\nabla_\mu\nabla_\nu\tilde{\xi}_1)A,
\end{equation}
which vanishes provided that \eqref{extracondition} is satisfied.  We then have that
\begin{align}
[\delta_\xi', \delta_{\tilde{\xi}}']A &= U^\mu D_\mu A + 2i\xi\tilde{\xi}\sigma A + \Delta\rho A - \Delta\alpha A, \nonumber \\
[\delta_\xi', \delta_{\tilde{\xi}}']\psi &= U^\mu D_\mu\psi + 2i\xi\tilde{\xi}\sigma\psi + \epsilon_{\mu\nu}\sigma^{\mu\nu}\psi + (\Delta + 1/2)\rho\psi + (1 - \Delta)\alpha\psi, \\
[\delta_\xi', \delta_{\tilde{\xi}}']F &= U^\mu D_\mu F + 2i\xi\tilde{\xi}\sigma F + (\Delta + 1)\rho F + (2 - \Delta)\alpha F. \nonumber
\end{align}
Thus the $R$-charges are fixed by $\Delta$.  Similar commutators hold for an antichiral multiplet.  These commutators can be written more succinctly in terms of the (gauge-covariant) Lie-Lorentz derivative \cite{Doroud:2012xw}.

Finally, the curved-space Chern-Simons action is given by
\begin{equation}
\mathcal{L}_\text{CS}|_{M^3} = \frac{k}{4\pi i}\Tr\left[\sqrt{g}^{-1}\epsilon^{\mu\nu\rho}\left(A_\mu\partial_\nu A_\rho - \frac{2i}{3}A_\mu A_\nu A_\rho\right) - 2\lambda\tilde{\lambda} + 2iD\sigma\right],
\label{LCS-curved}
\end{equation}
integrated against the standard volume element $\sqrt{g}\, d^3 x$.  We compute that
\begin{equation}
\delta'\mathcal{L}_\text{CS}|_{M^3} = \frac{k}{4\pi}\nabla_\mu\Tr[\sqrt{g}^{-1}\epsilon^{\mu\nu\rho}(\xi\gamma_\nu\tilde{\lambda} + \tilde{\xi}\gamma_\nu\lambda)A_\rho - 2(\xi\gamma^\mu\tilde{\lambda} - \tilde{\xi}\gamma^\mu\lambda)\sigma]
\label{LCS-curved-boundary}
\end{equation}
(compare to \eqref{LCS-boundary}).  Aside from possible boundary conditions, no special conditions need to be imposed on $\xi$ and $\tilde{\xi}$ to ensure SUSY'-invariance of the Chern-Simons action.

\section{Gauging SUSY NLSMs} \label{gaugingsusy}

Here, we describe how to gauge a global symmetry (under which the fields do not transform in a linear representation) at the nonlinear level while preserving global SUSY.

\subsection{K\"ahler Potential} \label{higherGamma}

We first consider a supersymmetric sigma model described by a K\"ahler potential.  We work in $0 + 1$ dimensions (the logic is the same in higher dimensions), namely $\mathcal{N} = 2$ quantum mechanics with K\"ahler target space, and illustrate the logic with the simplest example of a $\mathbb{CP}^1$ sigma model parametrized by a single complex scalar, with isometry group $SU(2)$ acting by linear fractional transformations (the example relevant for the main text).  Let $\Phi$ be a 1D $\mathcal{N} = 2$ chiral superfield.  In the patch containing the origin, $\mathbb{CP}^1$ has coordinates $\Phi, \Phi^\dag$ with K\"ahler potential $K = \log(1 + |\Phi|^2)$.  To gauge the action $S = \int dt\, d^2\theta\, K$, consider infinitesimal local $SU(2)$ transformations parametrized by chiral superfields $\Lambda_i$: $\delta_{SU(2)}\Phi = \Lambda_i X_i$.  The corresponding change in $K$ is given by \eqref{deltaK}.  We wish to cancel the terms involving $J_i$ in \eqref{deltaK}, leaving only a K\"ahler transformation, by means of an appropriate counterterm:
\begin{equation}
K\to K + \Gamma(\Phi, \Phi^\dag, V).
\end{equation}
The construction of $\Gamma$ satisfying $\delta_{SU(2)}\Gamma = i(\Lambda_i - \bar{\Lambda}_i)J_i$ is a special case of a more general problem.  An action invariant under $H$ can be promoted to an action invariant under $G\supset H$ by adding a counterterm constructed out of the original fields and fields $V$ parametrizing the coset $G/H$, with the condition that the counterterm vanishes when $V = 0$:
\begin{equation}
L_G(X, V) = L_H(X) + L_\text{ct}(X, V), \quad L_\text{ct}(X, 0) = 0.
\end{equation}
In particular, a vector superfield can be thought of as parametrizing $G_\mathbb{C}/G$, with gauge transformations corresponding to the action of $G_\mathbb{C}$ by left multiplication.

The interpretation of a vector superfield as a coset parameter proceeds as follows \cite{Wess:1992cp}.  Complexified gauge parameters implement gauge transformations in $G_\mathbb{C}$.  An arbitrary element of $G_\mathbb{C}$ can be written as
\begin{equation}
g = e^{v_i T_i}e^{iu_i T_i} = (\text{Hermitian})(\text{unitary})
\end{equation}
where $u_i$ and $v_i$ are real and $T_i$ are Hermitian ($u_i$ are coordinates on $G$, and $v_i$ are coordinates on $G_\mathbb{C}/G$).  The left cosets of $G$ in $G_\mathbb{C}/G$ are thus represented by
\begin{equation}
v = e^{v_i T_i}.
\end{equation}
$G_\mathbb{C}$ acts naturally on $G_\mathbb{C}/G$ by left multiplication, which can be written as a combination of $G$- and $G_\mathbb{C}/G$- transformations $v' = u_0 vu_0^{-1}$ and $v'^2 = v_0 v^2 v_0$ parametrized by
\begin{equation}
u_0 = e^{iu_{0i}T_i}\in G, \quad v_0 = e^{v_{0i}T_i}\in G_\mathbb{C}/G.
\end{equation}
The infinitesimal form is
\begin{equation}
\delta e^{2v_i T_i} = i(u_{0j} - iv_{0j})T_j e^{2v_i T_i} - ie^{2v_i T_i}(u_{0j} + iv_{0j})T_j\equiv i\epsilon e^{2v_i T_i} - ie^{2v_i T_i}\bar{\epsilon}.
\end{equation}
If we identify $\epsilon_i$ with the lowest component of $\Lambda_i$ and $v_i$ with the lowest component of $V_i$, then this transformation is simply the lowest component of the super gauge transformation
\begin{equation}
\delta e^{2V} = i\Lambda e^{2V} - ie^{2V}\bar{\Lambda} \Longleftrightarrow e^{2V'} = e^{i\Lambda}e^{2V}e^{-i\bar{\Lambda}}
\end{equation}
where $V = V_i T_i$ and $\Lambda = \Lambda_i T_i$.

In our case of interest, $G = SU(2)$ with $SU(2)_\mathbb{C} = SL(2, \mathbb{C})$.  On a function of $\Phi, \Phi^\dag, V$, a local $SU(2)$ variation can be written as
\begin{equation}
\delta_{SU(2)} = \Lambda_i X_i\partial_\Phi + \bar{\Lambda}_i \bar{X}_i\partial_{\Phi^\dag} + \delta_{SU(2)} V\partial_V = \operatorname{Re}(\Lambda_i)\mathcal{P}_i + i\operatorname{Im}(\Lambda_i)\mathcal{O}_i
\end{equation}
where (suppressing variations of $V$)
\begin{equation}
\mathcal{P}_i = X_i\partial_\Phi + \bar{X}_i\partial_{\Phi^\dag} + \cdots\equiv P_i + \cdots, \quad \mathcal{O}_i = X_i\partial_\Phi - \bar{X}_i\partial_{\Phi^\dag} + \cdots\equiv O_i + \cdots
\end{equation}
(Re and Im are shorthand for the appropriate linear combinations of $\Lambda_i$ and $\bar{\Lambda}_i$, which are not chiral superfields).  The $O_i$ satisfy $[O_i, O_j] = \epsilon_{ijk}O_k$ and $O_i J_j = O_j J_i = i(2J_i J_j - \frac{1}{2}\delta_{ij})$.

We focus on the $\mathcal{O}_i$ part in $\delta_{SU(2)}$ because the undesirable terms in \eqref{deltaK} involve $\operatorname{Im}(\Lambda_i)$.  Clearly, $K$ is invariant (up to a K\"ahler transformation) under gauge transformations with $\Lambda_i$ purely real (i.e., under $G\subset G_{\mathbb{C}}$).  With $\Lambda_i$ purely imaginary, gauge transformations (in the part of $G_{\mathbb{C}}$ not in $G$, namely $G_{\mathbb{C}}/G$) can be implemented by the $\mathcal{O}_i$.  Now we do a formal manipulation (forgetting that the parameters must be chiral superfields): let $\Lambda_i = iR_i$ with $R_i$ real to isolate $\mathcal{O}_i$; then up to K\"ahler transformations,
\begin{equation}
\delta_{SU(2)}K = -i(\Lambda_i - \bar{\Lambda}_i)J_i = 2R_i J_i = iR_i\mathcal{O}_i K = iR_i O_i K.
\end{equation}
To cancel this variation, we demand that $iR_i\mathcal{O}_i\Gamma = -2R_i J_i$, subject to the boundary condition $\Gamma(\Phi, \Phi^\dag, V = 0) = 0$.  Exponentiating, we want:
\begin{equation}
(e^{iR_i\mathcal{O}_i} - 1)\Gamma = \frac{e^{iR_i\mathcal{O}_i} - 1}{iR_j\mathcal{O}_j}(-2R_k J_k).
\end{equation}
If we take $R_i = V_i$, then $e^{iR_i\mathcal{O}_i}$ transforms $V$ to zero.  Thus by the boundary condition,
\begin{equation}
\Gamma = \frac{e^{iV_i\mathcal{O}_i} - 1}{iV_j\mathcal{O}_j}(2V_k J_k) = \frac{e^{iV_i O_i} - 1}{iV_j O_j}(2V_k J_k) = 2\int_0^1 d\alpha\, e^{i\alpha V_i O_i}V_j J_j
\end{equation}
where the derivatives in $\mathcal{O}$ do \emph{not} act on the $V$, thus justifying the replacement $\mathcal{O}\to O$.  For completeness, one should check that $\mathcal{P}_i\Gamma = 0$, which we do not do explicitly here.

\subsection{Higher-Derivative Terms} \label{higherGammap}

Our applications require gauging supersymmetric sigma models containing higher-derivative terms involving the Grassmann-odd superfield $D\Phi$.  Concretely, consider $K'$ in \eqref{Kprime}, associated to which are fermionic Noether currents as in \eqref{Kprimecurrents}, for which we wish to construct a counterterm $\Gamma'$ satisfying \eqref{Gammaprimecondition}.  First define the bosonic operators
\begin{equation}
O_i' = X_i\partial_\Phi - \bar{X}_i\partial_{\Phi^\dag} + DX_i\partial_{D\Phi} - (DX_i)^\dag\partial_{(D\Phi)^\dag} \implies O_i' K' = -2iJ_i'
\end{equation}
as well as the fermionic operators
\begin{equation}
P_i = X_i\partial_{D\Phi} \implies P_i K' = -iI_i, \mbox{ } P_i^\dag K' = iI_i^\dag
\end{equation}
(note that $D^\dag\Phi^\dag = -(D\Phi)^\dag$ and $(\partial_{D\Phi})^\dag = -\smash{\partial_{(D\Phi)^\dag}}$).  We may write
\begin{equation}
\delta_{SU(2)}K' = -i(\Lambda_i - \Lambda_i^\dag)J_i' - iD(\Lambda_i - \Lambda_i^\dag)I_i - iD^\dag(\Lambda_i - \Lambda_i^\dag)I_i^\dag
\end{equation}
by virtue of $D\Lambda_i^\dag = D^\dag\Lambda_i = 0$, so that if $\Lambda_i = i\xi_i$ with $\xi_i$ real, then
\begin{align}
\delta_{SU(2)}K' &= 2(\xi_i J_i' + D\xi_i I_i + D^\dag\xi_i I_i^\dag) \nonumber \\
&= i(\xi_i O_i' + 2D\xi_i P_i - 2D^\dag\xi_i P_i^\dag)K'.
\end{align}
To cancel this variation, we demand that
\begin{equation}
i(\xi_i O_i' + 2D\xi_i P_i - 2D^\dag\xi_i P_i^\dag)\Gamma' = -2(\xi_i J_i' + D\xi_i I_i + D^\dag\xi_i I_i^\dag),
\end{equation}
subject to the usual boundary condition.  This exponentiates to give
\[
(e^{i(\xi_i O_i' + 2D\xi_i P_i - 2D^\dag\xi_i P_i^\dag)} - 1)\Gamma' = -\frac{e^{i(\xi_i O_i' + 2D\xi_i P_i - 2D^\dag\xi_i P_i^\dag)} - 1}{i(\xi_j O_j' + 2D\xi_j P_j - 2D^\dag\xi_j P_j^\dag)}2(\xi_k J_k' + D\xi_k I_k + D^\dag\xi_k I_k^\dag).
\]
Setting $\xi_i = V_i$, the boundary condition yields
\begin{align}
\Gamma' &= \frac{e^{i(V_i O_i' + 2DV_i P_i - 2D^\dag V_i P_i^\dag)} - 1}{i(V_j O_j' + 2DV_j P_j - 2D^\dag V_j P_j^\dag)}2(V_k J_k' + DV_k I_k + D^\dag V_k I_k^\dag) \nonumber \\
&= 2\int_0^1 d\alpha\, e^{i\alpha(V_i O_i' + 2DV_i P_i - 2D^\dag V_i P_i^\dag)}(V_j J_j' + DV_j I_j + D^\dag V_j I_j^\dag).
\end{align}
Let us examine this result (and verify invariance under local $SU(2)$) in Wess-Zumino gauge.  First, we have
\begin{equation}
DV_i = -(D^\dag V_i)^\dag = \psi_i + \theta^\dag(A_i - i\dot{a}_i) + i\theta\theta^\dag\dot{\psi}_i \stackrel{\text{WZ}}{=} \theta^\dag A_i.
\end{equation}
Therefore, expanding gives
\begin{equation}
\Gamma' \stackrel{\text{WZ}}{=} 2(V_i J_i' + DV_i I_i + D^\dag V_i I_i^\dag) + DV_i(DV_j)^\dag(\delta_{ij} - 4J_i J_j + 2i\epsilon_{ijk}J_k),
\end{equation}
where we have used
\begin{equation}
iP_i\bar{I}_j = \frac{X_i\bar{X}_j}{(1 + |\Phi|^2)^2} = \frac{1}{4}\delta_{ij} - J_i J_j + \frac{i}{2}\epsilon_{ijk}J_k.
\end{equation}
Using the properties \eqref{deltaJ}, \eqref{gaugeV}, \eqref{deltaKprime}, $D\Lambda_i = -2i\theta^\dag\dot{\epsilon}_i$,
\begin{equation}
\delta_{SU(2)}I_i = -i(D\Lambda_j)^\dag J_j J_i - \frac{1}{2}\epsilon_{ijk}(D\Lambda_j)^\dag J_k + \frac{i}{4}(D\Lambda_i)^\dag + 2i\Lambda_j I_j J_i - 2i\bar{\Lambda}_j J_j I_i,
\end{equation}
and $I_i = -2i\epsilon_{ijk}J_j I_k$, we compute that $\delta_{SU(2)}\Gamma'$ is as expected, \eqref{Gammaprimecondition}.

\section{Details on the \texorpdfstring{$\mathcal{N} = 2$}{N = 2} Coadjoint Orbit}

\subsection{\texorpdfstring{$\mathbb{CP}^1$}{CP1} Sigma Model} \label{sigmadetails}

Our conventions for the geometry of $\mathbb{CP}^1$ are as follows.  Stereographic projection yields the relation $z = e^{i\varphi}/\tan(\theta/2)$ between the coordinate $z$ on $\mathbb{CP}^1$ and spherical coordinates $(\theta, \varphi)$ on $S^2$ (we use $\varphi$ to avoid confusion with $\phi$).  We thus get
\begin{equation}
ds^2 = \frac{\operatorname{Re}(dz\otimes d\bar{z})}{(1 + |z|^2)^2} = \frac{1}{4}(d\theta^2 + \sin^2\theta\, d\varphi^2),
\end{equation}
the round metric on $S^2$ of radius $1/2$.  In this setup, the projection is done from the north pole of the $S^2$ at $(0, 0, 1)^T$.  The adjoint action of $SU(2)$ on $\mathfrak{su}(2)\cong \mathbb{R}^3$ descends to an action on $S^2\subset \mathbb{R}^3$, giving rise to a two-to-one map $SU(2)\to SO(3)$:
\begin{equation}
\pm\left(\begin{array}{cc} a & b \\ -\bar{b} & \bar{a} \end{array}\right)\mapsto \left(\begin{array}{ccc} \operatorname{Re}(a^2 - b^2) & \operatorname{Im}(a^2 + b^2) & -2\operatorname{Re}(ab) \\ -\operatorname{Im}(a^2 - b^2) & \operatorname{Re}(a^2 + b^2) & 2\operatorname{Im}(ab) \\ 2\operatorname{Re}(a\bar{b}) & 2\operatorname{Im}(a\bar{b}) & |a|^2 - |b|^2 \end{array}\right).
\label{mapone}
\end{equation}
Alternatively, $SU(2)$ acts on $\mathbb{CP}^1$ by linear fractional transformations:
\begin{equation}
z\to \frac{az + b}{-\bar{b}z + \bar{a}}, \quad K\to K - \log(\bar{a} - \bar{b}z) - \log(a - b\bar{z}).
\end{equation}
Combined with stereographic projection, this results in the following map $SU(2)\to SO(3)$:
\begin{equation}
\pm\left(\begin{array}{cc} a & b \\ -\bar{b} & \bar{a} \end{array}\right)\mapsto \left(\begin{array}{ccc} \operatorname{Re}(a^2 - b^2) & -\operatorname{Im}(a^2 + b^2) & -2\operatorname{Re}(ab) \\ \operatorname{Im}(a^2 - b^2) & \operatorname{Re}(a^2 + b^2) & -2\operatorname{Im}(ab) \\ 2\operatorname{Re}(a\bar{b}) & -2\operatorname{Im}(a\bar{b}) & |a|^2 - |b|^2 \end{array}\right).
\label{maptwo}
\end{equation}
We use the latter convention.  The sign differences between \eqref{mapone} and \eqref{maptwo} have the following consequence.  The point $(\sin\theta\cos\varphi, \sin\theta\sin\varphi, \cos\theta)^T$ of $S^2$ transforms via the image of \eqref{infinitesimal} under the map \eqref{maptwo}, which coincides with the action of the standard rotation generators
\begin{equation}
\vec{D} = -i(-\sin\varphi\partial_\theta - \cos\varphi\cot\theta\partial_\varphi, \cos\varphi\partial_\theta - \sin\varphi\cot\theta\partial_\varphi, \partial_\varphi),
\label{dee}
\end{equation}
satisfying $[D_i, D_j] = i\epsilon_{ijk}D_k$, only after flipping the sign of $\epsilon_2$ in \eqref{infinitesimal}.  Finally, by ``Hopf map,'' we mean the map $SU(2)\to S^2$ that sends a given element of $SU(2)$ to the point to which it sends the north pole $(0, 0, 1)^T$ of $S^2$, according to \eqref{maptwo}.  Stereographic projection then allows us to identify an $SU(2)$ element with the point $z = -a/\bar{b}$ of $\mathbb{CP}^1$.

In writing \eqref{Lforg} as a sigma model to $\mathbb{CP}^1$, we identified $g = (\begin{smallmatrix} a & b \\ -\bar{b} & \bar{a} \end{smallmatrix}) = (\begin{smallmatrix} a & r \\ -r & \bar{a} \end{smallmatrix})$ with the complex scalar $\phi$ (playing the role of $z$ in the previous paragraph) via the Hopf map and an appropriate (partial) gauge fixing.  To see that the map \eqref{gtophi} is equivariant with respect to the action of $SU(2)$, consider an arbitrary $SU(2)$ transformation, either global or local:
\[
g\to \left(\begin{array}{cc} a_\ell & b_\ell \\ -\bar{b}_\ell & \bar{a}_\ell \end{array}\right)g = \left(\begin{array}{cc} a_\ell a - b_\ell r & a_\ell r + b_\ell\bar{a} \\ -\bar{a}_\ell r - \bar{b}_\ell a & \bar{a}_\ell\bar{a} - \bar{b}_\ell r \end{array}\right)
\]
(``$\ell$'' stands for ``left'').  To preserve the reality condition on $b$, as required by \eqref{gtophi}, we must again gauge away the phase of the off-diagonal components.  This is achieved by multiplying $a_\ell a - b_\ell r$ by $-e^{i\theta}$ and $a_\ell r + b_\ell\bar{a}$ by $-e^{-i\theta}$ where $\theta$ is the phase of $a_\ell r + b_\ell\bar{a}$ (the minus sign preserves the ``$(a, r) = (+, -)$'' convention), so that
\[
a\to a' = -\frac{(a_\ell a - b_\ell r)(a_\ell r + b_\ell\bar{a})}{|a_\ell r + b_\ell\bar{a}|}, \quad r\to r' = -|a_\ell r + b_\ell\bar{a}|, \quad \phi\to \phi' = \frac{a_\ell\phi + b_\ell}{-\bar{b}_\ell\phi + \bar{a}_\ell}
\]
under $SU(2)$.  Using \eqref{gtophi}, we indeed find that $(a', r') = (\phi', -1)/\sqrt{1 + |\phi'|^2}$, for which the minus sign in $r(\phi)$ is crucial.

Now, as in \eqref{infinitesimal}, let us drop the $\ell$ subscripts on the $SU(2)$ transformation parameters.  Under global $SU(2)$, the Wess-Zumino term $\mathcal{L}_0$ in \eqref{L0} picks up a total derivative,
\begin{equation}
\mathcal{L}_0\to \mathcal{L}_0 + i\partial_t\log\left(\frac{-\bar{b}\phi + \bar{a}}{-b\phi^\dag + a}\right),
\end{equation}
while $\mathcal{L}_A$ in \eqref{LA} is invariant.  Under local $SU(2)$, the variation of their sum $\mathcal{L} = \mathcal{L}_0 + \mathcal{L}_A$ (more conveniently written in terms of the $\epsilon_i$ in \eqref{infinitesimal}) is
\begin{equation}
\delta_{SU(2)}\mathcal{L} = \partial_t\left(\epsilon_3 - \frac{\epsilon_1 + i\epsilon_2}{2}\phi - \frac{\epsilon_1 - i\epsilon_2}{2}\phi^\dag\right).
\end{equation}
Thus the $SU(2)$ gauge invariance of the $g$ Lagrangian \eqref{Lforg} only holds up to total derivatives in the $\phi$ Lagrangian $\mathcal{L}$ due to the necessity of gauge-fixing the action of $U(1)$ on the right.  Upon promoting $\phi$ to the superfield $\Phi$, global $SU(2)$ acts as
\begin{equation}
\Phi\to \frac{a\Phi + b}{-\bar{b}\Phi + \bar{a}}\Longleftrightarrow (\phi, \psi, \dot{\phi})\to \bigg(\frac{a\phi + b}{-\bar{b}\phi + \bar{a}}, \frac{\psi}{(-\bar{b}\phi + \bar{a})^2}, \frac{\dot{\phi}}{(-\bar{b}\phi + \bar{a})^2}\bigg),
\end{equation}
under which the $\psi^\dag\psi$ term in $\tilde{\mathcal{L}}_0$ \eqref{K} and the $\dot{\phi}\dot{\phi}^\dag$ term in $\mathcal{L}'$ \eqref{Kprime} are invariant.

\subsection{Effective Action} \label{effactiondetails}

Upon changing variables to $\psi' = \psi/(1 + |\phi|^2)$, the path integral for $\tilde{\mathcal{L}}_\text{tot}$ acquires a Jacobian that renormalizes the action by $-2\tr\log(1 + |\phi|^2)$.  Integrating by parts, we have
\begin{equation}
\tilde{\mathcal{L}}_\text{tot}\supset j\mathcal{L} + \frac{i(\psi^\dag\dot{\psi} - \dot{\psi}^\dag\psi)}{2\mu(1 + |\phi|^2)^2} + \frac{(\mathcal{L}/\mu - j)\psi^\dag\psi}{(1 + |\phi|^2)^2} = j\mathcal{L} + \psi'^\dag\left(\frac{i\partial_t + \mathcal{L}}{\mu} - j\right)\psi'.
\end{equation}
Performing the path integral over $\psi$ thus generates the effective action
\begin{equation}
\tr\log\left(\frac{i\partial_t}{\mu} - j\right) + \sum_{n=1}^\infty \frac{(-1)^{n+1}}{n}\tr X^n
\end{equation}
where $X\equiv (i\partial_t - \mu')^{-1}\mathcal{L}$ and $\mu'\equiv j\mu$.  Inserting complete sets of states\footnote{Normalized as $\langle t|E\rangle = e^{-iEt} \implies \int \frac{dE}{2\pi}\, |E\rangle\langle E| = \int \frac{dt}{2\pi}\, |t\rangle\langle t| = 1$.} yields
\begin{align*}
\tr X^n &= \int \frac{dt_1\cdots dt_n\, dE_1\cdots dE_n}{(2\pi)^{2n-1}}\, \langle t_n|E_1\rangle\langle E_1|\left(\frac{\mathcal{L}(t_1)}{E_1 - \mu'}\right)|t_1\rangle\langle t_1|\cdots |E_n\rangle\langle E_n|\left(\frac{\mathcal{L}(t_n)}{E_n - \mu'}\right)|t_n\rangle \\
&= \int \frac{dE_1\cdots dE_{n-1}}{(2\pi)^{n-1}}\, G(E_1, \ldots, E_{n-1})\tilde{\mathcal{L}}(E_1)\cdots \tilde{\mathcal{L}}(E_{n-1})\tilde{\mathcal{L}}(-E_1 - \cdots - E_{n-1}) \vphantom{\frac{\tilde{F}}{E}}
\end{align*}
with $\tilde{\mathcal{L}}(E)\equiv \frac{1}{2\pi}\int dt\, e^{iEt}\mathcal{L}(t)$ and integral kernel
\begin{equation}
G(E_1, \ldots, E_{n-1}) = \int \frac{dE}{E(E + E_1)\cdots (E + E_1 + \cdots + E_{n-1})},
\end{equation}
whose $\mu'$-dependence drops out after suitably redefining $E$.  For $n > 1$, one can regulate this integral by shifting the contour of integration to $\mathbb{R}\pm i\epsilon$, in which case it obviously vanishes (the integrand falls off faster than $1/E$, and the sum of the residues vanishes).  When $n = 1$,
\begin{equation}
\tr X = \int \frac{dt\, dE}{2\pi}\, \langle t|E\rangle\langle E|\left(\frac{\mathcal{L}}{i\partial_t - \mu'}\right)|t\rangle = \frac{1}{2\pi}\int \frac{dE}{E}\int dt\, \mathcal{L}.
\end{equation}
In this case, the $E$ integral evaluates to $\pm i\pi$ depending on how we shift the pole away from the real line.  Hence the only relevant term in the one-loop effective action is
\begin{equation}
\tr X = \pm\frac{i}{2}\int dt\, \mathcal{L},
\end{equation}
namely the tadpole with one external $\mathcal{L}$ leg (we have ignored the vacuum energy).

\subsection{Canonical Quantization} \label{canonicalquantization}

\subsubsection{Bosonic System}

We begin by working in spherical coordinates $(\theta, \varphi)$ to make the connection to the monopole problem manifest.  It is instructive to quantize the system $L_B$ in arbitrary gauge.  For simplicity, we consider only longitudinal gauges for the monopole vector potential, parametrized by $\alpha$ in \eqref{LB}, that preserve the classical $U(1)$ symmetry manifest when $\alpha = 0$:
\begin{equation}
\vec{A} = \frac{(j - \alpha) + j\cos\theta}{r\sin\theta}\hat{\varphi}.
\end{equation}
The gauges $S$, $E$, and $N$ that we have defined by setting $\alpha = (0, j, 2j)$ are good near the south pole ($= 0$), equator, and north pole ($= \infty$), respectively.

At finite $\mu$, the phase space is $(2 + 2)$-dimensional.  The classical Lagrangian \eqref{LB} is
\[
L_B = ((j - \alpha) + j\cos\theta)\dot{\varphi} + \frac{\dot{\theta}^2}{2\mu} + \frac{\sin^2\theta\dot{\varphi}^2}{2\mu},
\]
from which we obtain the canonical momenta and classical Hamiltonian
\[
\pi_\theta = \frac{\dot{\theta}}{\mu}, \quad \pi_\varphi = (j - \alpha) + j\cos\theta + \frac{\sin^2\theta\dot{\varphi}}{\mu}, \quad H_\text{cl} = \frac{\mu}{2}\left[\left(\frac{\pi_\varphi - (j - \alpha) - j\cos\theta}{\sin\theta}\right)^2 + \pi_\theta^2\right].
\]
The ``good'' angular momentum operators $\smash{\vec{L}}$ in the presence of a monopole are well-known generalizations of the standard $\smash{\vec{D}}$ in \eqref{dee}, giving the quantum Hamiltonian \eqref{Hj} (with a subscript $j$ to emphasize the spin) where
\begin{equation}
\vec{L} = \vec{D} + \left(\frac{(j + (j - \alpha)\cos\theta)\cos\varphi}{\sin\theta}, \frac{(j + (j - \alpha)\cos\theta)\sin\varphi}{\sin\theta}, \alpha - j\right),
\label{Lquantums2withmu}
\end{equation}
which satisfy $[L_i, L_j] = i\epsilon_{ijk}L_k$.  The corresponding classical expressions $\smash{\vec{L}_\text{cl}}$, obtained simply by substituting $\pi_\theta$ and $\pi_\varphi$ for $-i\partial_\theta$ and $-i\partial_\varphi$ in \eqref{dee} and \eqref{Lquantums2withmu}, satisfy the expected Poisson brackets $[(L_i)_\text{cl}, (L_j)_\text{cl}]_\text{PB} = \epsilon_{ijk}(L_k)_\text{cl}$ with respect to $(\theta, \pi_\theta, \varphi, \pi_\varphi)$.\footnote{The $\vec{L}_\text{cl}$ can also be written in a gauge-independent manner in terms of the velocities $\dot{\theta}, \dot{\varphi}$:
\[
\vec{L}_\text{cl} = j(\sin\theta\cos\varphi, \sin\theta\sin\varphi, \cos\theta) + O(1/\mu).
\]
In this form, the $\mu\to\infty$ limit is manifest.}

At $\mu = \infty$, the phase space is $(1 + 1)$-dimensional and we have $\pi_\varphi = (j - \alpha) + j\cos\theta$, $\pi_\theta = 0$.  Hence there is a distinguished polarization in which $\varphi$ is the canonical coordinate:
\begin{equation}
(L_1)_\text{cl} = \cos\varphi\sqrt{j^2 - (L_3)_\text{cl}^2}, \quad (L_2)_\text{cl} = \sin\varphi\sqrt{j^2 - (L_3)_\text{cl}^2}, \quad (L_3)_\text{cl} = \pi_\varphi - (j - \alpha).
\end{equation}
The Poisson brackets on the reduced phase space $(\varphi, \pi_\varphi)$ take the same form as for finite $\mu$.  The corresponding quantum operators satisfying $[L_3, L_\pm] = \pm L_\pm$ and $[L_+, L_-] = 2L_3$ are
\begin{equation}
L_\pm = \sqrt{j\pm L_3}e^{\pm i\varphi}\sqrt{j\mp L_3}, \quad L_3 = -i\partial_\varphi - (j - \alpha),
\end{equation}
where we have set $L_\pm = L_1\pm iL_2$.

It is easy to recast the above statements in complex coordinates $(\phi, \phi^\dag)$.  At finite $\mu$, we immediately obtain the quantum angular momentum operators \eqref{Lquantumwithmu}.  Their classical counterparts $(L_i)_\text{cl}$, obtained simply by substituting $\pi_\phi$ for $-i\partial_\phi$ and $\pi_{\phi^\dag}$ for $-i\partial_{\phi^\dag}$, satisfy the expected Poisson brackets on the full phase space $(\phi, \pi_\phi, \phi^\dag, \pi_{\phi^\dag})$.  On the other hand, the classical canonical momenta following from \eqref{LB} are
\begin{equation}
\pi_\phi = (\pi_{\phi^\dag})^\dag = -\frac{ij\phi^\dag}{1 + |\phi|^2} + \frac{i\alpha}{2\phi} + \frac{2\dot{\phi}^\dag}{\mu(1 + |\phi|^2)^2}.
\label{momentawithmu}
\end{equation}
While canonical quantization does not fix the quantum representations of these momenta, we infer with the aid of \eqref{momentawithmu} that the quantization rules for the velocities are
\begin{align}
\frac{2i\dot{\phi}^\dag}{\mu} &\xrightarrow{q} (1 + |\phi|^2)^2\frac{\partial}{\partial\phi} - j\phi^\dag(1 + |\phi|^2) + \frac{\alpha}{2\phi}(1 + |\phi|^2)^2, \\
-\frac{2i\dot{\phi}}{\mu} &\xrightarrow{q} -(1 + |\phi|^2)^2\frac{\partial}{\partial\phi^\dag} - j\phi(1 + |\phi|^2) + \frac{\alpha}{2\phi^\dag}(1 + |\phi|^2)^2,
\end{align}
where ``$\xrightarrow{q}$'' means ``is represented quantumly by.''  After accounting for ordering ambiguities in the classical Hamiltonian, these rules indeed lead to the correct quantum Hamiltonian \eqref{Hj}.  At $\mu = \infty$, $\phi$ and $\phi^\dag$ are no longer independent canonical variables.  Letting the coordinate be $\phi$, we distinguish the corresponding momentum on the reduced phase space from that in \eqref{momentawithmu} with an extra subscript: $(\pi_\phi)_\text{red} = 2\lim_{\mu\to\infty} \pi_\phi$ (the factor of two comes from integration by parts), in terms of which
\begin{equation}
(L_+)_\text{cl} = -i\phi^2(\pi_\phi)_\text{red} + (2j - \alpha)\phi, \quad (L_-)_\text{cl} = i(\pi_\phi)_\text{red} + \frac{\alpha}{\phi}, \quad (L_3)_\text{cl} = i\phi(\pi_\phi)_\text{red} - (j - \alpha).
\label{Lclassicalnomu}
\end{equation}
Note that this holomorphic polarization differs from the $\varphi$-based polarization that we used in spherical coordinates, which treats $\phi$ and $\phi^\dag$ equally.  The Poisson brackets with respect to $(\phi, (\pi_\phi)_\text{red})$ are then as expected: $[(L_3)_\text{cl}, (L_\pm)_\text{cl}]_\text{PB} = \mp i(L_\pm)_\text{cl}$ and $[(L_+)_\text{cl}, (L_-)_\text{cl}]_\text{PB} = -2i(L_3)_\text{cl}$.  Representing $(\pi_\phi)_\text{red}$ by $-i\partial_\phi$ gives the correct quantum operators \eqref{Lquantumnomu}.

\subsubsection{Supersymmetric System}

Here, for simplicity, we set $\alpha = 0$ ($S$ gauge), keep $\mu$ finite, and work in complex coordinates $\phi, \phi^\dag$.  The canonical momenta for $L_B + L_F$ are
\begin{equation}
\pi_\phi = -\left[j + \frac{\psi^\dag\psi}{\mu(1 + |\phi|^2)^2}\right]\frac{i\phi^\dag}{1 + |\phi|^2} + \frac{2\dot{\phi}^\dag}{\mu(1 + |\phi|^2)^2}, \quad \pi_\psi = \frac{i\psi^\dag}{2\mu(1 + |\phi|^2)^2},
\end{equation}
where $\pi_{\phi^\dag} = \pi_\phi^\dag$ and $\pi_{\psi^\dag} = -\pi_\psi^\dag$ (note that $\psi$ and $\psi^\dag$ are not independent canonical coordinates).  Defining $\chi$ as in \eqref{chi}, the classical Hamiltonian can be written as
\begin{equation}
H_\text{cl}' = \frac{\mu}{2}(1 + |\phi|^2)^2\left[\pi_\phi + (j + \chi^\dag\chi)\frac{i\phi^\dag}{1 + |\phi|^2}\right]\left[\pi_{\phi^\dag} - (j + \chi^\dag\chi)\frac{i\phi}{1 + |\phi|^2}\right] + j\mu\chi^\dag\chi.
\end{equation}
To quantize, we impose the canonical (anti)commutation relations
\begin{equation}
[\phi, \pi_\phi] = [\phi^\dag, \pi_{\phi^\dag}] = i, \quad \{\chi, \chi^\dag\} = 1.
\end{equation}
The last relation introduces (in addition to the ambiguities already present when quantizing $H_\text{cl}$) fermion ordering ambiguities in $H_\text{cl}'$, which allow us to determine the quantum Hamiltonian corresponding to $H_\text{cl}'$ up to two constants $c_{1, 2}$:
\begin{equation}
H' = H_{j - c_1} + \frac{\mu}{2}(1 + |\phi|^2)\left[2j + \frac{(1 - 2c_1)|\phi|^2}{1 + |\phi|^2} - \left(\phi\frac{\partial}{\partial\phi} - \phi^\dag\frac{\partial}{\partial\phi^\dag}\right)\right]\chi^\dag\chi - c_2 j\mu.
\end{equation}
Hermiticity requires that $c_{1, 2}$ be real numbers, and the Dirac quantization condition further requires that $c_1$ be a half-integer.  It turns out that $H'$ can be diagonalized separately in the bosonic and fermionic sectors of the Hilbert space
\begin{equation}
(L^2(S^2, \mathbb{C})\otimes |0\rangle)\oplus (L^2(S^2, \mathbb{C})\otimes \chi^\dag|0\rangle)
\end{equation}
by the ``good'' angular momentum operators \eqref{Lquantumwithmu}:
\begin{align}
H' &= \left(\begin{array}{cc} H_{j - c_1} - c_2 j\mu & 0 \\ 0 & H_{j - c_1 + 1} + (1 - c_2)j\mu \end{array}\right) \\
&= \frac{\mu}{2}\left(\begin{array}{cc} \ell_b(\ell_b + 1) - (j - c_1)^2 - 2c_2 j & 0 \\ 0 & \ell_f(\ell_f + 1) - (j - c_1 + 1)^2 + 2(1 - c_2)j \end{array}\right),
\end{align}
where $\ell_b\geq j - c_1$ and $\ell_f\geq j - c_1 + 1$.  The constants $c_{1, 2}$ are then uniquely fixed by demanding that the quantum theory be supersymmetric, namely that each positive energy level have equal numbers of bosonic and fermionic states and that SUSY not be spontaneously broken.  Under these conditions, we find that $c_1 = c_2 + 1/4j = 1/2$, leading to precisely the quantum Hamiltonian stated in \eqref{Hpmatrix}.  In particular, $\ell_{b, f}$ are half-integers and the $2j$ bosonic ground states of zero energy occur at $\ell_b = j - 1/2$.  Having fixed $c_{1, 2}$, it is convenient to note that the ``fermionic'' monopole angular momenta
\begin{equation}
\vec{L}_f = \vec{L}|_{j + \chi^\dag\chi - 1/2}\equiv \vec{D} + (j + \chi^\dag\chi - 1/2)\left(\frac{\phi + \phi^\dag}{2}, \frac{\phi - \phi^\dag}{2i}, -1\right)
\label{Lfquantum}
\end{equation}
(compare to \eqref{Lquantums2withmu}), where
\begin{equation}
\vec{D} = \left(\frac{1 - \phi^2}{2}\frac{\partial}{\partial\phi} - \frac{1 - (\phi^\dag)^2}{2}\frac{\partial}{\partial\phi^\dag}, -\frac{1 + \phi^2}{2i}\frac{\partial}{\partial\phi} - \frac{1 + (\phi^\dag)^2}{2i}\frac{\partial}{\partial\phi^\dag}, \phi\frac{\partial}{\partial\phi} - \phi^\dag\frac{\partial}{\partial\phi^\dag}\right)
\end{equation}
in complex coordinates, diagonalize $H'$ as in \eqref{Hp}.

Classically, the supercharges follow from the Noether procedure:\footnote{Up to total derivatives, we have the \emph{local} SUSY variation $\delta(L_B + L_F) = -i\dot{\epsilon}Q - i\dot{\epsilon}^\dag Q^\dag$.}
\begin{equation}
Q = \frac{2i\dot{\phi}^\dag\psi}{\mu(1 + |\phi|^2)^2}, \quad Q^\dag = -\frac{2i\dot{\phi}\psi^\dag}{\mu(1 + |\phi|^2)^2}.
\end{equation}
Quantumly, the supercharges are represented by the nilpotent operators \eqref{Qquantum}, which are adjoints with respect to the Fubini-Study measure on the sphere:
\begin{equation}
\sin\theta\, d\varphi\wedge d\theta = \frac{2i\, d\phi\wedge d\phi^\dag}{(1 + |\phi|^2)^2}.
\label{FSmeasure}
\end{equation}
In deriving \eqref{Qquantum}, ordering ambiguities are fixed by demanding that $\frac{1}{2}\{Q, Q^\dag\} = H'$, where $H'$ is known from \eqref{Hpmatrix}.  One can verify that both $Q, Q^\dag$ annihilate the ground states of $H'$ and commute with the $(L_f)_i$ in \eqref{Lfquantum}.  Analysis of the supersymmetry transformations then shows that the velocities are represented by\footnote{The quantum supercharges generate the SUSY transformation of any $\mathcal{O} = \mathcal{O}(\phi, \phi^\dag, \psi, \psi^\dag)$ via $\delta\mathcal{O} = [\epsilon Q + \epsilon^\dag Q^\dag, \mathcal{O}]$.  The relations \eqref{qrule1} and \eqref{qrule2} ensure that the SUSY transformations of a chiral superfield are correctly realized, or specifically that $[\epsilon Q + \epsilon^\dag Q^\dag, \psi]$ and $[\epsilon Q + \epsilon^\dag Q^\dag, \psi^\dag]$ equate to $\delta\psi = -2i\epsilon^\dag\dot{\phi}$ and $\delta\psi^\dag = 2i\epsilon\dot{\phi}^\dag$, respectively.}
\begin{align}
\frac{2i\dot{\phi}^\dag}{\mu} &\xrightarrow{q} (1 + |\phi|^2)^2\frac{\partial}{\partial\phi} - (j + \chi^\dag\chi - 1/2)\phi^\dag(1 + |\phi|^2), \label{qrule1} \\
-\frac{2i\dot{\phi}}{\mu} &\xrightarrow{q} -(1 + |\phi|^2)^2\frac{\partial}{\partial\phi^\dag} - (j + \chi^\dag\chi - 1/2)\phi(1 + |\phi|^2). \label{qrule2}
\end{align}
These relations are consistent with the representations \eqref{Qquantum} of the supercharges as quantum operators if we choose the following ordering:
\begin{equation}
Q = \frac{\sqrt{\mu}\chi}{1 + |\phi|^2}\bigg(\frac{2i\dot{\phi}^\dag}{\mu}\bigg), \quad Q^\dag = \bigg({-\frac{2i\dot{\phi}}{\mu}}\bigg)\frac{\sqrt{\mu}\chi^\dag}{1 + |\phi|^2},
\label{Qordered}
\end{equation}
where the parenthesized expressions are understood as differential operators acting on the left.  This understanding allows us to identify the ordering prescription needed to directly quantize the classical Hamiltonian, written in terms of velocities as
\begin{equation}
H_\text{cl}' = H_\text{cl} + j\mu\chi^\dag\chi, \quad H_\text{cl} = \frac{2\dot{\phi}\dot{\phi}^\dag}{\mu(1 + |\phi|^2)^2}.
\label{Hpclassical}
\end{equation}
We have seen that the quantum Hamiltonian is
\begin{equation}
H' = \frac{1}{2}\{Q, Q^\dag\} = H_{j + \chi^\dag\chi - 1/2} + j\mu\chi^\dag\chi - \frac{\mu}{2}(j - 1/2),
\label{Hpquantum}
\end{equation}
where
\begin{equation}
H_j = -\frac{\mu}{2}\left[(1 + |\phi|^2)^2\frac{\partial}{\partial\phi}\frac{\partial}{\partial\phi^\dag} + j(1 + |\phi|^2)\left(\phi\frac{\partial}{\partial\phi} - \phi^\dag\frac{\partial}{\partial\phi^\dag}\right) - j^2|\phi|^2\right]
\end{equation}
in complex coordinates.  Applying \eqref{qrule1} and \eqref{qrule2} to the symmetrized expression
\begin{equation}
H_\text{cl} = \frac{\mu}{4}\left[\bigg(\frac{2i\dot{\phi}^\dag}{\mu}\bigg)\frac{1}{(1 + |\phi|^2)^2}\bigg({-\frac{2i\dot{\phi}}{\mu}}\bigg) + \bigg({-\frac{2i\dot{\phi}}{\mu}}\bigg)\frac{1}{(1 + |\phi|^2)^2}\bigg(\frac{2i\dot{\phi}^\dag}{\mu}\bigg)\right]
\label{Hordered}
\end{equation}
shows that $H_\text{cl} \xrightarrow{q} H_{j + \chi^\dag\chi - 1/2}$.  Further stipulating that
\begin{equation}
\chi^\dag\chi\xrightarrow{q} \chi^\dag\chi - \left(\frac{1}{2} - \frac{1}{4j}\right)
\label{qrulechi}
\end{equation}
reproduces precisely \eqref{Hpquantum} from \eqref{Hpclassical}.  In summary, the supersymmetric system $L_B + L_F$ in $S$ gauge can be quantized by applying the quantization rules \eqref{qrule1}, \eqref{qrule2}, \eqref{qrulechi} necessary to implement the supersymmetry algebra to the properly ordered classical expressions \eqref{Qordered} and \eqref{Hordered}.\footnote{Again, these quantization rules do not fix the quantum representations of the canonical momenta $\pi_\phi$ and $\pi_{\phi^\dag}$, but because we have chosen the boundary terms in the classical Lagrangian such that it is manifestly real, these representations are necessarily adjoints with respect to \eqref{FSmeasure}.}

Finally, judicious application of the aforementioned quantization rules\footnote{Namely, after symmetrization of the form $f(\phi)\dot{\phi}^\dag = \sqrt{f(\phi)}\dot{\phi}^\dag\sqrt{f(\phi)}$ and $f(\phi^\dag)\dot{\phi} = \sqrt{f(\phi^\dag)}\dot{\phi}\sqrt{f(\phi^\dag)}$.} shows that the classical expressions that, when quantized, give rise to the quantum operators $(L_f)_i$ are
\begin{align}
(\vec{L}_f)_\text{cl} &= \vec{V} + \frac{j + \chi^\dag\chi - 1/2}{1 + |\phi|^2}(\phi + \phi^\dag, -i(\phi - \phi^\dag), -(1 - |\phi|^2)), \label{Lfclassical} \\
\vec{V} &\equiv \frac{(i(1 - \phi^2)\dot{\phi}^\dag - i(1 - (\phi^\dag)^2)\dot{\phi}, -(1 + \phi^2)\dot{\phi}^\dag - (1 + (\phi^\dag)^2)\dot{\phi}, 2i(\phi\dot{\phi}^\dag - \phi^\dag\dot{\phi}))}{\mu(1 + |\phi|^2)^2}. \nonumber
\end{align}
Classically, the bosonic angular momenta \eqref{Lclassicalnomu} take the form
\begin{equation}
(L_1)_\text{cl} = j\left(\frac{\phi + \phi^\dag}{1 + |\phi|^2}\right), \quad (L_2)_\text{cl} = -ij\left(\frac{\phi - \phi^\dag}{1 + |\phi|^2}\right), \quad (L_3)_\text{cl} = -j\left(\frac{1 - |\phi|^2}{1 + |\phi|^2}\right),
\label{Lclassical}
\end{equation}
so that $(\vec{L}_f)_\text{cl}$ reduces to $\vec{L}_\text{cl}$ with $j - 1/2$ as $\mu\to\infty$ (note the differences in normalization and signs between the $J_i$ in \eqref{Noether} and the $(L_i)_\text{cl}$).

\section{Quantization of Chern-Simons Theory} \label{quantization}

Here, we review some basic aspects of the quantization of Chern-Simons theory with simple, compact $G$ that are relevant to our discussion in the main text, following \cite{Witten:1988hf, Elitzur:1989nr} (our conventions in Section \ref{gaugetransformations} entail some differences in formulas relative to those references).

\subsection{Generalities}

Let $\Sigma$ be an oriented Riemann surface, to which canonical quantization on $\Sigma\times \mathbb{R}$ associates a Hilbert space $\mathcal{H}_\Sigma$.  In temporal gauge $A_0 = 0$, the action \eqref{N0SCS} is $-\frac{k}{4\pi}\int_{M^3} A_1^a\dot{A}_2^a\, d^3 x$; we have the Poisson brackets and the (source-free) Gauss law constraint
\begin{equation}
[A_i^a(x), A_j^b(y)]_\text{PB} = \frac{4\pi}{k}\epsilon_{ij}\delta^{ab}\delta^{(2)}(x - y), \quad \epsilon^{ij}F_{ij}^a = 0.
\label{poissonbrackets}
\end{equation}
If we impose the constraints before quantizing, then the physical phase space is the moduli space of flat connections on $\Sigma$, modulo gauge transformations: $\mathcal{M} = \operatorname{Hom}(\pi_1(\Sigma), G)/G$.  In genus $g > 1$, $\dim\mathcal{M} = (2g - 2)\dim G$.  $\mathcal{M}$ inherits a symplectic structure from the Poisson brackets on the infinite-dimensional space of all connections $A_i^a(x)$ and has finite volume with respect to the symplectic volume element, so $\mathcal{H}_\Sigma$ is finite-dimensional.

Choosing a complex structure $J$ on $\Sigma$ induces a K\"ahler structure on $\mathcal{M}$, after which $\mathcal{H}_\Sigma$ has the interpretation as the space of holomorphic sections of $\mathcal{L}^k$ where $\mathcal{L}$ is a suitable holomorphic line bundle over $\mathcal{M}$ (e.g., for $SU(N)$, it is the determinant line bundle of the $\bar{\partial}$ operator on $\Sigma$).  The symplectic form on $\mathcal{M}$ is the curvature of $\mathcal{L}^k$.  One can show that the construction of $\mathcal{H}_\Sigma$ is canonically independent of $J$.

Here, we have assumed that $\partial\Sigma = 0$, so that $\mathcal{H}_\Sigma$ is the finite-dimensional vector space of conformal blocks of the $\smash{\widehat{G}_k}$ WZW model on $\Sigma$.  If instead $\partial\Sigma\neq 0$, then $\mathcal{H}_\Sigma$ is an infinite-dimensional representation of the chiral algebra of the 2D CFT, since there exist local degrees of freedom on the boundary that cannot be gauged away.

``Vertical'' Wilson lines correspond to marked points $P_i$ on $\Sigma$ with an irreducible representation $R_i$ of $G$ associated to each.  For $\mathcal{H}_{\Sigma; P_i, R_i}$ to be nontrivial, all representations $R_i$ must correspond to integrable representations of the loop group of $G$.  The appropriate reduced phase space $\mathcal{M}_{P_i, R_i}$ can be constructed with the aid of the Borel-Weil-Bott theorem.  Roughly speaking, it says the following.  Let $T$ be a maximal torus of $G$ and $r = \operatorname{rank} G$.  We have from the exact sequence in homotopy that $\pi_2(G/T) = \mathbb{Z}^r$ and $\pi_1(G/T) = 0$, hence $H_2(G/T) = \mathbb{Z}^r$ and there are $r$ nontrivial two-cycles in $G/T$.  To each two-cycle, we associate an exact ``unit'' two-form and specify a symplectic form on $G/T$ as an integral linear combination thereof, which is equivalent to specifying the highest weight of an irreducible representation $R$ of $G$ as an integral sum of $r$ fundamental weights.  The Hilbert space obtained by quantizing $G/T$ with this symplectic structure is the representation $R$.

This procedure can be phrased in the language of geometric quantization (see, e.g., \cite{Schottenloher:2008zz}); we will not do so here.  Let us instead make this abstract discussion concrete.

\subsection{Boundary Conditions} \label{boundaryconditions}

Suppose $\partial M^3\neq 0$.  The variation of \eqref{N0SCS} has both bulk and boundary components:\footnote{In coordinates, the variation of $\mathcal{L}_\text{CS}$ in \eqref{N0LCS} is
\[
\delta\mathcal{L}_\text{CS} = \epsilon^{\mu\nu\rho}\Tr(\delta A_\mu F_{\nu\rho}) + \partial_\nu[\epsilon^{\mu\nu\rho}\Tr(A_\mu\delta A_\rho)].
\]
For an infinitesimal gauge transformation, $\smash{\delta A_\mu^a} = D_\mu\theta^a$ and both terms reduce to boundary terms, which combine to give $\delta\mathcal{L}_\text{CS} = \frac{1}{2}\epsilon^{\mu\nu\rho}\partial_\nu(A_\rho^a\partial_\mu\theta^a)$.  This is indeed the infinitesimal form of $-i\epsilon^{\mu\nu\rho}\partial_\mu\Tr(A_\nu g^{-1}\partial_\rho g)$ from \eqref{gaugechange} (for an infinitesimal gauge transformation, the Pontryagin density term does not contribute).}
\begin{equation}
\delta S_\text{CS} = \frac{k}{2\pi}\int_{M^3} \Tr(\delta AF) + \frac{k}{4\pi}\int_{\partial M^3} \Tr(\delta AA).
\end{equation}
We will always choose boundary conditions such that there are no boundary corrections to the equations of motion.  This can be achieved by setting one component of the gauge field to zero at the boundary: for instance, $A_0$ in the case of a spatial boundary (e.g., $M^3 = \Sigma\times \mathbb{R}$ with $\partial\Sigma\neq 0$ and $\partial M^3 = \partial\Sigma\times \mathbb{R}$) and either $A_1$ or $A_2$ in the case of a temporal boundary (e.g., $M^3 = \Sigma\times (-\infty, 0]$ with $\partial\Sigma = 0$ and $\partial M^3 = \Sigma$).  These two cases are suited to the ``constrain, then quantize'' and ``quantize, then constrain'' approaches, respectively.

When $M^3 = \Sigma\times \mathbb{R}$, it is convenient to separate the temporal and spatial components of $d = dt\, \partial_t + \tilde{d}$ and $A = A_0 + \tilde{A}$, giving
\begin{equation}
S_\text{CS} = -\frac{k}{4\pi}\int_{M^3} \Tr(\tilde{A}\partial_t\tilde{A})\, dt + \frac{k}{2\pi}\int_{M^3} \Tr(A_0\tilde{F}) + \frac{k}{4\pi}\int_{\partial M^3} \Tr(A_0\tilde{A})
\label{effectiveCS}
\end{equation}
where $\tilde{F} = \tilde{d}\tilde{A} - i\tilde{A}^2$.  If $\partial\Sigma\neq 0$, then we impose $A_0|_{\partial M^3} = 0$, which kills the boundary term in \eqref{effectiveCS}.  This boundary condition implies that gauge transformations independent of time on the boundary are global (because, in an alternate quantization, they act nontrivially on the wavefunctions of physical states), while only those that reduce to the identity on the boundary are truly gauge.  Further integrating out $A_0$ in \eqref{effectiveCS} enforces $\tilde{F} = 0$, and we arrive at an effective action for $\tilde{A}$ satisfying this constraint by substituting such $\tilde{A}$ into the first term of \eqref{effectiveCS} (gauge-equivalent choices of $\tilde{A}$ yield the same action).

When $\partial M^3 = \Sigma$, it is natural to compute wavefunctions in the path integral formalism.  In the gauge $A_0 = 0$, the phase space coordinates are the two components of $\tilde{A}$ in the $\Sigma$ direction, one of which represents the canonical coordinate on which the wavefunction(al) depends.\footnote{Here, we have in mind a real polarization where the canonical coordinate is, e.g., $A_1$ or $A_2$ rather than $A_z$; we will discuss the holomorphic polarization later.}  Specifying nonzero values of $A_1$ on the boundary requires adding a term
\begin{equation}
\frac{k}{4\pi}\int_{\partial M^3} \Tr(A_1 A_2)
\label{boundaryterm}
\end{equation}
to \eqref{N0SCS} so that the boundary term in the variation of the total action is $\frac{k}{2\pi}\int_{\partial M^3} \Tr(\delta A_1 A_2)$, which vanishes because $\delta A_1|_{\partial M^3} = 0$.  Specifying $A_2$ requires a term of the opposite sign as in \eqref{boundaryterm}.  Aside from ensuring no boundary corrections to the equations of motion, the necessity of the boundary action \eqref{boundaryterm} follows from consistency of the canonical formalism in which $A_1, A_2$ are conjugate variables $q, p$: it is precisely the surface term that, when added to $S_\text{CS}|_{A_0 = 0}$, brings the action to the standard form $\propto \int p\dot{q}$.

\subsection{Real Polarization} \label{realpol}

In this subsection, we take $M^3 = \Sigma\times \mathbb{R}$.

\subsubsection{\texorpdfstring{$\Sigma = D^2$}{Sigma = D2}}

Because $D^2$ is simply connected, as is $G$ (by assumption), the flatness constraint $\tilde{F} = 0$ is solved by
\begin{equation}
\tilde{A} = -i\tilde{d}UU^{-1}
\end{equation}
where $U : M^3\to G$ is single-valued.  The change of variables $D\tilde{A}\, \delta(\tilde{F})\to DU$ in the path integral incurs no Jacobian \cite{Elitzur:1989nr}.  Setting $A_0|_{\partial M^3} = 0$, the effective action \eqref{effectiveCS} when written in terms of $U$ is
\begin{equation}
S_\text{eff} = kS_C^+(U)\equiv -\frac{k}{4\pi}\int_{\partial M^3} \Tr(U^{-1}\partial_\phi UU^{-1}\partial_t U)\, d\phi\, dt - \frac{k}{12\pi}\int_{M^3} \Tr(U^{-1}dU)^3
\label{SeffD2}
\end{equation}
where $\phi$ denotes the angular coordinate on $D^2$ and the chiral WZW (CWZW) action $S_C^+(U)$ depends only on the boundary values of $U$.  The action \eqref{SeffD2} is invariant under the following transformation on $\partial M^3$:
\begin{equation}
U(\phi, t)\to \tilde{V}(\phi)U(\phi, t)V(t).
\label{invariance}
\end{equation}
Since $\tilde{V}$ is a global symmetry, the Hilbert space is a representation of the loop group $LG$.  On the other hand, $V$ is a gauge symmetry.  The classical phase space is then $LG/G$, where $LG$ is the space of flat $G$-connections on $D^2$ modulo gauge transformations that reduce to the identity on $\partial D^2 = S^1$ and the quotient by $G$ fixes the $V$ gauge symmetry.  This space inherits a symplectic structure from the gauge-fixed Lagrangian.  Namely, given an action
\begin{equation}
I = \int dt\, A_i(\phi)\frac{d\phi^i}{dt}
\end{equation}
that is first-order in time derivatives, the symplectic form is $\omega = dA$ \cite{Witten:1983ar}.  Indeed, one can compute the Poisson brackets without choosing an explicit polarization of the phase space coordinates $\phi^i$: under an arbitrary variation $\phi^i\to \phi^i + \delta\phi^i$,
\begin{equation}
\delta I = \int dt\left(\frac{\partial A_j}{\partial\phi^i} - \frac{\partial A_i}{\partial\phi^j}\right)\delta\phi^i\frac{d\phi^j}{dt}\equiv \int dt\, F_{ij}\delta\phi^i\frac{d\phi^j}{dt} \implies [X, Y]_\text{PB} = \sum_{i, j} F^{ij}\frac{\partial X}{\partial\phi^i}\frac{\partial Y}{\partial\phi^j}
\label{symplectic}
\end{equation}
(we recover the usual definition by setting $\phi^i = q^i$, $A^i = p^i$).  In the case of a group-valued sigma model, $F$ can be constructed relative to a basis of Lie algebra-valued tangent vectors to the phase space, and it acts on both the Lie algebra index and the base space coordinates.  In our case, we compute that the variation of \eqref{SeffD2} is
\begin{equation}
\delta S_\text{eff} = \frac{k}{2\pi}\int_{\partial M^3} \Tr(U^{-1}\delta U\partial_\phi(U^{-1}\partial_t U))\, d\phi\, dt.
\end{equation}
Integrating by parts and using the recipe \eqref{symplectic}, we obtain
\begin{equation}
\omega = -\frac{k}{4\pi}\oint \Tr((U^{-1}\delta U)\partial_\phi(U^{-1}\delta U))
\end{equation}
where the $U^{-1}\delta U$ are tangent vectors to the phase space ($\delta$ can be thought of as an exterior derivative on phase space).  Further integrating by parts shows that $\omega$ is antisymmetric.

Our primary interest is in the case where the $D^2$ contains a source in the representation $\lambda$.  Its effect is modeled by adding to \eqref{effectiveCS} the coadjoint orbit action
\begin{equation}
i\int_{\mathbb{R}} \Tr(\lambda g^{-1}(\partial_t - iA_t)g)\, dt = \frac{i}{2\pi}\int_{\partial M^3} \Tr(\lambda g^{-1}(\partial_t - iA_t)g)\, d\phi\, dt,
\label{coadjointaction}
\end{equation}
where $g(t)\in G$ and $\lambda$ (which fixes a maximal torus $T\subset G$) is written in a basis of Cartan generators.  As discussed in Section \ref{loopsinCS}, the gauge invariance of \eqref{coadjointaction} under $g(t)\to g(t)h(t)$ with $h(t)\in T$ suffers from global anomalies unless $\lambda$ is quantized as a weight.\footnote{A weight $\lambda$ of $G$ is integral if for each $t\in \mathfrak{t}$ such that $\exp(it) = 1\in G$, $\lambda(t)\in 2\pi\mathbb{Z}$; for $G$ semisimple, the integral weights comprise a sublattice of the weight lattice, but these lattices coincide if $G$ is simply connected.}  By itself, \eqref{coadjointaction} describes the quantum mechanics of $g$ coupled to the background gauge field $A_t$, with classical phase space $G/T$ and symplectic structure $\Tr(\lambda(g^{-1}\delta g)^2)$.  Putting the source at the origin and integrating over $A_t$ in the total action \eqref{effectiveCS} plus \eqref{coadjointaction} yields the constraint
\begin{equation}
\frac{k}{2\pi}\tilde{F}(x, t) + g(t)\lambda g^{-1}(t)\delta^{(2)}(x)\, d^2 x = 0.
\label{constraint}
\end{equation}
At any given $t$, integrating \eqref{constraint} over a disk containing the origin immediately shows that the logarithm of the holonomy of the flat connection $\tilde{A}$ around the source is $-\smash{\frac{2\pi i}{k}}g(t)\lambda g^{-1}(t)$.  In other words, the conjugacy class of the holonomy of $\tilde{A}$ is determined by $\lambda$ to be that of $e^{-2\pi i\lambda/k}$.  Explicitly, \eqref{constraint} is solved by
\begin{equation}
\tilde{A} = -i\tilde{d}\tilde{U}\tilde{U}^{-1}, \quad \tilde{U}\equiv U\exp\left[\frac{i}{k}g(t)\lambda g^{-1}(t)\phi\right]
\label{flatwithsource}
\end{equation}
where at any given $t$, $U$ is single-valued on $D^2$ and its value at the origin $U(0, t)$ commutes with $g(t)\lambda g^{-1}(t)$.  Substituting \eqref{flatwithsource} into the total action \eqref{effectiveCS} plus \eqref{coadjointaction} and integrating out $g$, whose equation of motion imposes $[\lambda, g] = 0$, yields the effective action
\begin{equation}
S_\text{eff} = kS_C^+(U) - \frac{i}{2\pi}\int_{\partial M^3} \Tr(\lambda U^{-1}\partial_t U)\, d\phi\, dt.
\label{SeffD2withsource}
\end{equation}
The action \eqref{SeffD2withsource} is now invariant under \eqref{invariance} where $V$ commutes with $\lambda$, so the classical phase space is $LG/T$ with symplectic structure
\begin{equation}
\omega = -\frac{k}{4\pi}\oint \Tr((U^{-1}\delta U)\partial_\phi(U^{-1}\delta U)) - \frac{i}{2\pi}\oint \Tr(\lambda(U^{-1}\delta U)^2).
\end{equation}
The Hilbert space $\mathcal{H}_\Sigma$ is the integrable representation $\mathcal{H}_\lambda$ of $\smash{\widehat{G}_k}$.

\subsubsection{\texorpdfstring{$\Sigma = T^2$}{Sigma = T2}}

For $G$ connected and simply connected, the most general flat connection on $T^2$ is
\begin{equation}
\tilde{A} = -i\tilde{d}UU^{-1} + U\theta(t)U^{-1}
\label{flatonT2}
\end{equation}
where $U$ is single-valued and $\theta$ is a $\mathfrak{g}$-valued one-form representing the holonomies.  Since $\pi_1(T^2)$ is abelian, the two components of $\theta$ can be chosen to lie in a Cartan subalgebra $\mathfrak{t}$ with basis $\smash{\vec{H}}$: $\theta = \smash{\vec{\theta}\cdot\vec{H}}$ (in particular, $\theta^2 = 0$).  This description suffers from a gauge redundancy $\smash{\vec{\theta}}\sim \smash{\vec{\theta}} + 2\pi\vec{\alpha}$ where $\vec{\alpha}$ is a one-form valued in $\Lambda_R^\vee$.  Hence the classical phase space is
\begin{equation}
\mathcal{M} = \frac{T\times T}{\mathcal{W}}
\end{equation}
where $T$ is a maximal torus (with one copy for each component of $\theta$) and $\mathcal{W}$ acts diagonally.  Substituting \eqref{flatonT2} into \eqref{effectiveCS} gives
\begin{equation}
S_\text{eff} = -\frac{k}{4\pi}\int_{M^3} \epsilon^{ij}\Tr(\theta_i\partial_t\theta_j + U^{-1}\partial_t UU^{-1}\partial_i UU^{-1}\partial_j U)\, dx_1\, dx_2\, dt
\end{equation}
where the second term is proportional to the winding number of the map $U : T^2\times \mathbb{R}\to G$, which vanishes since $\pi_1(G) = 0$.  The change of variables $D\tilde{A}\, \delta(\tilde{F})\to DU D\theta = DU D\theta_1 D\theta_2$ entails no Jacobian \cite{Elitzur:1989nr}, so we are left with the following effective action for the holonomies $\theta_{1, 2}$, which are naturally interpreted as canonical coordinates and momenta:
\begin{equation}
S_\text{eff} = -\frac{k}{2\pi}\int_{M^3} \vec{\theta}_1\cdot\dot{\vec{\theta}}_2\, dt \implies [\theta_1^i, \theta_2^j] = \frac{2\pi i}{k}\delta^{ij}.
\end{equation}
Since the coordinates $\theta_1$ are compact, the momentum is quantized and momentum eigenstates are labeled by $\smash{\vec{\lambda}}$ in the weight lattice $\Lambda_W$ of $G$.  The momenta $\theta_2$ are also compact by virtue of the aforementioned gauge redundancy.  Finally, the action of the Weyl group leads to a further redundancy $\smash{\vec{\theta}}\sim \mathcal{W}(\smash{\vec{\theta}})$.  Thus we deduce that the Hilbert space in genus one is
\begin{equation}
\mathcal{H}_\Sigma\cong \frac{\Lambda_W}{\mathcal{W}\ltimes k\Lambda_R^\vee},
\end{equation}
where $\Lambda_R^\vee\cong \Lambda_R$ for simply laced $G$ (using our normalization conventions).  This coset space is precisely the space of integrable representations of $\smash{\widehat{G}_k}$ ($\mathcal{W}\ltimes \Lambda_R^\vee$ is the affine Weyl group).

In particular, for $\Sigma = T^2$, $\dim\mathcal{H}_\Sigma = t$ where $t$ is the number of integrable highest-weight representations of $LG$ at level $k$ (each corresponds to an irreducible representation $R_i$ of $G$, $i = 0, \ldots, t - 1$, $R_0\equiv \text{triv}$).  Every choice of homology basis ($a$ and $b$ cycles) yields a canonical basis (``Verlinde basis'') in $\mathcal{H}_\Sigma$: let $\Sigma$ be the boundary of a solid torus $U$ in which $a$ is contractible and place a Wilson line in the representation $R_i$ parallel to $b$; then the path integral over $U$ defines a vector $v_i\in \mathcal{H}_\Sigma$.  $\mathcal{H}_\Sigma$ has a natural metric, given by $g_{ij} = [R_j = \overline{R}_i]$ in this basis; diffeomorphisms $K$ of $\Sigma$ are represented by linear transformations $K_i{}^j$ on $\mathcal{H}_\Sigma$.

Our primary case of interest is $G = SU(2)$, for which $\mathcal{M} = \mathbb{CP}^1$.  The line bundle over $\mathbb{CP}^1$ of degree $k$ is obtained by gluing together the trivializations over the two standard patches by the transition function $z^k$ (in local coordinates), and the space of holomorphic sections is spanned by $1, z, \ldots, z^k$.  This space is identified with the first $k + 1$ characters ($j = 0, 1/2, \ldots, k/2$) of $\smash{\widehat{SU}(2)_k}$ (integrable representations at level $k$).  As another example, for $G = SU(N)$, the Hilbert space has dimension $\smash{\binom{k + N - 1}{N - 1}}$ by the Verlinde formula.  Indeed, the space of holomorphic sections of the degree-$k$ line bundle over $\mathbb{CP}^n$ is $\binom{k + n}{n}$-dimensional and spanned by monomials in $z_1, \ldots, z_n$ of degree at most $k$.

\subsection{Holomorphic Polarization} \label{holomorphicpol}

In this scheme, we first quantize and then constrain: the classical phase space is the infinite-dimensional space of all gauge fields $\mathcal{A}$ rather than the moduli space of flat connections $\mathcal{M}$.  Wavefunctions are derived by demanding that they satisfy Gauss's law rather than by evaluating the path integral.  This approach has the virtue of making explicit the identification of wavefunctions and conformal blocks of the corresponding 2D CFT on any $\Sigma$.

\subsubsection{Coherent States}

In holomorphic quantization, we separate out the holomorphic part of the wavefunction and regard the non-holomorphic part (namely, the exponentiated K\"ahler potential on $\mathcal{A}$ induced by a choice of complex structure on $\Sigma$) as part of the integration measure.  Coherent states furnish a resolution of the identity with respect to this measure.\footnote{Physically, one can justify these coherent states by turning on a small Yang-Mills interaction and making the usual quantum-mechanical analogy with a charged particle in an external magnetic field $\propto k$ \cite{Dunne:1998qy}: in the pure Chern-Simons (zero-mass) limit, all but the lowest Landau level are projected out (as in the monopole problem treated earlier), and the ground-state wavefunctionals are coherent states.}

In the gauge $A_0 = 0$, and having fixed a complex structure on $\Sigma$, we obtain from \eqref{poissonbrackets} the canonical commutation relations
\begin{equation}
[A_z^a(z_1), A_{\bar{z}}^b(z_2)] = \frac{2\pi i}{k}\delta^{ab}\delta^{(2)}(z_1 - z_2)
\end{equation}
where $A_z^a = \frac{1}{2}(A_1^a - iA_2^a)$.  Wavefunctions are holomorphic in $A_z$, with the action of $A_z^a$ and $A_{\bar{z}}^a$ being represented thereon by
\begin{equation}
\mathbf{A}_z^a = A_z^a, \quad \mathbf{A}_{\bar{z}}^a = -\frac{2\pi i}{k}\frac{\delta}{\delta A_z^a}.
\label{Arepresentations}
\end{equation}
The coherent state inner product is defined as
\begin{equation}
\langle\Psi_1|\Psi_2\rangle = \int DA_z^a(x)\, DA_z^a(x)^\ast\, e^{\frac{ik}{\pi}\int_\Sigma d^2 z \Tr(A_z A_z^\ast)}\Psi_1[A_z]^\ast\Psi_2[A_z]
\label{innerproduct}
\end{equation}
on the infinite-dimensional space of functionals of $A_z$, of which the physical Hilbert space $\mathcal{H}_\Sigma$ is the gauge-invariant subspace.

The following formulas are useful in constructing the physical wavefunctions.  We define the WZW actions
\begin{equation}
S^\pm[g]\equiv -\frac{1}{4\pi}\int_\Sigma d^2 z \Tr(g^{-1}\partial_z gg^{-1}\partial_{\bar{z}}g)\mp \frac{1}{12\pi}\int_{M^3} \Tr(g^{-1}dg)^3,
\end{equation}
where $\partial M^3 = \Sigma$.  The corresponding Polyakov-Wiegmann identities are
\begin{equation}
S^\pm[g_1 g_2] = S^\pm[g_1] + S^\pm[g_2] - \frac{1}{2\pi}\int_\Sigma d^2 z \begin{cases} \Tr(\partial_z g_2 g_2^{-1}g_1^{-1}\partial_{\bar{z}}g_1) & (+), \\ \Tr(g_1^{-1}\partial_z g_1\partial_{\bar{z}}g_2 g_2^{-1}) & (-), \end{cases}
\label{PW}
\end{equation}
from which we read off the variations
\begin{equation}
\delta S^\pm[g] = \frac{1}{2\pi}\int_\Sigma d^2 z \begin{cases} \Tr(g^{-1}\delta g\partial_z(g^{-1}\partial_{\bar{z}}g)) & (+), \\ \Tr(g^{-1}\delta g\partial_{\bar{z}}(g^{-1}\partial_z g)) & (-). \end{cases}
\label{WZWvariations}
\end{equation}
To begin, consider $\Sigma = S^2$.  For all $A_z$ (not necessarily flat) except in a subset of codimension one, it is possible to write
\begin{equation}
A_z = -i\partial_z UU^{-1}
\label{HodgeS2}
\end{equation}
where $U : S^2\to G_\mathbb{C}$ \cite{Elitzur:1989nr, Bos:1989kn}.  The unique physical state on $S^2$ has wavefunction
\begin{equation}
\Psi_0[A_z] = e^{2ikS^-[U]},
\end{equation}
up to normalization.  Indeed, we have by \eqref{WZWvariations} and \eqref{Arepresentations} that
\[
\delta\Psi_0[A_z] = \frac{k}{\pi}\left[\int_\Sigma d^2 z \Tr(\delta A_z\partial_{\bar{z}}UU^{-1})\right]\Psi_0[A_z] \implies \mathbf{A}_{\bar{z}}^a\Psi_0[A_z] = -i(\partial_{\bar{z}}UU^{-1})^a\Psi_0[A_z]
\]
(recall the conventions in Section \ref{gaugetransformations}).  Again in light of \eqref{Arepresentations}, the Gauss law constraint $\mathbf{F}_{z\bar{z}}\Psi_0[A_z] = 0$ is satisfied.  Under $A_z\to A_z^g$ (i.e., $U\to gU$), $\Psi_0[A_z]$ transforms by a phase (1-cocycle), which motivates us to represent the gauge transformation $g(x)$ by an operator $\mathbf{U}(g)$ that acts on arbitrary functionals as
\begin{equation}
\mathbf{U}(g)\Psi[A_z] = e^{-2ikS^+[g] - \frac{k}{\pi}\int_\Sigma d^2 z \Tr(A_z g^{-1}\partial_{\bar{z}}g)}\Psi[A_z^g].
\label{Ug}
\end{equation}
By \eqref{PW}, we deduce that $\mathbf{U}(g)\Psi_0[A_z] = \Psi_0[A_z]$, which is a restatement of the fact that the physical wavefunction is gauge-invariant.  $\mathbf{U}(g)$ is unitary with respect to the inner product \eqref{innerproduct} and, by \eqref{PW}, satisfies the composition law $\mathbf{U}(g_2)\mathbf{U}(g_1) = \mathbf{U}(g_1 g_2)$.

\subsubsection{\texorpdfstring{$\Sigma = T^2$}{Sigma = T2} Redux}

The Hodge decomposition in this case (analogous to \eqref{HodgeS2} on $S^2$), which holds for almost all $A_z$, is
\begin{equation}
A_z = -i\partial_z UU^{-1} + Ua_z U^{-1}
\end{equation}
where $U\in G_\mathbb{C}$ and $a_z$ lies in a fixed Cartan subalgebra $\mathfrak{t}\subset \mathfrak{g}$ (more precisely, $\mathfrak{t}_\mathbb{C}$).  Gauge invariance with respect to \eqref{Ug} restricts physical wavefunctions, which we preemptively label by a subscript $\lambda$, to take the form
\begin{equation}
\Psi_\lambda[A_z] = e^{2ikS^-[U] + \frac{k}{\pi}\int_\Sigma d^2 z \Tr(a_z U^{-1}\partial_{\bar{z}}U)}\psi_\lambda(a_z).
\label{ansatzT2}
\end{equation}
Using this parametrization in \eqref{innerproduct}, changing variables from $A_z$ to $(U, a_z)$ (which involves a nontrivial Jacobian), and integrating over $U$ leads to an effective quantum mechanics with coherent state inner product
\begin{equation}
\langle\psi_{\lambda_1}|\psi_{\lambda_2}\rangle = \int da_z\, da_z^\ast\, e^{\frac{i(k + h)}{\pi}\int_\Sigma d^2 z \Tr(a_z a_z^\ast)}\psi_{\lambda_1}^\text{eff}(a_z)^\ast\psi_{\lambda_2}^\text{eff}(a_z).
\label{qminnerproduct}
\end{equation}
The effective wavefunctions $\psi_\lambda^\text{eff}(a_z)$ are related to the $\psi_\lambda(a_z)$ in \eqref{ansatzT2} by
\begin{equation}
\psi_\lambda^\text{eff}(a_z) = e^{-\frac{ih}{2\pi}\int_\Sigma d^2 z \Tr a_z^2}\Pi(\bar{\tau}, u)\psi_\lambda(a_z),
\end{equation}
where $\tau$ denotes the complex structure and $u\equiv -\frac{\operatorname{Im}\tau}{\pi}a_z$ as in \eqref{wavefunctionsT2}.  In particular, for the $\psi_\lambda(a_z)$ in \eqref{wavefunctionsT2}, the corresponding effective wavefunctions
\begin{equation}
\psi_\lambda^\text{eff}(a_z) = e^{-\frac{(k + h)\operatorname{Im}\tau}{\pi}\Tr a_z^2}\Theta_{\lambda + \rho, k + h}^-(\bar{\tau}, u)
\label{qmwavefunctionsT2}
\end{equation}
are orthogonal with respect to \eqref{qminnerproduct}: $\langle\psi_{\lambda_1}|\psi_{\lambda_2}\rangle\propto \delta_{\lambda_1\lambda_2}$.  In writing \eqref{qmwavefunctionsT2}, we have used
\begin{equation}
\Pi(\tau, u)\chi_\lambda^{(k)}(\tau, u) = \Theta_{\lambda + \rho, k + h}^-(\tau, u)
\end{equation}
and the same conventions for the geometry of $\Sigma = T^2$ as \cite{Elitzur:1989nr}, in which the area is $-2i\operatorname{Im}\tau$; the functional determinant $\Pi$ and the Weyl-odd theta functions $\Theta^-$ are defined in \cite{Elitzur:1989nr} and references therein (in comparing Lie algebra conventions, note that we use $\Tr(T^a T^b) = \frac{1}{2}\delta^{ab}$, \cite{Elitzur:1989nr} uses $\Tr(T^a T^b) = -\delta^{ab}$, and \cite{Bos:1989kn} uses $\Tr(T^a T^b) = -\frac{1}{2}\delta^{ab}$).

The key point is the following.  Level-$k$ theta functions for $\Lambda_R$ are indexed by weights in $\Lambda_W/k\Lambda_R$ and transform by phases under large gauge transformations, which act as $u\to u + r_1 + \tau r_2$ for $r_{1, 2}\in \Lambda_R$ (hence the integration region for $a_z$ in \eqref{qminnerproduct} is the complex torus $T_\mathbb{C} = \mathfrak{t}_\mathbb{C}/\Lambda_R$).  The Weyl-invariant subspace of such theta functions is spanned by Weyl-Kac characters at level $k$.  However, Weyl-Kac characters at $k, \lambda$ are expressible in terms of theta functions at $k + h, \lambda + \rho$.  Therefore, rather than labeling the spectrum by $\lambda$ in the Weyl alcove $\Lambda_W/(\mathcal{W}\ltimes k\Lambda_R)$, one may equivalently label it by $\lambda$ in the \emph{interior} of the dilated Weyl alcove $\Lambda_W/(\mathcal{W}\ltimes (k + h)\Lambda_R)$.  While this statement is familiar from representation theory \cite{DiFrancesco:1997nk}, the relation between \eqref{wavefunctionsT2} and \eqref{qmwavefunctionsT2} gives it a physical interpretation.

\section{Surgery versus Localization} \label{surgvsloc}

It is amusing, and possibly even useful, that localization offers an alternative to traditional algebraic or surgery-based methods for the computation of certain knot invariants.  Let us compare localization for $\mathcal{N} = 2$ $SU(2)_{k+2}$ and surgery for $\mathcal{N} = 0$ $SU(2)_k$ on $S^3$ in a few examples.  The Chern-Simons observables that are accessible to localization on $S^3$ include links composed of Hopf fibers: their components are unknots with pairwise linking number one.  For such links, one can check that the matrix model results match those from surgery for small numbers of components or sufficiently small representations, but it is possible to obtain clean answers using localization even when the latter method becomes cumbersome.

The localization approach is as follows.  For $\mathcal{N} = 2$ $SU(2)_{k+2}$ with Cartan parametrized by $\operatorname{diag}(a, -a)$, the matrix model of \cite{Kapustin:2009kz}, written in \eqref{matrixmodelS3}, reduces to
\begin{equation}
\langle W_{j_1}\cdots W_{j_n}\rangle_{\mathcal{N} = 2} = -\frac{2}{Z_{\mathcal{N} = 2}}\int_{\mathbb{R}} da\, e^{-2\pi i(k + 2 - i\epsilon)a^2}\sinh(2\pi a)^2\Tr_{j_1}(e^{2\pi a})\cdots \Tr_{j_n}(e^{2\pi a})
\label{matrixmodelSU2}
\end{equation}
where $Z_{\mathcal{N} = 2} = e^{i\pi/4}e^{-i\pi/(k + 2)}Z_{\mathcal{N} = 0}$ with $Z_{\mathcal{N} = 0}$ in \eqref{ZN0}.  It is useful to define two equivalent expressions for $\Tr_j(e^{2\pi a})$, both before and after applying the Weyl character formula:
\begin{equation}
\Tr_j^{\exp}(e^{2\pi a})\equiv \sum_{m=-j}^j e^{4\pi ma}, \quad \Tr_j^{\sinh}(e^{2\pi a})\equiv \frac{\sinh(2\pi a(2j + 1))}{\sinh(2\pi a)}.
\end{equation}
Note that we label representations by their spin, not by their dimension.  Results are conveniently written in terms of $q\equiv e^{2\pi i/(k + 2)}$ and the quadratic Casimirs $C_2(j)\equiv j(j + 1)$ of the representations.  The supersymmetric framing in the $\mathcal{N} = 2$ $SU(2)_{k+2}$ theory leads to a phase of $q^{-C_2(j)}$ for each Wilson line relative to the $\mathcal{N} = 0$ $SU(2)_k$ theory:
\begin{equation}
\langle W_{j_1}\cdots W_{j_n}\rangle_{\mathcal{N} = 2} = q^{-C_2(j_1) - \cdots - C_2(j_n)}\langle W_{j_1}\cdots W_{j_n}\rangle_{\mathcal{N} = 0}.
\label{framing}
\end{equation}
From the integration measure in \eqref{matrixmodelSU2}, we see that although the integrand can always be written as a sum of Gaussians by inserting $\smash{\Tr_j^{\exp}(e^{2\pi a})}$, correlators of two or fewer Wilson loops are particularly simple because each insertion of $\smash{\Tr_j^{\sinh}(e^{2\pi a})}$ cancels a factor of $\sinh(2\pi a)$.  These are precisely the cases where the results have simple expressions in terms of the $S$-matrix elements \eqref{Smatrix}.

To describe the surgery approach, we follow \cite{Guadagnini:1994cc}, where our $q$ is denoted there by $q^{-1}$.  A link with $n$ components is conveniently regarded as the closure of a braid on $\geq n$ strands.\footnote{The following is a technical simplification of the original procedure of \cite{Witten:1988hf}, which involves passing from $S^2\times S^1$ to $S^3$ via braid traces with a ``spectator'' strand.}  Given a link with specified representations for its components, we write its expectation value $\langle\rangle_{\mathcal{N} = 0}$ in the $\mathcal{N} = 0$ $SU(2)_k$ theory as a braid group element enclosed in brackets $\langle\rangle_\#$, where $\#$ denotes the number of strands (if all components are in the same representation, then the brackets $\langle\rangle_\#$ are cyclic).  Strands are labeled from left to right, and braid moves are applied from bottom to top.  The braid group generator $g_k$ corresponds to crossing strand $k$ over strand $k + 1$, and the braid group relations are
\begin{equation}
g_i g_{i+1}g_i = g_{i+1}g_i g_{i+1}, \quad g_i g_j = g_j g_i \text{ ($|i - j|\geq 2$)}.
\label{braidrelations}
\end{equation}
Canonical framing is assumed, which means that each unit of writhe (self-intersection number) introduces a factor of $q^{-C_2(j)}$ for a line of spin $j$, where self-overcross corresponds to positive writhe.  The basic properties that allow us to compute link expectation values are the fusion property (OPE) for cabled unknots,\footnote{From \eqref{matrixmodelS3}, BPS Wilson loops in the $\mathcal{N} = 2$ theory manifestly satisfy this property as well \cite{Kapustin:2013hpk}.}
\begin{equation}
\langle W_{\rho_1}W_{\rho_2}\cdots\rangle_{\mathcal{N} = 0} = \langle W_{\rho_1\otimes \rho_2}\cdots\rangle_{\mathcal{N} = 0} = \sum_{\rho\in \rho_1\otimes \rho_2} \langle W_\rho\cdots\rangle_{\mathcal{N} = 0},
\end{equation}
and similar fusion properties for crossed lines (derivable from the braiding matrix), which hold for lines in arbitrary representations of any $G$.  Using these properties, one can derive inductively the results for the unknot and Hopf link in arbitrary representations of $SU(2)$:
\begin{align}
\langle W_j\rangle_{\mathcal{N} = 0} &= \langle 1\rangle_1 = \frac{q^{(2j + 1)/2} - q^{-(2j + 1)/2}}{q^{1/2} - q^{-1/2}}, \label{unknotSU2} \\
\langle W_{j_1}W_{j_2}\rangle_{\mathcal{N} = 0} &= \langle g_1^2\rangle_2 = \frac{q^{(2j_1 + 1)(2j_2 + 1)/2} - q^{-(2j_1 + 1)(2j_2 + 1)/2}}{q^{1/2} - q^{-1/2}}. \label{HopfSU2}
\end{align}
Let us, however, restrict our attention to lines in the fundamental of $SU(2)$ (the situation relevant to the Jones polynomial).  For such lines, the basic fusion properties (whose explicit forms we will not need) imply the familiar skein relation
\begin{equation}
q^{-1/4}L_+ - q^{1/4}L_- = (q^{-1/2} - q^{1/2})L_0
\label{skein}
\end{equation}
where $L_+$, $L_-$, and $L_0$ denote overcross, undercross, and no cross, respectively.  For fundamental lines, the skein relation \eqref{skein}, the writhe relations (factors of $q^{\mp 3/4}$ for each self-overcross and self-undercross), and the result $\langle W_{1/2}\rangle_{\mathcal{N} = 0}$ for the unknot suffice to determine all link expectation values \cite{Guadagnini:1994cc}.  Moreover, using \eqref{skein} rather than fusion allows us to consider only fundamental lines at all intermediate steps in the computation.

As an example of the use of \eqref{skein}, we compute for the fundamental trefoil knot that
\begin{align*}
\langle g_1^3\rangle_2 = q^{-7/4}(1 + q + q^2 - q^4).
\end{align*}
As a more relevant class of examples, consider $W_{1/2}^n$ where all components are understood to lie along Hopf fibers.  We may construct the corresponding braid on $n$ strands by linking the first with the rest, the second with the remainder, and so on:
\begin{equation}
\langle W_{1/2}^n\rangle_{\mathcal{N} = 0} = \langle(g_{n-1}^2)\cdots (g_2\cdots g_{n-2}g_{n-1}^2 g_{n-2}\cdots g_2)(g_1\cdots g_{n-2}g_{n-1}^2 g_{n-2}\cdots g_1)\rangle_n.
\end{equation}
Special cases are
\begin{align*}
\langle W_{1/2}\rangle_{\mathcal{N} = 0} &= q^{1/2} + q^{-1/2}, \\
\langle W_{1/2}^2\rangle_{\mathcal{N} = 0} &= q^{3/2} + q^{1/2} + q^{-1/2} + q^{-3/2}, \\
\langle W_{1/2}^3\rangle_{\mathcal{N} = 0} &= q^{-3} + q^{-2} + q^{-1} + 1 + 2q + 2q^2,
\end{align*}
where the first two expressions follow from \eqref{unknotSU2} and \eqref{HopfSU2} and the last expression is computed from $\smash{\langle W_{1/2}^3\rangle_{\mathcal{N} = 0}} = \langle g_2^2 g_1 g_2^2 g_1\rangle_3$ using \eqref{skein}, cyclicity of $\langle\rangle_\#$, and the first of the braid group relations \eqref{braidrelations}.  Note that when $n\geq 3$, the result for $\smash{\langle W_{1/2}^n\rangle_{\mathcal{N} = 0}}$ is not invariant under $q\leftrightarrow q^{-1}$.  An inductive argument might suffice to compute $\smash{\langle W_{1/2}^n\rangle}_{\mathcal{N} = 0}$ via surgery, or a representation-theoretic point of view might prove more useful, as in the case of torus knots (see \cite{Beasley:2009mb} and references therein).  Regardless, by inserting $\smash{\Tr_{1/2}^{\exp}(e^{2\pi a})^n}$ into the localization matrix model \eqref{matrixmodelSU2}, we compute with almost no effort that
\begin{equation}
\langle W_{1/2}^n\rangle_{\mathcal{N} = 2} = \frac{1}{2(1 - q)}\sum_{\ell=0}^n \binom{n}{\ell}q^{-(n - 2\ell)^2/4}(q^{n - 2\ell} + q^{-(n - 2\ell)} - 2q).
\end{equation}
Accounting for the framing discrepancy \eqref{framing}, we deduce that
\begin{equation}
\langle W_{1/2}^n\rangle_{\mathcal{N} = 2} = q^{-3n/4}\langle W_{1/2}^n\rangle_{\mathcal{N} = 0},
\end{equation}
from which we read off the Jones polynomial of all links whose components are Hopf fibers.\footnote{To get the Jones polynomial from $SU(2)$ Chern-Simons with fundamental Wilson loops, we must divide by the expectation value of the unknot, adjust by an overall power of $q$, and redefine $q$ slightly \cite{Guadagnini:1994cc}:
\[
\frac{q^{3w(L)/4}}{q^{1/2} + q^{-1/2}}\langle L\rangle \xrightarrow{q^{1/2}\to -q^{-1/2}} V(L),
\]
where $w(L)$ is the writhe of the link $L$.  For example, for the unknot ($w = 0$), Hopf link ($w = +2$), and trefoil ($w = +3$), we get 1, $-q^{-5/2} - q^{-1/2}$, and $-q^{-4} + q^{-3} + q^{-1}$, respectively.  This is a Laurent polynomial in $q^{1/2}$ (or, if the link has an odd number of components, in $q$).}

Finally, one can consider more general $SU(2)$ representations.  For example, the expectation value of a three-component Hopf link with each component in an arbitrary representation is easily computed via localization by inserting $\Tr_{J_1, J_2}^{\sinh}(e^{2\pi a})$ and $\Tr_{J_3}^{\exp}(e^{2\pi a})$ into \eqref{matrixmodelSU2}:
\begin{equation}
\langle W_{J_1}W_{J_2}W_{J_3}\rangle_{\mathcal{N} = 2} = \frac{1}{1 - q^{-1}}\sum_{\ell=-J_3}^{J_3} (q^{-(J_1 - J_2 + \ell)^2} - q^{-(1 + J_1 + J_2 + \ell)^2}).
\end{equation}
This expression is invariant under permutations of $\{J_1, J_2, J_3\}$.  To compare to surgery, one might hope to use a generalized skein relation, which is a linear relation between $N + 1$ crossings where $N$ is the number of irreducible representations in the decomposition of the tensor product of two lines.  However, skein relations only make sense for all lines in the same representation, and only for the fundamental do they alone suffice to determine knot invariant polynomials \cite{Guadagnini:1994cc}.  Hence deriving this result using surgery would require appealing to the underlying fusion properties, which is arguably more complicated than evaluating a one-dimensional Gaussian integral.

\nocite{*}
\bibliographystyle{utphys}
\bibliography{\jobname}

\providecommand{\href}[2]{#2}\begingroup\raggedright\begin{thebibliography}{10}

\bibitem{Birmingham:1991ty}
D.~Birmingham, M.~Blau, M.~Rakowski, and G.~Thompson, ``{Topological field
  theory},''
\href{http://dx.doi.org/10.1016/0370-1573(91)90117-5}{{\em Phys. Rept.}
  {\bfseries 209} (1991) 129--340}.

\bibitem{Kallen:2011ny}
J.~K{\"a}ll{\'e}n, ``{Cohomological localization of Chern-Simons theory},''
  \href{http://dx.doi.org/10.1007/JHEP08(2011)008}{{\em JHEP} {\bfseries 08}
  (2011) 008},
\href{http://arxiv.org/abs/1104.5353}{{\ttfamily arXiv:1104.5353 [hep-th]}}.

\bibitem{Chen:1992ee}
W.~Chen, G.~W. Semenoff, and Y.-S. Wu, ``{Two-loop analysis of non-Abelian
  Chern-Simons theory},''
  \href{http://dx.doi.org/10.1103/PhysRevD.46.5521}{{\em Phys. Rev.} {\bfseries
  D46} (1992) 5521--5539},
\href{http://arxiv.org/abs/hep-th/9209005}{{\ttfamily arXiv:hep-th/9209005
  [hep-th]}}.

\bibitem{Witten:1999ds}
E.~Witten, ``{Supersymmetric index of three-dimensional gauge theory},''
\href{http://arxiv.org/abs/hep-th/9903005}{{\ttfamily arXiv:hep-th/9903005
  [hep-th]}}.

\bibitem{Coleman:1985zi}
S.~R. Coleman and B.~R. Hill, ``{No More Corrections to the Topological Mass
  Term in QED in Three Dimensions},''
\href{http://dx.doi.org/10.1016/0370-2693(85)90883-4}{{\em Phys. Lett.}
  {\bfseries 159B} (1985) 184--188}.

\bibitem{Closset:2012vg}
C.~Closset, T.~T. Dumitrescu, G.~Festuccia, Z.~Komargodski, and N.~Seiberg,
  ``{Contact Terms, Unitarity, and F-Maximization in Three-Dimensional
  Superconformal Theories},''
  \href{http://dx.doi.org/10.1007/JHEP10(2012)053}{{\em JHEP} {\bfseries 10}
  (2012) 053},
\href{http://arxiv.org/abs/1205.4142}{{\ttfamily arXiv:1205.4142 [hep-th]}}.

\bibitem{Closset:2012vp}
C.~Closset, T.~T. Dumitrescu, G.~Festuccia, Z.~Komargodski, and N.~Seiberg,
  ``{Comments on Chern-Simons Contact Terms in Three Dimensions},''
  \href{http://dx.doi.org/10.1007/JHEP09(2012)091}{{\em JHEP} {\bfseries 09}
  (2012) 091},
\href{http://arxiv.org/abs/1206.5218}{{\ttfamily arXiv:1206.5218 [hep-th]}}.

\bibitem{Aharony:1997bx}
O.~Aharony, A.~Hanany, K.~A. Intriligator, N.~Seiberg, and M.~J. Strassler,
  ``{Aspects of N=2 supersymmetric gauge theories in three dimensions},''
  \href{http://dx.doi.org/10.1016/S0550-3213(97)00323-4}{{\em Nucl. Phys.}
  {\bfseries B499} (1997) 67--99},
\href{http://arxiv.org/abs/hep-th/9703110}{{\ttfamily arXiv:hep-th/9703110
  [hep-th]}}.

\bibitem{Elitzur:1989nr}
S.~Elitzur, G.~W. Moore, A.~Schwimmer, and N.~Seiberg, ``{Remarks on the
  Canonical Quantization of the Chern-Simons-Witten Theory},''
\href{http://dx.doi.org/10.1016/0550-3213(89)90436-7}{{\em Nucl. Phys.}
  {\bfseries B326} (1989) 108--134}.

\bibitem{Kapustin:2010mh}
A.~Kapustin, B.~Willett, and I.~Yaakov, ``{Tests of Seiberg-like Duality in
  Three Dimensions},''
\href{http://arxiv.org/abs/1012.4021}{{\ttfamily arXiv:1012.4021 [hep-th]}}.

\bibitem{Aharony:2015mjs}
O.~Aharony, ``{Baryons, monopoles and dualities in Chern-Simons-matter
  theories},'' \href{http://dx.doi.org/10.1007/JHEP02(2016)093}{{\em JHEP}
  {\bfseries 02} (2016) 093},
\href{http://arxiv.org/abs/1512.00161}{{\ttfamily arXiv:1512.00161 [hep-th]}}.

\bibitem{Mikhaylov:2014aoa}
V.~Mikhaylov and E.~Witten, ``{Branes And Supergroups},''
  \href{http://dx.doi.org/10.1007/s00220-015-2449-y}{{\em Commun. Math. Phys.}
  {\bfseries 340} no.~2, (2015) 699--832},
\href{http://arxiv.org/abs/1410.1175}{{\ttfamily arXiv:1410.1175 [hep-th]}}.

\bibitem{Schottenloher:2008zz}
M.~Schottenloher, ``{A mathematical introduction to conformal field theory},''
\href{http://dx.doi.org/10.1007/978-3-540-68628-6}{{\em Lect. Notes Phys.}
  {\bfseries 759} (2008) 1--237}.

\bibitem{Szabo:1996md}
R.~J. Szabo, ``{Equivariant localization of path integrals},''
\href{http://arxiv.org/abs/hep-th/9608068}{{\ttfamily arXiv:hep-th/9608068
  [hep-th]}}.

\bibitem{Pestun:2016zxk}
V.~Pestun {\em et~al.}, ``{Localization techniques in quantum field
  theories},'' \href{http://dx.doi.org/10.1088/1751-8121/aa63c1}{{\em J. Phys.}
  {\bfseries A50} no.~44, (2017) 440301},
\href{http://arxiv.org/abs/1608.02952}{{\ttfamily arXiv:1608.02952 [hep-th]}}.

\bibitem{Stone:1988fu}
M.~Stone, ``{Supersymmetry and the Quantum Mechanics of Spin},''
\href{http://dx.doi.org/10.1016/0550-3213(89)90408-2}{{\em Nucl. Phys.}
  {\bfseries B314} (1989) 557--586}.

\bibitem{Stanford:2017thb}
D.~Stanford and E.~Witten, ``{Fermionic Localization of the Schwarzian
  Theory},'' \href{http://dx.doi.org/10.1007/JHEP10(2017)008}{{\em JHEP}
  {\bfseries 10} (2017) 008},
\href{http://arxiv.org/abs/1703.04612}{{\ttfamily arXiv:1703.04612 [hep-th]}}.

\bibitem{Blau:1993tv}
M.~Blau and G.~Thompson, ``{Derivation of the Verlinde formula from
  Chern-Simons theory and the G/G model},''
  \href{http://dx.doi.org/10.1016/0550-3213(93)90538-Z}{{\em Nucl. Phys.}
  {\bfseries B408} (1993) 345--390},
\href{http://arxiv.org/abs/hep-th/9305010}{{\ttfamily arXiv:hep-th/9305010
  [hep-th]}}.

\bibitem{Blau:2006gh}
M.~Blau and G.~Thompson, ``{Chern-Simons theory on S1-bundles: Abelianisation
  and q-deformed Yang-Mills theory},''
  \href{http://dx.doi.org/10.1088/1126-6708/2006/05/003}{{\em JHEP} {\bfseries
  05} (2006) 003},
\href{http://arxiv.org/abs/hep-th/0601068}{{\ttfamily arXiv:hep-th/0601068
  [hep-th]}}.

\bibitem{Beasley:2005vf}
C.~Beasley and E.~Witten, ``{Non-Abelian localization for Chern-Simons
  theory},'' {\em J. Diff. Geom.} {\bfseries 70} no.~2, (2005) 183--323,
\href{http://arxiv.org/abs/hep-th/0503126}{{\ttfamily arXiv:hep-th/0503126
  [hep-th]}}.

\bibitem{Beasley:2009mb}
C.~Beasley, ``{Localization for Wilson Loops in Chern-Simons Theory},''
  \href{http://dx.doi.org/10.4310/ATMP.2013.v17.n1.a1}{{\em Adv. Theor. Math.
  Phys.} {\bfseries 17} no.~1, (2013) 1--240},
\href{http://arxiv.org/abs/0911.2687}{{\ttfamily arXiv:0911.2687 [hep-th]}}.

\bibitem{Witten:1988hf}
E.~Witten, ``{Quantum Field Theory and the Jones Polynomial},''
\href{http://dx.doi.org/10.1007/BF01217730}{{\em Commun. Math. Phys.}
  {\bfseries 121} (1989) 351--399}.

\bibitem{Closset:2012ru}
C.~Closset, T.~T. Dumitrescu, G.~Festuccia, and Z.~Komargodski,
  ``{Supersymmetric Field Theories on Three-Manifolds},''
  \href{http://dx.doi.org/10.1007/JHEP05(2013)017}{{\em JHEP} {\bfseries 05}
  (2013) 017},
\href{http://arxiv.org/abs/1212.3388}{{\ttfamily arXiv:1212.3388 [hep-th]}}.

\bibitem{Closset:2013vra}
C.~Closset, T.~T. Dumitrescu, G.~Festuccia, and Z.~Komargodski, ``{The Geometry
  of Supersymmetric Partition Functions},''
  \href{http://dx.doi.org/10.1007/JHEP01(2014)124}{{\em JHEP} {\bfseries 01}
  (2014) 124},
\href{http://arxiv.org/abs/1309.5876}{{\ttfamily arXiv:1309.5876 [hep-th]}}.

\bibitem{Ohta:2012ev}
K.~Ohta and Y.~Yoshida, ``{Non-Abelian Localization for Supersymmetric
  Yang-Mills-Chern-Simons Theories on Seifert Manifold},''
  \href{http://dx.doi.org/10.1103/PhysRevD.86.105018}{{\em Phys. Rev.}
  {\bfseries D86} (2012) 105018},
\href{http://arxiv.org/abs/1205.0046}{{\ttfamily arXiv:1205.0046 [hep-th]}}.

\bibitem{Dijkgraaf:1989pz}
R.~Dijkgraaf and E.~Witten, ``{Topological Gauge Theories and Group
  Cohomology},''
\href{http://dx.doi.org/10.1007/BF02096988}{{\em Commun. Math. Phys.}
  {\bfseries 129} (1990) 393}.

\bibitem{Kao:1995gf}
H.-C. Kao, K.-M. Lee, and T.~Lee, ``{The Chern-Simons coefficient in
  supersymmetric Yang-Mills Chern-Simons theories},''
  \href{http://dx.doi.org/10.1016/0370-2693(96)00119-0}{{\em Phys. Lett.}
  {\bfseries B373} (1996) 94--99},
\href{http://arxiv.org/abs/hep-th/9506170}{{\ttfamily arXiv:hep-th/9506170
  [hep-th]}}.

\bibitem{Redlich:1983kn}
A.~N. Redlich, ``{Gauge Noninvariance and Parity Violation of Three-Dimensional
  Fermions},''
\href{http://dx.doi.org/10.1103/PhysRevLett.52.18}{{\em Phys. Rev. Lett.}
  {\bfseries 52} (1984) 18}.

\bibitem{Redlich:1983dv}
A.~N. Redlich, ``{Parity Violation and Gauge Noninvariance of the Effective
  Gauge Field Action in Three Dimensions},''
\href{http://dx.doi.org/10.1103/PhysRevD.29.2366}{{\em Phys. Rev.} {\bfseries
  D29} (1984) 2366--2374}.

\bibitem{Pisarski:1985yj}
R.~D. Pisarski and S.~Rao, ``{Topologically Massive Chromodynamics in the
  Perturbative Regime},''
\href{http://dx.doi.org/10.1103/PhysRevD.32.2081}{{\em Phys. Rev.} {\bfseries
  D32} (1985) 2081}.

\bibitem{Kapustin:2009kz}
A.~Kapustin, B.~Willett, and I.~Yaakov, ``{Exact Results for Wilson Loops in
  Superconformal Chern-Simons Theories with Matter},''
  \href{http://dx.doi.org/10.1007/JHEP03(2010)089}{{\em JHEP} {\bfseries 03}
  (2010) 089},
\href{http://arxiv.org/abs/0909.4559}{{\ttfamily arXiv:0909.4559 [hep-th]}}.

\bibitem{Marino:2002fk}
M.~Mari{\~n}o, ``{Chern-Simons theory, matrix integrals, and perturbative three
  manifold invariants},''
  \href{http://dx.doi.org/10.1007/s00220-004-1194-4}{{\em Commun. Math. Phys.}
  {\bfseries 253} (2004) 25--49},
\href{http://arxiv.org/abs/hep-th/0207096}{{\ttfamily arXiv:hep-th/0207096
  [hep-th]}}.

\bibitem{Marino:2005sj}
M.~Mari{\~n}o, ``{Chern-Simons theory, matrix models, and topological
  strings},''
{\em Int. Ser. Monogr. Phys.} {\bfseries 131} (2005) 1--197.

\bibitem{Moore:1989yh}
G.~W. Moore and N.~Seiberg, ``{Taming the Conformal Zoo},''
\href{http://dx.doi.org/10.1016/0370-2693(89)90897-6}{{\em Phys. Lett.}
  {\bfseries B220} (1989) 422--430}.

\bibitem{Hosomichi:2014hja}
K.~Hosomichi, \href{http://dx.doi.org/10.1007/978-3-319-18769-3_10}{``{A Review
  on SUSY Gauge Theories on $S^3$},''} in {\em New Dualities of Supersymmetric
  Gauge Theories}, J.~Teschner, ed., pp.~307--338.
\newblock Springer, 2016.
\newblock
\href{http://arxiv.org/abs/1412.7128}{{\ttfamily arXiv:1412.7128 [hep-th]}}.
\newblock

\bibitem{Wess:1992cp}
J.~Wess and J.~Bagger, {\em {Supersymmetry and Supergravity}}.
\newblock Princeton, USA: Univ. Pr. (1992) 259 p,
1992.
\newblock

\bibitem{Dunne:1998qy}
G.~V. Dunne, ``{Aspects of Chern-Simons theory},'' in {\em {Proceedings, Les
  Houches Summer School on Topological Aspects of Low-Dimensional Systems}}.
\newblock 1998.
\newblock
\href{http://arxiv.org/abs/hep-th/9902115}{{\ttfamily arXiv:hep-th/9902115
  [hep-th]}}.
\newblock

\bibitem{Elitzur:1985xj}
S.~Elitzur, Y.~Frishman, E.~Rabinovici, and A.~Schwimmer, ``{Origins of Global
  Anomalies in Quantum Mechanics},''
\href{http://dx.doi.org/10.1016/0550-3213(86)90042-8}{{\em Nucl. Phys.}
  {\bfseries B273} (1986) 93--108}.

\bibitem{Noma:2009dq}
Y.~Noma, ``{Coadjoint Orbits and Wilson Loops in Five Dimensional Topological
  Gauge Theories},''
\href{http://arxiv.org/abs/0911.2386}{{\ttfamily arXiv:0911.2386 [hep-th]}}.

\bibitem{Ivanov:1991fn}
E.~A. Ivanov, ``{Chern-Simons matter systems with manifest N=2
  supersymmetry},''
\href{http://dx.doi.org/10.1016/0370-2693(91)90804-Y}{{\em Phys. Lett.}
  {\bfseries B268} (1991) 203--208}.

\bibitem{Willett:2016adv}
B.~Willett, ``{Localization on three-dimensional manifolds},'' in {\em
  Localization techniques in quantum field theories}.
\newblock 2016.
\newblock
\href{http://arxiv.org/abs/1608.02958}{{\ttfamily arXiv:1608.02958 [hep-th]}}.
\newblock

\bibitem{Aganagic:2002wv}
M.~Aganagic, A.~Klemm, M.~Mari{\~n}o, and C.~Vafa, ``{Matrix model as a mirror
  of Chern-Simons theory},''
  \href{http://dx.doi.org/10.1088/1126-6708/2004/02/010}{{\em JHEP} {\bfseries
  02} (2004) 010},
\href{http://arxiv.org/abs/hep-th/0211098}{{\ttfamily arXiv:hep-th/0211098
  [hep-th]}}.

\bibitem{Gaiotto:2007qi}
D.~Gaiotto and X.~Yin, ``{Notes on superconformal Chern-Simons-Matter
  theories},'' \href{http://dx.doi.org/10.1088/1126-6708/2007/08/056}{{\em
  JHEP} {\bfseries 08} (2007) 056},
\href{http://arxiv.org/abs/0704.3740}{{\ttfamily arXiv:0704.3740 [hep-th]}}.

\bibitem{Hama:2010av}
N.~Hama, K.~Hosomichi, and S.~Lee, ``{Notes on SUSY Gauge Theories on
  Three-Sphere},'' \href{http://dx.doi.org/10.1007/JHEP03(2011)127}{{\em JHEP}
  {\bfseries 03} (2011) 127},
\href{http://arxiv.org/abs/1012.3512}{{\ttfamily arXiv:1012.3512 [hep-th]}}.

\bibitem{Jafferis:2010un}
D.~L. Jafferis, ``{The Exact Superconformal R-Symmetry Extremizes Z},''
  \href{http://dx.doi.org/10.1007/JHEP05(2012)159}{{\em JHEP} {\bfseries 05}
  (2012) 159},
\href{http://arxiv.org/abs/1012.3210}{{\ttfamily arXiv:1012.3210 [hep-th]}}.

\bibitem{Pestun:2007rz}
V.~Pestun, ``{Localization of gauge theory on a four-sphere and supersymmetric
  Wilson loops},'' \href{http://dx.doi.org/10.1007/s00220-012-1485-0}{{\em
  Commun. Math. Phys.} {\bfseries 313} (2012) 71--129},
\href{http://arxiv.org/abs/0712.2824}{{\ttfamily arXiv:0712.2824 [hep-th]}}.

\bibitem{Tanaka:2012nr}
A.~Tanaka, ``{Comments on knotted 1/2 BPS Wilson loops},''
  \href{http://dx.doi.org/10.1007/JHEP07(2012)097}{{\em JHEP} {\bfseries 07}
  (2012) 097},
\href{http://arxiv.org/abs/1204.5975}{{\ttfamily arXiv:1204.5975 [hep-th]}}.

\bibitem{Hama:2011ea}
N.~Hama, K.~Hosomichi, and S.~Lee, ``{SUSY Gauge Theories on Squashed
  Three-Spheres},'' \href{http://dx.doi.org/10.1007/JHEP05(2011)014}{{\em JHEP}
  {\bfseries 05} (2011) 014},
\href{http://arxiv.org/abs/1102.4716}{{\ttfamily arXiv:1102.4716 [hep-th]}}.

\bibitem{Imamura:2011wg}
Y.~Imamura and D.~Yokoyama, ``{N=2 supersymmetric theories on squashed
  three-sphere},'' \href{http://dx.doi.org/10.1103/PhysRevD.85.025015}{{\em
  Phys. Rev.} {\bfseries D85} (2012) 025015},
\href{http://arxiv.org/abs/1109.4734}{{\ttfamily arXiv:1109.4734 [hep-th]}}.

\bibitem{Kapustin:2013hpk}
A.~Kapustin and B.~Willett, ``{Wilson loops in supersymmetric
  Chern-Simons-matter theories and duality},''
\href{http://arxiv.org/abs/1302.2164}{{\ttfamily arXiv:1302.2164 [hep-th]}}.

\bibitem{Gang:2009wy}
D.~Gang, ``{Chern-Simons theory on L(p,q) lens spaces and Localization},''
\href{http://arxiv.org/abs/0912.4664}{{\ttfamily arXiv:0912.4664 [hep-th]}}.

\bibitem{Closset:2016arn}
C.~Closset and H.~Kim, ``{Comments on twisted indices in 3d supersymmetric
  gauge theories},'' \href{http://dx.doi.org/10.1007/JHEP08(2016)059}{{\em
  JHEP} {\bfseries 08} (2016) 059},
\href{http://arxiv.org/abs/1605.06531}{{\ttfamily arXiv:1605.06531 [hep-th]}}.

\bibitem{Closset:2017zgf}
C.~Closset, H.~Kim, and B.~Willett, ``{Supersymmetric partition functions and
  the three-dimensional A-twist},''
  \href{http://dx.doi.org/10.1007/JHEP03(2017)074}{{\em JHEP} {\bfseries 03}
  (2017) 074},
\href{http://arxiv.org/abs/1701.03171}{{\ttfamily arXiv:1701.03171 [hep-th]}}.

\bibitem{Sugishita:2013jca}
S.~Sugishita and S.~Terashima, ``{Exact Results in Supersymmetric Field
  Theories on Manifolds with Boundaries},''
  \href{http://dx.doi.org/10.1007/JHEP11(2013)021}{{\em JHEP} {\bfseries 11}
  (2013) 021},
\href{http://arxiv.org/abs/1308.1973}{{\ttfamily arXiv:1308.1973 [hep-th]}}.

\bibitem{Yoshida:2014ssa}
Y.~Yoshida and K.~Sugiyama, ``{Localization of 3d $\mathcal{N}=2$
  Supersymmetric Theories on $S^1 \times D^2$},''
\href{http://arxiv.org/abs/1409.6713}{{\ttfamily arXiv:1409.6713 [hep-th]}}.

\bibitem{Pasquetti:2011fj}
S.~Pasquetti, ``{Factorisation of N = 2 Theories on the Squashed 3-Sphere},''
  \href{http://dx.doi.org/10.1007/JHEP04(2012)120}{{\em JHEP} {\bfseries 04}
  (2012) 120},
\href{http://arxiv.org/abs/1111.6905}{{\ttfamily arXiv:1111.6905 [hep-th]}}.

\bibitem{Beem:2012mb}
C.~Beem, T.~Dimofte, and S.~Pasquetti, ``{Holomorphic Blocks in Three
  Dimensions},'' \href{http://dx.doi.org/10.1007/JHEP12(2014)177}{{\em JHEP}
  {\bfseries 12} (2014) 177},
\href{http://arxiv.org/abs/1211.1986}{{\ttfamily arXiv:1211.1986 [hep-th]}}.

\bibitem{Fujitsuka:2013fga}
M.~Fujitsuka, M.~Honda, and Y.~Yoshida, ``{Higgs branch localization of 3d N =
  2 theories},'' \href{http://dx.doi.org/10.1093/ptep/ptu158}{{\em PTEP}
  {\bfseries 2014} no.~12, (2014) 123B02},
\href{http://arxiv.org/abs/1312.3627}{{\ttfamily arXiv:1312.3627 [hep-th]}}.

\bibitem{Benini:2013yva}
F.~Benini and W.~Peelaers, ``{Higgs branch localization in three dimensions},''
  \href{http://dx.doi.org/10.1007/JHEP05(2014)030}{{\em JHEP} {\bfseries 05}
  (2014) 030},
\href{http://arxiv.org/abs/1312.6078}{{\ttfamily arXiv:1312.6078 [hep-th]}}.

\bibitem{Festuccia:2011ws}
G.~Festuccia and N.~Seiberg, ``{Rigid Supersymmetric Theories in Curved
  Superspace},'' \href{http://dx.doi.org/10.1007/JHEP06(2011)114}{{\em JHEP}
  {\bfseries 06} (2011) 114},
\href{http://arxiv.org/abs/1105.0689}{{\ttfamily arXiv:1105.0689 [hep-th]}}.

\bibitem{Imamura:2011su}
Y.~Imamura and S.~Yokoyama, ``{Index for three dimensional superconformal field
  theories with general R-charge assignments},''
  \href{http://dx.doi.org/10.1007/JHEP04(2011)007}{{\em JHEP} {\bfseries 04}
  (2011) 007},
\href{http://arxiv.org/abs/1101.0557}{{\ttfamily arXiv:1101.0557 [hep-th]}}.

\bibitem{Benini:2015noa}
F.~Benini and A.~Zaffaroni, ``{A topologically twisted index for
  three-dimensional supersymmetric theories},''
  \href{http://dx.doi.org/10.1007/JHEP07(2015)127}{{\em JHEP} {\bfseries 07}
  (2015) 127},
\href{http://arxiv.org/abs/1504.03698}{{\ttfamily arXiv:1504.03698 [hep-th]}}.

\bibitem{Alday:2013lba}
L.~F. Alday, D.~Martelli, P.~Richmond, and J.~Sparks, ``{Localization on
  Three-Manifolds},'' \href{http://dx.doi.org/10.1007/JHEP10(2013)095}{{\em
  JHEP} {\bfseries 10} (2013) 095},
\href{http://arxiv.org/abs/1307.6848}{{\ttfamily arXiv:1307.6848 [hep-th]}}.

\bibitem{Drukker:2012sr}
N.~Drukker, T.~Okuda, and F.~Passerini, ``{Exact results for vortex loop
  operators in 3d supersymmetric theories},''
  \href{http://dx.doi.org/10.1007/JHEP07(2014)137}{{\em JHEP} {\bfseries 07}
  (2014) 137},
\href{http://arxiv.org/abs/1211.3409}{{\ttfamily arXiv:1211.3409 [hep-th]}}.

\bibitem{Alday:2012au}
L.~F. Alday, M.~Fluder, and J.~Sparks, ``{The Large N limit of M2-branes on
  Lens spaces},'' \href{http://dx.doi.org/10.1007/JHEP10(2012)057}{{\em JHEP}
  {\bfseries 10} (2012) 057},
\href{http://arxiv.org/abs/1204.1280}{{\ttfamily arXiv:1204.1280 [hep-th]}}.

\bibitem{Intriligator:2013lca}
K.~Intriligator and N.~Seiberg, ``{Aspects of 3d N=2 Chern-Simons-Matter
  Theories},'' \href{http://dx.doi.org/10.1007/JHEP07(2013)079}{{\em JHEP}
  {\bfseries 07} (2013) 079},
\href{http://arxiv.org/abs/1305.1633}{{\ttfamily arXiv:1305.1633 [hep-th]}}.

\bibitem{Benini:2016hjo}
F.~Benini and A.~Zaffaroni, ``{Supersymmetric partition functions on Riemann
  surfaces},'' {\em Proc. Symp. Pure Math.} {\bfseries 96} (2017) 13--46,
\href{http://arxiv.org/abs/1605.06120}{{\ttfamily arXiv:1605.06120 [hep-th]}}.

\bibitem{Guadagnini:1994cc}
E.~Guadagnini, {\em {The Link invariants of the Chern-Simons field theory: New
  developments in topological quantum field theory}}.
\newblock Berlin, Germany: de Gruyter (1993) 312 p. (de Gruyter expositions in
  mathematics, 10),
1994.
\newblock

\bibitem{Bos:1989kn}
M.~Bos and V.~P. Nair, ``{Coherent State Quantization of Chern-Simons
  Theory},''
\href{http://dx.doi.org/10.1142/S0217751X90000453}{{\em Int. J. Mod. Phys.}
  {\bfseries A5} (1990) 959}.

\bibitem{Dimofte:2017tpi}
T.~Dimofte, D.~Gaiotto, and N.~M. Paquette, ``{Dual Boundary Conditions in 3d
  SCFT's},'' \href{http://dx.doi.org/10.1007/JHEP05(2018)060}{{\em JHEP}
  {\bfseries 05} (2018) 060},
\href{http://arxiv.org/abs/1712.07654}{{\ttfamily arXiv:1712.07654 [hep-th]}}.

\bibitem{Drukker:1999zq}
N.~Drukker, D.~J. Gross, and H.~Ooguri, ``{Wilson loops and minimal
  surfaces},'' \href{http://dx.doi.org/10.1103/PhysRevD.60.125006}{{\em Phys.
  Rev.} {\bfseries D60} (1999) 125006},
\href{http://arxiv.org/abs/hep-th/9904191}{{\ttfamily arXiv:hep-th/9904191
  [hep-th]}}.

\bibitem{Kapustin:2012iw}
A.~Kapustin, B.~Willett, and I.~Yaakov, ``{Exact results for supersymmetric
  abelian vortex loops in 2+1 dimensions},''
  \href{http://dx.doi.org/10.1007/JHEP06(2013)099}{{\em JHEP} {\bfseries 06}
  (2013) 099},
\href{http://arxiv.org/abs/1211.2861}{{\ttfamily arXiv:1211.2861 [hep-th]}}.

\bibitem{Seiberg:2016rsg}
N.~Seiberg and E.~Witten, ``{Gapped Boundary Phases of Topological Insulators
  via Weak Coupling},'' \href{http://dx.doi.org/10.1093/ptep/ptw083}{{\em PTEP}
  {\bfseries 2016} no.~12, (2016) 12C101},
\href{http://arxiv.org/abs/1602.04251}{{\ttfamily arXiv:1602.04251
  [cond-mat.str-el]}}.

\bibitem{Verlinde:1988sn}
E.~P. Verlinde, ``{Fusion Rules and Modular Transformations in 2D Conformal
  Field Theory},''
\href{http://dx.doi.org/10.1016/0550-3213(88)90603-7}{{\em Nucl. Phys.}
  {\bfseries B300} (1988) 360--376}.

\bibitem{Faizal:2016skd}
M.~Faizal, Y.~Luo, D.~J. Smith, M.-C. Tan, and Q.~Zhao, ``{Gauge and
  supersymmetry invariance of $\mathcal N=2$ boundary Chern-Simons theory},''
  \href{http://dx.doi.org/10.1016/j.nuclphysb.2016.11.020}{{\em Nucl. Phys.}
  {\bfseries B914} (2017) 577--598},
\href{http://arxiv.org/abs/1601.05429}{{\ttfamily arXiv:1601.05429 [hep-th]}}.

\bibitem{DiVecchia:1984nyg}
P.~Di~Vecchia, V.~G. Knizhnik, J.~L. Petersen, and P.~Rossi, ``{A
  Supersymmetric Wess-Zumino Lagrangian in Two Dimensions},''
\href{http://dx.doi.org/10.1016/0550-3213(85)90554-1}{{\em Nucl. Phys.}
  {\bfseries B253} (1985) 701--726}.

\bibitem{Fuchs:1988gm}
J.~Fuchs, ``{More on the Super {WZW} Theory},''
\href{http://dx.doi.org/10.1016/0550-3213(89)90634-2}{{\em Nucl. Phys.}
  {\bfseries B318} (1989) 631--654}.

\bibitem{Bernatska:2008}
J.~Bernatska and P.~Holod,
  \href{http://dx.doi.org/10.7546/giq-9-2008-146-166}{``{Geometry and topology
  of coadjoint orbits of semisimple Lie groups},''} in {\em {Geometry,
  Integrability, and Quantization IX}}, pp.~146--166.
\newblock {Softex, Sofia}, 2008.
\newblock \href{http://arxiv.org/abs/0801.2913}{{\ttfamily arXiv:0801.2913}}.

\bibitem{Assel:2015oxa}
B.~Assel and J.~Gomis, ``{Mirror Symmetry And Loop Operators},''
  \href{http://dx.doi.org/10.1007/JHEP11(2015)055}{{\em JHEP} {\bfseries 11}
  (2015) 055},
\href{http://arxiv.org/abs/1506.01718}{{\ttfamily arXiv:1506.01718 [hep-th]}}.

\bibitem{Doroud:2012xw}
N.~Doroud, J.~Gomis, B.~Le~Floch, and S.~Lee, ``{Exact Results in D=2
  Supersymmetric Gauge Theories},''
  \href{http://dx.doi.org/10.1007/JHEP05(2013)093}{{\em JHEP} {\bfseries 05}
  (2013) 093},
\href{http://arxiv.org/abs/1206.2606}{{\ttfamily arXiv:1206.2606 [hep-th]}}.

\bibitem{Marino:2011nm}
M.~Mari{\~n}o, ``{Lectures on localization and matrix models in supersymmetric
  Chern-Simons-matter theories},''
  \href{http://dx.doi.org/10.1088/1751-8113/44/46/463001}{{\em J. Phys.}
  {\bfseries A44} (2011) 463001},
\href{http://arxiv.org/abs/1104.0783}{{\ttfamily arXiv:1104.0783 [hep-th]}}.

\bibitem{Witten:1983ar}
E.~Witten, ``{Nonabelian Bosonization in Two Dimensions},''
\href{http://dx.doi.org/10.1007/BF01215276}{{\em Commun. Math. Phys.}
  {\bfseries 92} (1984) 455--472}.

\bibitem{DiFrancesco:1997nk}
P.~Di~Francesco, P.~Mathieu, and D.~S{\'e}n{\'e}chal,
  \href{http://dx.doi.org/10.1007/978-1-4612-2256-9}{{\em {Conformal Field
  Theory}}}.
\newblock Graduate Texts in Contemporary Physics. Springer-Verlag, New York,
1997.
\newblock

\bibitem{Hori:2003ic}
K.~Hori, S.~Katz, A.~Klemm, R.~Pandharipande, R.~Thomas, C.~Vafa, R.~Vakil, and
  E.~Zaslow, {\em {Mirror Symmetry}}.
\newblock AMS, Providence, USA,
2003.
\newblock

\end{thebibliography}\endgroup

\end{document}